\documentclass[cernpreprint, UKenglish, texmf, orcidlogo]{atlasdoc}
 
\usepackage{atlaspackage}
\usepackage[block=none]{atlasbiblatex}
 
\usepackage{atlasphysics}

\graphicspath{{logos/}{figures/}}
 
\usepackage{ANA-DAPR-2021-01-PAPER-defs}

\AtlasTitle{Luminosity determination in $pp$ collisions at $\sqrt{s}=13$\,\TeV\
using the ATLAS detector at the LHC}
 
\AtlasAbstract{
The luminosity determination for the ATLAS detector at the LHC during
Run~2 is presented, with $pp$ collisions at a centre-of-mass energy
$\sqrt{s}=13$\,\TeV.
The absolute
luminosity scale is determined using van der Meer beam separation scans
during dedicated running periods in each year, and extrapolated to the physics
data-taking regime using complementary measurements from several
luminosity-sensitive detectors. The total uncertainties in the integrated
luminosity for each individual year of data-taking range from 0.9\% to 1.1\%,
and are partially correlated between years. After standard data-quality
selections, the full Run~2 $pp$ data sample corresponds to an integrated
luminosity of $\totintl\pm\etotintl$\,fb$^{-1}$, i.e.\ an uncertainty of \totlfrac\%.
A dedicated sample of low-pileup data recorded in 2017--18 for precision
Standard Model physics measurements is analysed separately, and has
an integrated luminosity of $\lowmuintl\pm\lowmuetot$\,pb$^{-1}$.
 
}
 
\author{The ATLAS Collaboration}
 
\AtlasRefCode{DAPR-2021-01}

\AtlasJournal{EPJC}
\PreprintIdNumber{CERN-EP-2022-281}
 
\AtlasJournalRef{Eur. Phys. J. C 83 (2023) 982}
\AtlasDOI{10.1140/epjc/s10052-023-11747-w}

\hypersetup{pdftitle={ATLAS document},pdfauthor={The ATLAS Collaboration}}
 
\begin{document}
 
\maketitle
 
\tableofcontents

% The next lines are included from the .//intro.tex input file
\section{Introduction}\label{s:intro}
 
A precise measurement of the integrated luminosity is a key component
of the ATLAS physics programme at the CERN Large Hadron Collider (LHC),
in particular for cross-section
measurements where it is often one of the leading sources of uncertainty.
Searches for new physics phenomena beyond those predicted by the Standard Model
also often require accurate estimates of the luminosity to determine
background levels and sensitivity. This paper describes the measurement
of the luminosity of the proton--proton ($pp$)
collision data sample delivered to the
ATLAS detector at a centre-of-mass energy of \sxyt\ during Run~2 of
the LHC in the years 2015--2018.
The measurement builds on the experience and techniques developed for Run~1 and
documented extensively in Refs.~\cite{DAPR-2010-01,DAPR-2011-01,DAPR-2013-01},
where uncertainties in the total integrated luminosities of
$\Delta{\cal L}/{\cal L}=1.8$\% for \sxwt\ and 1.9\% for \sxvt\ were achieved.
 
The luminosity measurement is based on an absolute calibration of the
primary luminosity-sensitive detectors in low-luminosity runs with
specially tailored LHC conditions using the van der Meer (vdM)
method \cite{vdmisr,lumibib}. A calibration transfer procedure was then used to
transport this calibration to the physics data-taking regime at high
luminosity. The vdM  calibration was performed in dedicated fills once per
year during Run~2, and relative comparisons of the luminosities measured by
different detectors were used to set limits on any possible change in the
calibration during the year. Finally, the integrated luminosity and
uncertainty for the whole Run~2 data-taking period was derived, taking into
account correlations between the uncertainties in each of the component years.
 
The structure of this paper mirrors the calibration process.
After a brief introduction
to the methodology and the Run~2 dataset in Section~\ref{s:meth}, the various
luminosity-sensitive detectors are described in Section~\ref{s:lumidet},
the absolute vdM calibration in Section~\ref{s:vdmcal}, the
calibration transfer procedure in Sections~\ref{s:calt} and~\ref{s:calsyst},
the long-term stability studies in Section~\ref{s:stab} and the correlations
and combination in Section~\ref{s:unccorl}. The latter also contains
a summary (Table~\ref{t:unc}) of the contributing uncertainties in all
individual years and the combination. A dedicated luminosity calibration
for special datasets recorded for precision $W/Z$ physics with low numbers
of $pp$ interactions per bunch crossing is described in Section~\ref{s:lowmu}.
Conclusions are given in Section~\ref{s:conc}.

% End of text imported from the .//intro.tex input file

% The next lines are included from the .//meth.tex input file
\section{Methodology and datasets}\label{s:meth}
 
The instantaneous luminosity \lbun\ produced by a single pair of colliding
bunches can be expressed as
\begin{equation}\label{e:lbungen}
\lbun=\frac{R}{\sigma}\ ,
\end{equation}
where $R$ is the rate of a particular process with cross-section $\sigma$.
In the case of inelastic $pp$ collisions
with cross-section \sigmainel, Eq.~(\ref{e:lbungen}) becomes
\begin{equation}\label{e:lbuninel}
\lbun=\frac{\mu f_\mathrm{r}}{\sigmainel}\ ,
\end{equation}
where the pileup parameter $\mu$ is the average number of inelastic interactions
per bunch crossing, and $f_\mathrm{r}$ is the LHC bunch revolution frequency
(11246\,Hz for protons).
The total instantaneous luminosity is given by
\begin{equation}\nonumber
\linst=\sum_{b=1}^{\nbun} \lbun = \nbun\langle{\lbun}\rangle = \nbun \frac{\meanmu f_\mathrm{r}}{\sigmainel}\ ,
\end{equation}
where the sum runs over the \nbun\ bunch pairs colliding at the interaction
point (IP), $\langle\lbun\rangle$ is the mean per-bunch luminosity and
\meanmu\ is the pileup parameter averaged over all colliding bunch pairs.
 
The instantaneous luminosity can be measured by monitoring \muvis, the visible
interaction rate per bunch-crossing for a particular luminosity algorithm,
based on a chosen luminosity-sensitive detector.\footnote{The term `luminosity
algorithm' refers
to the measurement of a quantity proportional to the instantaneous luminosity,
e.g.\ a counting rate or electrical current, using a procedure or set of
selection criteria specific to a particular luminosity detector.}
The per-bunch luminosity can be written in analogy with Eqs.~(\ref{e:lbungen})
and~(\ref{e:lbuninel}) as
\begin{equation}\label{e:lbunvis}
\lbun=\frac{\muvis f_\mathrm{r}}{\sigmavis}\ ,
\end{equation}
where the algorithm-specific
visible cross-section is a calibration constant which represents the
absolute luminosity calibration of the given algorithm, and can be determined
via the vdM calibration method discussed in Section~\ref{s:vdmcal}.
Once \sigmavis\ is known, the pileup parameter $\mu$ can be determined
from $\mu=\muvis\sigmainel/\sigmavis$. Since \sigmainel\ is not precisely
known, a reference value of $\sigmainel=80$\,mb is used by convention
by all LHC experiments for
$pp$ collisions at \sxyt. The values of $\mu$ and its bunch-averaged
counterpart \meanmu\ are important parameters for characterising the
performance of the LHC machine and detectors, but the luminosity calibration
based on \sigmavis\ is independent of the chosen reference value of \sigmainel.
 
Provided the calibration constant \sigmavis\
is stable in time, and does not depend on the LHC conditions
(e.g.\ the number of bunches or value of \meanmu), the measurement of
\muvis\ at any moment is sufficient to determine the per-bunch
instantaneous luminosity at that time. In practice, no luminosity algorithm
is perfectly stable in time and independent of LHC conditions, and comparisons
between many different algorithms and detectors, each with their own strengths
and weaknesses, are essential to produce a precise luminosity measurement with
a robust uncertainty estimate.
 
In ATLAS, the rate \muvis\ is measured over a finite time interval
called a luminosity block (LB), during which data-taking conditions (including
the instantaneous luminosity) are assumed to remain stable. The start and
end times of each LB are defined by the ATLAS central trigger processor, and
the normal duration during Run~2 data-taking was one minute, sometimes cut short
if a significant change of conditions such as a detector problem occurred.
The instantaneous luminosity for each LB was calculated from the sum of
per-bunch luminosities, each derived using Eq.~(\ref{e:lbunvis}) and the
per-bunch \muvis\
values measured within the LB, modified by the calibration transfer
corrections discussed in Section~\ref{s:calt}. The integrated luminosity
was then obtained by multiplying \linst\ by the duration of the LB, and these
values were stored in the ATLAS conditions database \cite{conddb}.
The data-taking is organised into a series of `runs' of the ATLAS data
acquisition system, each of which normally corresponds to an LHC fill,
including the periods outside stable beam collisions such as LHC injection,
acceleration and preparation for collisions.
ATLAS physics analyses are based on a `good runs list' (GRL), a list of runs
and LBs within them with stable beam collisions and
where all the ATLAS subdetectors were functioning correctly according to
a standard set of data-quality requirements \cite{DAPR-2018-01}.
The integrated luminosity corresponding to the GRL can be  calculated as
the sum of the integrated luminosities of all LBs in the GRL, after correcting
each LB for readout dead-time. If the trigger used for a particular analysis
is prescaled, the effective integrated luminosity is reduced accordingly
\cite{TRIG-2019-04}.
 
The data-taking conditions evolved significantly during Run~2, with the LHC
peak instantaneous luminosity at the start of fills (\lpeak) increasing from
5 to $19\times 10^{33}\,\mathrm{cm}^{-2}\mathrm{s}^{-1}$
as the number of colliding bunch pairs \nbun\ and average current per bunch
were increased during the course of each year. In addition,
progressively stronger focusing in the
ATLAS and CMS interaction regions (characterised by the $\beta^*$ parameter,
the value of the $\beta$ function at the interaction point
\cite{weidermann}) was used in each successive year,
leading to smaller transverse beam sizes and higher per-bunch
instantaneous luminosity.  Table~\ref{t:lhcpar} shows
an overview of typical best parameters for each year, together with the
total delivered integrated luminosity. All this running took place with
long `trains' of bunches featuring 25\,ns bunch spacing within the trains,
except for the second part of 2017, where a special filling pattern
with eight filled bunches separated by 25\,ns followed by a four bunch-slot
gap (denoted `8b4e') was used. This bunch pattern mitigated the enhancement of
electron-cloud induced instabilities caused by a vacuum incident.
The luminosity was levelled by partial beam separation at
the beginning of such 8b4e LHC fills to give a maximum pileup parameter
of $\meanmu\approx 60$, whereas the maximum \meanmu\ achieved in standard
25\,ns running was about 55 in 2018. These values are significantly larger
than the maximum \meanmu\ of around~40 achieved with 50\,ns bunch spacing
in Run~1 \cite{DAPR-2013-01}, and posed substantial challenges to the detectors.
 
\begin{table}[tp]
\caption{\label{t:lhcpar} Selected LHC parameters for $pp$ collisions at
\sxyt\ in 2015--2018. The values shown are representative of the best
accelerator performance during normal physics operation. In 2017, the LHC
was run in two modes: standard 25\,ns bunch train operation with long trains,
and `8b4e', denoting a pattern of eight bunches separated by 25\,ns followed by a four bunch-slot gap. Values are given
for both configurations. The instantaneous luminosity was levelled by beam
separation to about $\lpeak=16\times 10^{33}\,\mathrm{cm^{-2}s^{-1}}$ for part of the 8b4e
period. The $0.1\,\ifb$ of physics data delivered during 2015 with 50\,ns
bunch spacing is not included.}
\centering
 
\begin{tabular}{l|cccc}\hline
Parameter & 2015 & 2016 & 2017 & 2018 \\
\hline
Maximum number of colliding bunch pairs  ($n_b$) & 2232 & 2208 & 2544/1909 & 2544 \\
Bunch spacing [ns] & 25 & 25 & 25/8b4e & 25 \\
Typical bunch population [$10^{11}$ protons] & 1.1 & 1.1 & 1.1/1.2 & 1.1 \\
$\beta^{*}$ [m] & 0.8 & 0.4 & 0.3 &  0.3--0.25 \\
Peak luminosity \lpeak\ [$10^{33}\,\mathrm{cm^{-2}s^{-1}}]$ & 5 & 13 & 16 & 19 \\
Peak number of inelastic interactions/crossing (\meanmu) & $\sim 16 \phantom{\sim\ }$ & $\sim 41 \phantom{\sim\ }$ & $\sim 45/60 \phantom{\sim}$ & $\sim 55 \phantom{\sim\ }$ \\
Luminosity-weighted mean inelastic interactions/crossing & 13 & 25 & 38 & 36 \\
Total delivered integrated luminosity [\ifb] & 4.0 & 39.0 & 50.6 & 63.8 \\
\hline
\end{tabular}
\end{table}

% End of text imported from the .//meth.tex input file

% The next lines are included from the .//lumidet.tex input file
\section{Luminosity detectors and algorithms}\label{s:lumidet}
 
The ATLAS detector \cite{PERF-2007-01,atlasibl,IBLTDR,ATL-SOFT-PUB-2021-001}
consists of an
inner tracking  detector surrounded by a thin superconducting solenoid
producing a 2\,T axial magnetic field, electromagnetic and hadronic
calorimeters, and an external muon spectrometer incorporating three
large toroidal magnet assemblies.\footnote{ATLAS
uses a right-handed coordinate system with its origin at
the nominal interaction point in the centre of the detector, and the $z$ axis
along the beam line. The $x$ axis points from the nominal interaction point
to the centre of the LHC ring, and the $y$ axis points upwards. The two
sides of the detector are denoted A (corresponding to $z>0$) and C ($z<0$).
Pseudorapidity is defined in terms of the polar angle
$\theta$ as $\eta=-\ln\tan{\theta/2}$, and transverse momentum and energy
are defined relative to the beam line as $\pt=p\sin\theta$ and
$\et=E\sin\theta$. The azimuthal angle around the beam line is denoted by
$\phi$.}
The ATLAS luminosity measurement relies on multiple redundant luminosity
detectors and algorithms, which have complementary capabilities and
different systematic uncertainties. For LHC Run~2, the primary bunch-by-bunch
luminosity measurement was provided by the LUCID2 Cherenkov
detector \cite{lucid2} in the far forward region,
upgraded from its Run~1 configuration and referred to hereafter as
LUCID. This was complemented by bunch-by-bunch measurements from the
ATLAS beam conditions monitor (BCM) diamond detectors, and from offline
measurements of the
multiplicity of reconstructed charged particles in randomly selected
colliding-bunch crossings (track counting). The ATLAS calorimeters provided
bunch-integrated measurements (i.e.\ summed over all bunches)
based on quantities proportional to
instantaneous luminosity: liquid-argon (LAr) gap currents in the case of
the endcap electromagnetic (EMEC) and forward (FCal) calorimeters,
and photomultiplier currents from the scintillator-tile hadronic
calorimeter (TileCal).
All these measurements are discussed in more detail below.
 
\subsection{LUCID Cherenkov detector}\label{ss:lucid}
 
The LUCID detector contains 16~photomultiplier tubes (PMTs) in each forward arm
of the ATLAS detector (side A and side C), placed around the beam pipe in
different azimuthal $\phi$ positions at approximately
$z=\pm 17$\,m from the interaction point and covering the pseudorapidity
range $5.561<|\eta|<5.641$. Cherenkov light is produced in the quartz windows
of the PMTs, which are coated with $^{207}$Bi radioactive sources that provide
a calibration signal. Regular calibration runs were performed between LHC fills,
allowing the  high voltage applied to the PMTs to be adjusted so as to keep
either the charge or amplitude of the calibration pulse constant over time.
The LUCID detector was read out with dedicated electronics which provided
luminosity counts for each of the 3564 nominal LHC bunch slots.
These counts were integrated
over the duration of each luminosity block, and over shorter time periods
of 1--2 seconds
to provide real-time feedback to the LHC machine for beam optimisation.
 
Several algorithms were used to convert the raw signals from the PMTs to
luminosity measurements, using a single PMT or combining the information
from several PMTs in various ways. The simplest algorithm uses a single
PMT, and counts an `event' if there is a signal in the PMT above a given
threshold (a `hit'), corresponding to one or more inelastic $pp$ interactions
detected in a given bunch crossing.
Assuming that the number of inelastic $pp$ interactions
in a bunch crossing follows a Poisson distribution, the probability for
such an event $P_\mathrm{evt}$ is given in terms of the single-PMT
visible interaction rate $\muvis=\epsilon\mu$ by
\begin{equation*}
P_\mathrm{evt}=\frac{N_\mathrm{evt}}{N_\mathrm{BC}}=1-\mathrm{e}^{-\muvis}\ ,
\end{equation*}
where $N_\mathrm{evt}$ is the number of events counted in the luminosity block
and $N_\mathrm{BC}$ is the number of bunch crossings sampled (equal for a single
colliding bunch pair to
$f_\mathrm{r}\Delta t$ where $\Delta t$ is the duration of the luminosity block).
The \muvis\
value is then given by $\muvis=-\ln(1-P_\mathrm{evt})$, from which the
instantaneous luminosity can be calculated via Eq.~(\ref{e:lbunvis}) once
\sigmavis\ is known.
Several PMTs can be combined in an `EventOR' (hereafter abbreviated to EvtOR)
algorithm by counting an event if
any  of a group of PMTs registers a hit in a given bunch crossing.
Such an algorithm has a larger efficiency and \sigmavis\ than a single-PMT
algorithm, giving a smaller statistical uncertainty at low bunch luminosity.
However, at moderate single-bunch luminosity ($\mu>20$--30), it already suffers
from `zero starvation', when for a given colliding-bunch pair,
all bunch crossings in the LB considered contain at least one hit. In this
situation, there are no bunch crossings left without an event, making
it impossible to determine \muvis.
LUCID EvtOR algorithms were therefore not used for standard high-pileup
physics running during Run~2. The `HitOR' algorithm provides an alternative
method for combining PMTs at high luminosity. For $N_\mathrm{PMT}$ PMTs, the
average probability $P_\mathrm{hit}$ to have a hit in any given PMT
during the $N_\mathrm{BC}$ bunch crossings of one luminosity block is inferred
from the total number of hits summed over all PMTs $N_\mathrm{hit}$ by
\begin{equation}\label{e:phit}
\phit = \frac{N_\mathrm{hit}}{N_\mathrm{BC}N_\mathrm{PMT}} = 1-\mathrm{e}^{-\muvis} ,
\end{equation}
which leads for the HitOR algorithm to $\muvis=-\ln(1-\phit)$.
Per-bunch LB-averaged event and hit counts from a variety of PMT combinations
were accumulated online in the LUCID readout electronics.
 
Both the configuration of LUCID and the LHC running conditions evolved
significantly over the course of Run~2, and different algorithms
gave the best LUCID offline luminosity measurement in each data-taking year.
These algorithms are  summarised in Table~\ref{t:lucidalg}, together with
the corresponding visible cross-sections derived from the vdM calibration,
the peak \meanmu\ value and the corresponding fraction of bunch crossings
which do not have a PMT hit. The latter fell below 1\% at the highest
instantaneous luminosities achieved in~2017 and~2018.
Where possible, HitOR algorithms
were used, as the averaging over multiple PMTs reduces systematic uncertainties
due to drifts in the calibration of individual PMTs, and also evens out
the asymmetric response of PMTs in different $\phi$ locations due to the
LHC beam crossing angle used in bunch train running. The BiHitOR algorithm,
which combines four bismuth-calibrated PMTs on the A-side of LUCID with four
more on the C-side using the HitOR method,
was used in both 2016 and 2017. Some PMTs were changed
during each winter shutdown, so the sets of PMTs used in the two years
were not exactly the same. In 2015, nearly all PMTs suffered from long-term
timing drifts
due to adjustments of the high-voltage settings, apart from one PMT on
the C-side (C9), and the most stable luminosity measurement (determined from
comparisons with independent measurements from other detectors) was obtained
from the single-PMT algorithm applied to PMT C9.
In 2015 and 2016, the bismuth calibration
signals were used to adjust the PMT gains so as to keep the mean charge
recorded by the PMT constant over the year. However, the efficiencies of the
hit-counting algorithms depend primarily on the pulse amplitude. Additional
offline corrections were therefore applied to the LUCID data in these years to
correct
for this imperfect calibration. In 2017 and 2018, the gain adjustments were
made using the amplitudes derived from the calibration signals, and so these
corrections were not needed. In 2018, a significant number of PMTs
stopped working during the course of the data-taking year, and a single PMT
on the C-side (C12) was used for the final offline measurement, as
it showed good stability throughout the year and gave results similar to those
of a more complicated offline HitOR-type combination of the remaining seven
working PMTs. In all years, other LUCID algorithms were also available, and
were studied as part of the vdM and stability analyses. These algorithms include
Bi2HitOR, an alternative to BiHitOR using an independent set of
eight PMTs with four on each side of the detector, single-sided HitOR and EvtOR
algorithms, EvtAND algorithms requiring a coincidence of hits on both sides,
and single-PMT algorithms based on individual PMTs.

\begin{table}[tp]
\caption{\label{t:lucidalg}LUCID algorithms used for the baseline luminosity
determination in each year of Run~2 data-taking, together with the visible
cross-sections \sigmavis\ determined from the absolute vdM calibration
(see Section~\ref{s:vdmcal}),
the peak \meanmu\ value taken from Table~\ref{t:lhcpar}, and the fraction of
bunch crossings $f_\mathrm{no-hit}$ which do not have a PMT hit
at this \meanmu\ value. For the HitOR algorithms, this fraction
represents an average over all the contributing PMTs.}
\centering
 
\begin{tabular}{clrrr}\hline
Year & Algorithm & \sigmavis\,[mb] & Peak \meanmu\ & $f_\mathrm{no-hit}\,[\%]$ \\
\hline
2015 & PMT C9  & 6.540 & 16 & 27.0 \\
2016 & BiHitOR & 6.525 & 41 & 3.5 \\
2017 & BiHitOR & 6.706 & 60 & 0.6 \\
2018 & PMT C12 & 6.860 & 55 & 0.9 \\
\hline
\end{tabular}
\end{table}
 
\subsection{Beam conditions monitor}\label{ss:bcm}
 
The BCM detector consists of four $8\times 8\,\mathrm{mm}^2$ diamond sensors
arranged around the beam pipe in a cross pattern at $z=\pm 1.84$\,m on each
side of the ATLAS interaction point \cite{bcmdet}.
The BCM mainly provides beam conditions
information and beam abort functionality to protect the ATLAS inner detector,
but it also gives a bunch-by-bunch luminosity signal with
sub-nanosecond timing resolution. Various luminosity algorithms are available,
combining hits from individual sensors in different ways with EvtOR and EvtAND
algorithms in the same way as for LUCID. The BCM provided the primary ATLAS
luminosity measurement during most of Run~1 \cite{DAPR-2011-01,DAPR-2013-01}
but performed much less well in the physics data-taking conditions of Run~2,
with 25\,ns bunch spacing and higher pileup, due to a combination of radiation
damage, charge-pumping effects \cite{DAPR-2011-01} and
bunch--position-dependent
biases along bunch trains. Its use in the Run~2 analysis was therefore
limited to consistency checks during some vdM scan periods with isolated bunches
and low instantaneous luminosity, and luminosity measurement in
heavy-ion collisions.

\subsection{Track-counting algorithms}\label{ss:trkcnt}
 
The track-counting luminosity measurement determines the per-bunch visible
interaction
rate \muvis\ from the mean number of reconstructed tracks per bunch crossing
averaged over a luminosity block. The measurement was derived from
randomly sampled colliding-bunch crossings, where only the data from the
silicon tracking detectors (i.e. the SCT
and pixel detectors, including a new innermost layer, the `insertable B-layer' (IBL) \cite{atlasibl,IBLTDR} added before Run~2)
were read out, typically at 200\,Hz during normal physics data-taking and at
much higher rates during vdM scans and other special runs. These events
were saved in a dedicated event stream, which was then reconstructed
offline using special track reconstruction settings, optimised for luminosity
monitoring. All reconstructed tracks
used in the track-counting luminosity measurement were required to
satisfy the TightPrimary requirements of Ref.~\cite{ATL-PHYS-PUB-2015-051},
and to have transverse momentum $\pt>0.9$\,GeV and a track impact parameter
significance of
$|d_0|/\sigma_{d_0}<7$, where $d_0$ is the impact parameter of the
track with respect to the beamline in the transverse plane,
and $\sigma_{d_0}$ the uncertainty
in the measured $d_0$, including the transverse spread of the luminous region.
Several different track selection working points,
applying additional criteria on top of these basic requirements, were
used, as summarised in Table~\ref{t:trkwp}.
The baseline selection~A uses tracks only in the
barrel region of the inner detector ($|\eta|<1.0$), requires
at least nine silicon hits \nsihit\ (counting both pixel and SCT hits), and
requires at most one pixel `hole' (\npixhole), i.e.\ a missing pixel hit
where one is expected, taking into account known dead modules.
Selection~B extends the acceptance into the endcap
tracking region, with tighter hit requirements for $|\eta|>1.65$, and
does not allow any pixel holes. Selection~C is based on
selection~A, with a tighter requirement of at least ten silicon hits.
Selection~A was used as the baseline track-counting luminosity measurement,
and the other selections were used to study systematic uncertainties.
 
\begin{table}[tp]
\caption{\label{t:trkwp}Track-selection criteria for the different working
points used in the track-counting luminosity measurement, applied in addition
to the basic TightPrimary selection of Ref.~\cite{ATL-PHYS-PUB-2015-051}.
Selection A was used for the baseline track-counting luminosity measurement.}
\centering
 
\begin{tabular}{l|ccc}\hline
Criterion & Selection A & Selection B & Selection C \\
\hline
\pt[GeV] & $>0.9$ & $>0.9$ & $>0.9$ \\
$|\eta|$ & $<1.0$ & $<2.5$ & $<1.0$ \\
\nsihit & $\geq 9$ & $\geq 9$ if $|\eta|<1.65$ & $\geq 10$ \\
& & else $\geq 11$ & \\
\npixhole & $\leq 1$ & $=0$ & $\leq 1$ \\
$|d_0|/\sigma_{d_0}$ & $<7$ & $<7$ & $<7$ \\
\hline
\end{tabular}
\end{table}
 
The statistical precision of the track-counting luminosity measurement is
limited  at low luminosity, making it impractical to calibrate it directly
using vdM scans. Instead, the working points were calibrated to agree with LUCID
luminosity measurements in the same LHC fills as the vdM scans, utilising
periods with stable, almost constant
luminosity where the beams were colliding head-on in ATLAS, typically while
vdM scans were being performed at the CMS interaction point.
The average number of selected tracks
per \sxyt\ inelastic $pp$ collision was about 1.7 for selections~A and~C,
and about 3.7 for selection~B, which has a larger acceptance in $|\eta|$.

\subsection{Calorimeter-based algorithms}\label{ss:calo}
 
The electromagnetic endcap calorimeters  are sampling calorimeters
with liquid argon as the active medium and lead/stainless-steel absorbers,
covering the region $1.5<|\eta|<3.2$ on each side of the ATLAS detector.
Ionisation electrons produced by charged particles crossing the LAr-filled
gaps between the absorbers draw a current through the high-voltage (HV) power
supplies that is proportional to the instantaneous luminosity, after
subtracting the electronics pedestal determined from the period without
collisions at the start of each LHC fill. A subset of the EMEC HV lines
are used to produce a luminosity measurement, independently from the
A- and C-side EMEC detectors. A similar measurement is available from the
forward calorimeters that cover the region $3.2<|\eta|<4.9$ with longitudinal
segmentation into three modules. The first module has copper absorbers,
optimised for the detection of electromagnetic showers, and the HV currents from
the 16~azimuthal sectors are combined to produce a luminosity measurement,
again separately from the A- and C-side calorimeters. Since both the
EMEC and FCal luminosity measurements rely on `slow' ionisation currents,
they cannot resolve individual bunches and therefore give bunch-integrated
measurements
which are read out every few seconds and then averaged per luminosity block.
They have insufficient sensitivity at low luminosity to be calibrated during
vdM scans, so they were cross-calibrated to track-counting measurements in
high-luminosity physics fills as discussed in Section~\ref{s:stab}.
 
The event-by-event `energy flow' through the cells of the LAr calorimeters,
as read out by the standard LAr pulse-shaping electronics, provides another
potential luminosity measurement. However, the long LAr drift time and bipolar
pulse shaping \cite{LARG-2009-02} wash out any usable signal during running
with bunch trains. Only the EMEC and FCal cells have short enough drift
times to yield useful measurements in runs with isolated bunches
separated by at least 500\,ns. Under these conditions, the total energy
deposited in a group of LAr cells and averaged over a luminosity block is
proportional to the instantaneous luminosity (after pedestal subtraction). A
dedicated data stream reading out only the LAr calorimeter data in
randomly sampled bunch crossings was used in certain runs with isolated bunches
to derive a LAr energy-flow luminosity measurement. This was exploited
to constrain potential non-linearity in the track-counting luminosity
measurement as discussed in Section~\ref{ss:caltladder}.
 
The TileCal hadronic calorimeter uses steel absorber plates interleaved with
plastic scintillators as the active medium, read out by wavelength-shifting
fibres connected to PMTs \cite{TCAL-2010-01}.
It is divided into a central barrel section covering
$|\eta|<1.0$, and an extended barrel on each side of the detector (A and C)
covering $0.8<|\eta|<1.7$. Both the barrel and extended barrels are segmented
azimuthally into 64 sectors, and longitudinally into three sections.
The current drawn by each PMT (after pedestal correction) is proportional to
the total number of particles traversing the corresponding TileCal cell, and
hence to the instantaneous luminosity. In principle, all TileCal cells can be
used for luminosity measurement, but the D-cells, located at the largest radius
in the last longitudinal sampling, suffer least from variations in response
over time due to radiation damage. The D6 cells, at the highest pseudorapidity
in the extended barrels on the A- and C-sides, were used for the primary
TileCal luminosity
measurements in this analysis, with other D-cells at lower pseudorapidity
being used for systematic comparisons. Changes in the gains of the PMTs with
time were monitored between physics runs
using a laser calibration system which injects pulses
directly into the PMTs, and the response of the scintillators was also
monitored once or twice per year using $^{137}$Cs radioactive sources
circulating through the calorimeter cells during shutdown periods. The TileCal
D-cells have insufficient sensitivity to be calibrated during vdM scans, so
they were cross-calibrated to track-counting in the same way as for the EMEC
and FCal measurements.
 
The TileCal system also includes the E1 to E4 scintillators, installed
in the gaps between the barrel and endcap calorimeter assemblies, and designed
primarily to measure energy loss in this region. Being much more exposed to
particles from the interaction point than the rest of the TileCal scintillators
(which are shielded by the electromagnetic calorimeters)
the E-cells (in particular E3, and E4 which is closest to the beamline) are
sensitive enough to make precise luminosity measurements during vdM fills.
They play a vital role in constraining
the calibration transfer uncertainties, as discussed in Section~\ref{s:calsyst}.
However, their response changes rapidly over time due to radiation damage,
so they cannot be used for long-term stability studies. The high currents
drawn by the E-cell PMTs also cause the PMT gains to increase slightly
at high luminosity, resulting in an overestimation of the actual luminosity.
This effect was monitored during collision data-taking by firing the
TileCal calibration laser periodically during the LHC abort gap---a time window
during each revolution of the LHC beams, in which no bunches pass through the ATLAS interaction point, and which is left free for the duration of the
rise time of the LHC beam-dump kicker magnets.
By comparing, in 2018 data, the response of the E-cell PMTs  to the laser pulse
with that of the D6-cell PMTs (the gain of which remains stable thanks to the
much lower currents), a correction to the E-cell PMT gain was derived, reducing
the reported luminosity in high-luminosity running by up to 1\% for the E4
cells, and somewhat less for E3.
This correction is applied to all the TileCal E-cell data shown in this paper.

% End of text imported from the .//lumidet.tex input file

% The next lines are included from the .//vdm.tex input file
\section{Absolute luminosity calibration in vdM scans}\label{s:vdmcal}
 
The absolute luminosity calibration of LUCID and BCM, corresponding
to the determination of the visible cross-section \sigmavis\ for each of
the LUCID and BCM algorithms, was derived using dedicated vdM scan sessions
during special LHC fills in each data-taking year. The calibration
methodology and main sources of uncertainty are similar to those in
Run~1 \cite{DAPR-2011-01,DAPR-2013-01},
but with significant refinements in the light of Run~2
experience and additional studies. The improvements especially concern
the scan-curve fitting, the treatment of beam--beam and emittance
growth effects, and the
potential effects of non-linearity from magnetic hysteresis in the LHC
steering corrector magnets used to move the beams in vdM scans.
These aspects are given particular attention below.
The uncertainties related to each aspect of the vdM calibration are discussed
in the relevant subsections, and listed for each data-taking year
in Table~\ref{t:unc} in Section~\ref{s:unccorl}.
 
\subsection{vdM formalism}\label{ss:vdmform}
 
The instantaneous luminosity for a single colliding bunch pair, \lbun, is given
in terms of LHC beam parameters by
\begin{equation}\label{e:vdm}
\lbun = \frac{f_\mathrm{r} n_1 n_2}{2\pi \capsigx \capsigy}\ ,
\end{equation}
where $f_\mathrm{r}$ is the LHC revolution frequency,
$n_1$ and $n_2$ are the numbers of protons in the beam-1 and beam-2 colliding
bunches, and \capsigx\ and \capsigy\ are the convolved beam sizes
in the horizontal and vertical (transverse) planes \cite{lumibib}.
In the vdM method, the uncalibrated instantaneous luminosity or counting rate
$R(\Delta x)$ is measured as a function of the nominal separation
$\Delta x$ between the two beams in the horizontal plane, given by
\begin{equation}\nonumber
\Delta x=\xpos{1}-\xpos{2}\ ,
\end{equation}
where \xpos{i} is the displacement of beam $i$ from its initial position
(corresponding to $\xpos{i}=0$) with approximately head-on collisions
(i.e.\ no
nominal transverse separation). These initial positions were established by
optimising the luminosity as a function of separation in each plane
before starting the first vdM scan in a session. The separation was then
varied in steps by displacing the two beams in opposite directions
($\xpos{2}=-\xpos{1}$), starting with the maximum negative separation,
moving through the peak and onward to maximum positive separation,
and measuring the instantaneous luminosity $R(\Delta x)$ at each step.
The procedure was then repeated in the vertical plane (varying $\Delta y$ and
keeping $\Delta x=0$) to measure $R(\Delta y)$,
completing an $x$--$y$ scan pair.
The quantity \capsigx\ is then given by
\begin{equation}\label{e:capsig}
\capsigx = \frac{1}{\sqrt{2\pi}} \frac{\int R(\Delta x)\, \mathrm{d}\Delta x}{R(\dxmax)}\ ,
\end{equation}
i.e.\ the ratio of the integral over the horizontal scan to the
instantaneous luminosity $R(\dxmax)$ on the peak of the scan
at $\dxmax\approx 0$, and similarly for \capsigy\ with a vertical
scan.
 
If the form of $R(\Delta x)$ is Gaussian, \capsigx\ is equal to the
standard deviation of the distribution, but the method is valid for any
functional form of $R(\Delta x)$.  However, the formulation does assume that
the particle densities in each bunch can be factorised into independent
functions of $x$ and $y$. The effect of violations of this assumption
(i.e.\ of non-factorisation) is quantified in Section~\ref{ss:nonfact} below.
Since the normalisation of $R(\Delta x)$
cancels out in Eq.~(\ref{e:capsig}), any quantity proportional to luminosity
can be used to determine the scan curve. The calibration of a given algorithm
(i.e.\ its \sigmavis\ value) can then be determined by combining
Eqs.~(\ref{e:lbunvis}) and~(\ref{e:vdm}) to give
\begin{equation}\label{e:sigcal}
\sigmavis = \muvismax \frac{2\pi \capsigx\capsigy}{n_1 n_2}\ ,
\end{equation}
where $\muvismax$ is the visible interaction rate per bunch crossing
at the peak of the scan curve. In practice, the vdM scan curve was analysed
in terms of the specific visible interaction rate \mubvis, the visible
interaction rate \muvis\ divided  by the bunch population product $n_1n_2$
measured at each scan step, as discussed in Section~\ref{ss:bunch}.
This normalisation accounts for the small
decreases in bunch populations during the course of each scan. The specific
interaction rate in head-on collisions \mubvismax\ was determined from the
average of the rates fitted at the peaks of the $x$ and $y$ scan curves
\mubvismaxi{x}\ and \mubvismaxi{y}, so Eq.~(\ref{e:sigcal}) becomes
\begin{equation}\label{e:sigcalav}
\sigmavis=\frac{1}{2}\left( \mubvismaxi{x}+\mubvismaxi{y} \right)\cdot 2\pi\capsigx\capsigy\ .
\end{equation}
A single pair of $x$--$y$ vdM scans
suffices to determine \sigmavis\ for each algorithm active during the
scan. Since the quantities entering Eq.~(\ref{e:sigcalav}) may be different
for each colliding bunch pair, it is essential to perform a bunch-by-bunch
analysis
to determine \sigmavis, limiting the vdM absolute calibration
in ATLAS to the LUCID and BCM luminosity algorithms. Track-counting cannot
be used since only a limited rate of bunch crossings
can be read out for offline analysis,\footnote{At \sxvt, track-counting
algorithms were calibrated in vdM scans \cite{DAPR-2013-01}, but trigger-rate
and dataflow limitations restricted this calibration to a small subset
of the colliding bunch pairs. A more precise calibration can be achieved
by cross-calibrating to LUCID algorithms as discussed in
Section~\ref{ss:trkcnt}.}
and the calorimeter algorithms only measure bunch-integrated luminosity.
 
\subsection{vdM scan datasets}\label{ss:scandata}
 
In order to minimise both instrumental and accelerator-related uncertainties
\cite{lumibib}, \sxyt\ vdM scans were not performed in standard physics
conditions, but once per year in dedicated low-luminosity vdM scan fills using
a special LHC configuration. Filling schemes with 44--140 isolated
bunches in each beam were
used, so as to avoid the parasitic encounters between incoming and outgoing
bunches that occur in normal bunch train running. This allowed the beam
crossing angle to be set to zero, reducing orbit drift and beam--beam-related
uncertainties. The parameter $\beta^*$ was set to 19.2\,m rather than the
0.25--0.8\,m used during normal physics running, and the injected
beam emittance was increased to 3--4\,$\mu$m-rad. These changes
increased the RMS transverse size of the luminous region to about $60\,\mu$m,
thus reducing vertex-resolution uncertainties in evaluating the
non-factorisation corrections discussed below.
The longitudinal size of the luminous region was typically around
45\,mm, corresponding to an RMS bunch length of 64\,mm.
Special care was taken in the LHC injector chain to produce beams with
Gaussian-like transverse profiles \cite{bunchvdm}, a procedure that also
minimises non-factorisation effects in the scans.
Finally, bunch currents were reduced to $\sim 0.8\times 10^{11}$ protons/bunch,
minimising bunch-current normalisation uncertainties. When combined with
the enlarged emittance, this also reduces beam--beam biases.
These configurations typically resulted in
$\meanmu\approx 0.5$ at the peak of the scan curves, whilst maintaining
measurable rates in the tails of the scans at up to  $6\signom$
separation, where $\signom\approx 100\,\mu$m is the nominal single-beam size
at the interaction point.
 
The vdM scan datasets are listed in Table~\ref{t:vdmds}. Several pairs of
on-axis $x$--$y$ scans were performed in each session, spaced
through one or more fills in order to study reproducibility. One or two pairs of
off-axis scans, where the beam was scanned in $x$ with an offset in $y$ of
e.g.\ $300\,\mu$m (and vice versa) were also included,
to study the beam profiles in the tails
in order to constrain non-factorisation effects. A diagonal scan varying
$x$ and $y$ simultaneously was also performed in 2016.
A typical scan pair took $2\times 20$ minutes for $x$ and $y$, scanning
the beam separation between $\pm 6\signom$ in 25 steps of $0.5\signom$.
Length scale
calibration (LSC) scans were also performed in the vdM or nearby fills,
as described in Section~\ref{ss:lsc}. A vdM fill lasted up to one day,
with alternating groups of scans in ATLAS and CMS, and the beams colliding
head-on in the experiment which was not performing scans.
 
\begin{table}[tp]
\caption{\label{t:vdmds}Summary of the \sxyt\ $pp$ vdM scan fills used in the
ATLAS Run~2 vdM analysis, showing the dates, LHC fill numbers, numbers of
bunches in each beam (\nbeam), numbers of bunch pairs colliding in ATLAS
(\nbun), and the individual $x$--$y$ scan pairs performed in each fill
(S1, S2, etc., including off-axis scans denoted `Sn-off', a
diagonal $xy$ scan in 2016 denoted `S5-diag' and length scale
calibration scans denoted LSC). All fills
used the dedicated vdM optics with $\beta^*=19.2$\,m, with zero beam crossing
angle and only isolated bunches.}
\centering
 
\begin{tabular}{lc|cc|l}\hline
Date & Fill & \nbeam & \nbun & Scans \\
\hline
24/8/2015 & 4266 & 44 & 30 & S1, S2, S3-off \\
25/8/2015 & 4269 & 51 & 8 & S4, LSC \\
\hline
17/5/2016 & 4937 & 55 & 11 & LSC \\
18/5/2016 & 4945 & 52 & 32 & S1,S2, S3-off, S4, S5-diag \\
27/5/2016 & 4954 & 52 & 32 & S6 \\
\hline
28/7/2017 & 6016 & 52 & 32 & S1, S2, S3-off, S4, S5, LSC \\
\hline
29/6/2018 & 6864 & 70 & 58 & LSC \\
30/6/2018 & 6868 & 140 & 124 & S1, S2, S3-off, S4, S5-off, S6 \\
\hline
\end{tabular}
\end{table}
 
\subsection{Scan curve fitting}\label{ss:scanfit}
 
\begin{figure}[tp]
\parbox{83mm}{\includegraphics[width=76mm]{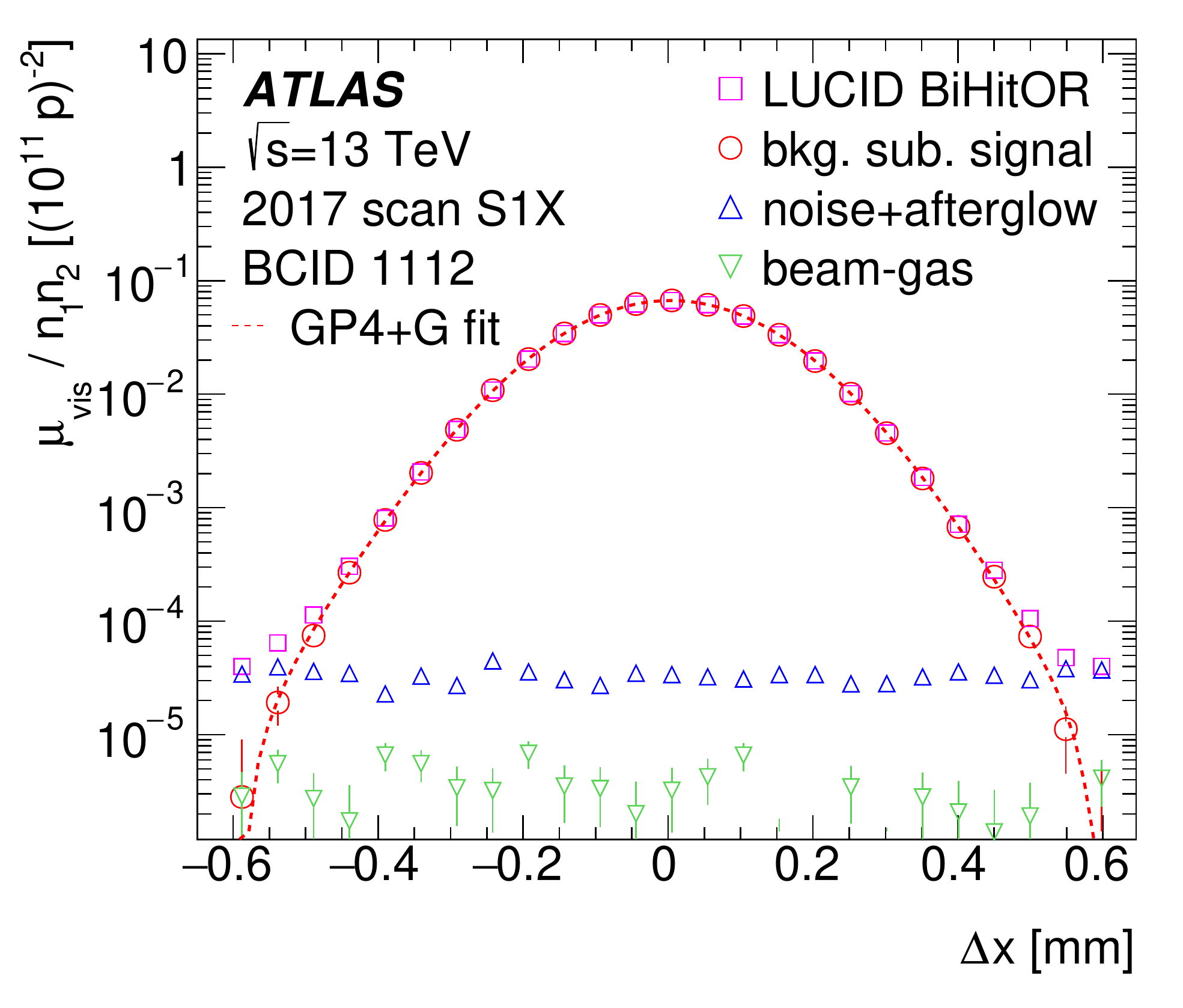}\vspace{-6mm}\center{(a)}}
\parbox{83mm}{\includegraphics[width=76mm]{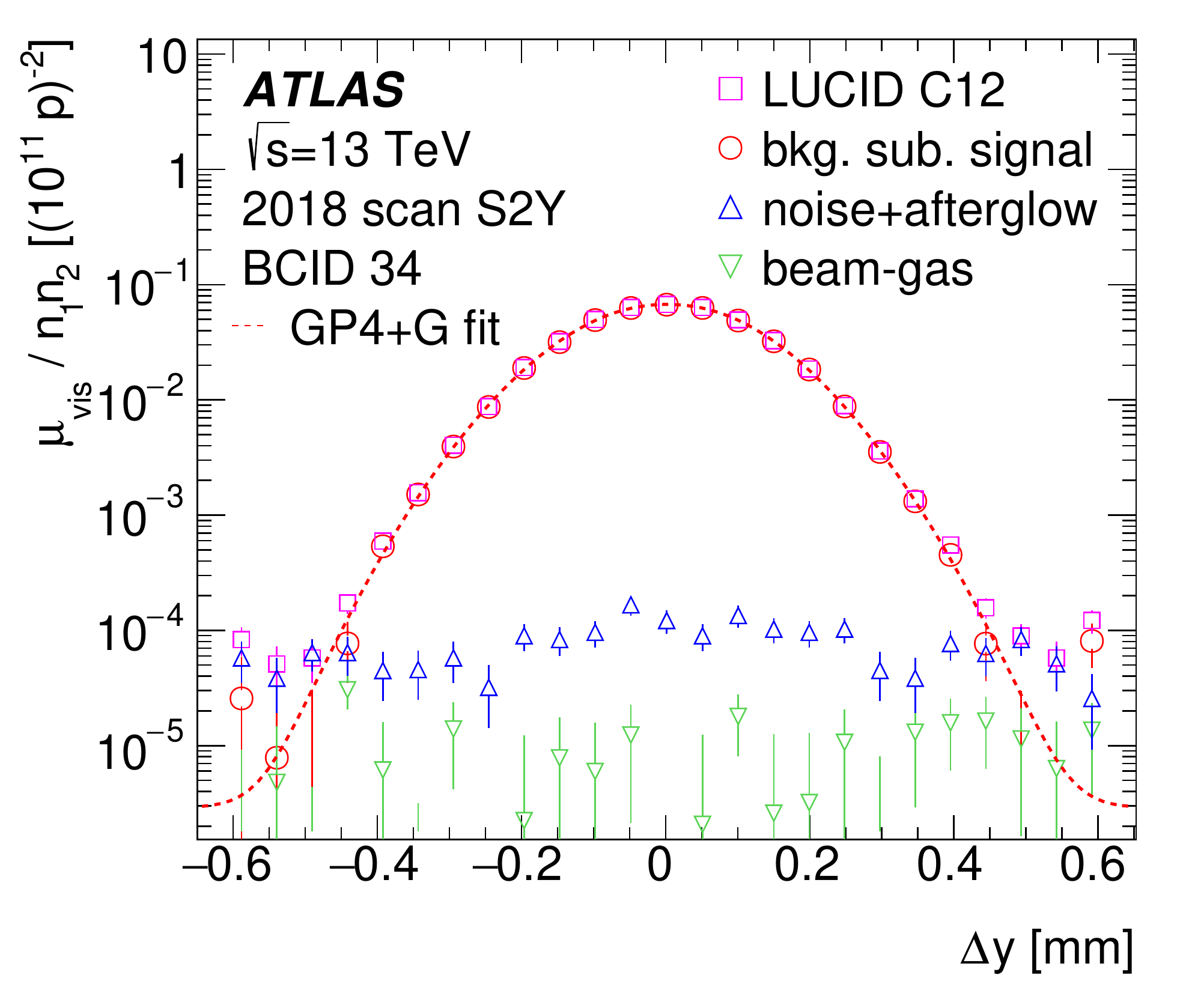}\vspace{-6mm}\center{(b)}}
\caption{\label{f:vdmcurve}Visible interaction rate \muvis\ per unit bunch
population product $n_1n_2$ (i.e.\ \mubvis)
vs. beam separation in (a) the horizontal plane
for bunch slot (BCID) 1112 in scan~1 of 2017, using the LUCID BiHitOR algorithm,
and (b) the vertical plane for BCID 34 in scan~2 of 2018, using the
LUCID C12 PMT algorithm. The measured rate is shown by the pink squares,
the background-subtracted rate by the red circles, the estimated background
from noise and afterglow by the blue upward-pointing triangles and that from
beam--gas interactions by the green downward-pointing triangles.
The error bars show the statistical uncertainties, which are in some cases
smaller than the symbol sizes.
The fits to a Gaussian function multiplied by a fourth-order polynomial (GP4)
plus an additional Gaussian function in the central region and a constant term
(see text) are shown by the dashed red lines.}
\end{figure}

Typical vdM scan curves from the 2017 and 2018 datasets using the LUCID BiHitOR
and C12 algorithms are shown in Figure~\ref{f:vdmcurve}.
The \mubvis\ distributions
were fitted using an analytic function (e.g.\ a Gaussian function multiplied
by a polynomial), after subtraction of backgrounds. For the LUCID algorithms,
the background is dominated by noise from the $^{207}$Bi calibration
source, which was estimated by subtracting the counting rate measured
in the preceding bunch slot. The latter was always empty in the LHC filling
patterns used for vdM scans, which only include isolated bunches separated
by at least 20 unfilled bunch slots.
This procedure also subtracts the smaller
background due to `afterglow' from photons produced in the decay of nuclei
produced in hadronic showers initiated by the $pp$ collisions. This background
scales with
the instantaneous luminosity at each scan point, and decays slowly over the
bunch slots following each filled bunch slot. A further small background
comes from the interaction of the protons in each beam with residual gas
molecules in the beam pipe, and scales with the bunch population
$n_1$ or $n_2$. It was estimated using the counting rate in the
bunch slots corresponding to `unpaired' bunches,
where a bunch from one beam passes through the ATLAS interaction point
without meeting a bunch travelling in the opposite direction.
 
Since the visible interaction rate is only sampled at a limited number of scan
points, the choice of fit function is crucial in order to ensure unbiased
estimates of \muvismax\ and \capsigxy. Fitting the scan curves to a Gaussian
function multiplied by a fourth-order polynomial (referred to as GP4)
gives a good general description
of the shape, but was found to systematically overestimate the peak of the
scan curves by about 0.5\%, biasing \mubvismax. Better results were achieved
by adding a second Gaussian function to the GP4 function for the five
central scan points, flattening the scan curve at the peak.\footnote{The second
Gaussian function introduces discontinuities into the fitting function just
outside the five central points, but the results were found not to depend on
whether the discontinuities were placed close to the central points or close to
the next two scan points at larger $|\Delta x|$.} This GP4+G function
was used for the baseline \sigmavis\ results. The statistical uncertainties
from the fit are smaller than 0.1\% in all years, and are listed as
`Statistical uncertainty' in Table~\ref{t:unc}.
A Gaussian function multiplied by a
sixth-order polynomial (GP6) was found to give a similar goodness-of-fit
to GP4+G, and the difference between the \sigmavis\ obtained from these two
functions was used to define the `Fit model' uncertainty shown for each
data-taking year in Table~\ref{t:unc}.
In all cases, a constant term was also included in the fit function, and the
integrals to determine \capsigxy\ (Eq.~(\ref{e:capsig})) were truncated at the
limits of the scan range. The constant term improves the fit stability and the
description of the tails.
 
To estimate the uncertainty due to the background subtraction procedure,
alternative fits were performed to the data without subtracting
the estimates of the noise and beam--gas components, instead using the constant
term to account for their contributions, and
implicitly assuming that the backgrounds do not depend on beam separation.
In this case, the constant term
was not included in the evaluation of \capsigxy\ and \muvismax.
The resulting difference in \sigmavis\ was used to define the `Background
subtraction' uncertainty in Table~\ref{t:unc}.
 
The vdM scan curve fits were performed for all available LUCID and BCM
algorithms, and the results are compared in Section~\ref{ss:vdmcon}. However,
most of the systematic uncertainty studies detailed below were only performed
with a single `reference' algorithm for each year of data-taking.
For 2016--18, this is the algorithm
listed in Table~\ref{t:lucidalg} and used for the baseline luminosity
determination; however for 2015, the LUCID BiEvtORA algorithm, using
four PMTs on the A-side of LUCID in an EvtOR combination, was used as the
reference algorithm instead.
 
\subsection{Bunch population measurement}\label{ss:bunch}
 
The determination of the bunch populations $n_1$ and $n_2$
was based on the LHC beam instrumentation, using
the methods discussed in detail in Ref.~\cite{DAPR-2013-01}.
The total intensity in each beam was measured by the LHC DC current
transformers (DCCT), which have an absolute precision of better than 0.1\% but
lack the ability to resolve individual bunches. The per-bunch intensities
were determined from the fast beam-current transformers (FBCT), which
can resolve the current in each of the 3564 nominal 25\,ns bunch slots in
each beam. The FBCT measurements were  normalised to the DCCT measurement at
each vdM scan step, and
supplemented by measurements from the ATLAS beam-pickup timing
system (BPTX), which can also resolve the relative populations in each bunch.
 
Both the FBCT and BPTX systems suffer from non-linear behaviour, which
can be parameterised as a constant offset between the measured and
true beam current, assuming the measurement is approximately linear near
the average bunch currents $\bar{n}_1$ and $\bar{n}_2$ for the two beams.
For the BPTX measurements, these offsets were determined
for each $x$--$y$ scan pair from the data by minimising the differences between
the corrected \sigmavis\ values measured for all bunches, defining a
$\chi^2$ by
\begin{equation}\nonumber
\chi^2(\sigbarvis,b_1,b_2)=\sum_i \left(\frac{S(\sigbarvis,b_1,b_2,n^i_1,n^i_2)-\siguncvisi}{\Delta\siguncvisi}\right)^2\ ,
\end{equation}
where $\siguncvisi$ is the uncorrected measured visible cross-section for
colliding bunch pair $i$ and $\Delta\siguncvisi$ its statistical error,
and $S$ represents the modified visible cross-section
which would be measured for bunch pair $i$, given a true average visible cross-section $\sigbarvis$ and bunch current offsets $b_1$ and $b_2$ for the two beams.
The quantity $S$ is given by
\begin{equation}\nonumber
S(\sigbarvis,b_1,b_2,n^i_1,n^i_2)=\sigbarvis\cdot\left(\frac{n^i_1+b_1(n^i_1-\bar{n}_1)}{n^i_1}\right) \left(\frac{n^i_2+b_2(n^i_2-\bar{n}_2)}{n^i_2}\right)\ ,
\end{equation}
where the two terms in brackets represent the bunch-dependent biases in
the measured \sigmavis\ values caused by the non-linearity in the BPTX bunch
current measurements. The values of $b_1$ and $b_2$ were determined separately
for each scan in each year by minimising the $\chi^2$ as a function of
$b_1$, $b_2$ and $\sigbarvis$ \cite{anders}, and have typical sizes of a
few percent.
In all vdM datasets, this correction reduces the apparent dependence of the
\sigmavis\ measured for each bunch pair on the BPTX bunch-population product
$n_1n_2$.
 
The FBCT uses independent electronics channels with separate scales and
offsets for the even- and odd-numbered bunch slots, making it difficult to apply
the above procedure to the FBCT measurements. Instead, the uncorrected FBCT
measurements in each scan were fitted to a linear function of the
offset-fit-corrected BPTX measurements, deriving parameters $p_0^j$
and $p_1^j$, where $j=1$ for odd and~2 for even bunch pairs.
These parameters, derived from separate fits to the odd and even
bunches within each scan, were used to determine corrected FBCT populations
$n_1^{i,j,\mathrm{FBCT-corr}}$ from the uncorrected measurements
$n_1^{i,j,\mathrm{FBCT-uncorr}}$ via
\begin{equation}\nonumber
n_1^{i,j,\mathrm{FBCT-corr}} = (n_1^{i,j,\mathrm{FBCT-uncorr}}-p_0^j)/p_1^j \ ,
\end{equation}
and similarly for $n_2^{i,j,\mathrm{FBCT-corr}}$.
This correction procedure reduced the $n_1n_2$ dependence of the \sigmavis\
values measured using the FBCT, with residual variations contributing
to the bunch-by-bunch consistency uncertainty discussed in
Section~\ref{ss:vdmcon} below.
The FBCT bunch population measurements (corrected
using the BPTX offset fit) were used for the baseline \sigmavis\ results,
with the alternative BPTX-only results used to define the `FBCT bunch-by-bunch
fractions' uncertainty in Table~\ref{t:unc}, which is below 0.1\% for all years.
The changes in \sigmavis\ resulting from these corrections amount to about
+0.4\% in 2015 and less than $\pm 0.05$\% in all other years.
 
The DCCT measurement used to normalise the BPTX and FBCT results summed over
all filled bunches is also sensitive to additional contributions to the
total circulating beam currents from
ghost charge and satellite bunches. Ghost charge refers to circulating protons
present in nominally unfilled bunch slots, which are included in the
DCCT reading but are typically below the FBCT or BPTX threshold.
Their contribution was measured by comparing the rate of beam--gas vertices
reconstructed by the LHCb experiment in nominally empty bunch slots with the
rate in those bunch slots containing an unpaired bunch \cite{lhcblumi}.
 
Satellite bunches are composed of protons present in
a nominally filled bunch slot, but which are captured in a radio-frequency (RF)
bucket at least one RF period (2.5\,ns) away from the nominally filled bucket.
Their current is included in the FBCT reading, but because collisions between
satellite bunches in each beam
make a negligible contribution to the nominal luminosity signal
at the interaction point, their contributions to the FBCT measurements
must be estimated and subtracted. Satellite populations were measured
using the LHC longitudinal density monitor \cite{boccardi}.
At zero crossing angle, longitudinally displaced collisions between a
satellite bunch and the opposing
in-time nominal bunch can contribute to the measured luminosity signal in the
corresponding bunch slot. The magnitude of this additional contribution
is proportional to the sum of the fractional satellite populations in the two
beams, and also depends on the emittance of the satellite bunch, on the
number of RF buckets by which it is offset from the nominal bunch, and
on the longitudinal location and time acceptance of the luminosity detector
considered.  The fractional satellite contribution
summed over both beams is at most 0.02\%, 0.03\%, 0.02\% and 0.09\%
in each of the datasets from 2015 to 2018, and these values were conservatively
taken as uncertainties in \sigmavis\ due to the additional luminosity produced
by nominal--satellite collisions.
 
The combined effects of ghost and satellite charges on the bunch-population
measurement lead to positive corrections
to \sigmavis\ of 0.1--0.4\%, depending on the vdM fill in question, with
uncertainties that are an order of magnitude smaller. The corresponding
uncertainty in Table~\ref{t:unc} also includes the effect of nominal--satellite
collisions.
 
A further uncertainty of 0.2\% arises from the absolute calibration of the
DCCT current measurements, derived from a precision current source. The
uncertainty takes into account residual non-linearities, long-term stability
and dependence on beam conditions, and was derived using the procedures
described in Ref.~\cite{dcctcalib}.

\subsection{Orbit drift corrections}\label{ss:odc}
 
Gradual orbit drifts of up to $O(10\,\mu\mathrm{m})$ in the positions of
one or both beams have
been observed during the course of a single vdM scan. These drifts change the
actual beam separation from the nominal one produced by the LHC steering
corrector magnets, and were monitored using two independent beam position
monitor (BPM) systems. The DOROS BPM system\footnote{Named for the  DOROS
(Diode ORbit and Oscillation System) front-end electronics, which can
separate the close-in-time signals from the bunches circulating in opposite
directions in the same beam pipe close to the IP.}
uses pickups located at the inner
(i.e.\ closest to the IP) ends of the LHC final-focus quadrupole triplet
assemblies
at $z=\pm 21.7$\,m on each side of the IP, allowing the beam position at the IP
to be determined by averaging the measurements from each side. Since these
BPMs are located closer to ATLAS than the steering correctors used to move the
beams during vdM scans, they also see the beam displacements during scans.
The arc BPMs are located at intervals throughout the LHC magnetic arcs, and
the measurements from those close to ATLAS can be extrapolated to give the
beam position at the IP. The arc BPMs do not see the displacement during scans,
apart from any residual effects leaking into the LHC arcs due to non-closure
of the orbit bumps around the IP. For both BPM systems, the beam positions in
both planes were measured at the zero nominal beam separation points
immediately before and after each single scan ($x$ or $y$), and any changes
were interpolated linearly as a function of time within the scan
to give corrections $\delta^{x,y}_1(t)$ and $\delta^{x,y}_{2}(t)$ for beams~1
and~2.
 
The dominant effect on the vdM analysis comes from in-plane drifts,
defined as horizontal
(vertical) drifts during a horizontal (vertical) scan. They distort
the beam-separation scale in Figure~\ref{f:vdmcurve}, changing the
values of \capsigxy\ obtained from Eq.~(\ref{e:capsig}). These drifts were
taken into account by correcting the separation at each scan step to include
the  interpolated orbit drifts:
\begin{equation}\nonumber
\Delta x^\mathrm{corr}=\Delta x^\mathrm{nominal}+\delta^x_1-\delta^x_2\ ,
\end{equation}
and similarly for $y$, before fitting the scan curves.
 
A subdominant effect is caused by out-of-plane drifts between the peaks of the
$x$ and $y$ scans (including during any pause between the two scans).
A horizontal orbit drift between the $x$-scan peak (where the beams are
perfectly centred horizontally by definition) and the $y$-scan peak (where
they may no longer be), would lead to an underestimate of the visible
interaction rate at the peak of the $y$ scan, and hence of \mubvismaxi{y}\ in
Eq.~(\ref{e:sigcalav}).
Similarly, a vertical orbit drift causes an underestimate
of the peak interaction rate in the $x$ scan.\footnote{Strictly speaking,
what matters is the cumulative effect on the collision rate of the change
in horizontal (vertical) beam separation, relative to that at the peak
of the $x$ ($y$) scan, from the beginning to the end of the $y$ ($x$) scan.
If the orbit drift rate is constant in time during the individual $x$ and $y$
scans, the cumulative effect on \mubvismax\ is equivalent to that evaluated
from the  scan peaks alone.}
Approximate corrections for out-of-plane drifts were made using the
differences in horizontal and vertical orbit drifts
$\delta x^0$ and $\delta y^0$ between the peaks of the $x$ and $y$ scan curves:
\begin{eqnarray*}
\delta x^0 & = &
\delta^x_1(t^0_y)-\delta^x_2(t^0_y)
-\delta^x_1(t^0_x)+\delta^x_2(t^0_x) \ ,\\
\delta y^0 & = &
\delta^y_1(t^0_x)-\delta^y_2(t^0_x)
-\delta^y_1(t^0_y)+\delta^y_2(t^0_y) \ ,
\end{eqnarray*}
where the central (zero nominal separation) points of the $x$ and $y$ scan
curves occur at times $t^0_x$
and $t^0_y$. It can be shown \cite{andersthesis} that the
visible cross-section defined in Eq.~(\ref{e:sigcalav}) and corrected for
out-of-plane orbit drifts is  given by
\begin{equation}\nonumber
\sigmavis = 4\pi\capsigx\capsigy\frac{{\cal G}_x(\Delta x^\mathrm{max}){\cal G}_y(\Delta y^\mathrm{max})}{{\cal G}_x(\delta x^0)+{\cal G}_y(\delta y^0)} \ ,
\end{equation}
where ${\cal G}_x(\Delta x)$  and ${\cal G}_y(\Delta y)$ denote the functions
of nominal separation fitted to the normalised vdM scan curves
$\mubvis(\Delta x)$ and $\mubvis(\Delta y)$, and
$\Delta x^\mathrm{max}$ and $\Delta y^\mathrm{max}$ are the fitted nominal
beam separations at the peaks of these functions, i.e.\
${\cal G}_x(\Delta x^\mathrm{max})=\mubvismaxi{x}$ and
${\cal G}_y(\Delta y^\mathrm{max})=\mubvismaxi{y}$. This formalism corrects
for both out-of-plane orbit drifts and any imperfect initial alignment
in the non-scanning plane.
 
A third effect is associated with the potential change in \capsigx\ due to
drifts in the $y$ (i.e.\ non-scanning) plane during the $x$ scan, and similarly
for \capsigy\ due to drifts in $x$ during the $y$ scan. This effect is distinct
from the effects of out-of-plane drifts on \mubvismax\ discussed above. However,
since the beams were reasonably well-centred on each other before each scan
started (as ensured by the scan protocol), these effects are negligible for
the orbit drifts observed in the non-scanning plane during on-axis scans.
 
Large orbit drift corrections were seen for scan S4 in 2015
(shifting \sigmavis\ by 2.6\%), S1 in 2017 (1.1\%) and S4 in 2018 ($-0.8$\%),
mainly due to drift in the horizontal plane. All other scans
have orbit drift corrections at the level of $\pm 0.5$\% or less. Applying the
corrections always improved the scan-to-scan consistency within each year. The
baseline corrections were evaluated using the DOROS BPM system, and using
the arc BPMs gave similar results. The `Orbit drift correction'
uncertainties in Table~\ref{t:unc} were calculated as the fractional
scan-averaged difference between the \sigmavis\ values determined using
the DOROS and arc BPM corrections, and are smaller than 0.1\% in all years.
 
Short-term movements of the beams during the 30\,s duration of a single vdM
scan step may lead to fluctuations in the luminosity, distorting the measured
scan curve. The potential effect of this beam position jitter was characterised
by studying the distribution of the individual arc BPM beam position
extrapolations, which were sampled every 5--7\,s,  i.e.\ several times
per scan step. All measurements in a single $x$ or $y$ scan were fitted as a
function of time to a second-order polynomial, and the RMS of the
residuals of the individual measurements with respect to this fit were
taken to be representative of the random beam jitter. Data from the $y$ scan
(where the beams are stationary in $x$) were used to measure the jitter in the
$x$-direction, and vice versa,
giving RMS values in the range 0.5--1.0\,$\mu$m, typically larger in the
horizontal plane than the vertical.
One thousand simulated replicas of each real data scan were created,
applying Gaussian smearing to the beam positions (and hence the separation) at
each step according to the RMS of the arc BPM fit residuals, coherently
for the data from all colliding bunch pairs. The RMS of the \sigmavis\ values
obtained from these fits, averaged over all scans in a year, is typically
around 0.2\%. No correction was made for this effect, but the RMS
was used to define the `Beam position jitter' uncertainty in Table~\ref{t:unc}.

\subsection{Non-factorisation effects}\label{ss:nonfact}
 
The vdM formalism described by Eqs.~(\ref{e:vdm}--\ref{e:sigcal}) assumes that
the particle density distributions in each bunch can be factorised into
independent horizontal and vertical components, such that the term
$1/2\pi\capsigx\capsigy$ in Eq.~(\ref{e:vdm}) fully describes the overlap
integral of the two beams. Evidence of non-factorisation was clearly
seen during Run~1 \cite{DAPR-2013-01}, especially when dedicated beam
tailoring in the LHC injector chain was not used. Such beam tailoring, designed
to produce Gaussian-like beam profiles in $x$ and $y$,
was used for all \sxyt\ vdM scan sessions during Run~2, and information
derived from the distribution of primary collision vertices reconstructed
by the ATLAS inner detector in both on- and off-axis scans was used
to constrain the possible residual effect of non-factorisation on the \sigmavis\
determination, following the procedure of Ref.~\cite{DAPR-2013-01}.
 
\begin{figure}[tp]
\parbox{83mm}{\includegraphics[width=76mm]{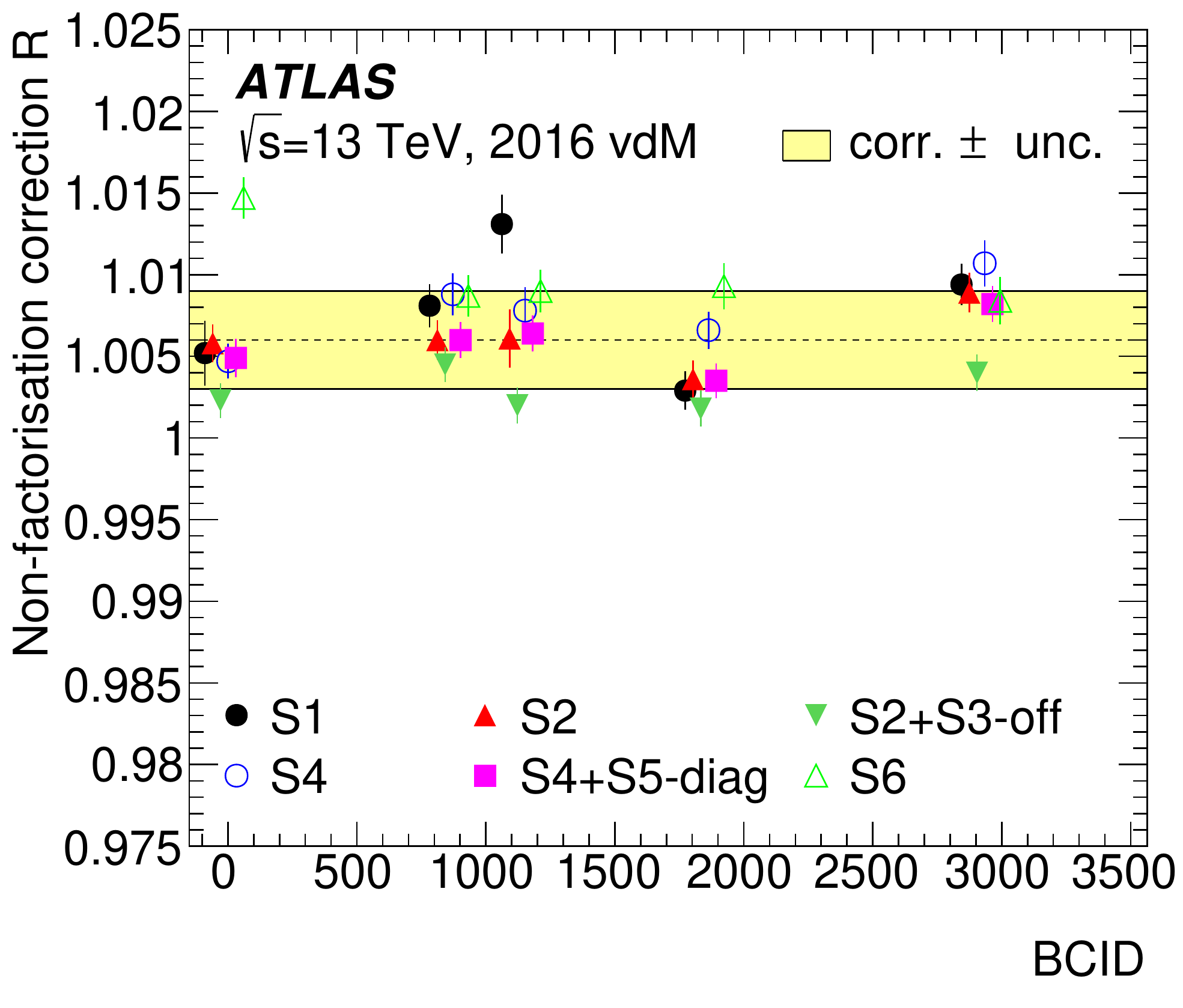}\vspace{-6mm}\center{(a)}}
\parbox{83mm}{\includegraphics[width=76mm]{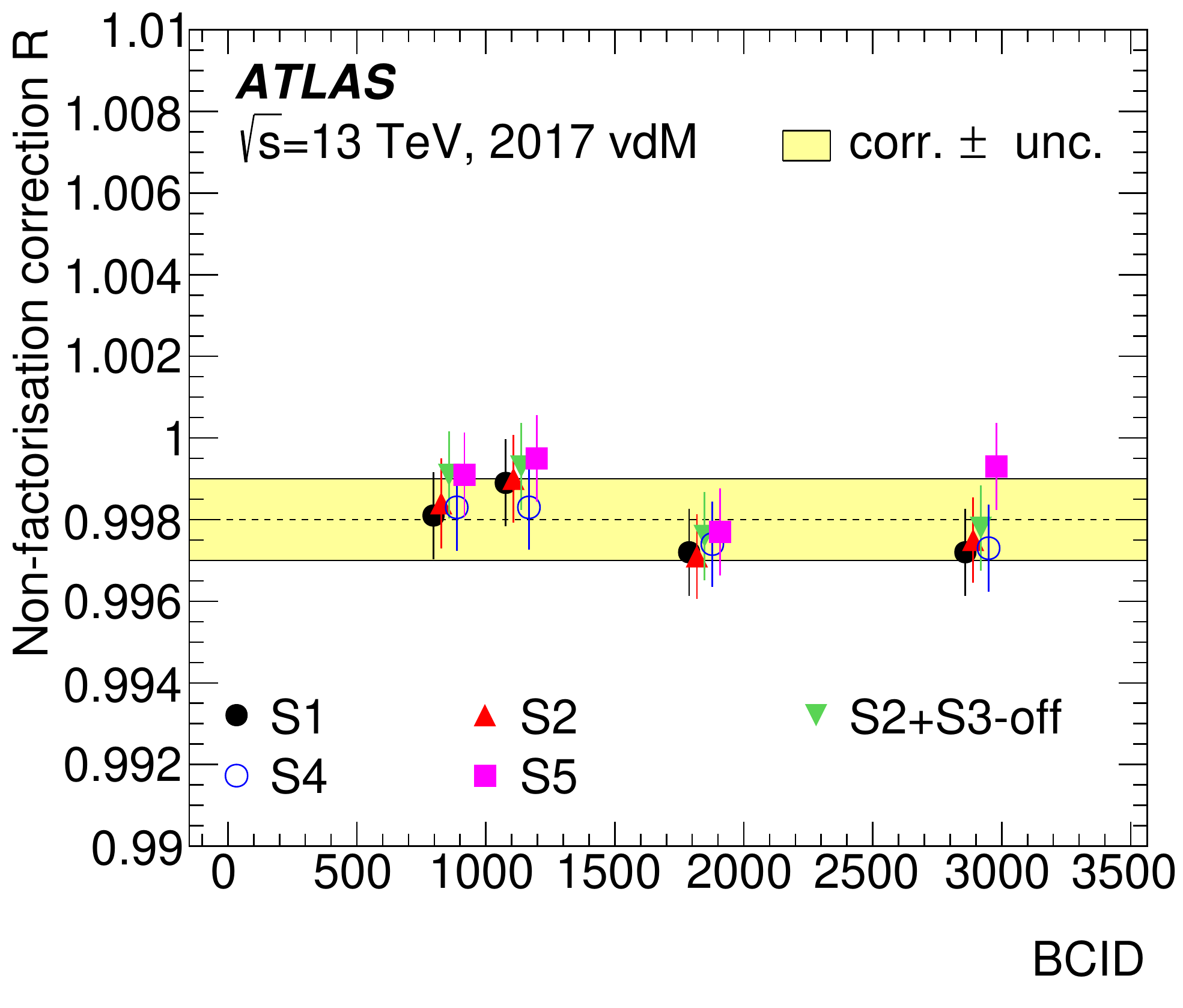}\vspace{-6mm}\center{(b)}}
\caption{\label{f:nonfact}Non-factorisation correction factors $R$ for several
colliding bunch pairs (BCID) and vdM scan sets in the (a) 2016 and (b) 2017
vdM sessions. The results were extracted from combined fits to the vdM
luminosity scan curves and reconstructed primary vertex data, using either
on-axis scans alone or combined fits to an on-axis scan and an off-axis or
diagonal scan. The uncertainties are statistical.
The dashed horizontal lines show the error-weighted
mean corrections,
and the yellow bands the uncertainties assigned from the RMS of the measured
$R$ values. The results from the different scans for each colliding bunch pair
are slightly offset on the horizontal axis for clarity.}
\end{figure}

In more detail, combined fits were performed to the beam-separation dependence
of both the LUCID luminosity measurements and the position, orientation and
shape of the luminous region, characterised by the three-dimensional (3D)
spatial distribution of the primary collision vertices formed from tracks
reconstructed in the inner detector (i.e.\ the beamspot) \cite{PERF-2015-01}.
Since this procedure requires a large number of
reconstructed vertices per scan step, it was carried out for only a
handful of colliding bunch pairs, whose tracking information was read out
at an enhanced rate. Non-factorisable single-beam profiles (each composed of a
weighted sum of three 3D Gaussian distributions $G(x,y,z)$ with arbitrary
widths and orientations)
were fitted to these data. These profiles were then  used in simulated vdM
scans to determine the ratio $R$ of the apparent luminosity scale which
would be extracted using the standard factorisable vdM formalism
to the true luminosity scale determined from the full 3D beam overlap
integral. This procedure was carried out both for each on-axis $x$--$y$
scan in each vdM dataset, and with fits combining each off-axis (and in 2016,
diagonal) scan with the closest-in-time on-axis scan. The resulting
$R$ values for the 2016 and 2017 datasets are shown in Figure~\ref{f:nonfact}.
In general, the results from on-axis and combined fits are similar,
although there is significant scatter between bunches and scans.
 
The ratio $R$ was used to correct the visible cross-section measured by the
standard factorised vdM analysis for the measured non-factorisation:
$\sigviscorr=\sigmavis/R$. Since the evaluation of $R$ only sampled a small
fraction of the colliding bunch pairs,
the error-weighted average of $R$ over this subset of bunches was
taken as representative of the complete set, and applied to
the bunch-averaged \sigmavis, giving corrections of $R=1.006\pm 0.003$ for
2016, $0.998\pm 0.001$ for 2017 and $1.003\pm 0.003$ for 2018, where the
uncertainties were defined as the RMS of the individual $R$ values for all
available bunches and scans in each year. In 2015, the same method gave
a central value of $R=0.995$ but with a poor fit to the offset scan
S3-off. An alternative method, fitting a non-factorisable function of both
$x$ and $y$ to the $x$ and $y$ scan curves simultaneously, but without the
use of beamspot data, gave a central value of $R=1.006$. For the 2015 dataset,
a value of $R=1.000\pm 0.006$ was used, i.e.\ no correction but an uncertainty
that spans the range of central values from the two methods.

\subsection{Beam--beam interactions}\label{ss:bbcorr}
 
The mutual electromagnetic interaction between colliding bunches produces
two effects that may bias their overlap integral, and that depend on the beam
separation: a transverse deflection that induces a non-linear distortion of the
intended separation, and a defocusing of one beam by the other that not only
modulates the optical demagnification from the LHC arcs to the interaction
point, but also modifies the shape of the transverse bunch profiles.
 
\begin{figure}[tp]
\parbox{83mm}{\includegraphics[width=76mm]{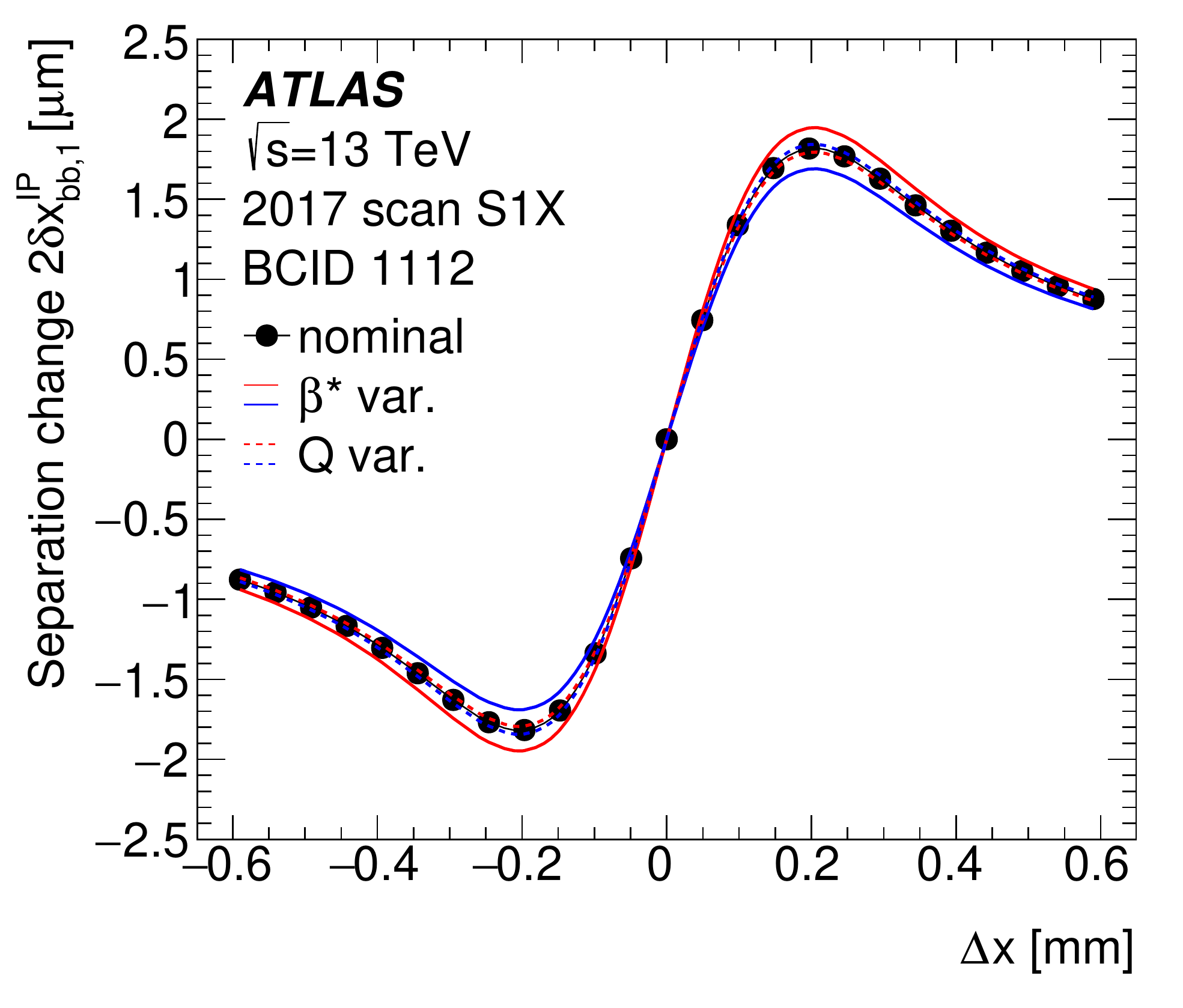}\vspace{-6mm}\center{(a)}}
\parbox{83mm}{\includegraphics[width=76mm]{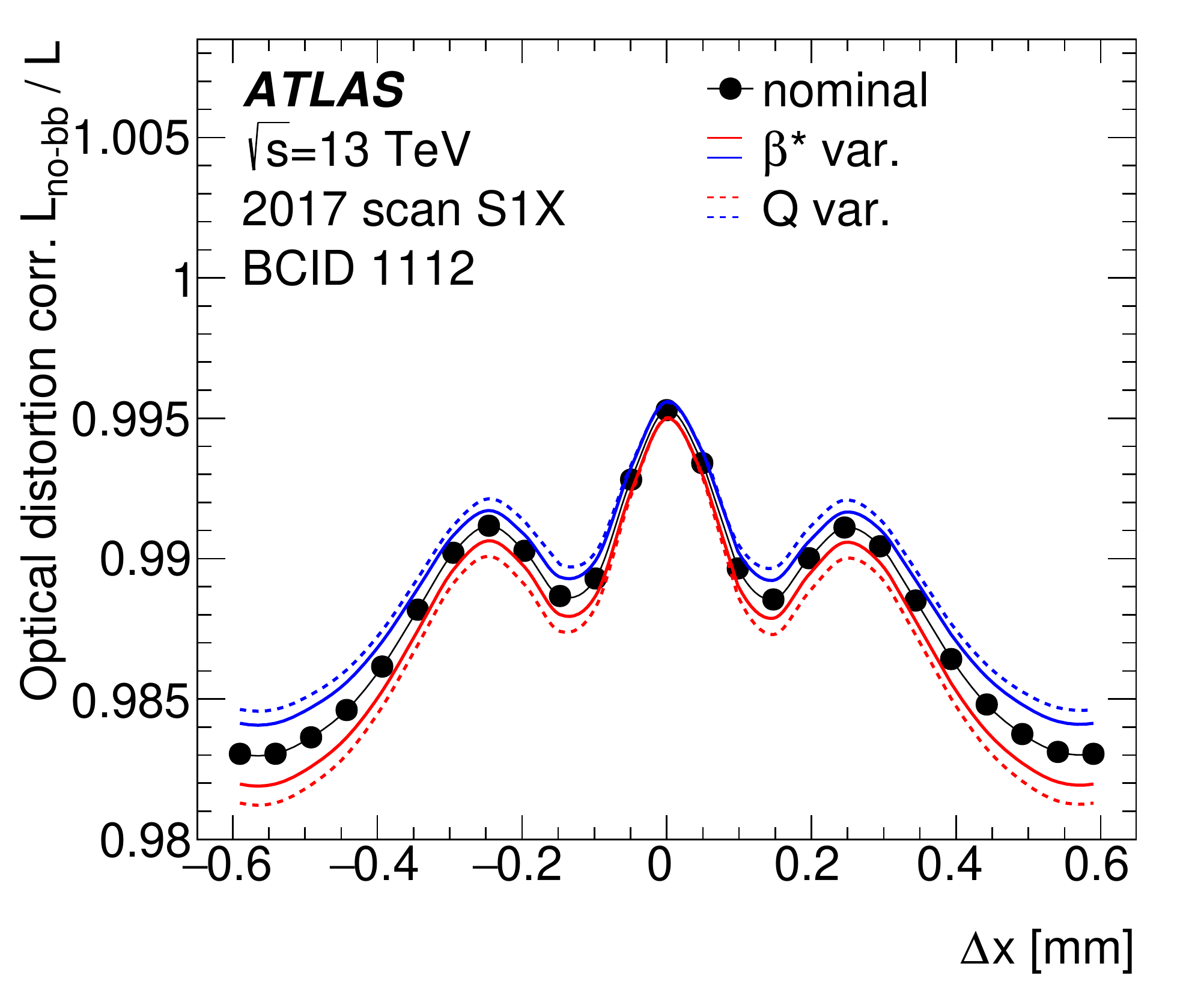}\vspace{-6mm}\center{(b)}}
\caption{\label{f:bbdef}(a) Calculated change in beam separation due to the
beam--beam deflection (orbit shift) and (b) optical-distortion correction, as
functions of the nominal beam separation in the horizontal plane, for bunch
slot (BCID) 1112 in the horizontal scan~1 of the 2017 vdM session. The nominal
values are shown by the black points and curve. The solid and dashed lines
illustrate the impact of varying the $\beta^*$ values by $\pm 10$\% or the
non-colliding tunes $Q$ by $\pm 0.002$, in each case for both beams and
both planes simultaneously.}
\end{figure}

The beam--beam deflection can be modelled as a kick produced by a small dipole
magnet, i.e.\ for a horizontal scan as an angular deflection $\theta_{x,i}$ that
results in a
horizontal beam displacement or orbit shift \dxbb{i}\ of each beam $i$ at the
IP. In the round-beam approximation, the deflection of a beam-1 bunch during a
horizontal on-axis scan is given by \cite{bambade}
\begin{equation}\label{e:bbang}
\theta_{x,1}=\frac{2r_{p} n_2}{\gamma_1\Delta x}\left[1-\exp(-\Delta x^2/(2\capsigx^2))\right];\ \theta_{x,2} \approx -\theta_{x,1}\ ,
\end{equation}
where $r_{p}=e^2/(4\pi\epsilon_{0}\,m_{p}\,c^2)$ is the classical proton radius,
$\gamma_1=E_1/m_{p}$ is the Lorentz factor of the beam-1 particles
with energy $E_1$ and mass $m_{p}$,
$n_2$ is the charge of the beam-2 bunch, $\Delta x$ is the nominal beam
separation produced by the steering corrector magnets, and \capsigx\ is the
transverse convolved beam size defined in Eq.~(\ref{e:capsig}). The induced
single-beam displacement at the interaction point is proportional to
$\theta_{x,i}$, and is given by
\begin{equation}\nonumber
\dxbb{i}=\theta_{x,i}\frac{\beta^*_x}{2\tan(\pi Q_x)};\
\dxbb{2}\approx -\dxbb{1} \ \ ,
\end{equation}
where $\beta^*_x$ is the value of the horizontal $\beta$-function at the IP and
$Q_x$ is the horizontal tune.\footnote{The tune of a storage ring is defined
as the betatron phase advance per turn, equivalent to the number of betatron
oscillations over one full ring circumference.}  Similar formulae apply to a
vertical scan. The beam--beam deflection at each scan point, and its impact on
the corresponding beam separation, can be calculated analytically from the
measured bunch currents and \capsigxy\ values using the Bassetti--Erskine
formula for elliptical beams~\cite{bassetti}. An example for one
colliding bunch pair in the 2017 scan S1X (i.e.\ the $x$-scan of S1)
is shown in Figure~\ref{f:bbdef}(a).
The change of separation is largest for $|\Delta x|\approx 200\,\mu$m, where it
amounts to almost $\pm 2\,\mu$m, and decreases to around $\pm 1\,\mu$m in the
tails of the scan.
Beam--beam deflections were taken into account on a bunch-by-bunch basis, by
correcting the nominal separation at each scan step by the calculated orbit
shifts for both beams
\begin{equation}\nonumber
\Delta x^\mathrm{corr}=\Delta x+\dxbb{1}-\dxbb{2}\ ,
\end{equation}
then refitting the vdM-scan curves and updating \capsigxy.
This orbit-shift correction
causes a 1.7--2.2\% increase in \sigmavis, depending on the bunch parameters.
 
In the Run-1 luminosity calibrations \cite{DAPR-2011-01,DAPR-2013-01}, the
defocusing effect was estimated using the MAD-X optics code \cite{madx},
giving a negative optical-distortion correction of 0.2--0.3\% in \sigmavis, and
thereby cancelling out a small fraction of the orbit-shift correction.
However, this calculation implicitly neglected the fact that the gradient of
the beam--beam force is different for the particles in the core of the bunch
than it is for those in the tails.
Recent studies~\cite{newbb} using two independent
multiparticle simulation codes, B*B \cite{bstarb} and COMBI \cite{combi},
have shown that the linear-field approximation used by MAD-X is inadequate,
and that the actual optical-distortion correction is larger than previously
considered.
 
Starting from the assumption of two initially round Gaussian proton bunches,
with equal populations ($n_1=n_2$) and sizes
($\sigma_{x,1}=\sigma_{x,2}=\sigma_{y,1}=\sigma_{y,2}$), that collide at a single
interaction point with zero crossing angle and the same $\beta$-function
($\beta^*_x = \beta^*_y = \beta^*$), Ref.~\cite{newbb} demonstrates that the
optical-distortion correction at each nominal beam separation $\Delta x$ can be
characterised by a reduction in luminosity $\lref/L$ parameterised as a function
of $\Delta x$, the non-colliding tunes\footnote{The non-colliding tunes refer
to the tunes in the absence of collisions at {\em any} of the four LHC IPs.}
$Q_{x,y}$ and the round-beam equivalent beam--beam parameter \xir:
\begin{equation}\nonumber
\frac{\lref}{L}\left(\Delta x\right)  =  f(\Delta x, Q_x, Q_y, \xir)\ ,
\end{equation}
with
\begin{equation}\label{e:xir}
\xir  =  \frac{r_p\bar{n}\beta^*}{2\pi\gamma\capsigx\capsigy} \ .
\end{equation}
At each given nominal separation $\Delta x$, $\lref$ is the luminosity in the
absence of (or equivalently, after correction for) optical-distortion effects,
$L$ is the uncorrected (i.e.\ measured) luminosity,
$\bar{n} = (n_1 + n_2)/2$ is the average number of protons per bunch,
$r_p$ is the common classical particle radius, and $\gamma$ is the common
Lorentz factor. For the Run~2 \sxyt\ vdM scans, the parameter \xir\ varies from
$3\times 10^{-3}$ to $4\times 10^{-3}$. Reference~\cite{newbb} provides a
polynomial parameterisation of the correction $\lref/L$ obtained from the full
B*B and COMBI simulations, valid for the range of beam parameters typical
of LHC vdM scans.
 
An additional complication arises from the effects of collisions at
`witness IPs', i.e.\ at interaction points other than that of ATLAS.
Studies with COMBI simulations \cite{newbb}
showed that the effect on the beam--beam corrections
can be adequately modelled by an ad-hoc shift $\Delta Q_{x,y}$ in the
non-colliding tune values that are input to the single-IP parameterisation
\begin{equation*}
\Delta Q_{x,y}= p_1(\nwit)\,\xir\ ,
\end{equation*}
that accounts for the average beam--beam tune shift induced by collisions at
interaction points other than the one where the scan is taking place.
Here, \nwit\ is the number of witness IP collisions for the colliding bunch
pair in question, incremented by one half for each additional collision that
each separate bunch experiences around the LHC ring, and $p_1(\nwit)$ is
a first-order polynomial function of \nwit.
Since all bunches that collide in ATLAS also collide in CMS, $\nwit\geq1$.
Depending on the LHC filling pattern, some bunches also experience collisions
in either LHCb or ALICE, leading to some bunch pairs with
$\nwit=1.5$ or $\nwit=2$, especially in 2018.
 
\begin{figure}[tp]
\centering
 
\includegraphics[width=90mm]{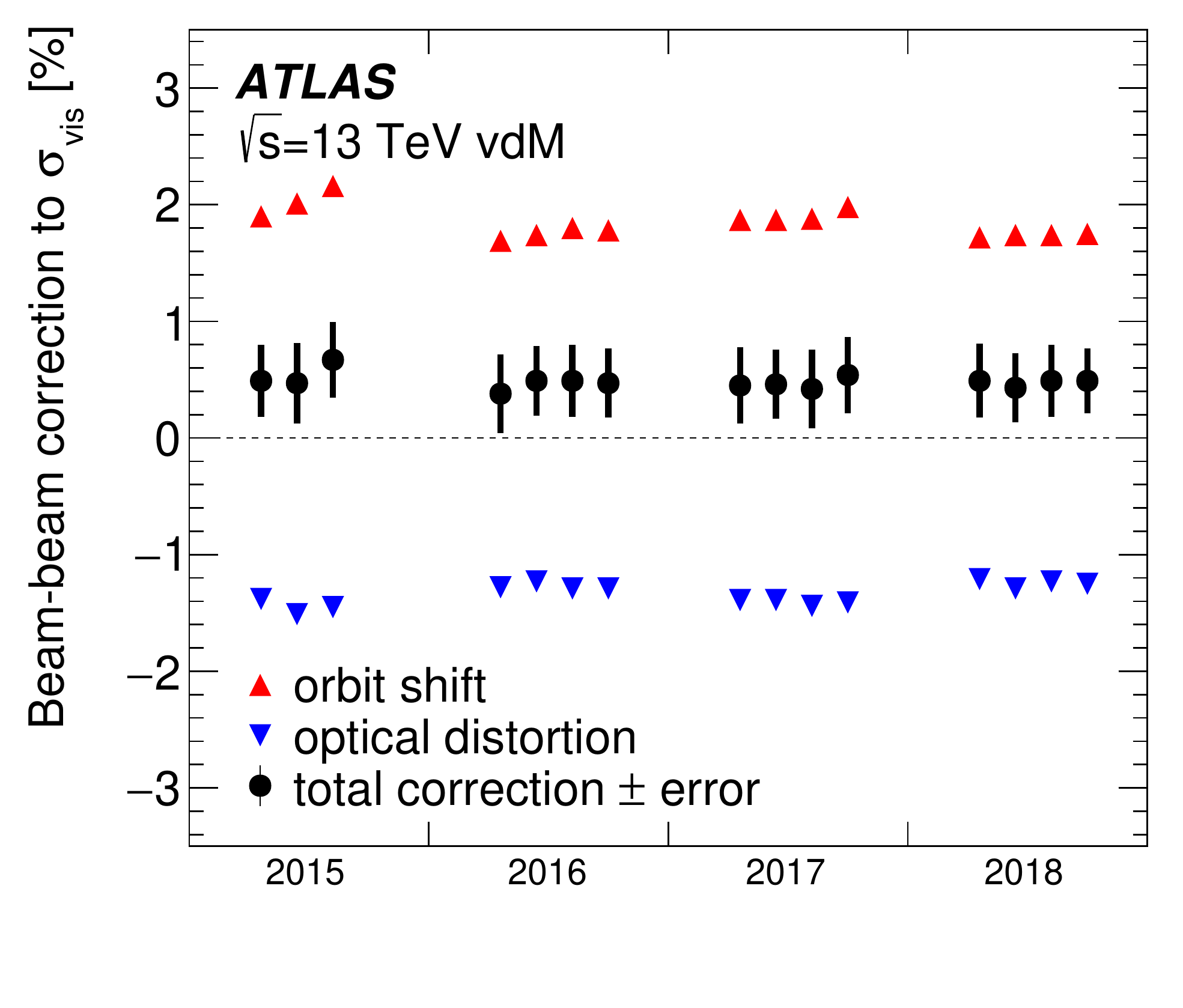}
\caption{\label{f:bbcorr}Bunch-averaged corrections to \sigmavis\ due to
the orbit shift and optical distortion caused by the mutual electromagnetic
interaction of the two beams, calculated for all Run~2 \sxyt\ on-axis vdM scans,
showing the two effects separately, and their total. The error bars show
the uncertainties of the total correction.}
\end{figure}

The resulting optical-distortion correction for one particular colliding bunch
pair in the 2017 scan S1X is shown in Figure~\ref{f:bbdef}(b); the luminosity
is reduced by 0.5\% at the peak and almost 2\% in the tails, depending slightly
on the $\beta^*$ and tune values. This correction was applied to the measured
luminosities (and hence \mubvis\ values) at each scan point, reducing
\sigmavis\ by around 1.5\%.
The values of the orbit-shift, optical-distortion and total beam--beam
corrections, averaged over all colliding bunch
pairs, are shown for each scan in each year in Figure~\ref{f:bbcorr}.
The revised optical-distortion correction cancels out a much larger fraction of
the orbit-shift correction than predicted by the original linear treatment.
 
The systematic uncertainty affecting the magnitude of the beam--beam
corrections~\cite{newbb} is dominated by the tunes.
An uncertainty of 0.002 is assigned to the non-colliding tunes, correlated
between beams and planes; an additional  uncertainty of 0.001 in the mean tunes
input to the parameterisation accounts for the approximate modelling of the
effect of witness-IP collisions on the unperturbed-tune\footnote{The unperturbed
tunes are the values of the horizontal and vertical tunes with the beam--beam
interaction `switched off' at the IP where the scans are taking place
(whether or not the beams are colliding at other IPs). Physically, these
correspond to the mean tune values that would be measured, for the bunch pair
under study, before the beams are brought into collision at this IP.} spectra.
The uncertainty in the actual value of $\beta^*$ (which is difficult to measure
precisely) is conservatively taken as  10\%, uncorrelated between the two beams
but correlated between $x$ and $y$; it directly translates into an uncertainty
in the value of \xir\ defined in Eq.~(\ref{e:xir}). A number of additional
uncertainties, detailed below,
arise due to the assumptions made in the evaluation of the
combination of the orbit-shift and optical-distortion corrections \cite{newbb}.
An uncertainty of 0.1\% results from
potential non-Gaussian tails in the initial
unperturbed transverse beam profiles. Other effects, such as the departure from
round beams (i.e. $\capsigx\neq\capsigy$), unequal transverse beam sizes, limitations in the parameterisation of optical distortions, uncertainties in the
modelling of witness-IP effects,
magnetic lattice non-linearity, and the effects of a residual beam
crossing angle, each lie below 0.1\%.
Most of the uncertainties are either correlated
or anticorrelated between the orbit-shift and optical-distortion corrections;
their effects were therefore evaluated taking this into account.
The total beam--beam uncertainty in \sigmavis\ in
Table~\ref{t:unc} is less than 0.3\% for all years.

\subsection{Emittance growth corrections}\label{ss:emit}
 
The determination of \sigmavis\ from Eq.~(\ref{e:sigcalav}) implicitly assumes
that the convolved beam sizes \capsigxy\ (and therefore the transverse
emittances \cite{lumibib} of the two beams)
remain constant both during a single $x$ or
$y$ scan and between the peaks of the two scans. The result
may be biased if the beam sizes (and hence also the peak interaction
rates) change during or between $x$ and $y$ scans.
 
The evolution of the emittances during the vdM scan sessions was studied by
comparing the values of \capsigxy\ obtained from the different on-axis scans
in each vdM fill. These studies showed that \capsigx\ generally
increases through the fill at rates of 0.2--0.6\,$\mu$m/hour, whereas
\capsigy\ decreases at around 0.8--1.3\,$\mu$m/hour, due to
synchrotron-radiation damping \cite{weidermann}.
The peak interaction rates (after correcting for out-of-plane
orbit drifts as discussed in Section~\ref{ss:odc}) also increase with
time, consistent with the larger change of emittance in the $y$ plane
(as a decrease in emittance should correspond to an increase in peak rate).
The effect of emittance growth on the estimate of \capsigxy\ from
a single $x$ or $y$ scan was studied by fitting simulated scan curves,
described by Gaussian distributions whose widths changed during the scan
according to the observed rates. The largest bias was found to be 0.03\%,
with typical values being an order of magnitude smaller, and this effect was
thus neglected.
 
Much larger potential biases in \sigmavis\ arise from
the evaluation of the $x$- and $y$-related
quantities in Eq.~(\ref{e:sigcalav}) at different times. Making
the time-dependence explicit, Eq.~(\ref{e:sigcalav}) can be rewritten as
\begin{equation*}
\sigmavis(t_x,t_y)=\frac{1}{2}\left( \mubvismaxi{x}(t_x)+\mubvismaxi{y}(t_y) \right)\cdot 2\pi\capsigx(t_x)\capsigy(t_y)\ ,
\end{equation*}
where the $x$-related quantities are evaluated at time $t_x$ and the $y$-related
quantities at time $t_y$. These quantities can be translated to a common
time $\tmid=(t^0_x+t^0_y)/2$ midway between the peaks of the $x$ and $y$ scans
at $t^0_x$ and $t^0_y$, using
linear fits of the evolution of \capsigxy\ and peak rates with time
(this procedure is only possible for
fills which have more than one on-axis scan). The emittance growth correction
\cee\ can then be defined as
\begin{equation}\nonumber
\cee = \frac{\sigmavis(\tmid,\tmid)}{\sigmavis(t^0_x,t^0_y)}-1 \ ,\\ \nonumber
\end{equation}
correcting the measured visible cross-section for the bias caused by evaluating
the $x$- and $y$-scan quantities at different times:
$\sigviscorr=(1+\cee)\,\sigmavis$. Since the visible
cross-section should not depend on the choice of \tmid, \cee\ was also
evaluated by evolving all quantities to $t=t^0_x$ or $t=t^0_y$, taking the
largest difference as the uncertainty in \cee\ due to the linear evolution
model. This uncertainty also accounts for the small inconsistencies observed
between the evolution of the measured \mubvismax\ and that predicted from the
measured evolution of \capsigxy.
 
\begin{figure}[tp]
\centering
 
\includegraphics[width=90mm]{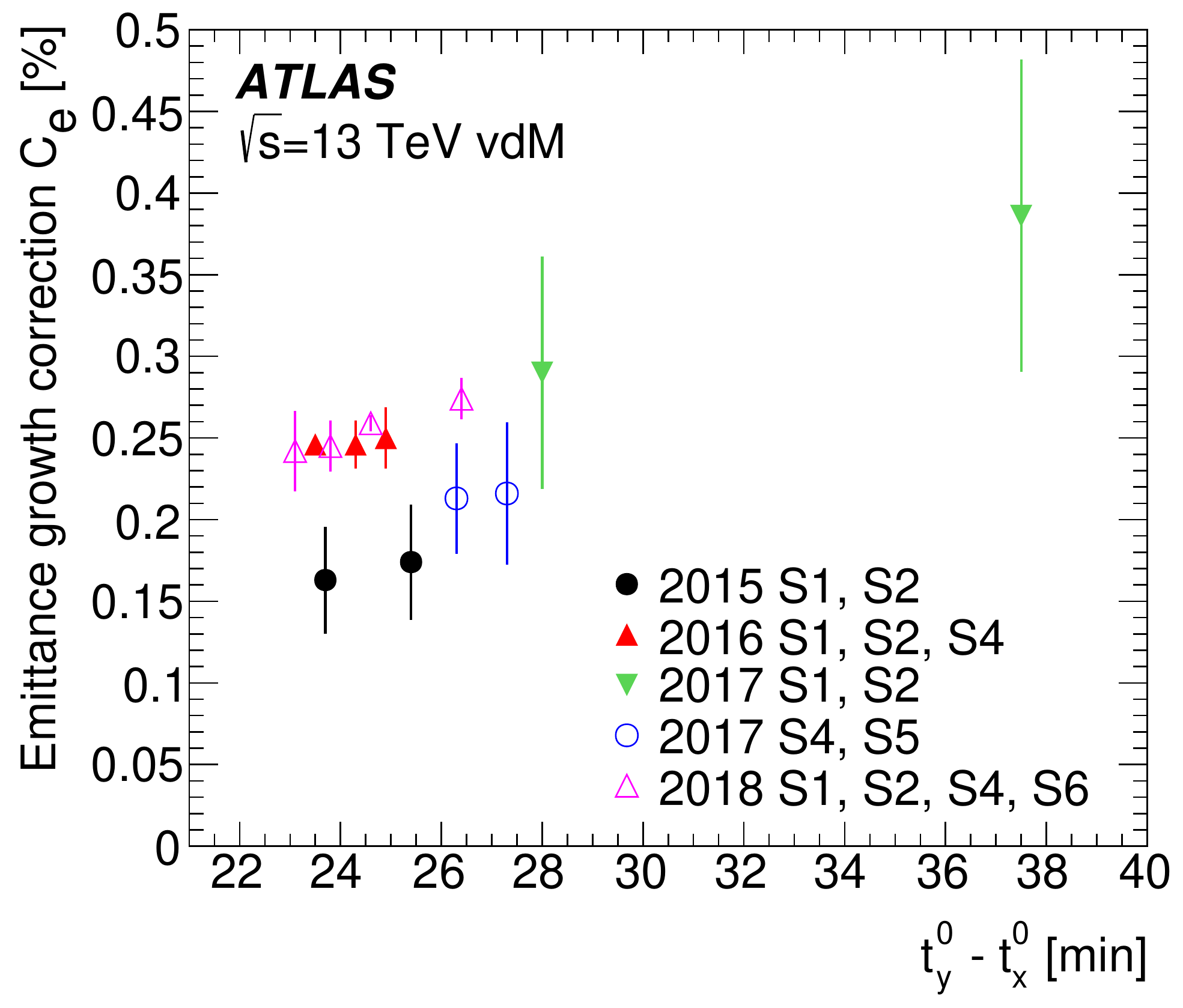}
\caption{\label{f:egrowth}Emittance growth correction \cee\ for each vdM scan
in fills with more than one scan, shown as a function of the elapsed time
between the peaks of the $x$ and $y$ scans. The corrections
were evaluated by determining the rate of change
of emittances during the sequences of scans given in the legend, dividing
the 2017 scans into two groups. The error bars show the total uncertainties,
including the statistical uncertainties from the linear fits and the systematic
uncertainty due to the mismatch between the \capsigxy\ and peak-rate evolution.}
\end{figure}
 
The corrections for each scan recorded in a fill with more than one
on-axis scan are shown as a function of the elapsed time between
the $x$ and $y$ scan peaks in Figure~\ref{f:egrowth}, together with the
systematic uncertainty from the fit model combined with the statistical
uncertainties from the linear fits. The corrections increase \sigmavis,
typically by 0.15--0.3\%, except for scan S1 in 2017, where there was a delay
between the $x$ and $y$ scans, and hence a larger correction of 0.39\%. Although
all four 2017 scans were performed in the same fill, there was a six-hour
gap between S1+S2 and S4+S5. The two pairs of scans have different
emittance growth rates, so they were analysed separately. Scans S4 in 2015 and
S6 in 2016 were performed in separate LHC fills (see Table~\ref{t:vdmds})
with only a single on-axis scan, so they were assigned the average correction
for other scans in that year. In all cases, the corrections were calculated
using the bunch-averaged \capsigxy\ and \mubvismax, but a bunch-by-bunch
analysis showed very similar results, the emittance growing at similar rates
in all bunches. The final uncertainties due to emittance growth corrections
shown in Table~\ref{t:unc} are smaller than 0.1\% for all years.
 
\subsection{Length scale determination}\label{ss:lsc}
 
\begin{figure}[tp]
\parbox{83mm}{\includegraphics[width=76mm]{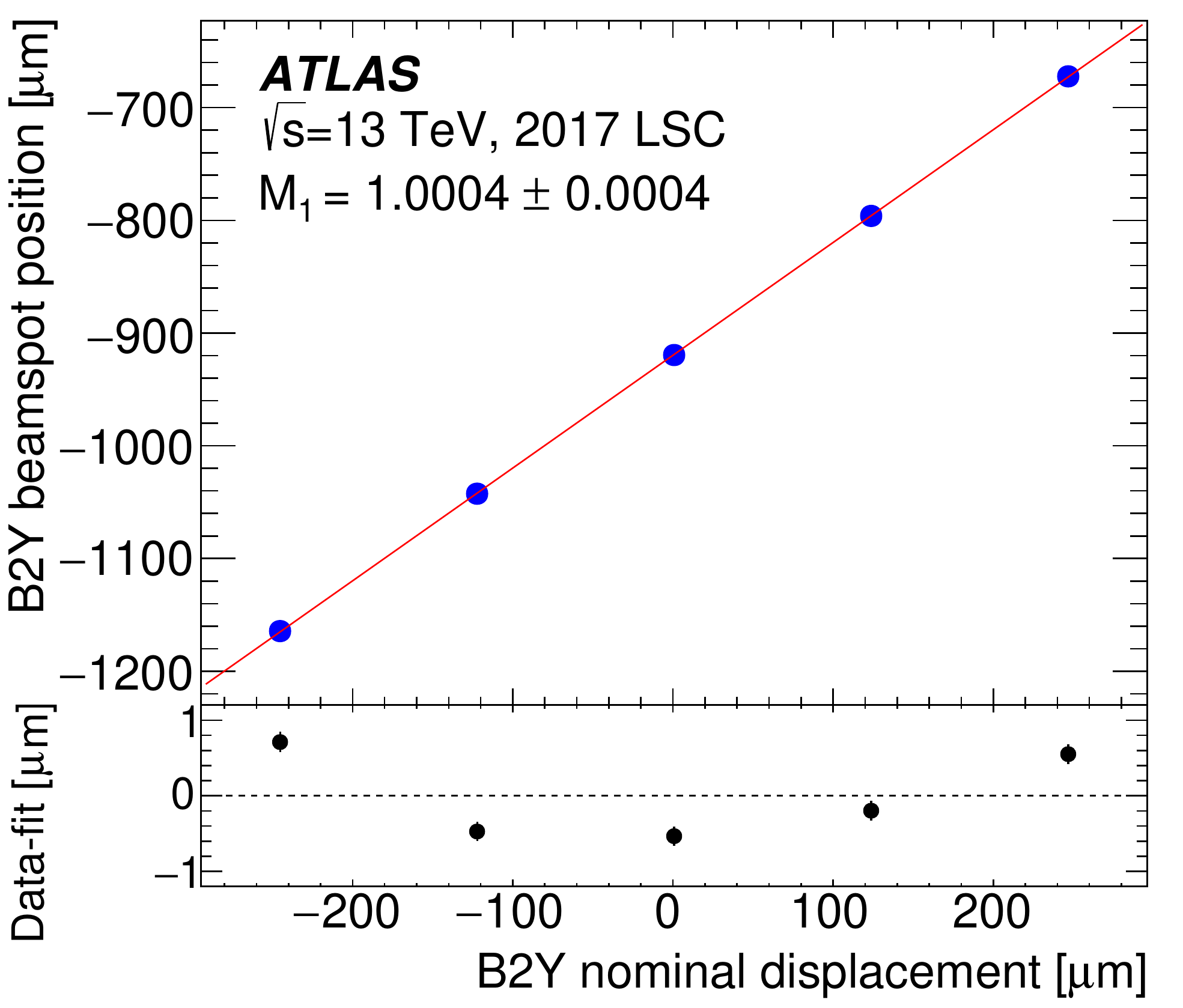}\vspace{-6mm}\center{(a)}}
\parbox{83mm}{\includegraphics[width=76mm]{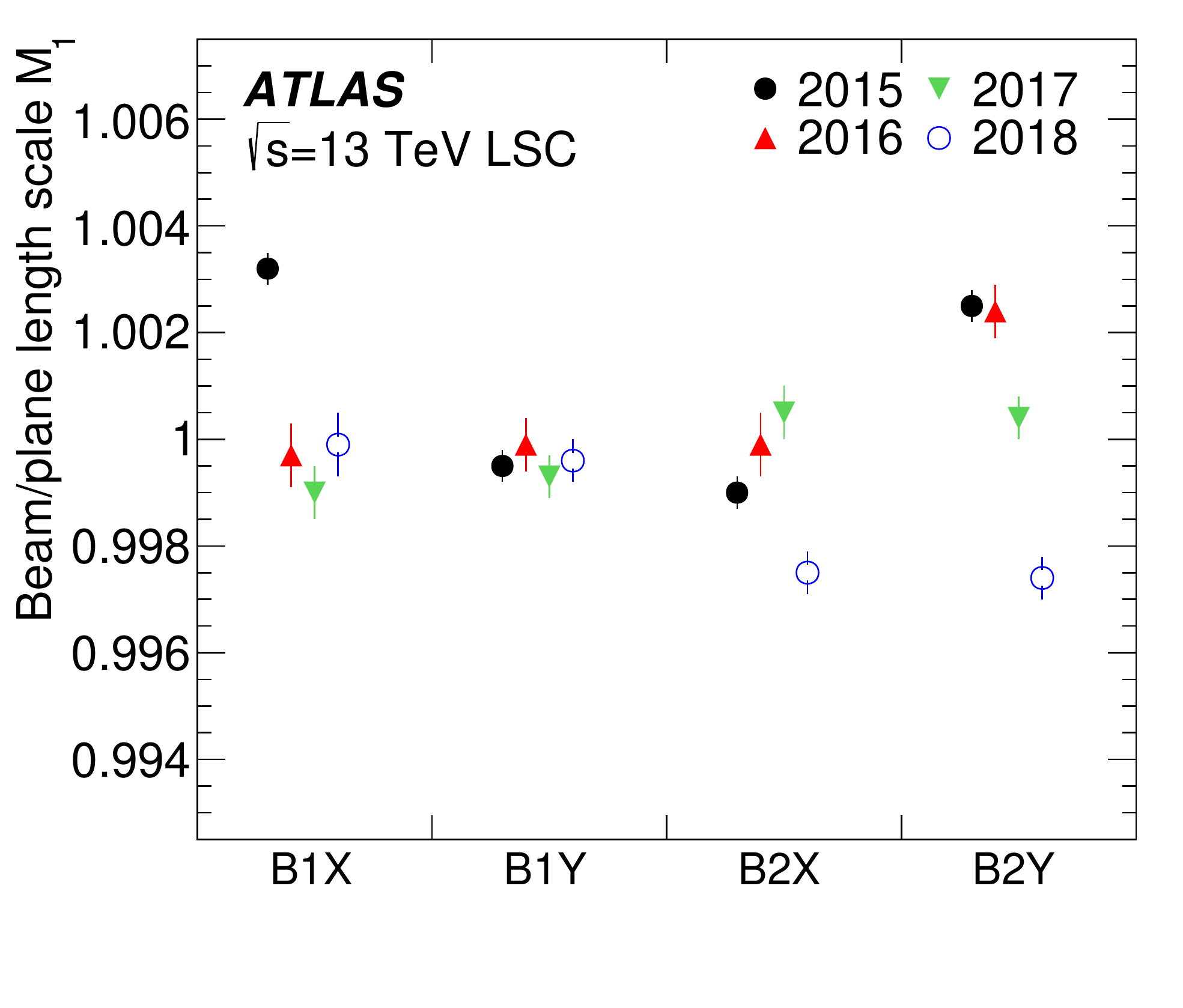}\vspace{-6mm}\center{(b)}}
\caption{\label{f:lsclin}(a) Length scale calibration for the $y$ direction
of beam~2 in the 2017 measurement, showing the beamspot position vs. the nominal
displacement of the target beam,
together with a linear fit giving the slope parameter $M_1$. The
lower plot shows the fit residuals, with the error bars indicating the
statistical uncertainties of the beamspot positions. These residuals
are discussed further in Section~\ref{ss:lscmag}. (b) Measurements of the
linear length scale $M_1$ for each beam and plane measured in the
calibration for each year of data-taking.}
\end{figure}

The \capsigxy\ measurements require knowledge of the length scale, i.e.\
the actual beam displacements (and hence beam separation) produced by
the settings of the LHC steering magnets intended to produce a given nominal
displacement. This was determined using length scale calibration (LSC) scans,
separately for each beam and plane ($x$ and $y$). In each scan, the target
beam being calibrated was moved successively to five equally spaced positions
within $\pm 3\signom$, and its position measured using the
luminous centroid (or beamspot) position fitted from primary collision vertices
reconstructed
in the ATLAS inner detector when the two beams were in head-on collision. Since
the alignment of the inner detector is very well understood
\cite{IDTR-2019-05}, the beamspot
gives a measurement of the beam displacement that is accurate at the 0.1\%
level over the scanning range. In practice, the requirement for the two
beams to be in head-on collision was satisfied by performing a three-point
miniscan of the non-target beam around the target beam nominal position at
each step, fitting the resulting curve of luminosity vs.\ beamspot position,
and interpolating the beamspot position to that of maximum luminosity and beam
overlap.
As shown in Figure~\ref{f:lsclin}(a), the slope of a linear fit of the beamspot
position vs. the nominal position gives the linear length scale $M_1$, i.e.\ the
ratio of actual to nominal beam movement for each beam and plane (the offset of
this fit depends on the relation between the LHC beam and ATLAS coordinate
systems). Orbit drifts can influence the measured
values of $M_1$, and were corrected using arc BPM data, interpolating
the difference in beam positions measured directly before and after each
scan linearly in time through the scan, in a procedure similar to that
described for the vdM scans in Section~\ref{ss:odc}.
 
The four length scales for B1X (i.e.\ beam~1 in the $x$-direction),
B1Y, B2X and B2Y were measured in turn in the LSC
scans for each year, which were always performed with the same machine
optics as, and close in time to, the vdM scans. The resulting values of $M_1$
for all four quantities are shown in Figure~\ref{f:lsclin}(b), and are always
within $\pm 0.4$\% of unity. However, apart from B1Y, the length scales
differ significantly between years, despite the same nominal machine
configuration being used. In 2015--17, all four scans were performed from
negative to positive displacement, meaning that beam~1 moved in the same
direction as in the vdM scans, but beam~2 moved in the opposite direction.
In 2018, the beam-2 LSC scans were performed from positive to negative
displacement, so both beams were calibrated by moving them in the same
direction as in the vdM scans.
 
The measured \capsigxy\ values from
Eq.~(\ref{e:capsig}) must be corrected by the average of the beam-1 and
beam-2 length scales $M_1$ in each plane, leading to a correction \lsp\ to the
visible cross-section $\sigviscorr=\lsp\sigmavis$, where the length scale
product \lsp\ is given by
\begin{equation*}
\lsp=(M_1^{x,1}+M_1^{x,2})(M_1^{y,1}+M_1^{y,2})/4 \ ,
\end{equation*}
and $M_1^{x|y,i}$ is the measured linear length scale for the $x$ or $y$ plane
and beam $i$. The statistical uncertainty from the beamspot position
measurements is shown as the `Length scale calibration' uncertainty in
Table~\ref{t:unc}, and is below 0.1\%. A further uncertainty of 0.12\%
arises from uncertainties in the absolute length scale of the inner detector,
determined from simulation studies of various realistic misalignment scenarios
as described in Ref.~\cite{DAPR-2011-01}. The studies were updated to use
scenarios appropriate to the Run~2 detector, including the impact of the
precise measurements from the innermost pixel layer, the IBL.

\subsection{Magnetic non-linearity}\label{ss:lscmag}
 
\begin{figure}[tp]
\parbox{83mm}{\includegraphics[width=76mm]{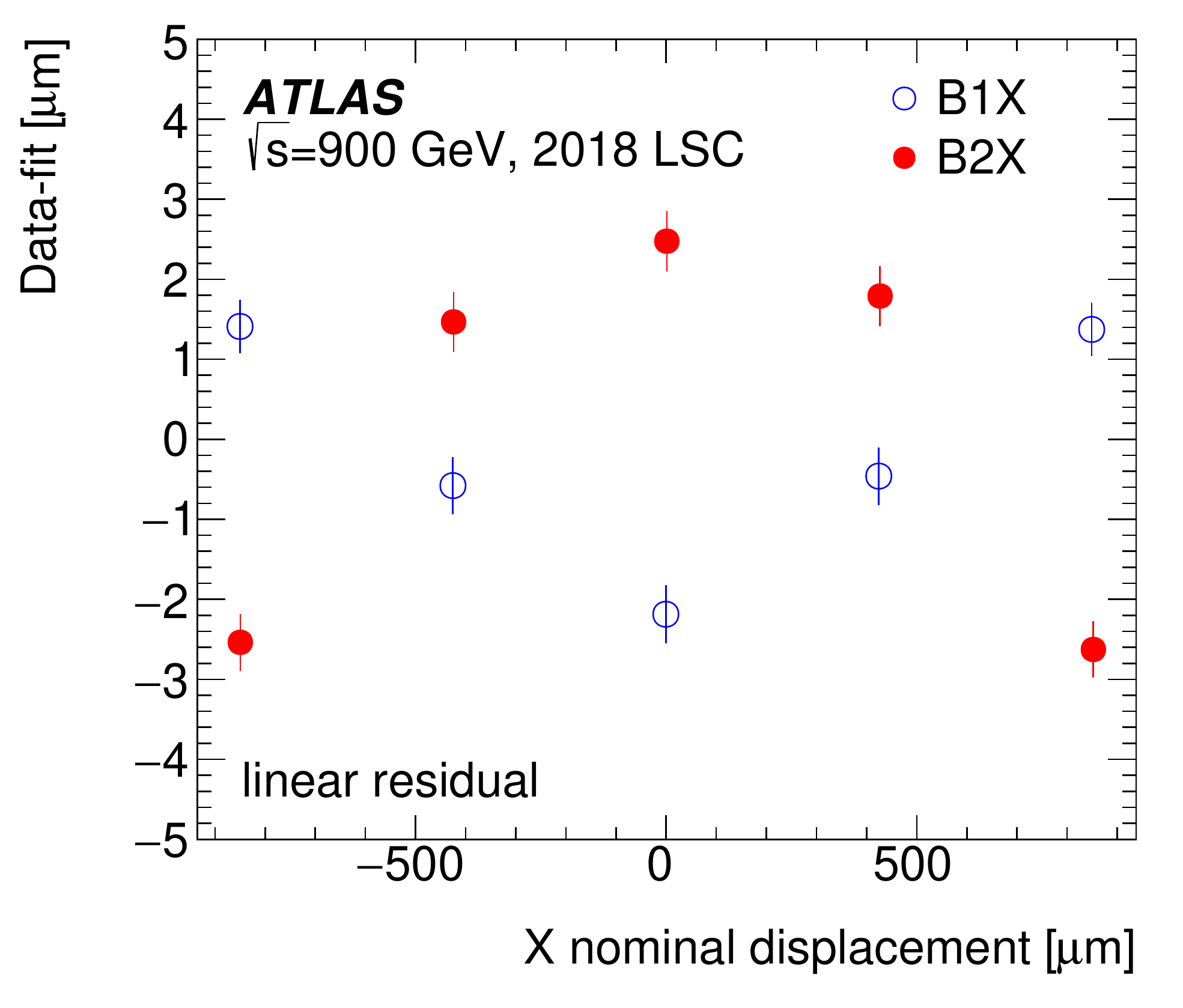}\vspace{-6mm}\center{(a)}}
\parbox{83mm}{\includegraphics[width=76mm]{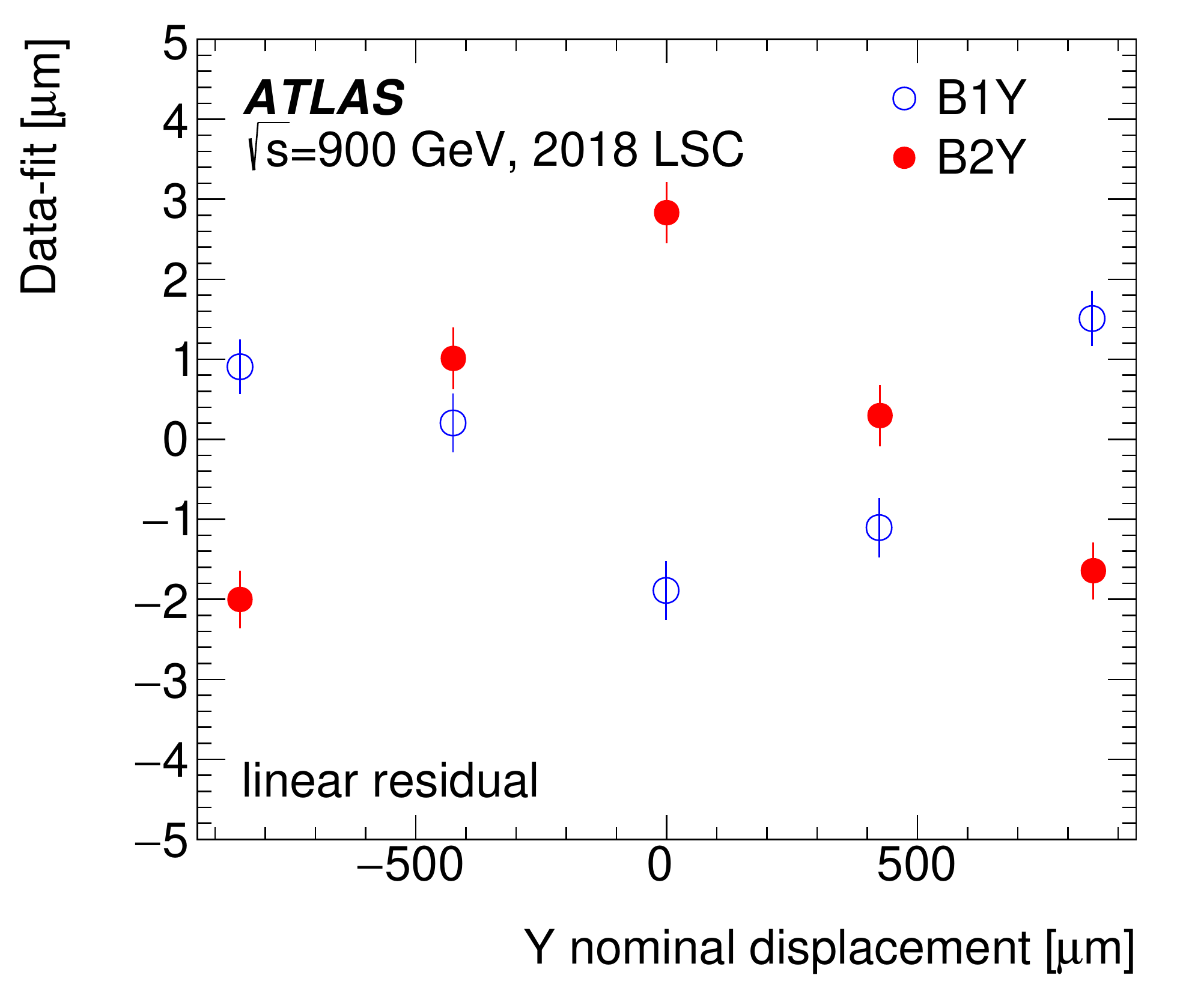}\vspace{-6mm}\center{(b)}}
\caption{\label{f:lscres900}Residuals of the beamspot position with respect
to a linear fit of the beamspot position vs. nominal displacement of the target
beam in the length scale
calibration of the 2018 \sxit\ vdM session, for (a) the $x$ and (b) the $y$
plane, for beam~1 (blue open points) and beam~2 (red filled points).
The error bars show the statistical uncertainties from the beamspot
position measurements.}
\end{figure}

The fit residuals shown in the lower panel of Figure~\ref{f:lsclin}(a) suggest
that, in this particular scan, the actual beam movement is not perfectly
linear with respect to the
nominal position, at the level of $1\,\mu$m in $\pm 300\,\mu$m of displacement.
The residuals from some of the other \sxyt\ LSC scans show similar hints of
systematic non-linear behaviour; some are more scattered, and some appear
consistent with a purely linear dependence. The \sxvt\ LSC data from 2012
are also suggestive of non-linear behaviour \cite{DAPR-2013-01}.
More recent data are available to study these effects further.
For five days in October~2018,
the LHC delivered $pp$ collisions at \sxit\ in support of the forward-physics
programme. As part of this run, a vdM scan session, including a length
scale calibration, took place with $\beta^*=11$\,m, a configuration
in which the transverse beam sizes (and hence scan ranges) are around three
times larger than in the vdM optics used at \sxyt.
As well as the length scale calibration,
a total of seven $x$--$y$ vdM scans were performed over four fills spanning a
three-week period, as shown in the top part of Table~\ref{t:vdm900}. The
residuals from the length scale calibration are shown in
Figure~\ref{f:lscres900}, and show a clear deviation from linearity,
of up to $\pm 3\,\mu$m over a $\pm 900\,\mu$m scan range. The shape of
the residual curve is approximately inverted for beam~2 compared to beam~1
in both planes. As the two beams were moved in opposite directions during the
LSC, this suggests the non-linearity may come from hysteresis effects in
the steering corrector magnets.
 
\begin{table}[tp]
\caption{\label{t:vdm900}Summary of the \sxit\ vdM scan fills used in the study
of magnetic non-linearity, showing the dates, LHC fill numbers, total numbers of
bunches in each beam (\nbeam) and colliding in ATLAS (\nbun), and the individual
scans performed in each fill. In the first group of scans performed in 2018,
each scan consisted of an $x$--$y$ vdM scan pair (S1, S3 etc.), an off-axis scan
(S2-off) or a length scale calibration scan (LSC). The second group
of scans was performed during the LHC beam test in October 2021, and individual
$x$ or $y$ scans are listed separately, as parallel (par), separation (sep)
or single-beam (B1 or B2) scans.}
\centering
 
\begin{tabular}{lc|cc|l}\hline
Date & Fill & \nbeam & \nbun & Scans \\
\hline
14/10/2018 & 7299 & 152 & 150 & S1, S2-off, S3 \\
14/10/2018 & 7300 & 152 & 150 & LSC, S4 \\
5/11/2018 & 7406 & 152 & 150 & S5 \\
5/11/2018 & 7407 & 152 & 150 & S6, S7-off \\
\hline
28/10/2021 & 7516 & 4 & 2 & S1xsep, S2xpar, S3ypar, S4ysep, S5ypar, S6xB1, S7xB2, S8ypar \\
30/10/2021 & 7524 & 3 & 0 & S1xpar, S2xpar, S3xsep, S4xpar, S5ypar, S6ypar, S7ysep, S8ypar \\
30/10/2021 & 7525 & 3 & 3 & S1xpar, S2xsep, S3ypar, S4ysep, S5yB1, S6yb2, S6xB1, S7xB2 \\
\hline
\end{tabular}
\end{table}
 
The beamspot positions measured during the \sxit\ LSC scans provide
unambiguous evidence of non-linearity, but have limited granularity
(only five points per scan) and do not address reproducibility, as only
one scan was performed per beam/plane. The multiple vdM scans per session
have 25~points per scan and so can potentially address these issues, but since
the beams move in opposite directions, the beamspot position remains
approximately stationary, only moving slightly if the two beams are of
unequal sizes. The DOROS BPMs measure the displacements of each beam separately
during vdM scans, and can be used to study magnetic non-linearity effects.
However, they suffer from short-term variations in their calibration,
and must also be corrected for the effects of beam--beam deflections.
 
\begin{figure}[tp]
\parbox{83mm}{\includegraphics[width=76mm]{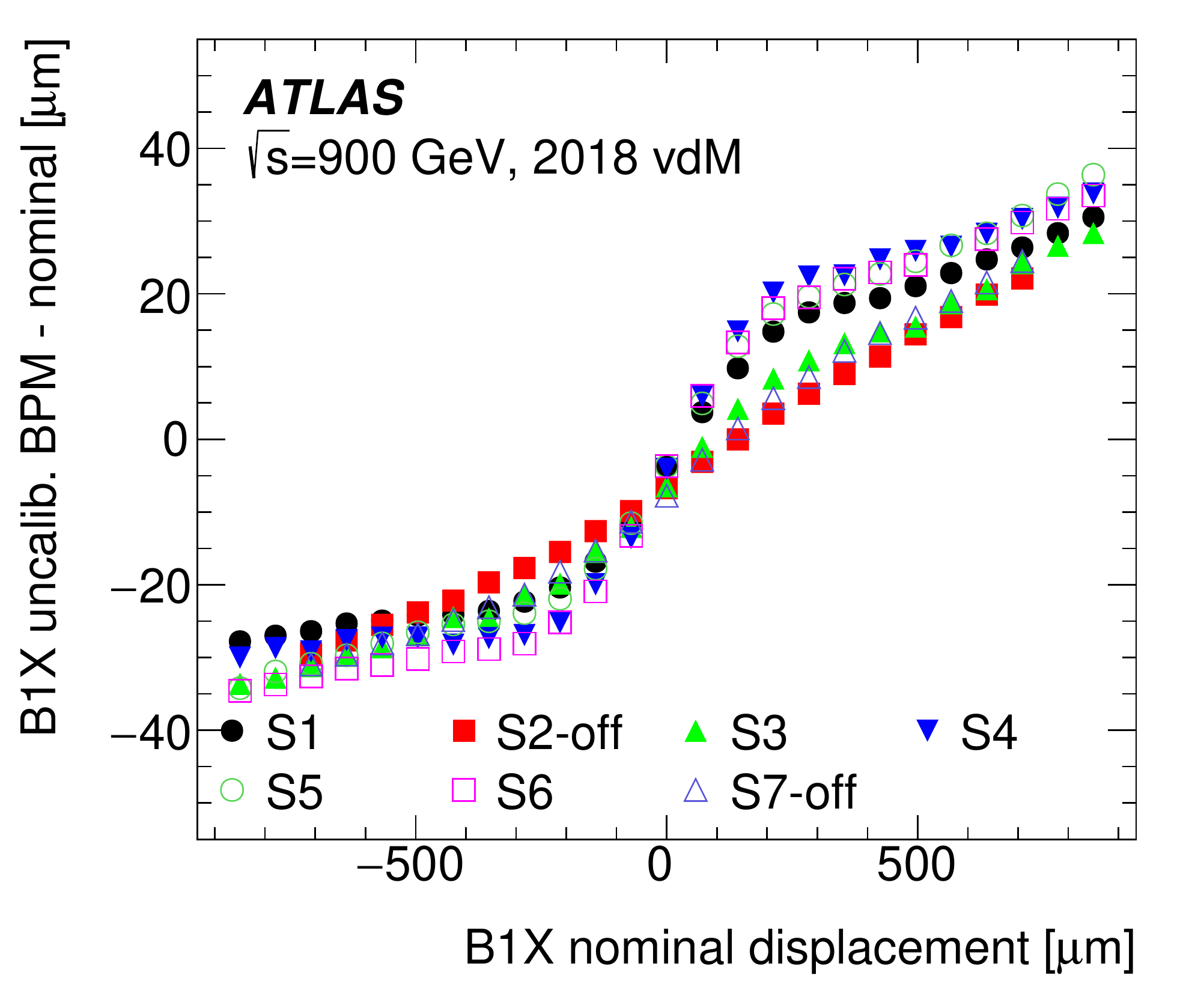}\vspace{-6mm}\center{(a)}}
\parbox{83mm}{\includegraphics[width=76mm]{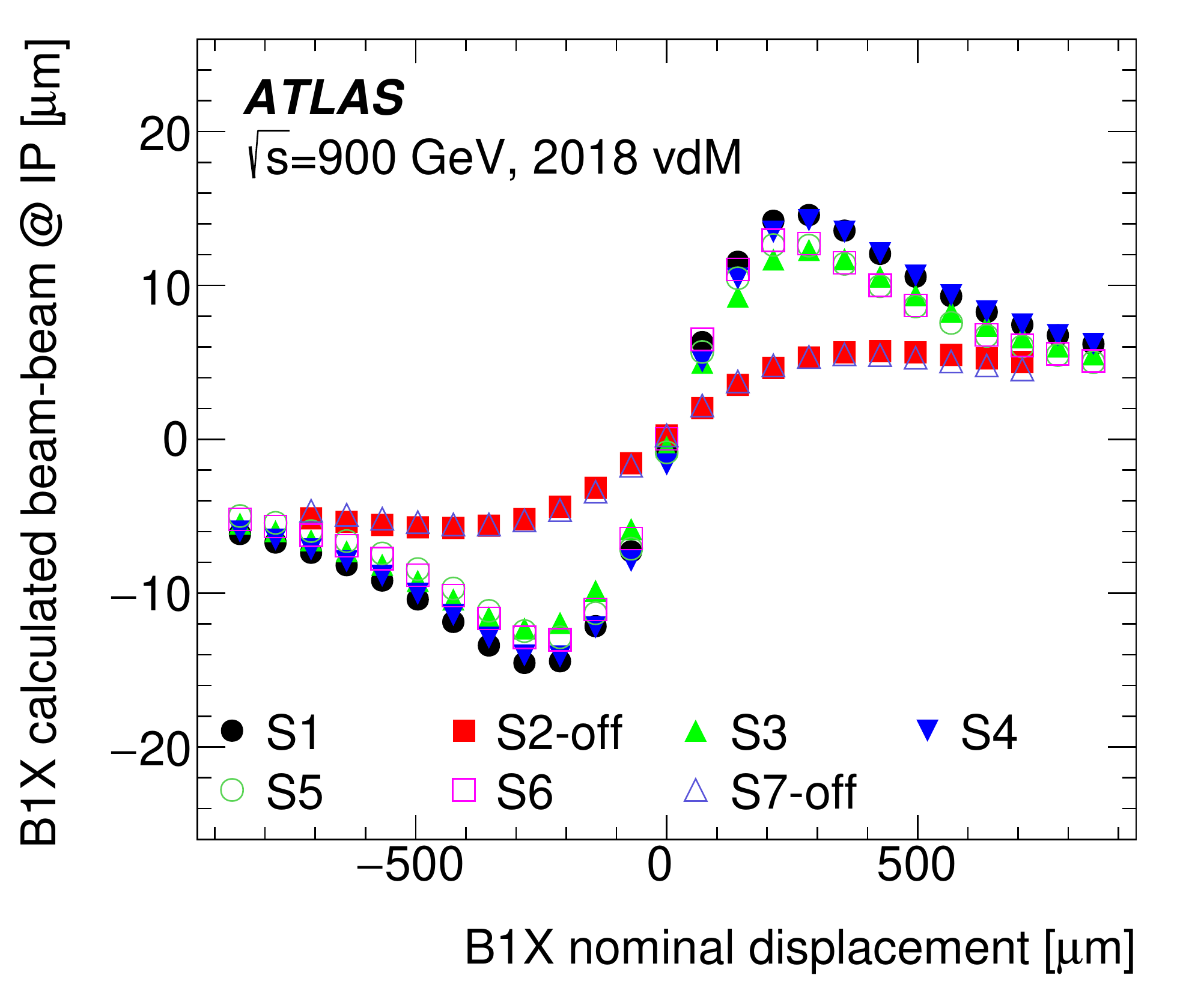}\vspace{-6mm}\center{(b)}}
\caption{\label{f:vdMraw900}Details of the analysis of beam-1 DOROS BPM
residuals in the $x$ scans
of the 2018 \sxit\ vdM scan session: (a) residuals between the uncalibrated
BPM measurements and the nominal beam displacement as a function of the
nominal displacement; (b) apparent beam--beam-induced displacement at the IP as
seen by the DOROS BPMs, calculated using the measured beam parameters.}
\end{figure}

The differences between the beam position at the interaction point determined
from the uncalibrated DOROS BPM measurements, and the nominal
beam-1 displacements, are shown  during the \sxit\ $x$ scans from~2018 in
Figure~\ref{f:vdMraw900}(a). The residuals are shown after subtracting an
offset from the BPM
such that the residual with the beam at the nominal head-on position,
immediately before the start of each scan, is zero. A significant slope is
visible, indicating a miscalibration of the length scale of the BPM.
The residuals are slightly negative at the nominal central point of the scan,
suggesting magnetic hysteresis. The curves also have an `S' shape,
due to the beam--beam deflection already discussed in
Section~\ref{ss:bbcorr}. As well as the true beam displacement at the
interaction point, the DOROS BPMs also measure an additional apparent
displacement from half the angular kick
$\theta_{x,i}$ projected over the distance $L_\mathrm{doros}=21.7$\,m between
the IP and the BPM.\footnote{The factor of one-half arises because in the new
closed orbit, both the incoming and outgoing beams are deviated from their initial trajectories by $\theta_{x,1}/2$, giving a total angular deflection of
$\theta_{x,1}$ at the IP.}  With $\theta_{x,i}$ given by
Eq.~(\ref{e:bbang}), the apparent displacement \dxbbd{i} of beam~$i$ at the IP
as measured by the DOROS BPMs is
\begin{equation}\label{e:bbdeld}
\dxbbd{1}=\theta_{x,1}\left(\frac{\beta^*}{2\tan(\pi Q_x)}
+\frac{L_\mathrm{doros}}{2}\right);\ \dxbbd{2}=-\dxbbd{1} \ .
\end{equation}
The expected apparent horizontal beam--beam displacements for beam~1 in
these scans, calculated
using the measured \capsigxy\ and beam currents, are shown in
Figure~\ref{f:vdMraw900}(b) and reach $\pm 15\,\mu$m at $\pm 250\,\mu$m
single-beam displacement (corresponding to a separation $\Delta x=500\,\mu$m).
The displacements for the off-axis scans S2 and S7 are smaller, and were
calculated using a generalisation of Eq.~(\ref{e:bbang}) that also takes into
account the constant beam separation in the non-scanning plane.
 
The miscalibration and offset of the DOROS BPM measurement visible in
Figure~\ref{f:vdMraw900}(a) can be corrected by replacing the initial
DOROS estimate  of the beam position at the IP \doros{x}{i}\ by
$\dcal{x}{i}\doros{x}{i}-\dofs{x}{i}$, where $\dcal{x}{i}\approx 1$
is the DOROS length scale and $\dofs{x}{i}$ the offset at the central scan
point.  Using the calibrated
DOROS measurement, together with the nominal beam displacement \setpos{x}{i}\
scaled by its length scale $M_{1,i}$ and the estimate of the apparent
beam--beam deflection \dxbbd{i}\ from Eq.~(\ref{e:bbdeld}) scaled by a parameter
\bbds, the residual between the calibrated DOROS measurement and the
nominal beam displacement for beam $i$ at a particular scan point is given
by
\begin{equation}\label{e:dresid}
\resid{doros}{x}{i}=\dcal{x}{i}\doros{x}{i}-\dofs{x}{i} -M_{1,i}\setpos{x}{i}-\bbds\,\dxbbd{i} \ .
\end{equation}
These residuals were used to define a $\chi^2$ summed over all the
scan points in a single scan or set of scans for one beam and plane:
\begin{equation}\label{e:chidoros}
\chi^2_\mathrm{doros}=\sum \left(\frac{\resid{doros}{x}{i}}{\sdd\,\sigma_\mathrm{doros}}\right)^2\ ,
\end{equation}
where the statistical uncertainties in the DOROS BPM measurements
$\sigma_\mathrm{doros}$ were determined from the spread of individual
measurements over the duration of one LB, and \sdd\ is a scale factor with
nominal value $\sdd=1$. The
$\chi^2$ was minimised, allowing the values of \dcal{x}{i}, \dofs{x}{i}\ and
\bbds\ to be determined. The length scale $M_1$ cannot be determined
from this fit (as it is almost degenerate with \dcal{x}{i}), but was determined
within the same framework by using the residuals between the beamspot and
nominal beam positions in the LSC scan to define an additional
$\chi^2$ term constraining $M_1$:
\begin{equation}\label{e:chilsc}
\chi^2_\mathrm{lsc}=\sum \left(\frac{\xbs{}-M_{1,i}\setpos{x}{i}}{\exbs{}}\right)^2\ ,
\end{equation}
where \xbs{} is the beamspot position measurement at maximum beam overlap
derived from each LSC miniscan and \exbs{} its uncertainty,
and the sum is taken over the five LSC scan points.\footnote{This $\chi^2$
minimisation to determine $M_1$ is equivalent to the linear fit shown in
Figure~\ref{f:lsclin}(a).}
A combined fit to all seven vdM scans and the LSC was performed separately for
each beam and plane to determine \dcal{x}{i}, \dofs{x}{i}, $M_1$ and \bbds.
Separate values of \dcal{x}{i}\ and \dofs{x}{i}\ were fitted in each scan,
and these were found to differ scan-to-scan with RMS values of 0.3--0.4\%
(the DOROS BPM responses are not expected to be stable in time, due to
temperature variations in their electronics). For each beam and plane,
a common \bbds\ value was fitted for all scans, and these values differed
from unity by up to 17\%.
 
\begin{figure}[tp]
\parbox{83mm}{\includegraphics[width=76mm]{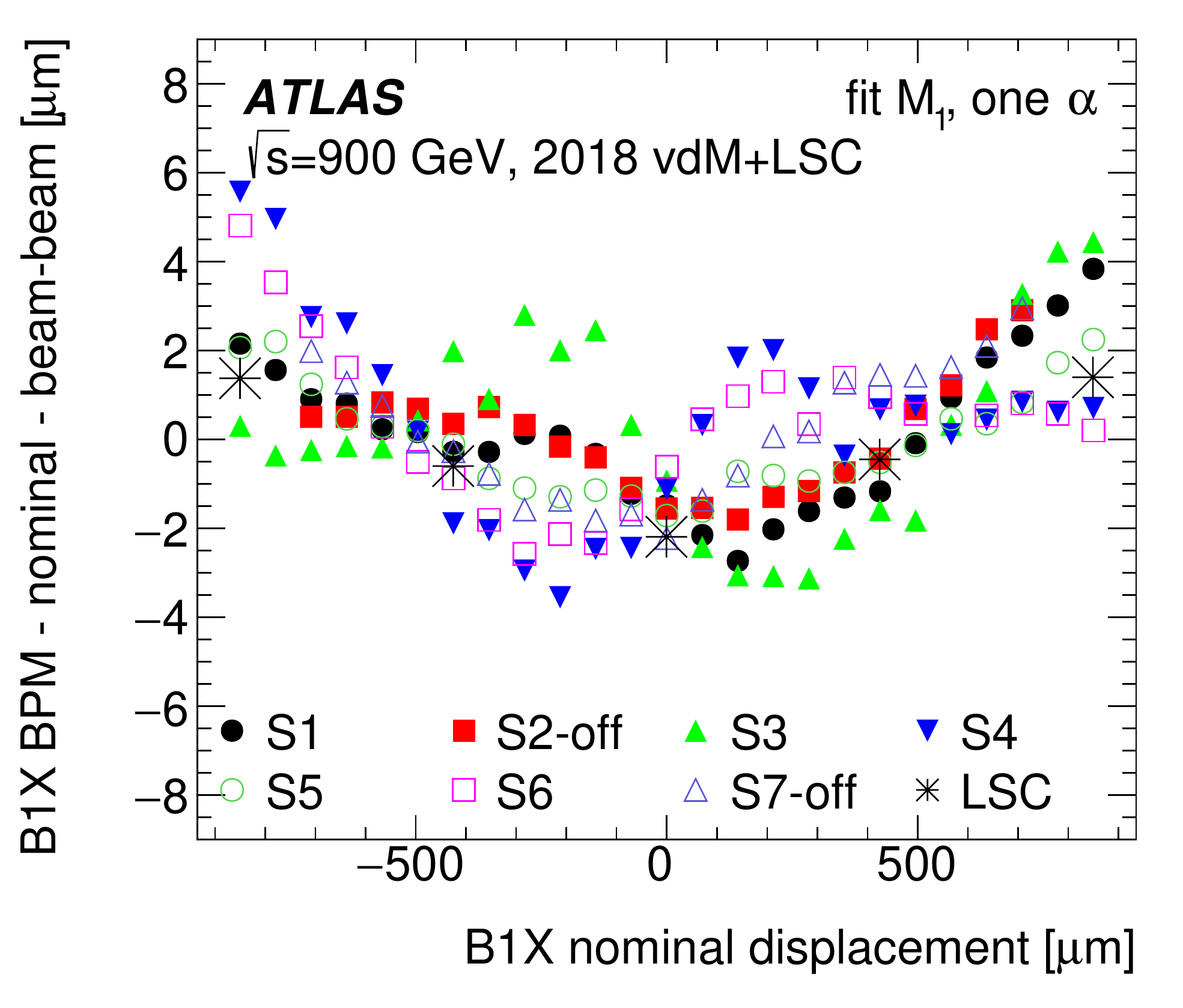}\vspace{-6mm}\center{(a)}}
\parbox{83mm}{\includegraphics[width=76mm]{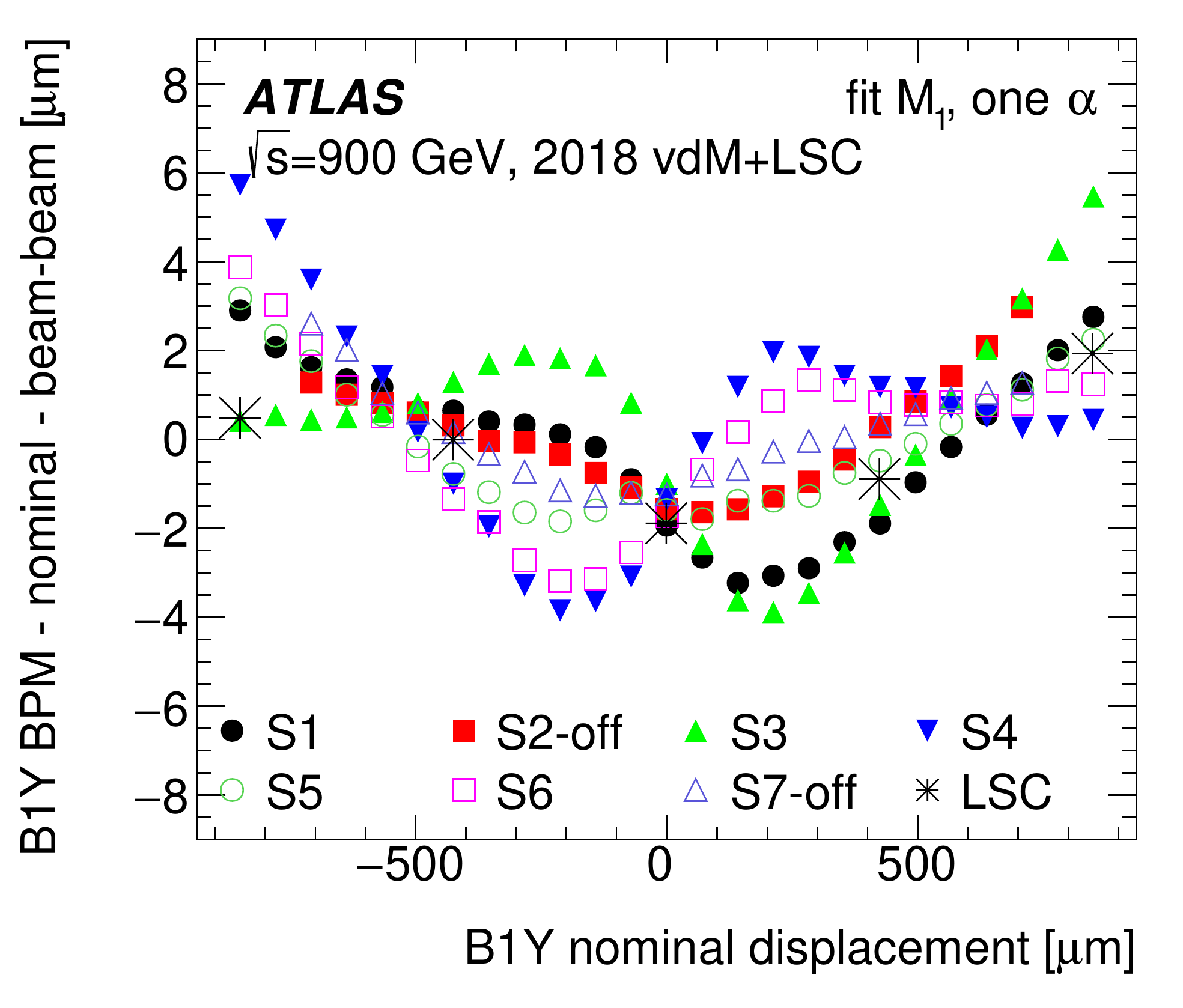}\vspace{-6mm}\center{(b)}}
\parbox{83mm}{\includegraphics[width=76mm]{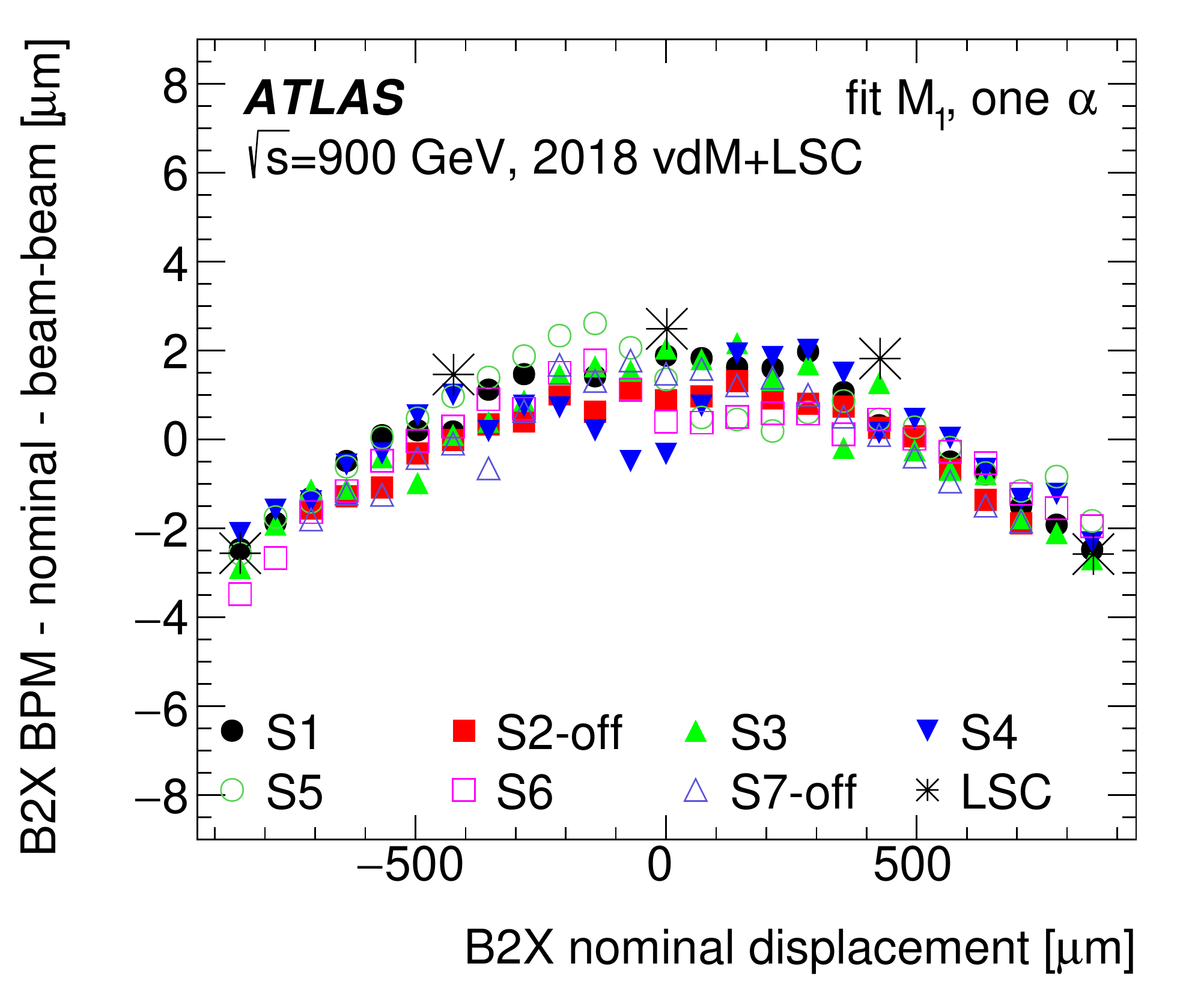}\vspace{-6mm}\center{(c)}}
\parbox{83mm}{\includegraphics[width=76mm]{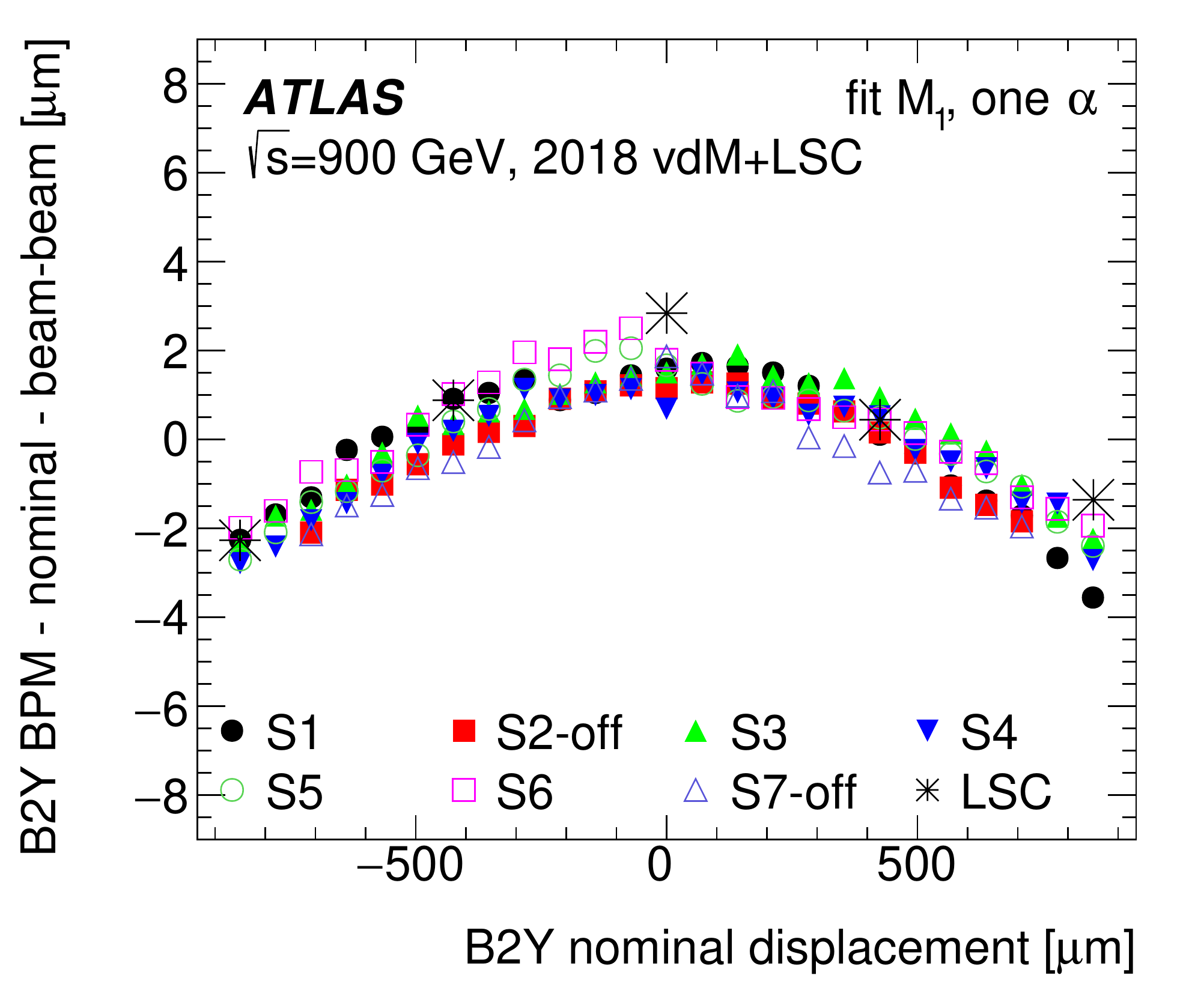}\vspace{-6mm}\center{(d)}}
\caption{\label{f:vdMres900}Residuals between the beam displacement
at the interaction point inferred from the calibrated DOROS BPM measurement
corrected for the beam--beam-induced displacement,
and the nominal beam displacement, during vdM scans at \sxit\ in 2018,
using a fit with a linear magnetic model and a single beam--beam deflection
scaling parameter \bbds\ per beam and plane. The four
plots show (a) B1X, (b) B1Y, (c) B2X and (d) B2Y. The residuals
for the seven scans are shown separately by the different coloured markers.
The black stars show the differences between the beamspot position and
the nominal beam displacement measured in the length scale calibration scan and
plotted on the same axes.}
\end{figure}

The resulting DOROS residuals from Eq.~(\ref{e:dresid})
are shown for each beam and plane in Figure~\ref{f:vdMres900}, together with the
beamspot residuals from the LSC scan that were also shown in
Figure~\ref{f:lscres900}.  In both $x$ and $y$ planes for beam~2,
the residuals from all scans are similar, and close to those from
the LSC scan, suggesting a non-linearity that is largely reproducible between
scans, and present whether the beam displacements are measured with DOROS BPM
or beamspot data. The situation for beam~1 appears more complex: the DOROS
residuals
broadly follow the beamspot data, but several scans exhibit residual S-shaped
wiggles even after subtracting the scaled beam--beam displacement, suggesting
that the magnitude of the apparent beam--beam displacement has been
overestimated or underestimated.
Allowing the value of \bbds\ to be different for each
scan (as well as for each plane) significantly improved the fits, giving
residuals for beam~1 that resemble the beam-2 curves, although inverted
due to the opposite scan direction. However, the fitted \bbds\ values
scatter between 0.5 and 1.2, typically being smaller for beam~1 than for beam~2.
This large range of fitted $\alpha$ values
suggests that although the separation-dependence of the beam--beam
displacement seen by
the DOROS BPMs is well predicted by Eqs.~(\ref{e:bbang}) and~(\ref{e:bbdeld}),
its magnitude is not.
 
The non-linearity between the nominal and actual beam displacements \xset{i}\ and \xpos{i}\ was parameterised as
\begin{equation}\label{e:lscnonlin}
\xpos{i}=M_{1,i}\xset{i}+M_{2,i}r_i^2+M_{3,i}r_i^3\ ,
\end{equation}
where $M_{1,i}$ is the linear length scale, and $M_{2,i}$ and $M_{3,i}$ are the
coefficients of quadratic and cubic terms in the reduced beam separation
$r=\xset{i}/x_\mathrm{max}$, where $x_\mathrm{max}=850\,\mu$m is the maximum
beam displacement in a \sxit\ scan. The corresponding non-linear terms
were added to Eqs.~(\ref{e:dresid}) and~(\ref{e:chilsc}), and the fits repeated,
also allowing \bbds\ to be different in each scan. The resulting residuals
are typically smaller than $0.5\,\mu$m and without significant structure,
demonstrating that the non-linearity seen in both DOROS BPM and beamspot
measurements can be reasonably well parameterised by Eq.~(\ref{e:lscnonlin}).
The fitted values of $M_1$, $M_2$ and $M_3$ for both beams and planes are
shown in Table~\ref{t:mag900}, together with the $M_1$ values from linear
fits to the LSC scan only, analogous to that shown in Figure~\ref{f:lsclin}.
The significant quadratic terms of $|M_2|\approx 4\,\mu$m, with opposite signs
for the two beams, account for the symmetric distortions seen in
Figure~\ref{f:vdMres900}. The cubic $M_3$ terms account for any distortions that
are asymmetric between positive and negative beam displacements, and their
fitted values are all smaller than $1\,\mu$m and consistent with zero.
However, the $M_3$ and
$M_1$ terms are anticorrelated, and the $M_1$ values also change
slightly between the linear and non-linear fits.
 
\begin{table}[tp]
\caption{\label{t:mag900}Fitted values of the length scale $M_1$ from
a linear fit to the LSC scan data only (left column) and the parameters
$M_1$, $M_2$ and $M_3$ from the non-linear
fit to vdM and LSC scan data (right columns), for each beam and plane
in the 2018 \sxit\ vdM scan session.}
\centering
 
\begin{tabular}{l|c|crr}\hline
& $M_1$ (linear LSC) & $M_1$ & $M_2$ [$\mu$m] & $M_3$ [$\mu$m] \\
\hline
B1X & $1.0044\pm 0.0003$ & $1.0046\pm 0.0006$ & $3.7\pm 0.2$ & $-0.2\pm 0.5$ \\
B1Y & $1.0064\pm 0.0003$ & $1.0067\pm 0.0003$ & $4.5\pm 0.1$ & $-0.3\pm 0.2$ \\
B2X & $1.0022\pm 0.0003$ & $1.0024\pm 0.0006$ & $-4.2\pm 0.2$ & $-0.1\pm 0.5$ \\
B2Y & $1.0012\pm 0.0003$ & $1.0014\pm 0.0003$ & $-4.2\pm 0.2$ & $-0.2\pm 0.2$ \\
\hline
\end{tabular}
\end{table}
 
The effect of the non-linear terms on the length scale product was estimated
by modelling the scan curves as single Gaussian functions with RMS equal to the
\capsigxy\ averaged over all scans, and calculating the change in the
integrals of Eq.~(\ref{e:capsig}) from the distortions in $\Delta x$ and
$\Delta y$ when using the non-linear instead of the linear length scale. For the
2018 \sxit\ scans \lsp\ changes from $1.0071\pm 0.0003$ to $1.0075\pm 0.0004$,
a change of only 0.04\%. Although the $M_2$ terms are large, they have no
effect on \lsp, because e.g.\ a decrease of the integral for $\Delta x<0$ is
compensated by an equal increase for $\Delta x>0$, and only the small changes
in $M_1$ and $M_3$ have any effect.
 
Since the non-linear effects seen in the \sxit\ scans were unexpected, two
additional studies were made to probe the interpretation in terms of
non-linearity of the steering magnets. Firstly, the magnetic fields produced by
two spare LHC corrector dipoles were precisely measured on a test bench, with
both standard magnetic cycles covering the full operating range, and test
cycles replaying the powering history during vdM scans at \sxyt\ from fill 6016
in 2017 \cite{magmeas}. These studies showed that, in the laboratory
environment, the magnetic fields are highly reproducible
(to better than $10^{-4}$),
but that there is a significant hysteresis effect depending on whether the
magnet current is being increased or decreased (and hence on the direction
of beam movement in a vdM scan).
 
\begin{figure}[tp]
\parbox{83mm}{\includegraphics[width=76mm]{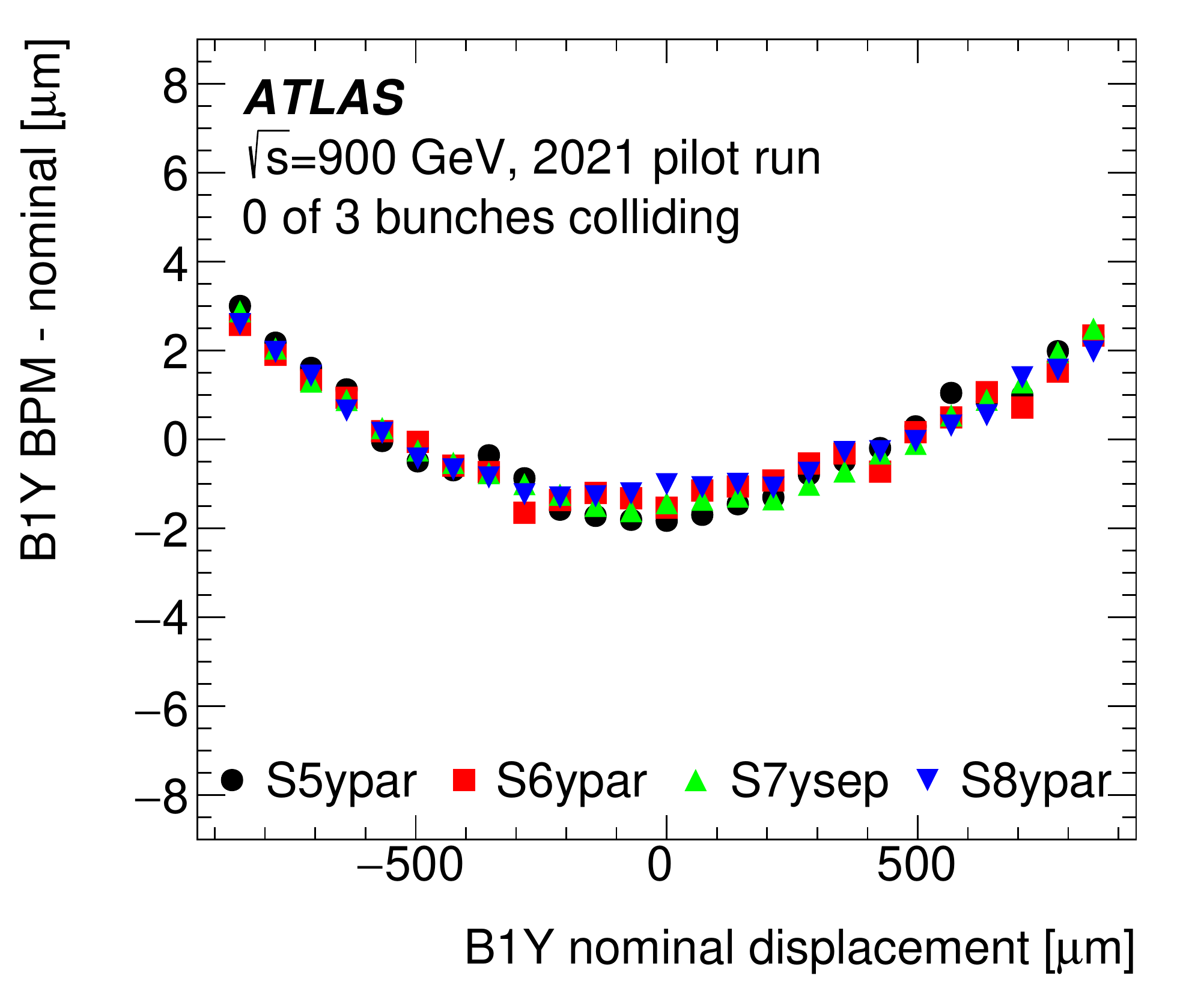}\vspace{-6mm}\center{(a)}}
\parbox{83mm}{\includegraphics[width=76mm]{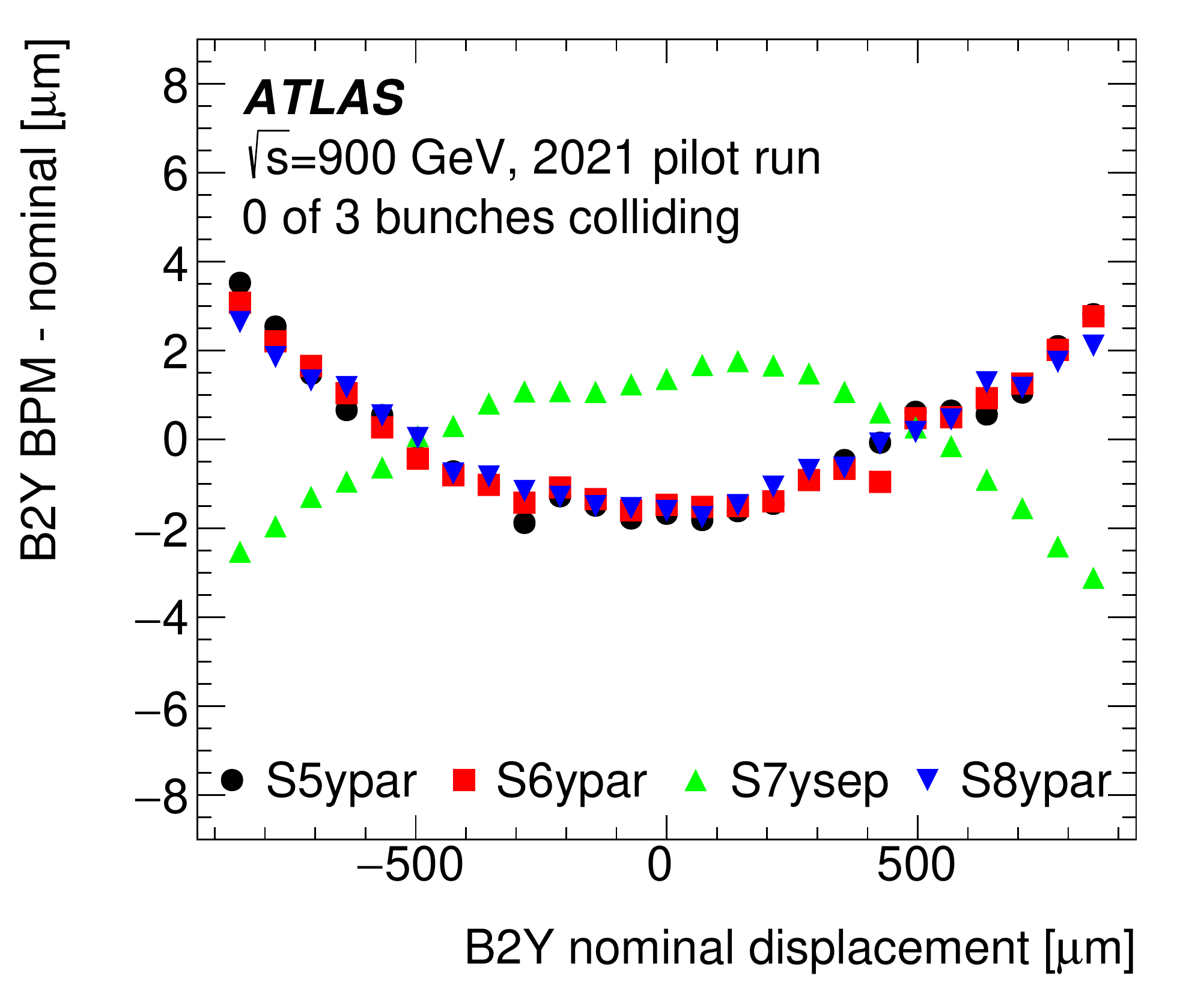}\vspace{-6mm}\center{(b)}}
\parbox{83mm}{\includegraphics[width=76mm]{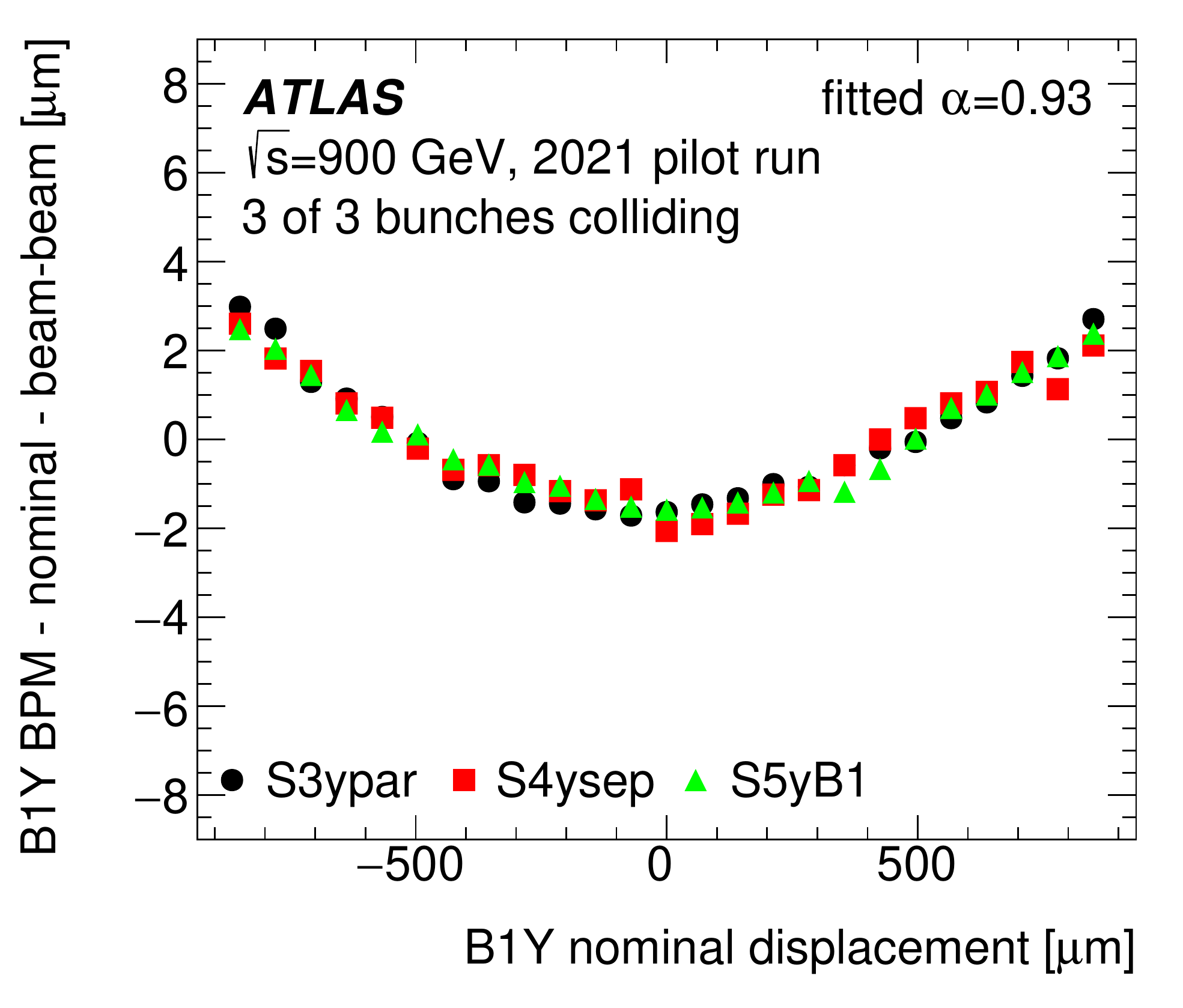}\vspace{-6mm}\center{(c)}}
\parbox{83mm}{\includegraphics[width=76mm]{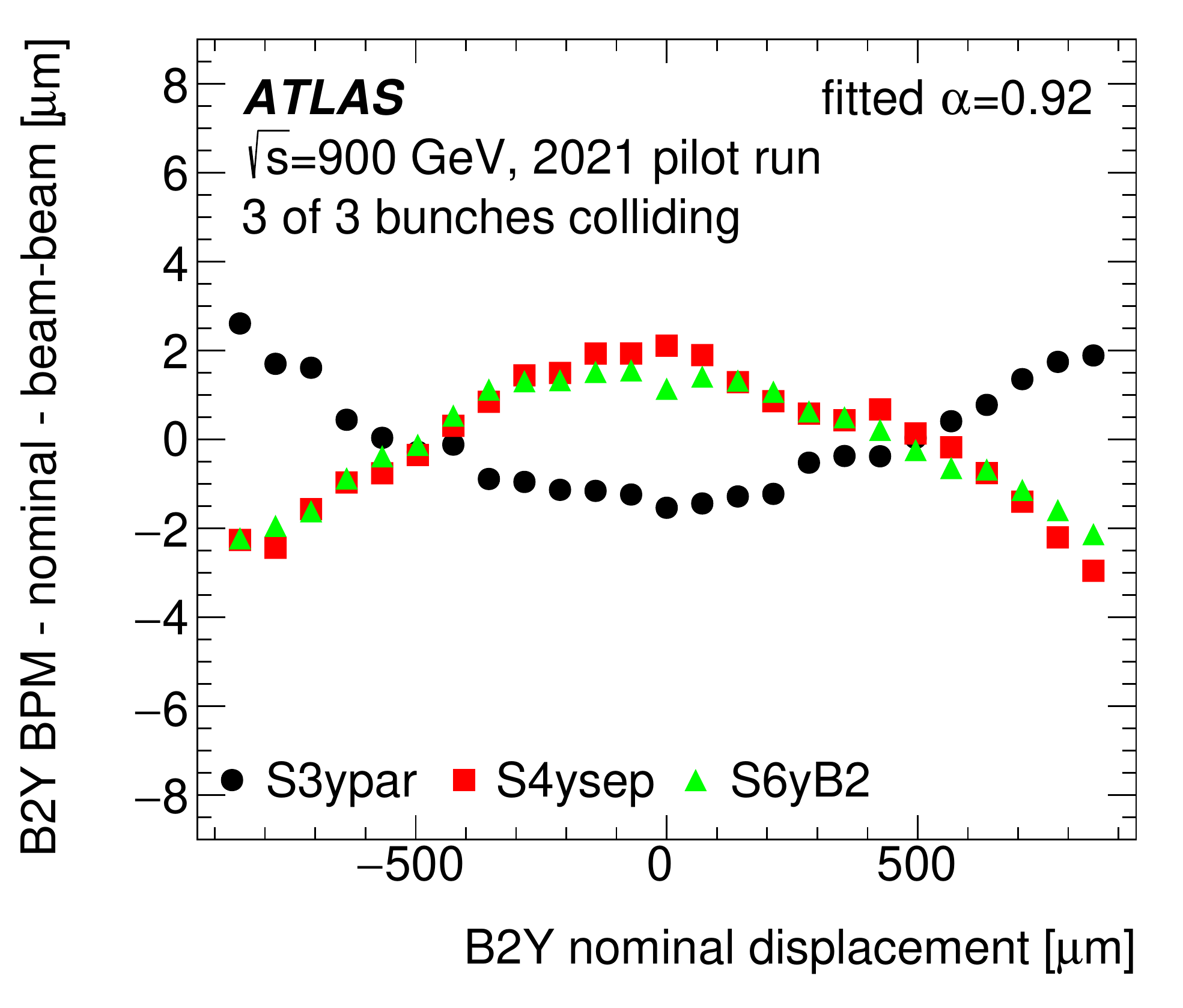}\vspace{-6mm}\center{(d)}}
\parbox{83mm}{\includegraphics[width=76mm]{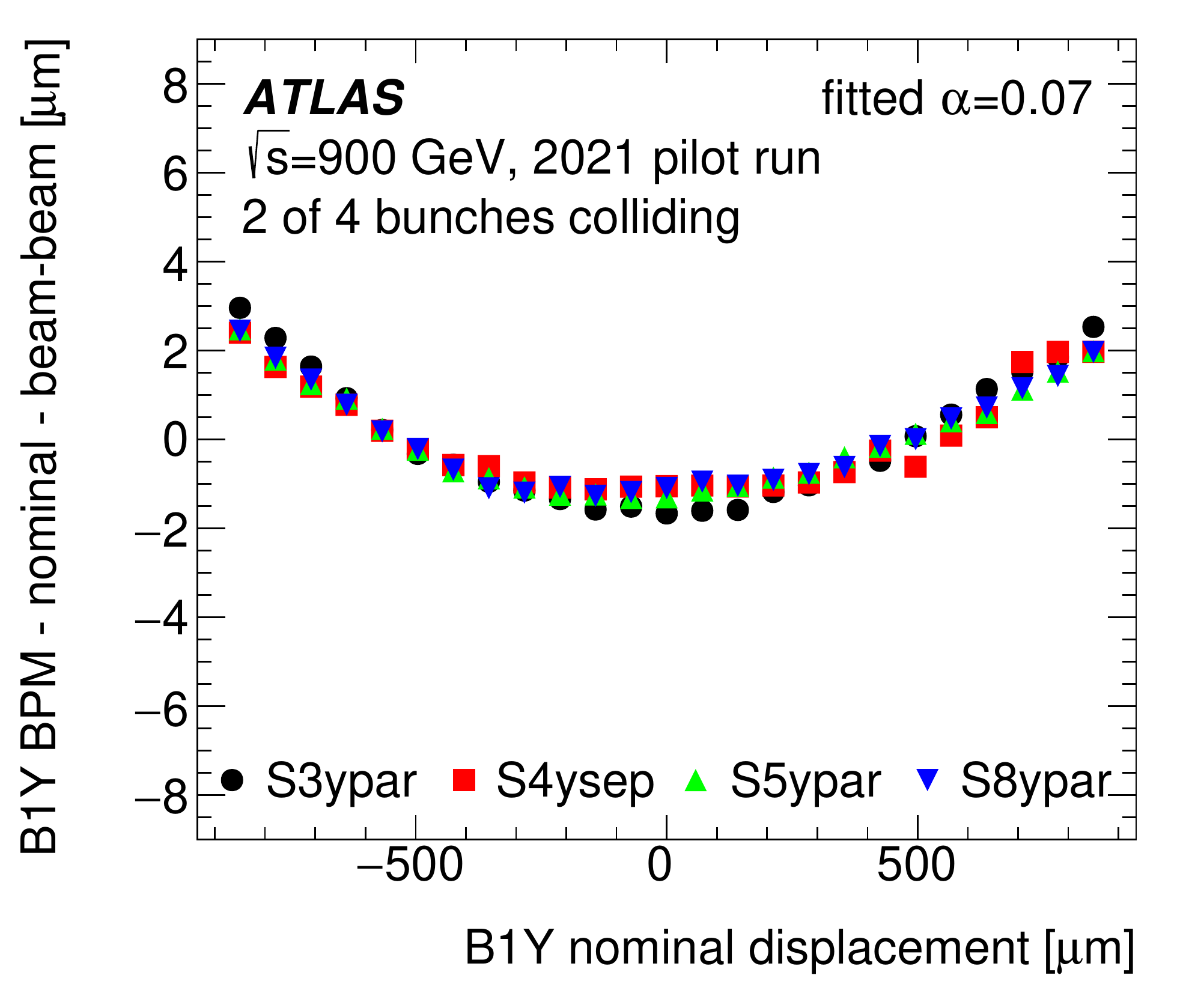}\vspace{-6mm}\center{(e)}}
\parbox{83mm}{\includegraphics[width=76mm]{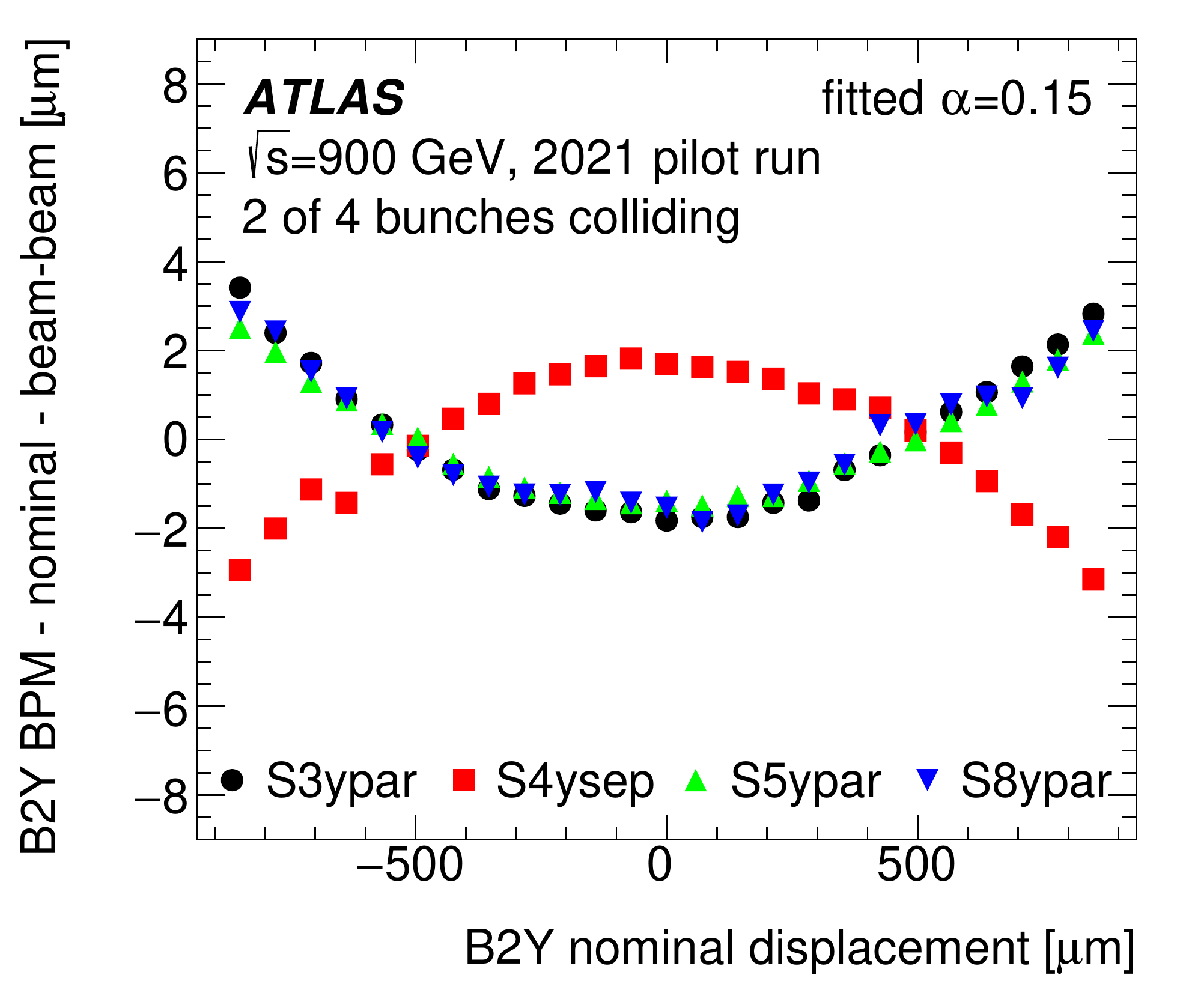}\vspace{-6mm}\center{(f)}}
\caption{\label{f:pilot}Residuals between the beam displacement at the
interaction point inferred from the calibrated DOROS BPM measurement and the
nominal beam displacement, during scans at \sxit\ in the October 2021 LHC pilot
run. The left column shows B1Y and the right column B2Y; the results in the
$x$-plane are similar. The three rows show (a, b) a fill with no bunches
colliding at the ATLAS IP, (c, d) all three bunches colliding at ATLAS,
and (e, f) a `mixed' configuration with two out of four bunches colliding
at ATLAS. The beam--beam displacement was subtracted in plots (c--f), scaled by
the parameter \bbds\ whose fitted values are given in the legends.
}
\end{figure}

Secondly, a series of test scans
were performed at \sxit\ during the October 2021 LHC pilot run, as shown
in the lower part of Table~\ref{t:vdm900}. The DOROS BPM data were analysed
in the same way as for the 2018 \sxit\ scans, but keeping $M_1=1$ and
$M_{2,3}=0$, and without additional constraints from beamspot data.
Figures~\ref{f:pilot}(a) and~\ref{f:pilot}(b) show the residuals from the two
beams in a series
of $y$-plane scans in fill~7524. In this fill, the three bunches in each
beam were positioned around the LHC ring such that none collided in ATLAS,
eliminating beam--beam deflections.
Two consecutive parallel scans were performed, in which both beams were moved
from negative to positive displacement. These were followed by a separation
scan in which beam~2 was moved in the opposite direction (positive to negative)
to beam~1, as in a vdM scan, and then a further parallel scan. The offsets
between the DOROS BPM and nominal displacements were determined separately for
each scan, so that they are zero at the nominal zero displacement
just before each scan started. These residuals are not affected by beam--beam
displacements, and clearly show that the non-linearity at \sxit\
is highly reproducible
from scan to scan, with the opposite sign for beam~2 in the separation scan
where it moves in the opposite direction.
 
Figures~\ref{f:pilot}(c) and~\ref{f:pilot}(d) show results from fill 7525,
where all three bunches were colliding in ATLAS, and parallel and separation
scans were followed by scans where only one beam was moved and the other
remained  stationary (and is hence not plotted). The beam--beam displacement
was calculated
using the \capsigy\ value obtained from the separation scan S4ysep, and
subtracted from the residuals after being scaled by a fitted parameter \bbds,
the same for all scans but different for the two beams.
The fitted values of \bbds\ are within 10\%
of unity, validating the modelling of the beam--beam displacement as seen by
the DOROS BPMs for the case where all bunches are colliding in ATLAS. The
residuals are again similar for all three scans for beam~1, and for the
separation and single-beam scans in beam~2 where it moves in the same direction.
 
Figures~\ref{f:pilot}(e) and~\ref{f:pilot}(f)  show results from fill 7516,
with four bunches
in each beam, and only two colliding in ATLAS. This was fitted in the same
way as for fill 7525, but the fitted \bbds\ values are much smaller, 0.07
for beam~1 and 0.15 for beam~2. After subtracting this scaled beam--beam
displacement, the residuals are again
very reproducible between scans and similar to those in the other fills.
In such a fill, the DOROS BPMs see a mixture of colliding bunches that are
deviated by the beam--beam interaction, and non-colliding bunches that are not
deviated. The response of the DOROS electronics is known to depend both
on the time structure of the bunch pattern, and the relative intensity
of the bunches, and in this case the sensitivity to the beam--beam displacement
is reduced by much more than the $\bbds=0.5$ which would be expected from
the fraction of bunches that were colliding.
 
Qualitatively similar results were obtained in the $x$-plane, though with
a larger scatter of residuals. These studies confirm that the non-linearity
in the beam displacements appears to originate from hysteresis in the LHC
steering magnets, and is highly reproducible from scan to scan. The effect
of beam--beam deflections can be modelled by Eqs.~(\ref{e:bbang})
and~(\ref{e:bbdeld}), after empirical scaling by the factor \bbds\
which must be determined from data.
 
\begin{figure}[tp]
\centering
 
\includegraphics[width=160mm]{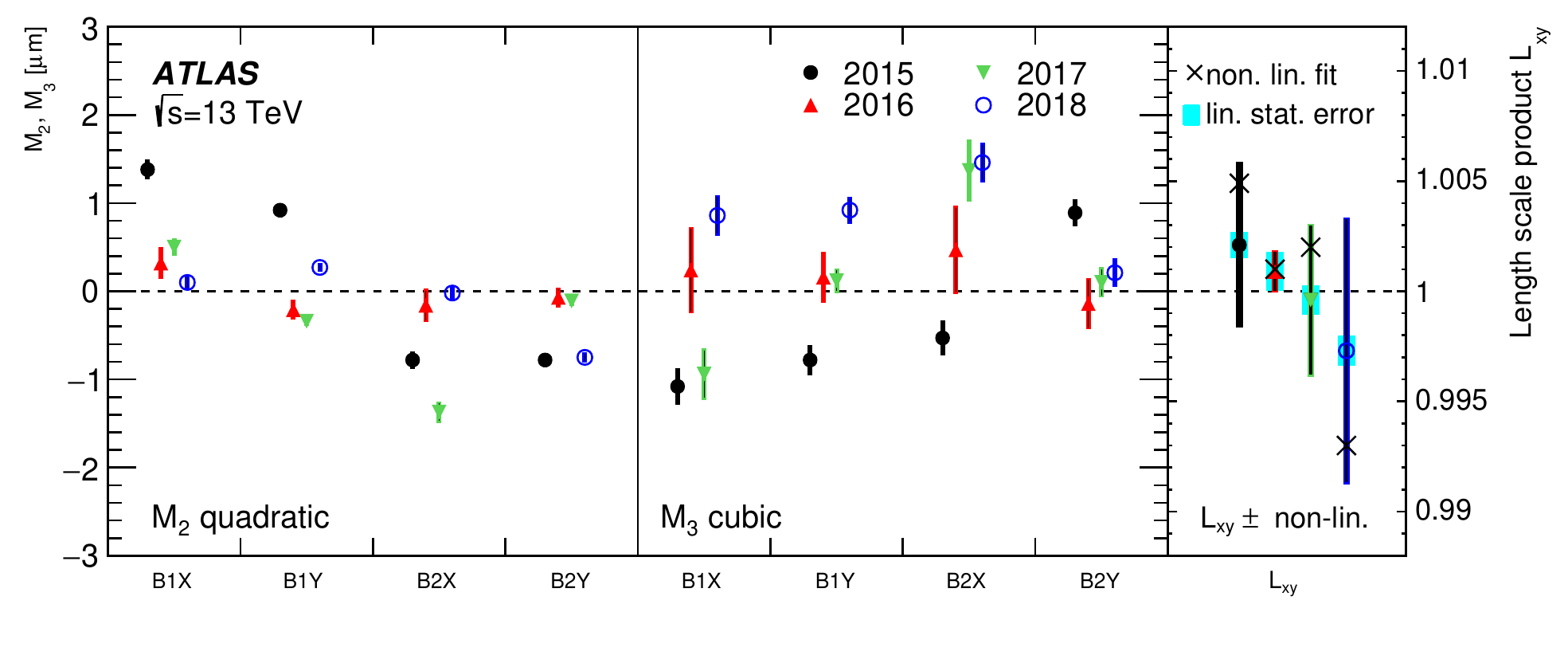}
\caption{\label{f:magsum}Summary of the non-linear length scale determination
for the \sxyt\ vdM+LSC datasets. The left and middle panels show the
quadratic $M_2$ and cubic $M_3$ terms fitted for each beam and plane from
the combined vdM and LSC data for each data-taking year, with error bars
indicating the statistical uncertainties. The right panel shows the linear
length scale product \lsp\ for each year, together with its statistical
error (cyan band), the central value of the baseline non-linear fit
incorporating the $M_2$ and $M_3$ terms (black cross), and the total uncertainty
in \lsp\ encompassing the central values of the baseline and alternative
non-linear fits (see text).}
\end{figure}

To estimate the potential effect of magnetic non-linearity on the length scale
at \sxyt, the data from each year's vdM and LSC scans were fitted in the same
way as at \sxit, fitting $M_{1,2,3}$ and \bbds\ for each beam and plane
separately in each year. However, due to the smaller beam displacements compared
to the BPM and beamspot resolution, the results were less conclusive.
The residual distributions for some scan sessions showed shapes reminiscent
of the non-linearities seen at \sxit, and which could be eliminated by
adding $M_2$ and $M_3$ terms to the basic linear length scale. Others,
particularly 2016, showed no significant non-linearity, and some scans
in 2015 showed evidence of higher-order structure. In most datasets,
the fitted values of \bbds\ are close to the naive expectation of
$\bbds\approx \nbun/\nbeam$, assuming the DOROS position readings correspond
to an unweighted average over all bunches.
The resulting values of $M_2$ and $M_3$
for each beam, plane and year are shown in Figure~\ref{f:magsum}. Significant
$M_2$ and $M_3$ terms were fitted for many of the datasets, but with magnitudes
that typically do not exceed $1\,\mu$m. For a given beam and plane, the results
for different years are often very different, for both $M_2$ and $M_3$. This
may reflect the fact that the LHC orbit was set up independently in each year,
leading to potentially different corrector magnet currents and hence hysteresis
effects for the same scan patterns in different years.
 
The baseline fit was performed with a common value of \bbds\ for all scans
for each individual beam and plane, and with the factor \sdd\ of
Eq.~(\ref{e:chidoros}) set to $\sdd=2$, rather than $\sdd=1$ as used at
\sxit. This was required to get $\chi^2_\mathrm{doros}$ values close to one
per degree of freedom, suggesting that the DOROS BPM errors
estimated from the spread within each luminosity block do not fully capture
the measurement uncertainties. Alternative fits with $\sdd=1$, or
allowing \bbds\ to vary per scan, were also considered. Since in the LSC scans
of 2015--2017, beam~2 was moved in the opposite direction
to that in the vdM scans, the baseline fits were performed with the signs
of $M_2$ and $M_3$ reversed in the LSC compared to the vdM scans, reflecting
the scan direction dependence suggested by Figures~\ref{f:lscres900},
\ref{f:vdMres900} and~\ref{f:pilot}. An alternative fit without these sign
reversals was also considered in these cases.
 
Given the lack of conclusive evidence for magnetic non-linearity at \sxyt,
the baseline \lsp\ values were taken from the linear fits described in
Section~\ref{ss:lsc}, and the largest deviations from any of the
non-linear fits were considered to define a (symmetric) systematic uncertainty
added in quadrature to the statistical uncertainty from the linear fits.
This procedure assumes that any significant non-linearity can be
adequately described by a cubic parameterisation determined from
the combined fit to vdM and LSC scans, but does not attempt to correct it.
The resulting \lsp\ values are shown in the rightmost panel of
Figure~\ref{f:magsum}, together with the \lsp\ values from the baseline
non-linear fits. The magnetic non-linearity uncertainty varies from 0.1\% in
2016 to 0.6\% in 2018, reflecting the fitted values of $M_3$ shown in
Figure~\ref{f:magsum}.
 
\subsection{Consistency checks}\label{ss:vdmcon}
 
\begin{figure}[tp]
\parbox{83mm}{\includegraphics[width=76mm]{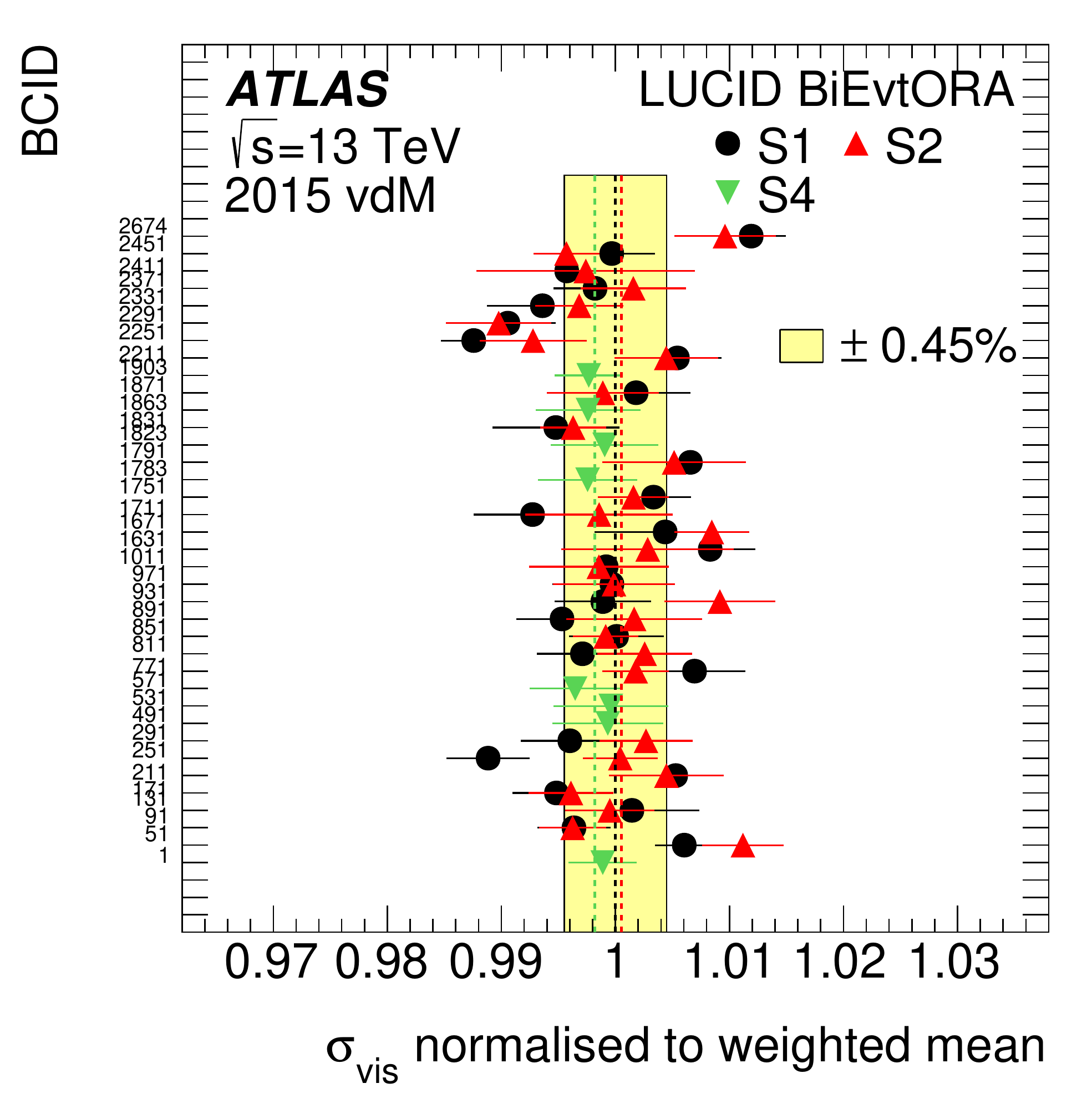}\vspace{-6mm}\center{(a)}}
\parbox{83mm}{\includegraphics[width=76mm]{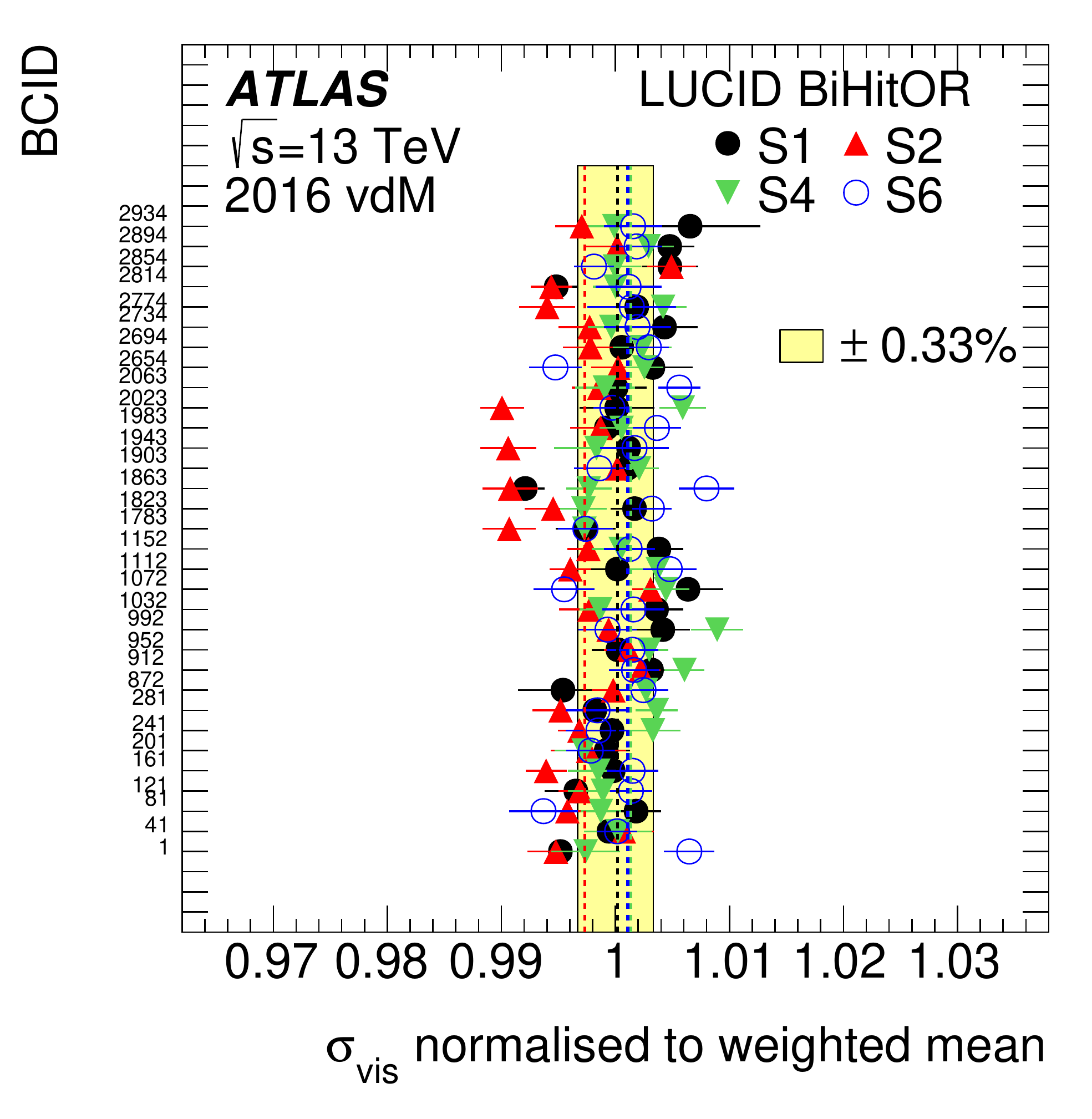}\vspace{-6mm}\center{(b)}}
\parbox{83mm}{\includegraphics[width=76mm]{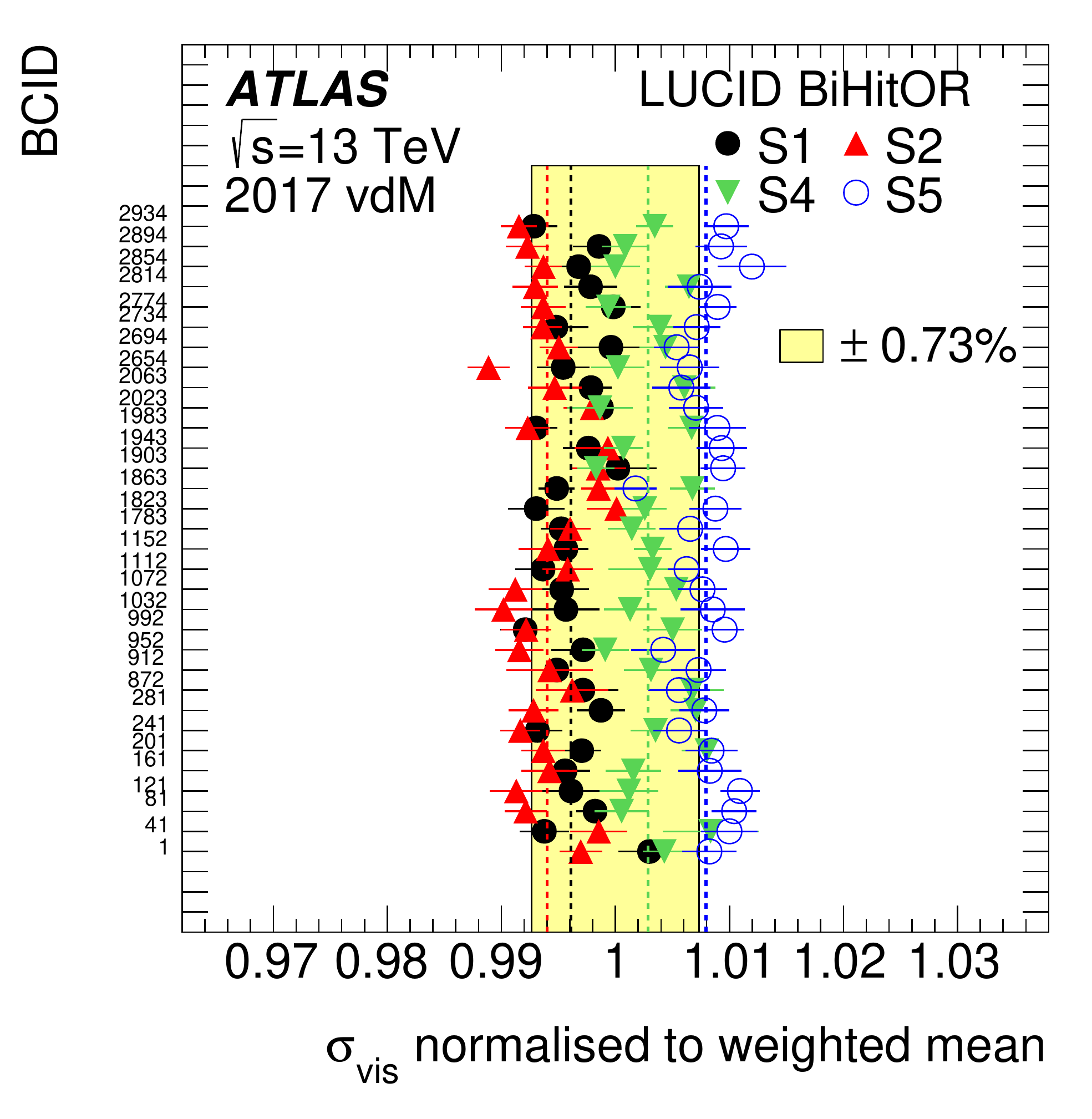}\vspace{-6mm}\center{(c)}}
\parbox{83mm}{\includegraphics[width=76mm]{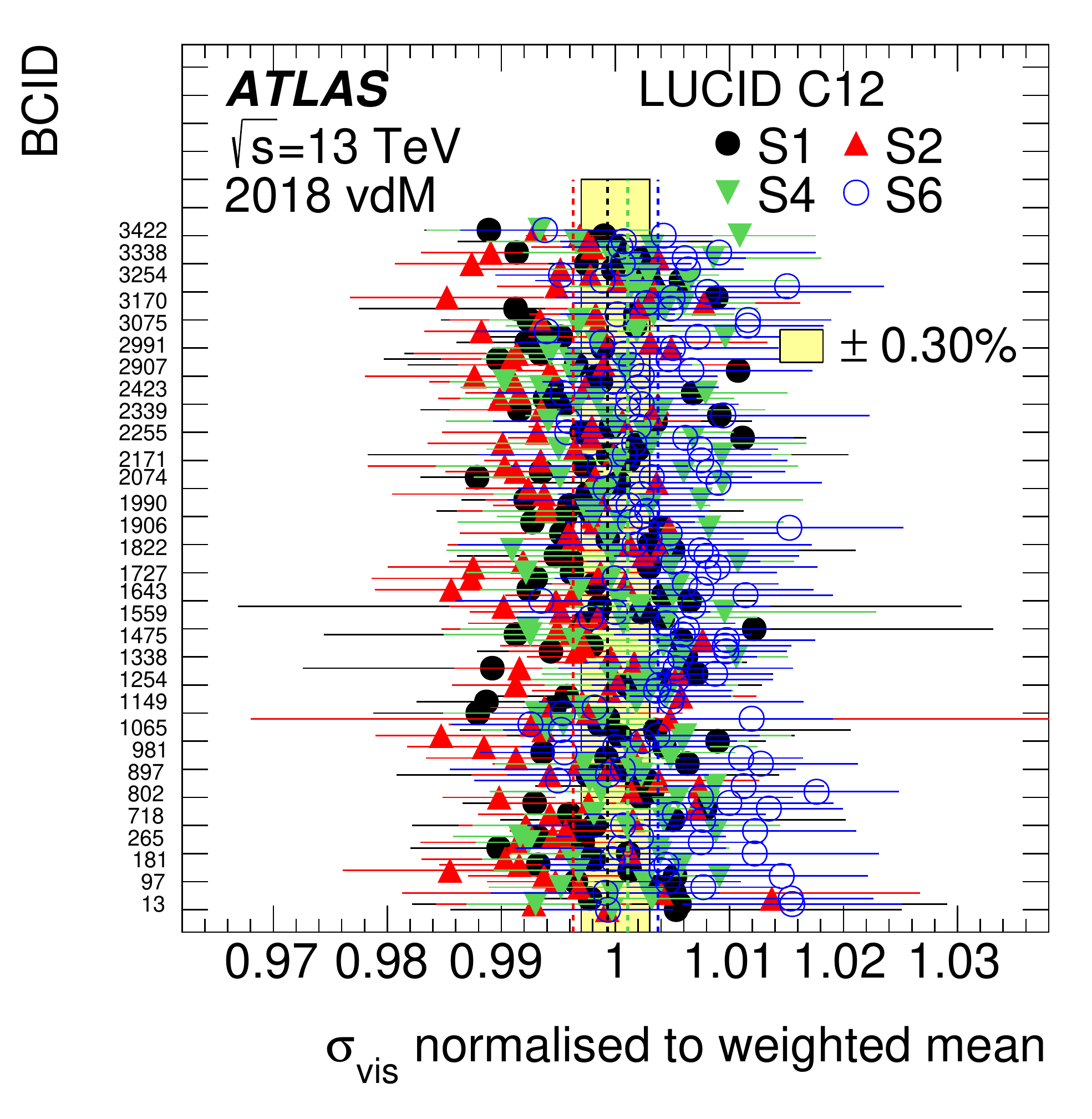}\vspace{-6mm}\center{(d)}}
\caption{\label{f:vdMconst}Ratio of bunch-by-bunch visible cross-sections
\sigmavis\ to the weighted mean of \sigmavis\ for all colliding bunch pairs
in all on-axis scans in the vdM scan sessions from each year of data-taking,
using the reference LUCID algorithm in each case. The vertical dashed lines show
the mean \sigmavis\ for all bunches in each individual scan set.
The uncertainties are statistical, and are larger for 2018 due to the use
of the C12 single-PMT algorithm rather than a multi-PMT OR algorithm as used
in 2015--2017. The yellow bands show the combination of bunch-to-bunch and
scan-to-scan consistency uncertainties in each year, with the numerical
values indicated in the legends.}
\end{figure}

Since the value of \sigmavis\ for a given luminosity algorithm should not depend
on beam parameters, the ability to determine it separately from each colliding
bunch pair and $x$--$y$ scan set allows the stability and reproducibility of the
calibrations to be checked. Figure~\ref{f:vdMconst} shows the \sigmavis\
values measured for each bunch pair and scan in each vdM scan session. The
values are shown after all corrections have been applied, normalised to the
error-weighted mean over all bunch pairs and scans in each year.
Within one scan, the spread of \sigmavis\ values over all bunch pairs
(after subtracting in quadrature the expected spread given the average
statistical error of the individual measurements) is a measure of
potential systematic bunch-to-bunch inconsistencies in the absolute luminosity
scale. The largest such value for each year was taken as the
`Bunch-by-bunch \sigmavis\ consistency' uncertainty in Table~\ref{t:unc}.
These uncertainties
are at most 0.4\% (in 2015), and zero in 2018, where the spread was consistent
with the expected bunch-to-bunch statistical fluctuations, which are larger than
in other years due to the use of a single LUCID PMT as reference algorithm.
 
The differences in bunch-averaged \sigmavis\ between scans give an indication
of the reproducibility of the calibration on a timescale of hours, or longer
in 2015 and 2016 where the scans were distributed over several LHC fills
(Table~\ref{t:vdmds}). The sampling-corrected standard deviation of these values
was used to define the `Scan-to-scan reproducibility' uncertainty in
Table~\ref{t:unc}, which is 0.3\% or less in all years except in 2017, where
significant inconsistencies between scans are apparent, reflected in the larger
uncertainty of 0.7\% that dominates the vdM calibration uncertainty for
that dataset. The average of the \sigmavis\ values from all scans in one year
was finally used as the central value of the calibration result for that year.
 
\begin{figure}[tp]
\includegraphics[width=140mm]{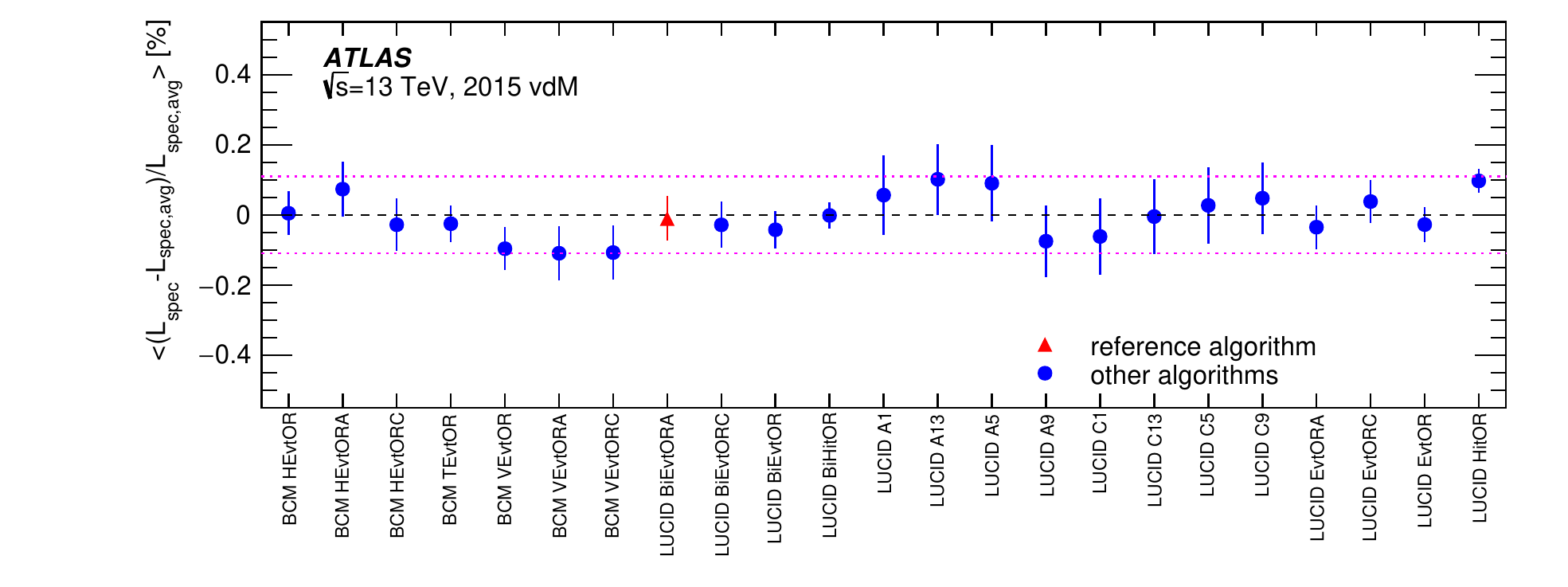}
\includegraphics[width=140mm]{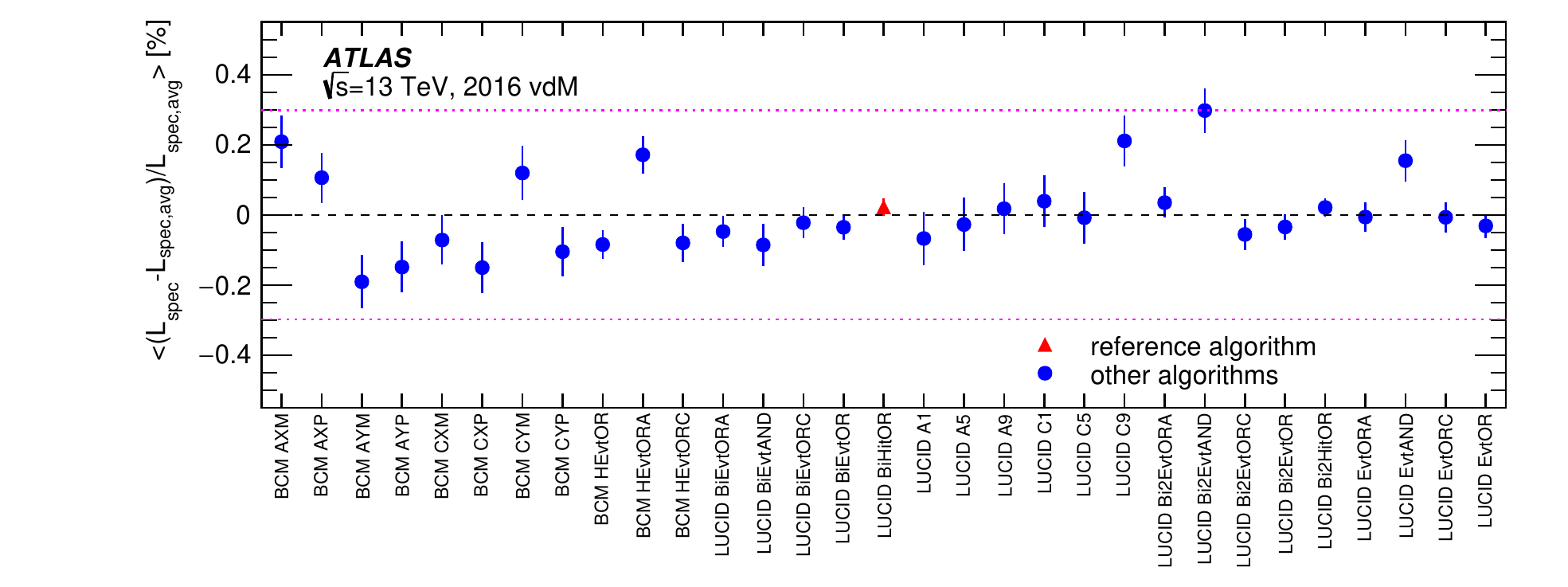}
\includegraphics[width=140mm]{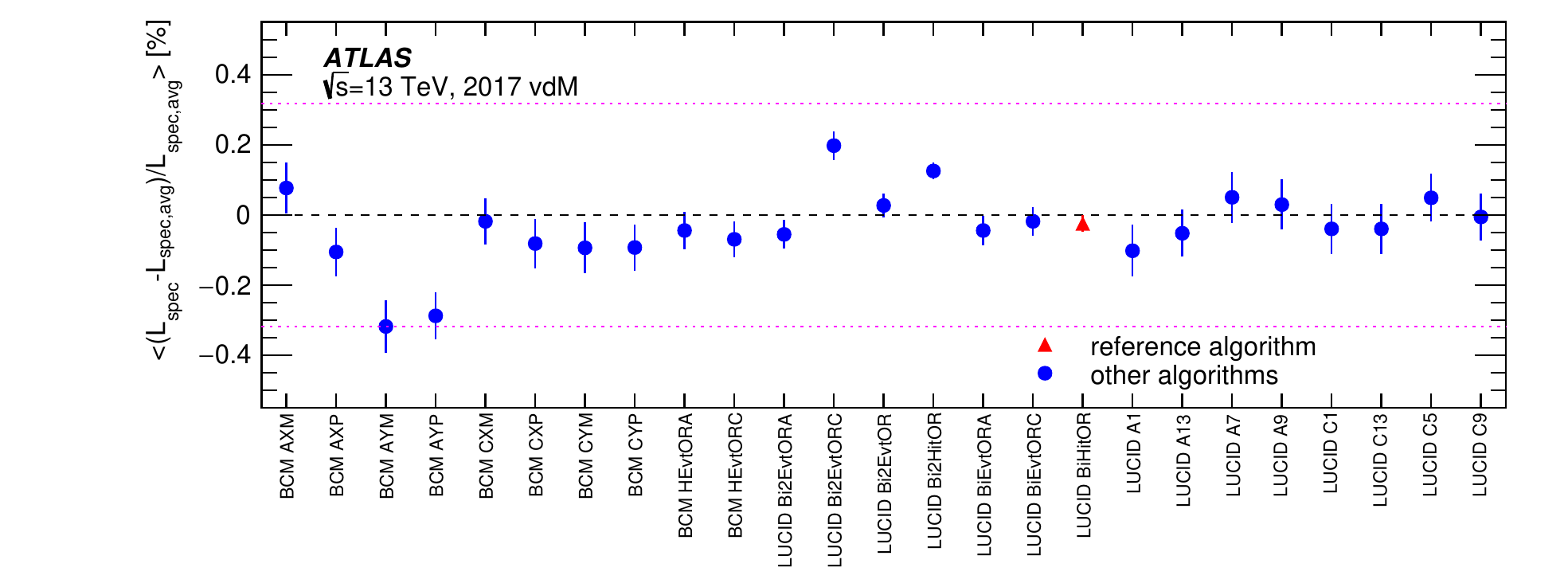}
\includegraphics[width=140mm]{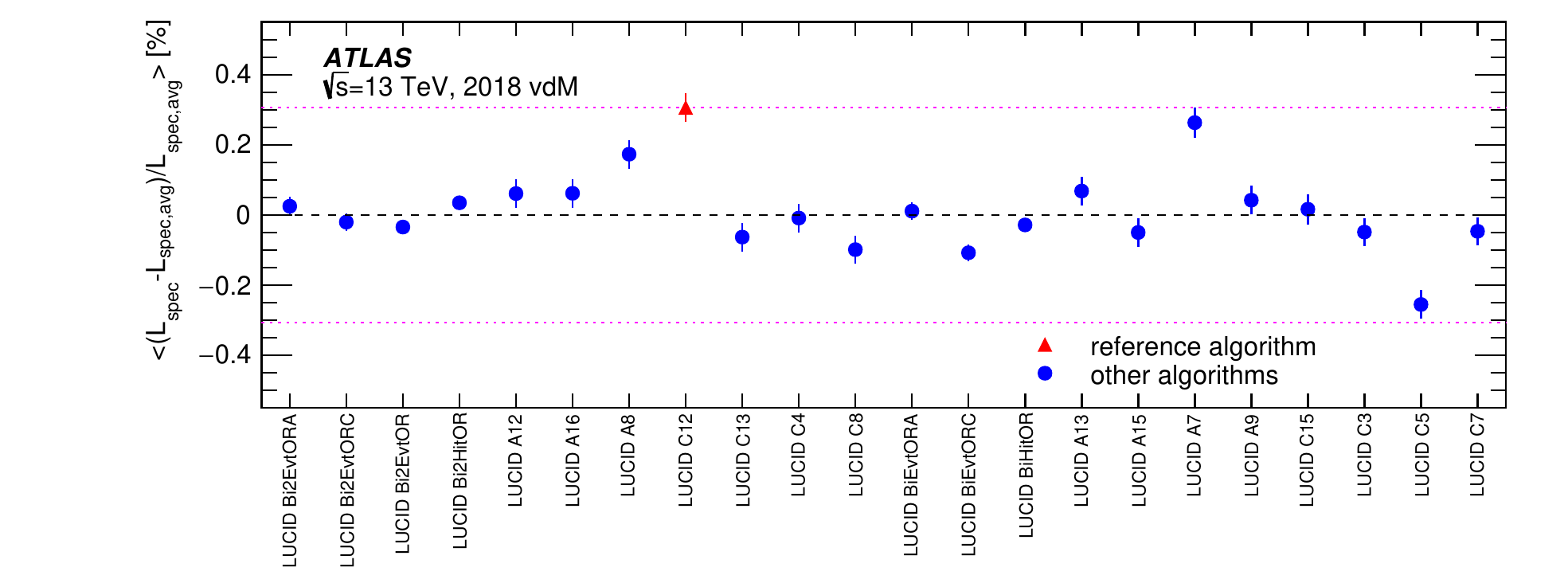}
\caption{\label{f:lspec}Relative deviations of the specific luminosity
measured by each algorithm \lspec\ from the average over all algorithms
$L_\mathrm{spec,avg}$, averaged over all colliding bunch pairs and scans
in the vdM session of each year. The uncertainties indicated by the error bars
are statistical, and the
symmetric uncertainty bands encompassing all algorithms in each year are shown
by the purple dotted lines. The reference algorithm for each year is indicated
by the red triangle.}
\end{figure}

In each year, the vdM calibration procedure was used to determine \sigmavis\
for all available LUCID and BCM algorithms. Since the acceptance of each
algorithm is different, the \sigmavis\ values are also different, but
all algorithms should agree on the values of \capsigxy\ and
hence on the specific luminosity \lspec\ for each colliding bunch pair, given by
\begin{equation*}
\lspec = \frac{\lbun}{n_1 n_2} = \frac{f_\mathrm{r}}{2\pi \capsigx \capsigy}\ ,
\end{equation*}
and obtained from the vdM scan curves using the luminosity measurements
of each individual algorithm. The consistency of the different algorithms
was quantified by calculating the per-bunch-pair and per-scan relative deviation
of \lspec\ for each algorithm compared to the average over all algorithms,
and then averaging this deviation over all bunch pairs and scans in each year.
Systematic deviations from this average are sensitive to problems with
individual algorithms
that may bias \capsigxy, such as detector non-linearity, short-term efficiency
drifts or background subtraction issues. The resulting relative deviations of
\lspec\ for all considered LUCID and BCM algorithms in each year are shown in
Figure~\ref{f:lspec}, with the reference algorithm shown in red. The set of
available algorithms evolved during Run~2, as the LUCID configuration changed
and progressively more PMTs were equipped with bismuth calibration sources.
The BCM algorithms were not considered for the 2018 analysis, as they showed
significant instabilities during the vdM fill due to radiation-ageing effects
caused by the accumulated dose. The largest deviation from the average
in each year was used to determine the `Reference specific luminosity'
uncertainty in Table~\ref{t:unc}, which is at most 0.31\%. The reference
algorithm has an \lspec\  close to the
average of all algorithms in all years except 2018, where the single-PMT
algorithm C12 has an \lspec\ 0.31\% higher than average. However, this algorithm
was chosen for the baseline luminosity measurement due to its availability and
stability throughout the year, as discussed in Section~\ref{ss:lucid}. The full
difference between the C12 and average \lspec\ values is covered by the
2018 reference specific luminosity uncertainty.
 
The total uncertainties in the absolute vdM calibration are listed in
the `Subtotal vdM calibration' row of Table~\ref{t:unc}, and range
from 0.7\% to 1.0\% across the years of Run~2 data-taking.
The largest individual
uncertainties come from a variety of sources---non-factorisation effects in
2015 and 2016, scan-to-scan reproducibility in 2017 and magnetic non-linearity
in 2018. All other individual sources are smaller than 0.5\% in all years.

% End of text imported from the .//vdm.tex input file

% The next lines are included from the .//caltrans.tex input file
\section{Calibration transfer to high-luminosity running}\label{s:calt}
 
The procedures described above provide the absolute calibration of the LUCID and
BCM luminosity measurements for low-$\mu$ data-taking with
a limited number of isolated bunches, where the instantaneous luminosities
are at least three orders of magnitude smaller than those typical
of normal physics running with bunch trains. The LUCID detector suffers
from significant non-linearity, and requires a downward correction
of $O(10\%)$ in the physics data-taking regime. This correction was derived
using track-counting luminosity measurements and validated with calorimeter
measurements, in particular those from the TileCal E-cell scintillators.
The calibration transfer procedures, and their
corresponding uncertainties, are discussed in detail below.
 
\subsection{Comparison of luminosity measurements at high pileup}\label{ss:algcomp}
 
\begin{figure}[tp]
\parbox{83mm}{\includegraphics[width=76mm]{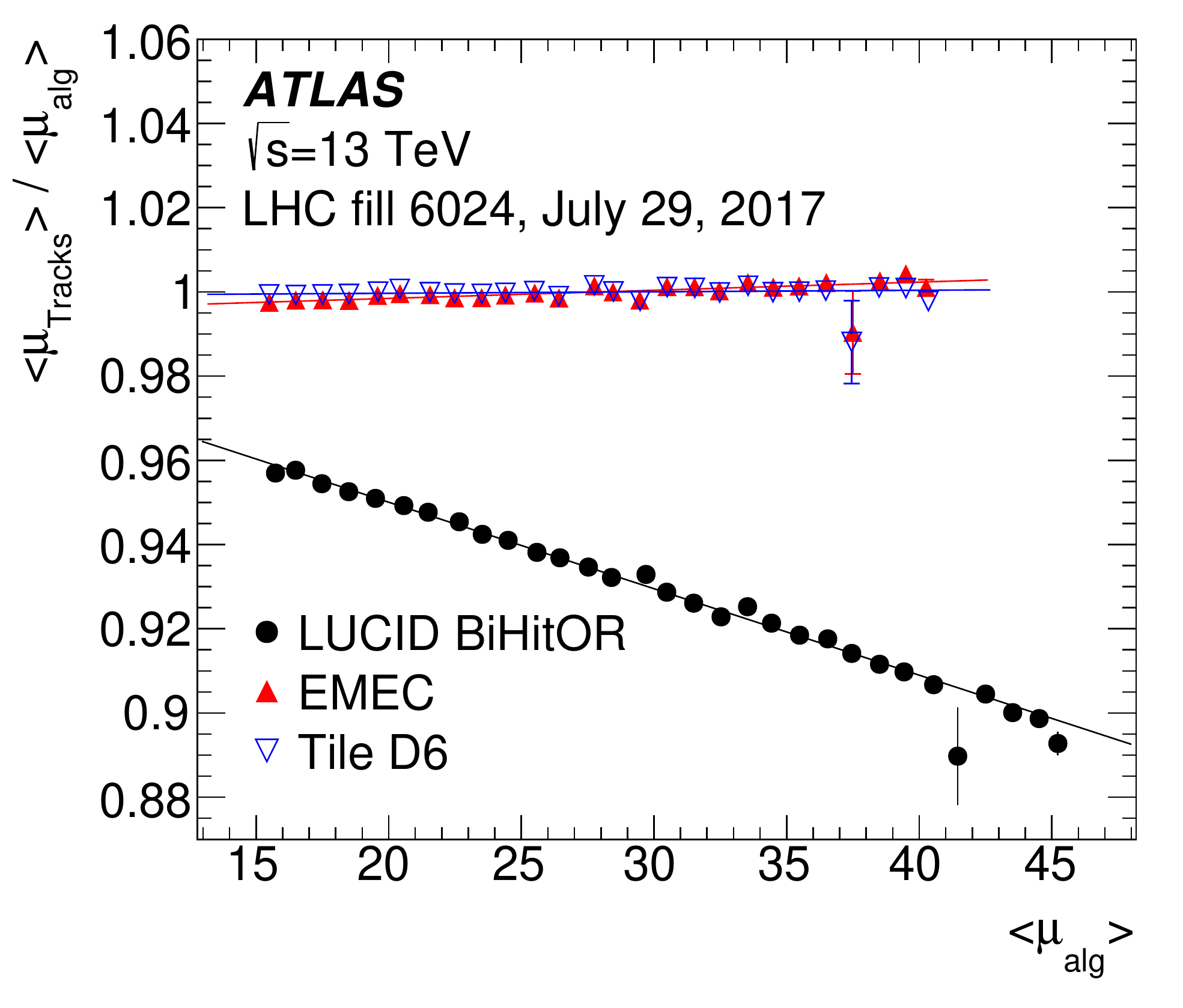}\vspace{-7mm}\center{(a)}}
\parbox{83mm}{\includegraphics[width=76mm]{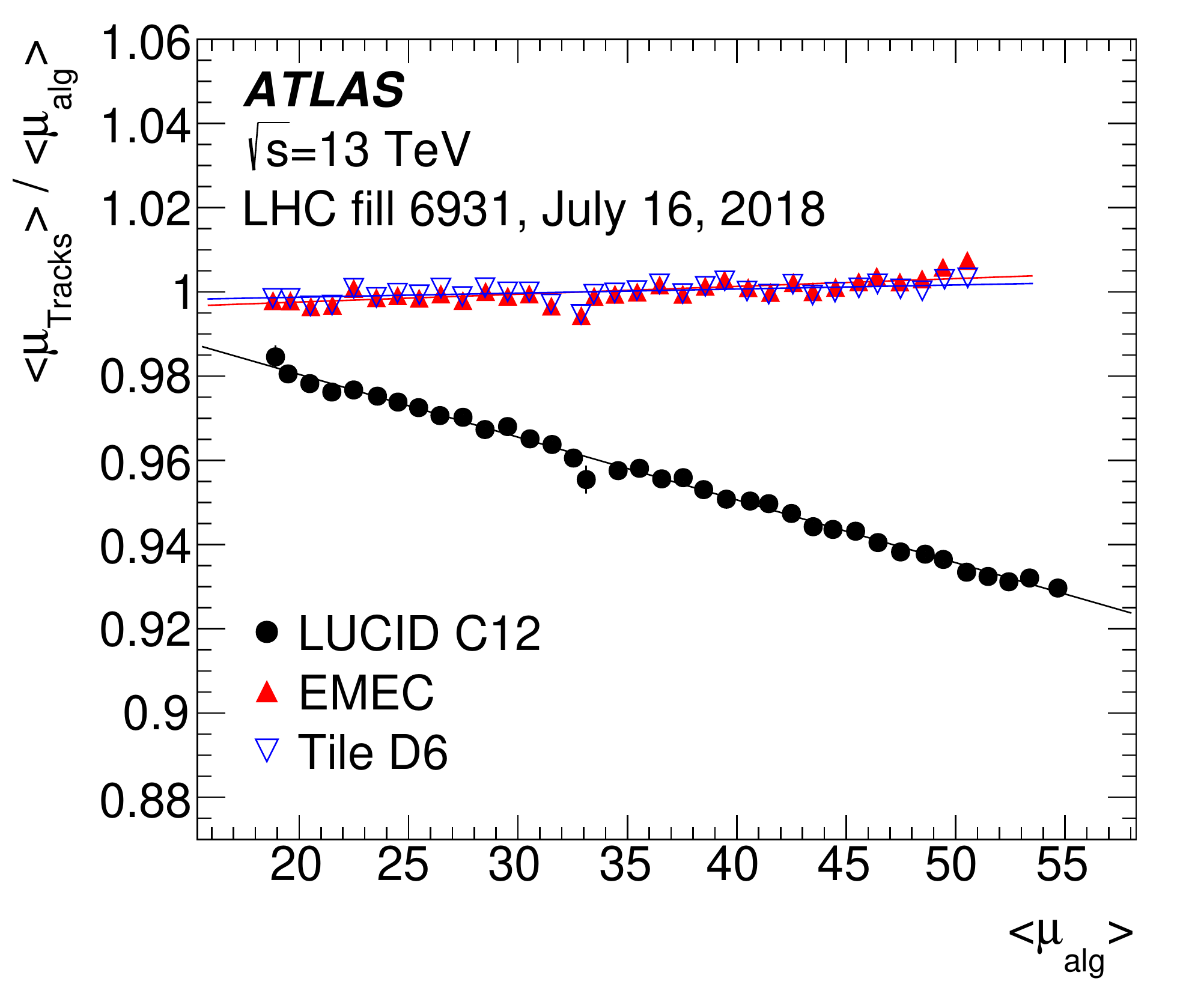}\vspace{-7mm}\center{(b)}}
\caption{\label{f:mudecayphys}Ratios of \meanmu\ (equivalent to ratios of
instantaneous luminosity)
measured by track-counting to that measured by LUCID (black points),
EMEC (red filled triangles) and TileCal D6 (blue open inverted triangles) as
a function of the mean number of interactions per bunch crossing \meanmu\
measured by the latter algorithms in two long LHC physics fills in (a) 2017
and (b) 2018. The track-counting luminosity measurements are normalised to
those from LUCID in the head-on parts of the corresponding vdM runs
at low luminosity, and the calorimeter measurements are normalised to
give the same integrated luminosity as track-counting in the two physics
fills. The lines show linear fits through each set of points. The error bars
show the statistical uncertainties in the track-counting measurements. Time
progresses from right to left in these plots as \meanmu\ decreases through
the fill.}
\end{figure}

The natural luminosity decay during a long LHC physics fill, where the
luminosity, and hence bunch-averaged pileup \meanmu, decrease by a factor of two
or three, provides a way to study the relative responses of the different
luminosity measurements as a function of \meanmu. Figure~\ref{f:mudecayphys}
shows the ratios of track-counting luminosity to several other measured
luminosities in two long physics fills in 2017 and 2018, plotted as a function
of \meanmu, which decreases steadily as the fills proceed. The ratios
of track-counting to TileCal D6-cell luminosities change by at
most 0.3\% as the instantaneous luminosity decreases by a factor of three. The
EMEC measurements include a contribution from material activation, which
builds up over the first few hours of the fill to eventually contribute about
0.5\% of the luminosity signal, leading to a somewhat larger decrease
in the track-counting to EMEC ratio of about 0.7\% over the course of
these long fills. In contrast, the track-counting/LUCID ratios are
$O(10\%)$ smaller than unity at the start of physics fills, and exhibit a
strong dependence on \meanmu\ which is well-described by a linear function,
with different slopes in each year of data-taking. Although the track-counting
measurement was normalised to agree with LUCID in the head-on parts of the
corresponding vdM fills in each year, extrapolating the track-counting/LUCID
ratios to $\meanmu=0.5$ from the fits shown in Figure~\ref{f:mudecayphys} gives
ratios which differ from unity by $O(1\%)$, suggesting that the relative
response of the two detectors may also be affected by running with bunch
trains, and with a crossing angle between the colliding beams. The level
of agreement between the luminosities measured by track-counting, EMEC and
TileCal, and the strong disagreement between track-counting and LUCID,
suggest that residual $\mu$-dependencies in the track-counting, TileCal
and EMEC measurements are at the percent level or less, and that LUCID
overestimates the luminosity in physics conditions by $O(10\%)$, with a strong
pileup dependence.
 
\begin{figure}[tp]
\parbox{83mm}{\includegraphics[width=76mm]{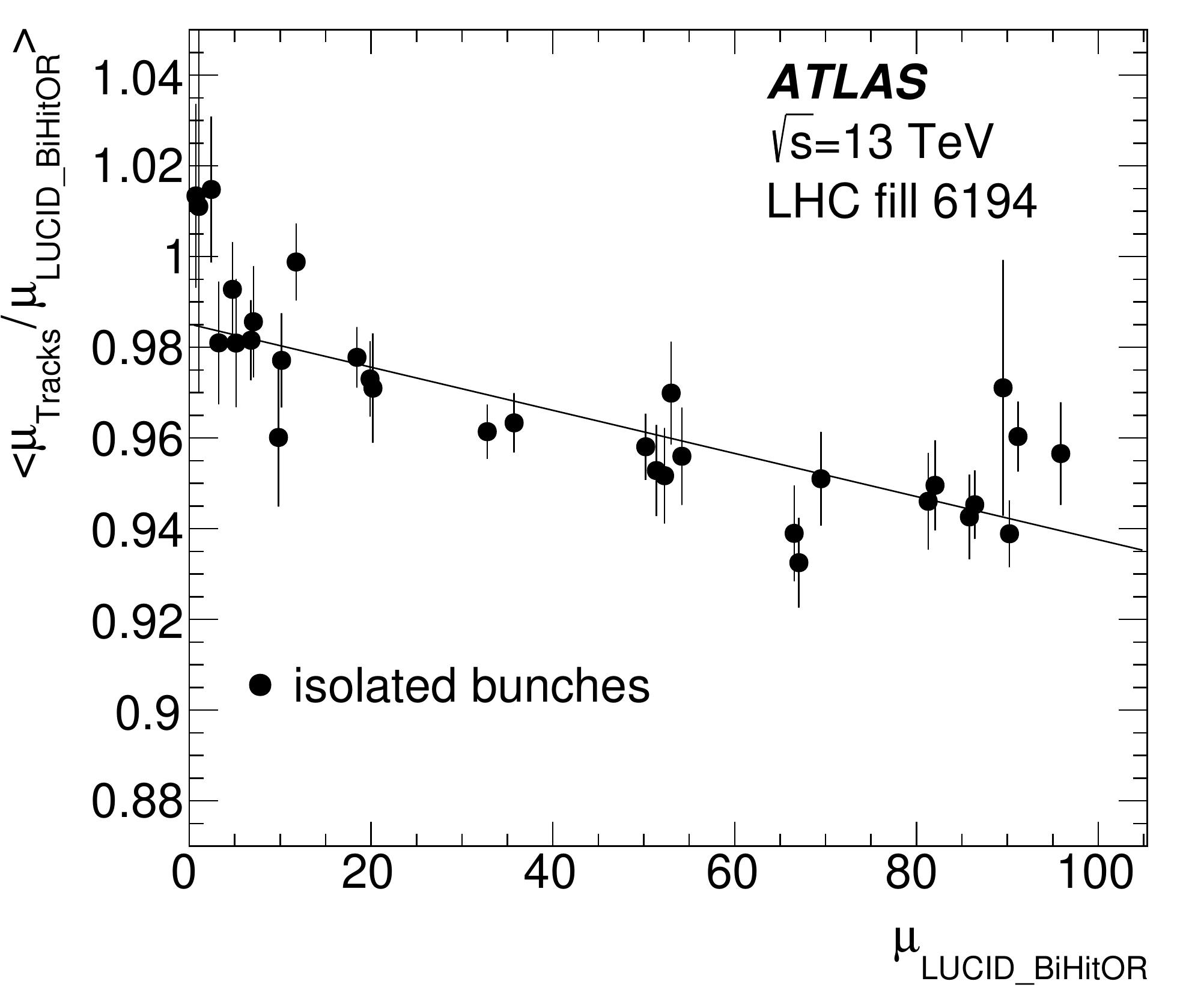}\vspace{-6mm}\center{(a)}}
\parbox{83mm}{\includegraphics[width=76mm]{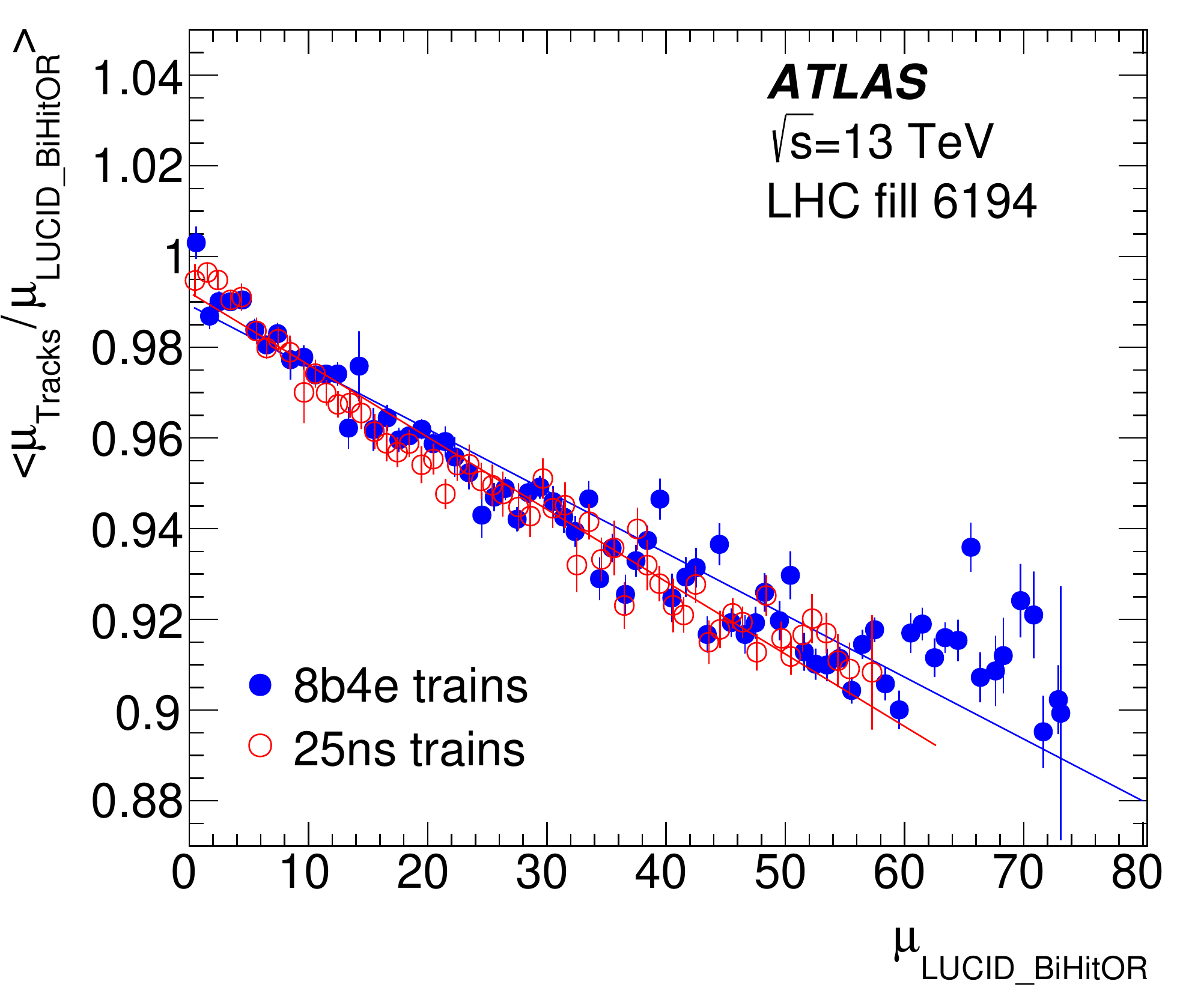}\vspace{-6mm}\center{(b)}}
\caption{\label{f:lcdtrkbgroup}Mean values of the bunch-by-bunch ratio of
$\mu$ measured by track-counting and LUCID as a function of the $\mu$
measured by LUCID from the $\mu$-scan in LHC fill 6194 with a mixture of
isolated bunches, 25\,ns and 8b4e trains. The ratios are shown separately for
(a) isolated bunches and (b) the two types of bunch trains. The lines show
linear fits through each set of points. The error bars show the uncertainties
in the track-counting measurements.}
\end{figure}

The relative dependence of the LUCID and track-counting luminosity measurements
was further studied by performing $\mu$-scans, where the luminosity was
varied between zero and the maximum achievable, by partially
separating the beams in the transverse plane at the ATLAS IP.
Since both LUCID and track-counting measure
$\mu$ for individual bunches, the performance as a function of the LHC bunch
pattern and of the position of individual bunches within a bunch train
could  also be investigated.
Figure~\ref{f:lcdtrkbgroup} shows results from a special fill (6194) recorded
in 2017, where the LHC filling pattern contained three isolated bunches,
two 48-bunch trains with 25\,ns spacing and two 56-bunch 8b4e trains with a
repeating pattern of eight bunches separated by 25\,ns followed by four empty
bunch-slots. During the $\mu$-scan, the ATLAS luminosity block boundaries
were synchronised to the steps in LHC beam separation, such that each LB
corresponds to a period with constant instantaneous luminosity at a different
$\mu$ value. The LBs where the luminosity was varying as the beam separation
was changed were discarded. Figure~\ref{f:lcdtrkbgroup}(a) shows the
ratio of $\mu$ from track-counting to that from LUCID averaged over the
three isolated bunches, as a function of $\mu_\mathrm{LUCID}$.
There is a significant negative slope (indicating that LUCID overestimates
the luminosity compared to track-counting at high pileup), but the slope
is several times smaller than those seen in Figure~\ref{f:mudecayphys}.
Figure~\ref{f:lcdtrkbgroup}(b) shows the same ratio averaged over all the bunches
in the 25\,ns and 8b4e trains separately. Here, the slopes are larger,
demonstrating a larger $\mu$-dependence in train bunches, which is slightly
stronger for 25\,ns than 8b4e bunch trains. Deviations from the linear slope
are also visible at the lower and upper ends of the $\mu$ range. Due to the
spread in the bunch intensities and emittances produced by the LHC injector
chain, the $\mu$ range sampled for 8b4e trains is slightly larger than
that for 25\,ns trains, and the isolated bunches extend almost to $\mu=100$,
much larger values than those encountered in normal LHC fills during Run~2.
 
\begin{figure}[tp]
\parbox{83mm}{\includegraphics[width=76mm]{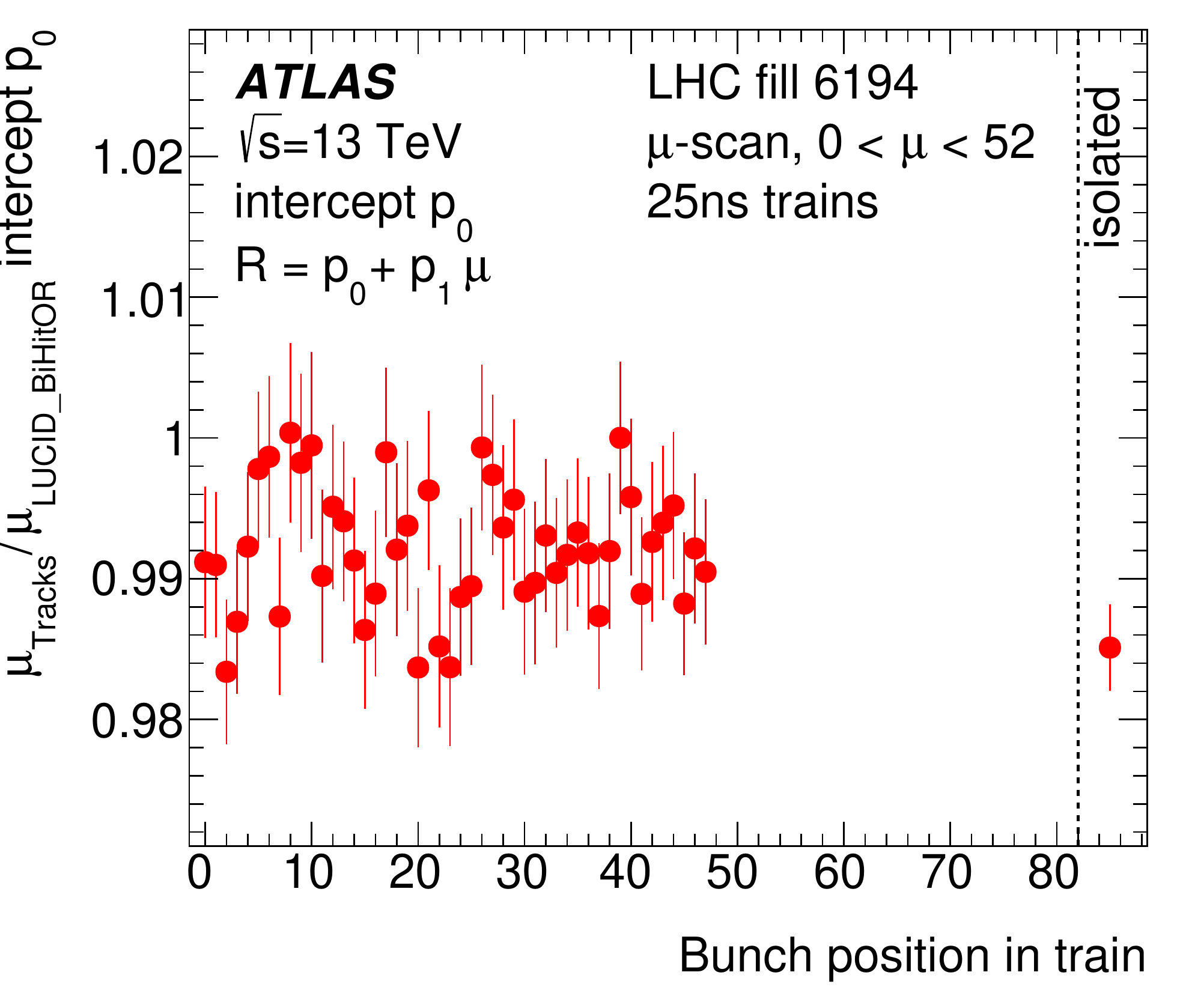}\vspace{-6mm}\center{(a)}}
\parbox{83mm}{\includegraphics[width=76mm]{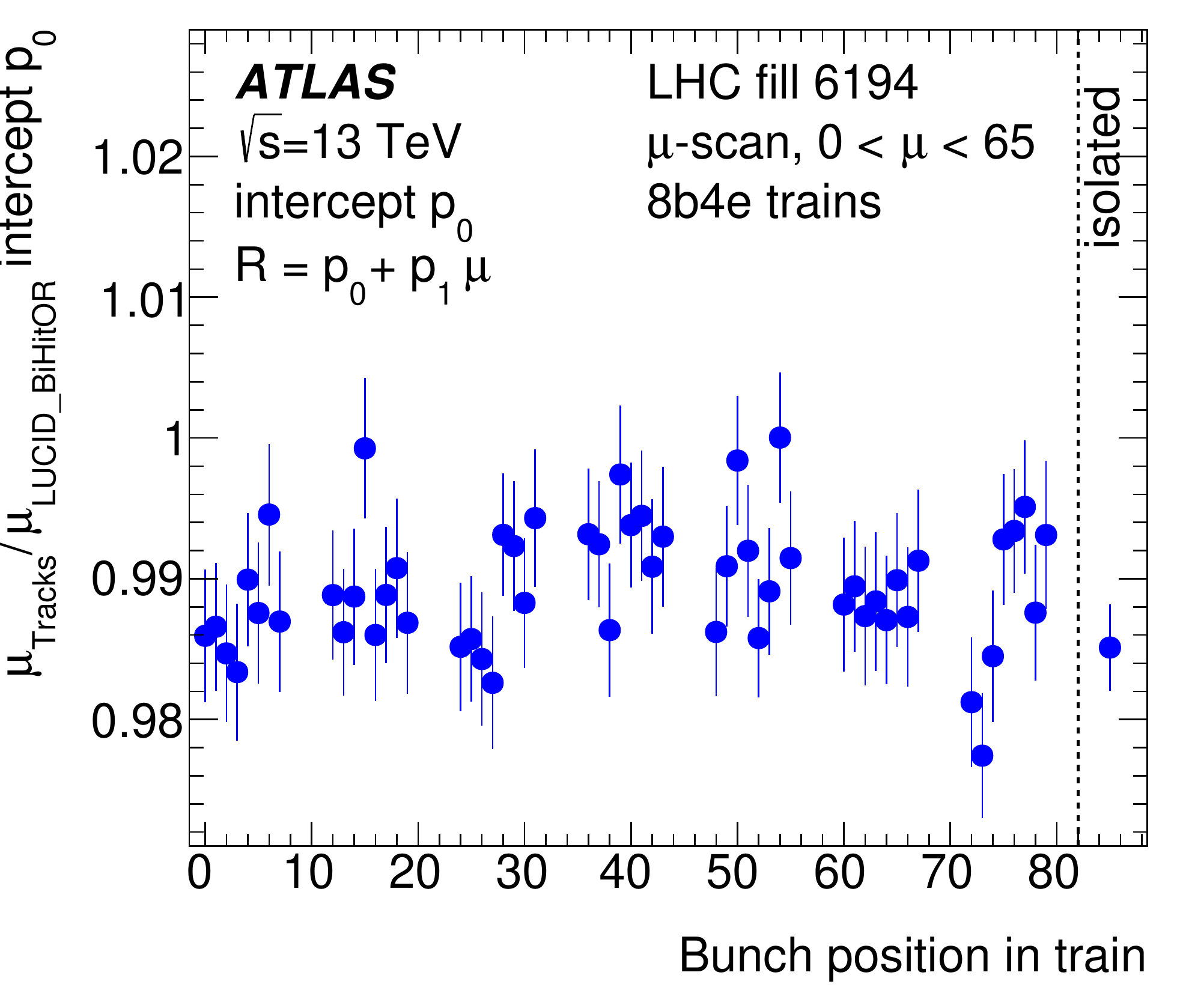}\vspace{-6mm}\center{(b)}}
 
\parbox{83mm}{\includegraphics[width=76mm]{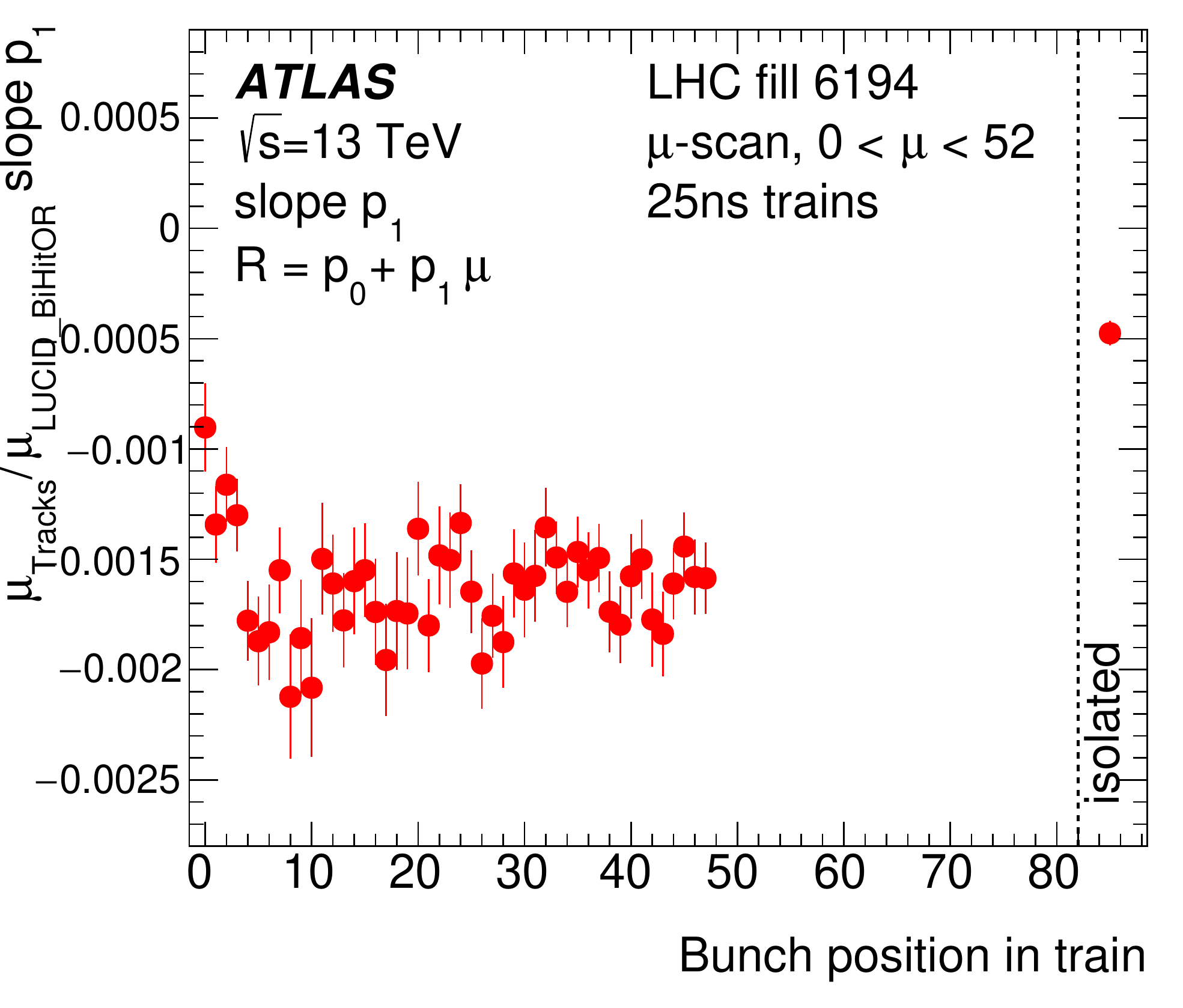}\vspace{-6mm}\center{(c)}}
\parbox{83mm}{\includegraphics[width=76mm]{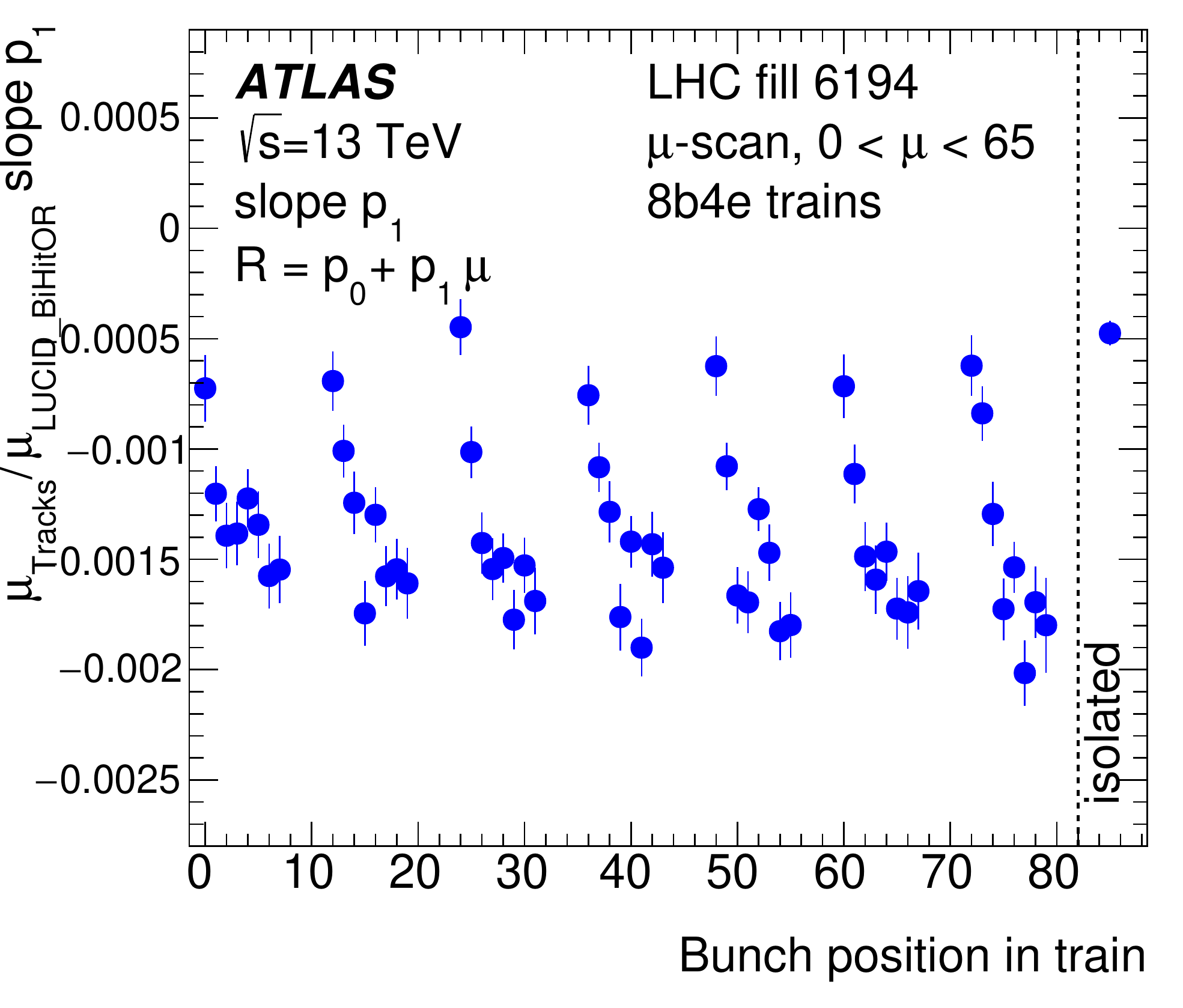}\vspace{-6mm}\center{(d)}}
\caption{\label{f:lcdtrktrain}Fitted intercept $p_0$ (upper plots) and slope
parameter $p_1$ (lower plots) for the track-counting/LUCID $\mu$ ratios in
LHC fill 6194 as a function of position within the bunch train, counting the
first bunch in the train as position zero. The left plots show the results for
the 48-bunch 25\,ns trains, and the right plots those for the 56-bunch 8b4e
trains with eight colliding bunches separated by 25\,ns followed by a four
bunch-slot gap. The right-most points in each plot show the corresponding values
for isolated bunches. The error bars show the statistical uncertainties from
the track-counting measurements.}
\end{figure}

The $\mu$-ratios between track-counting and LUCID were also analysed as a
function of bunch position $b$ within the train, fitting the ratio $R^b=\mu^b_\mathrm{Tracks}/\mu^b_\mathrm{LUCID}$ as a function of $\mu_\mathrm{LUCID}$ to a
linear function with intercept $p_0^b$ and slope $p_1^b$.
The resulting fitted parameters are shown separately for the 25\,ns and 8b4e
bunches in Figure~\ref{f:lcdtrktrain}. Whilst the intercepts $p_0$ are similar in
isolated bunches and both types of train bunches, and do not depend significantly
on train position, the slopes $p_1$ show significant structure. The first bunch
of a 25\,ns train has a similar $p_1$ (and hence $\mu$-dependence) to that
of isolated bunches. The subsequent bunches show increasingly negative
$p_1$ (and larger $\mu$-dependence), reaching a plateau about six bunches
into the train, after which there is a slight recovery to a constant value
through the bulk of the train. The 8b4e trains show a similar initial behaviour,
but the four bunch-slot gap is enough for the ratio to `recover', and the
first bunch of the next eight-bunch subtrain again behaves similarly to
an isolated bunch. This structure is consistent with the slightly smaller
$\mu$-dependence for 8b4e compared to 25\,ns trains seen in
Figure~\ref{f:lcdtrkbgroup}.
 
Since the calorimeter measurements cannot resolve the luminosity of individual
bunches, the bunch-by-bunch stability of track-counting measurements
cannot be validated except by `internal' comparisons of different track-counting
working points (see Section~\ref{ss:trkperf} below). However, given the small
relative $\mu$-dependence between the bunch-integrated track-counting
measurements and the EMEC and TileCal calorimeter measurements
shown in Figure~\ref{f:mudecayphys}, it is reasonable to attribute
the bulk of the effects seen in Figures~\ref{f:lcdtrkbgroup}
and~\ref{f:lcdtrktrain} to LUCID. A likely contributor is `signal migration'
\cite{lucid2}, the overlap of several below-threshold signals in the same LUCID
PMT from different
$pp$ interactions, which combine to produce an above-threshold signal
which gets counted as a hit, causing deviations from the Poisson assumption in
Eq.~(\ref{e:phit}). These effects become more likely for higher $\mu$ values,
leading to an overestimate of the luminosity at high pileup. The structures
seen in Figure~\ref{f:lcdtrktrain} also suggest that particles from previous
bunch crossings can contribute to this signal migration, making the overestimate
worse for bunches further from the start of the train.

\subsection{Performance of track-counting algorithms}\label{ss:trkperf}
 
\begin{figure}[tp]
\parbox{83mm}{\includegraphics[width=78mm]{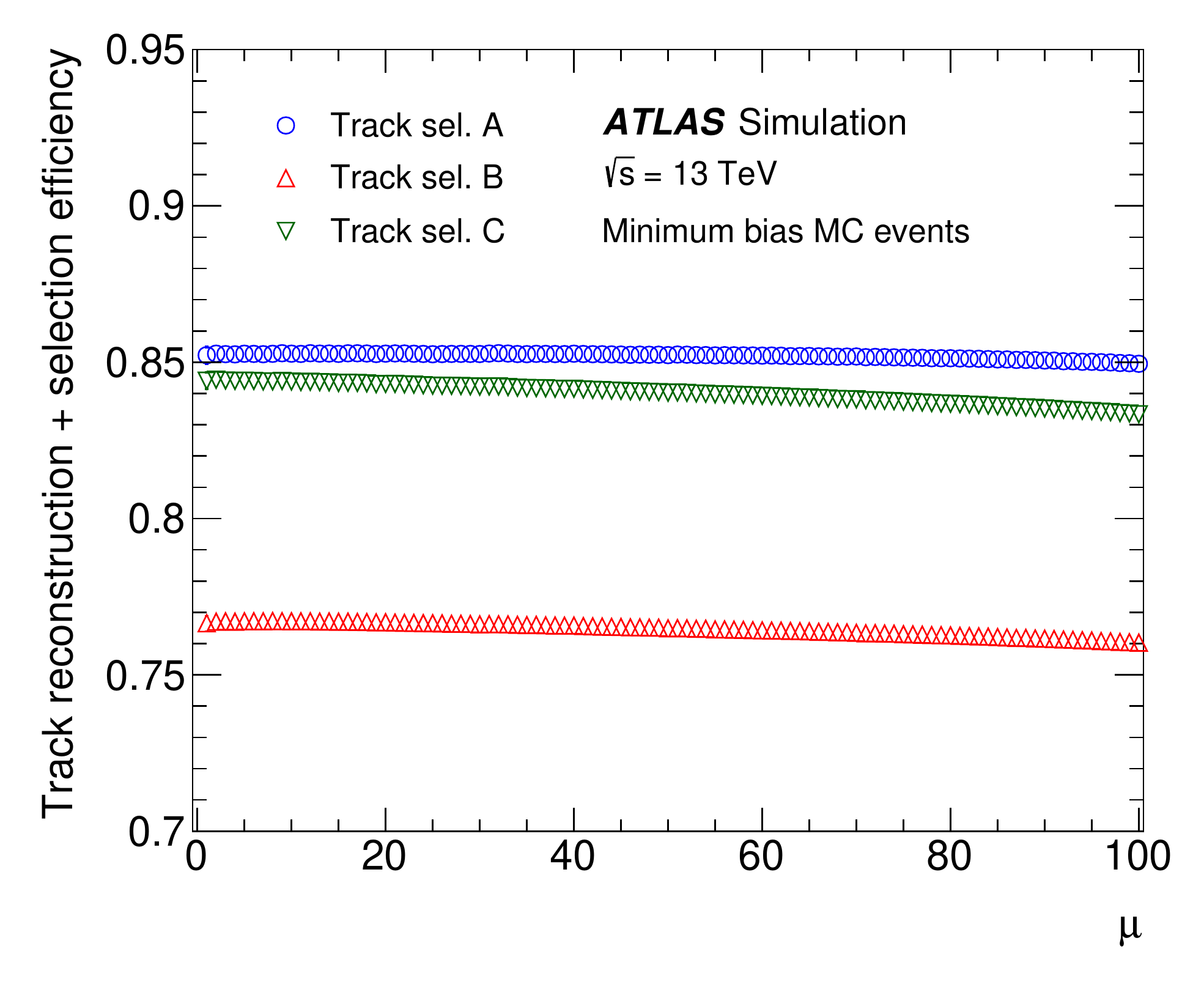}\vspace{-6mm}\center{(a)}}
\parbox{83mm}{\includegraphics[width=78mm]{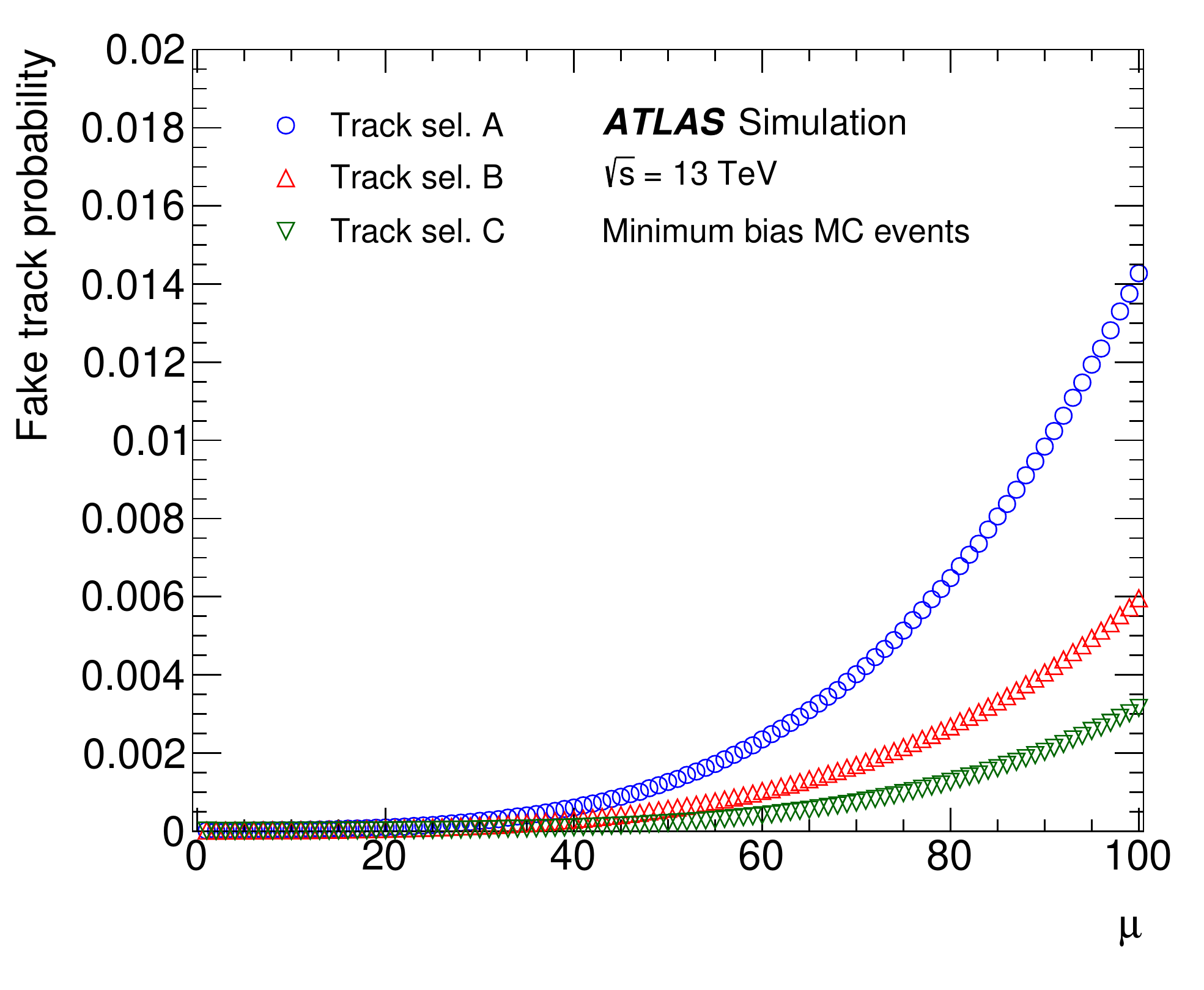}\vspace{-6mm}\center{(b)}}
\caption{\label{f:trkselmc}Performance of the different track-selection working
points for the track-counting luminosity measurement in simulation as a
function of $\mu$, showing (a) combined track reconstruction and
selection efficiency, and (b) fake-track probability.}
\end{figure}

The performance of the track-counting luminosity measurements was studied
using the standard ATLAS detector simulation \cite{SOFT-2010-01} based
on {\scr Geant4} \cite{geant4}. A sample of events containing multiple overlaid
inelastic $pp$ collisions was generated with {\scr Pythia 8.186}
\cite{pythia8} and the A3 set of tuned parameter values
\cite{ATL-PHYS-PUB-2016-017},
with a flat $\mu$ profile in the range $0<\mu<100$.
The sample included the simulation
of bunch trains, with 53~trains each containing 48 colliding-bunch pairs
with 25\,ns spacing, simulating some of the effects of in-time and out-of-time
pileup. These events were processed using the same dedicated track
reconstruction settings as used for the track-counting event stream in
real data, only making use of information from the pixel and SCT detectors.
Figure~\ref{f:trkselmc}(a) shows the efficiency to reconstruct a track
and select it using each of the working points (normalised to the
true number of tracks satisfying the \pt\ and $|\eta|$ requirements for
each working point  shown in Table~\ref{t:trkwp}), as a function of $\mu$.
For the baseline selection~A, the efficiency decreases by only about 0.3\%
between $\mu=0$ and $\mu=100$. The efficiencies for the other two working
points decrease faster, reflecting the more stringent hit requirements
(see Table~\ref{t:trkwp}).
Figure~\ref{f:trkselmc}(b) shows the fraction of fake tracks
(tracks which cannot be matched to a generator-level track with a hit-based
matching probability $P_\mathrm{match}>0.5$ as defined in
Ref.~\cite{ATL-PHYS-PUB-2015-051}), which is around 0.1\% at $\mu=50$ for
selection~A,  rising non-linearly to about 1\% at $\mu=90$ as the
detector occupancy increases. Here, the alternative selections show
smaller fake-track rates, again reflecting the more stringent hit requirements.
The simulation includes some of the effects which lead to a loss of performance
at high pileup, such as the `01X' hit pattern definition in the SCT, which
requires that there be no hit on a strip in the immediately preceding
bunch-crossing for a valid hit to be read out. However, it does not include
an inefficiency in the pixel detector due to time-over-threshold effects,
where a pixel which is hit again whilst the decaying signal from an initial hit
is still over the readout threshold will not record the second hit. This
leads to an inefficiency which may last for tens of bunch-crossings, and
particularly affects selection~B, which rejects
tracks with a missing pixel hit where one is expected.
 
\begin{figure}[tp]
\parbox{83mm}{\includegraphics[width=78mm]{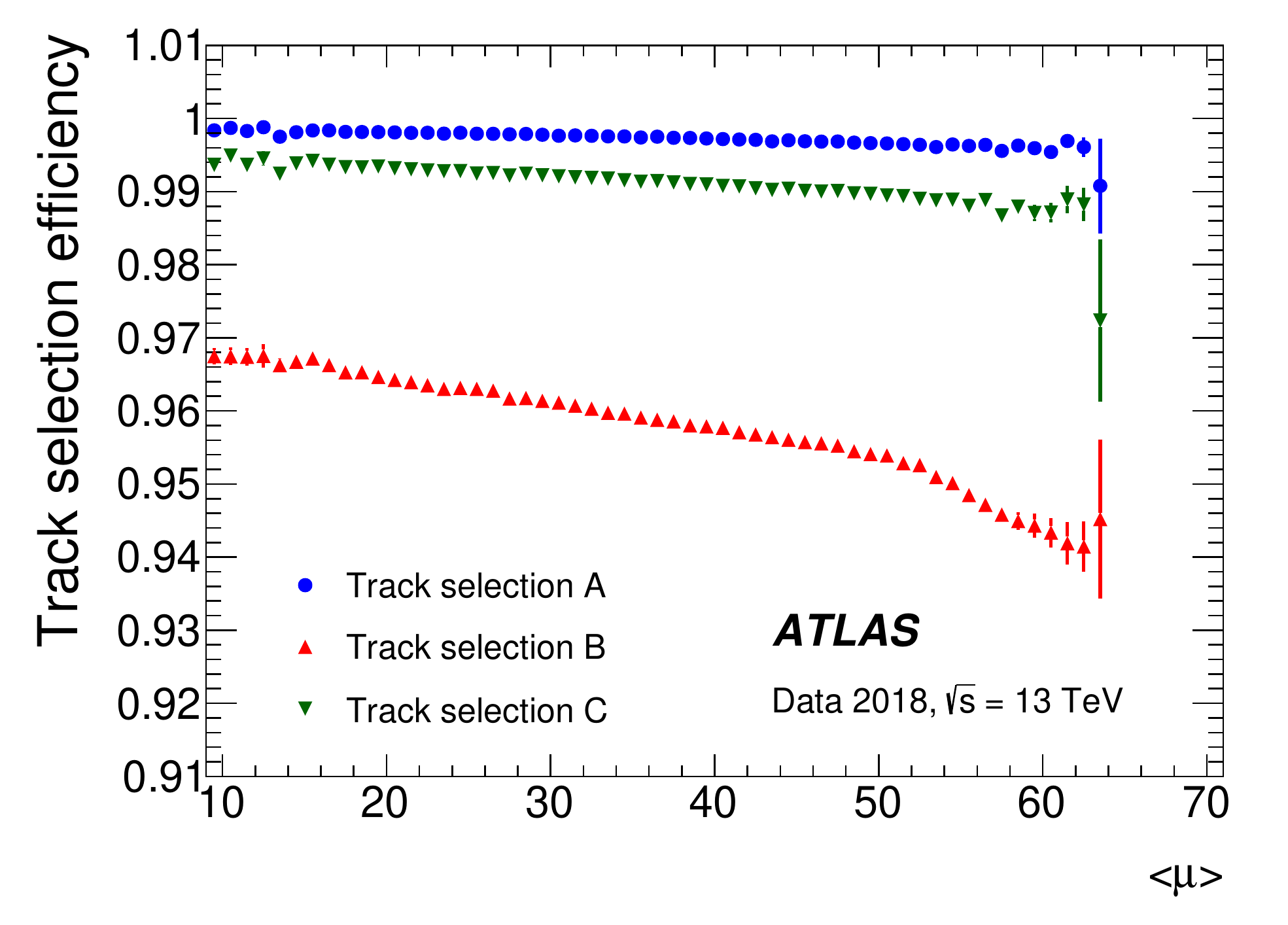}\vspace{-6mm}\center{(a)}}
\parbox{83mm}{\includegraphics[width=78mm]{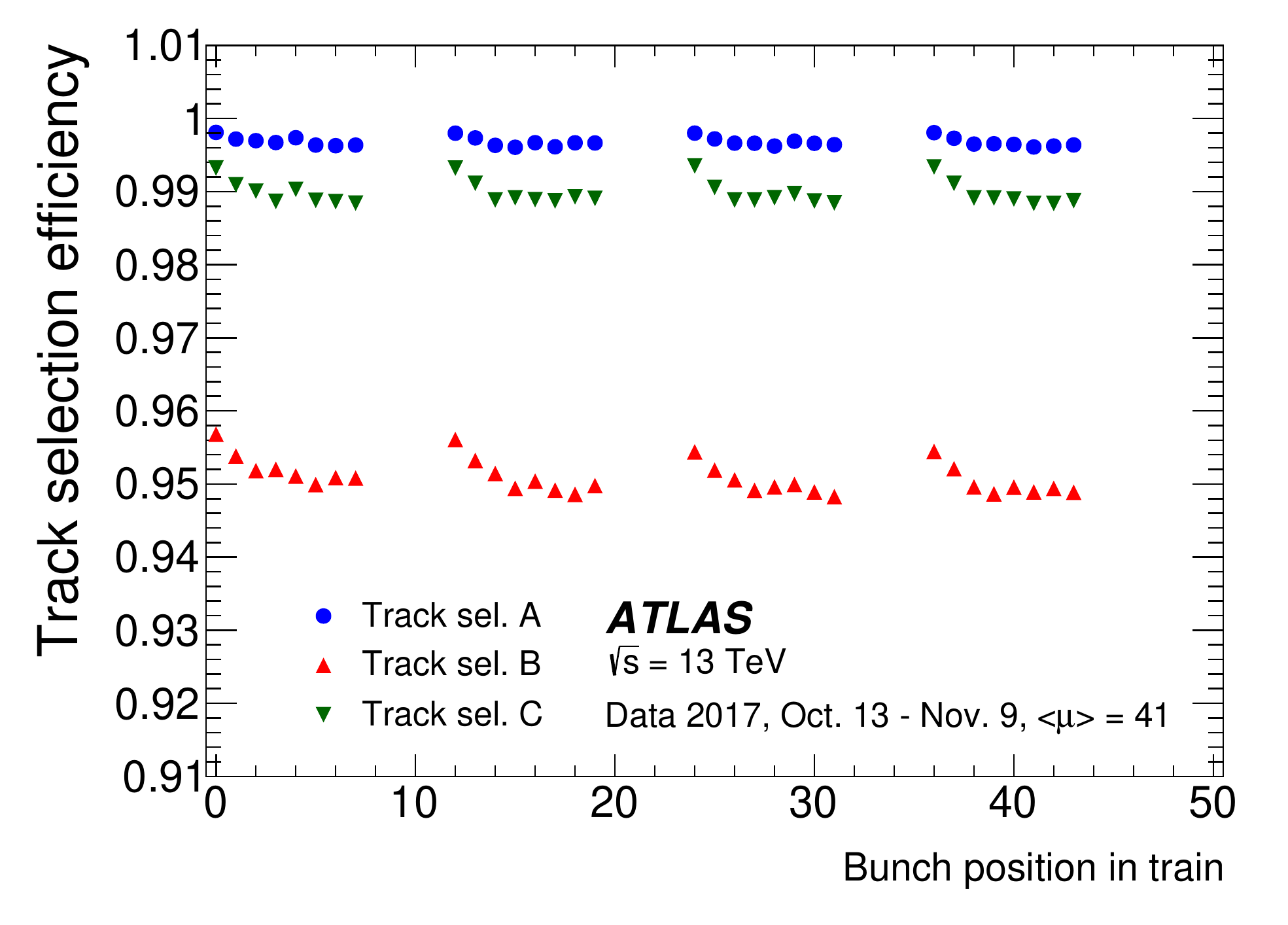}\vspace{-6mm}\center{(b)}}
\caption{\label{f:trkselmubun}Track-selection efficiency for the
track-counting luminosity measurement determined using $Z\rightarrow\mu\mu$
events in data: (a) Efficiencies for the different selections as a
function of \meanmu\ in 2018 data; (b) Efficiencies
for data recorded during part of the 8b4e collision period as a function of the
position of the bunch within the 8b4e train, numbering the first bunch
as position zero.}
\end{figure}

The efficiencies of the track-selection working points were also studied using
$Z\rightarrow\mu\mu$ events in data, where both muons were reconstructed
as isolated `combined' muons with $\pt>20$\,\GeV, consisting of matched tracks
in the ATLAS inner detector
and muon spectrometer \cite{MUON-2018-03}. The fraction of the inner detector
tracks associated with these muons that pass the track-counting selections,
which only rely on criteria related to the inner detector, gives
a measure of the inner detector track selection efficiency relative to the
inclusive initial selection. The latter just requires a very loose reconstructed
track passing $\pt$ and $|\eta|$ requirements, together with minimal
requirements on the number of inner detector hits to ensure the track can be
reconstructed \cite{PERF-2015-10}.  Figure~\ref{f:trkselmubun}(a)
shows the resulting efficiencies measured in 2018 data for the different
working points. The efficiency for selection~A
is reasonably stable with $\mu$, decreasing by about 0.4\% between
$\mu=0$ and $\mu=60$, a decrease which is however larger than that seen in
simulation. The efficiencies of the other selections show much stronger
dependence on $\mu$, especially for selection~B. Figure~\ref{f:trkselmubun}(b)
shows the efficiencies measured in part of the 2017 8b4e data-taking period
as a function of
the bunch position along the train. Again, the selection~A working point
is  rather stable, decreasing by about 0.2\% along the eight-bunch subtrains,
whilst selection~B shows $O(1\%)$ effects.
Although these efficiencies are measured for muons with $\pt>20$\,\GeV,
simulation studies showed that they also capture the main effects for the
track-counting selections, which are dominated by hadrons with lower \pt.
 
\begin{figure}[tp]
\centering
 
\includegraphics[width=150mm]{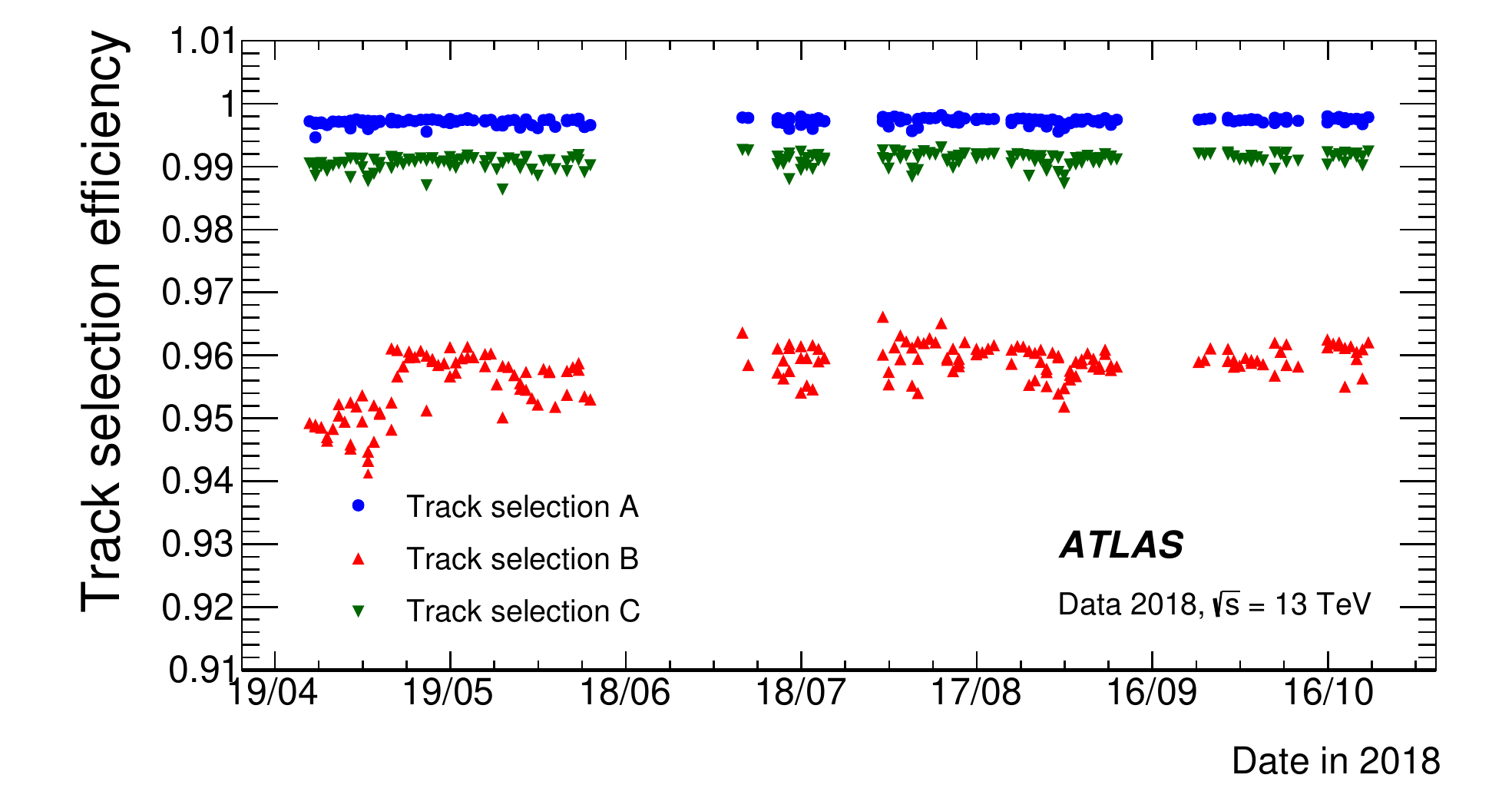}
\caption{\label{f:trkseltime}Track-selection efficiency for the track-counting
luminosity measurement determined using $Z\rightarrow\mu\mu$ events as a
function  of time within the entire 2018 high-luminosity physics dataset.
The efficiencies of the different track selection  working points are shown
separately.}
\end{figure}

Figure~\ref{f:trkseltime} shows the selection efficiencies measured
using $Z\rightarrow\mu\mu$ events  as a function of
time over the entire 2018 dataset. Selections~A and~C are very
stable, changing by less than 0.2\% during the course of the year.
Selection~B is much less stable, with about a 1\% lower efficiency in the
first few weeks. This was traced to an incorrect masking of inactive pixel
modules in the track reconstruction, leading to more tracks with unexpected
pixel holes. Since this early data also had atypically
high \meanmu\ values, this effect also causes the decrease in efficiency
for $\meanmu>50$ visible for selection~B in Figure~\ref{f:trkselmubun}(a).
Similar levels of stability for the selection~A working point
were seen for the datasets from other data-taking years within Run~2. Given
these observations, no attempt was made to use the $Z$-based efficiency
measurements to correct the track-counting
luminosity as a function of time during each year.
 
These results using both simulation and real collision data demonstrate that
the track-counting luminosity measurement with the selection~A working point is
much more robust than the LUCID luminosity measurements against variations of
$\mu$ and bunch position. The track-counting measurement also shows
excellent stability in time, and can thus be used as a reliable relative
reference to calibrate other luminosity algorithms, once its absolute
calibration has been determined by normalisation to LUCID in the vdM run.

\subsection{LUCID correction strategy}\label{ss:lucidcorr}
 
Exploiting the excellent stability of the track-counting luminosity measurement,
the non-linearity in LUCID was corrected as a function of \meanmu\ using one
or more long high-luminosity reference fills in each year of data-taking. Each
reference fill was used to derive a correction of the form
\begin{equation}\label{e:mucorr}
\mmucorr = p_0 \mmuuncorr + p_1 (\mmuuncorr)^2\ ,
\end{equation}
where \mmuuncorr\ is the uncorrected and \mmucorr\ the corrected LUCID \meanmu\
value. The parameters $p_0$ and $p_1$ were obtained from a linear fit to
the ratio of \meanmu\ values measured by track-counting and LUCID,
$R=\langle\mu_\mathrm{Tracks}\rangle/\langle\mu_\mathrm{LUCID}\rangle$, as a
function of \mmuuncorr, as shown in Figure~\ref{f:mudecayphys}. The track-counting
luminosity was first normalised to the absolute luminosity measured by LUCID
in parts of the vdM fill with stable head-on collisions at ATLAS, ensuring
that $R=1$ at low pileup ($\mu\approx 0.5$) with isolated bunches.
This correction also absorbs the $O(1\%)$ change in the LUCID calibration caused
by the non-zero LHC beam crossing angle used in physics runs, and the
$O(0.2\%)$ afterglow contribution to the LUCID luminosity signal
in bunch train running. In principle,
Eq.~(\ref{e:mucorr}) could also be written in terms of the bunch-by-bunch
$\mu$, correcting the LUCID measurement as a function of the instantaneous
luminosity in each individual bunch and potentially compensating for the
bunch-to-bunch variations seen in Figure~\ref{f:lcdtrktrain}. However, since
ATLAS physics analyses make use of the data from all LHC bunches, an
average correction based on \mmuuncorr\ is sufficient, and a more complex
bunch-position-dependent correction scheme was not used.
 
\begin{figure}[tp]
\parbox{83mm}{\includegraphics[width=78mm]{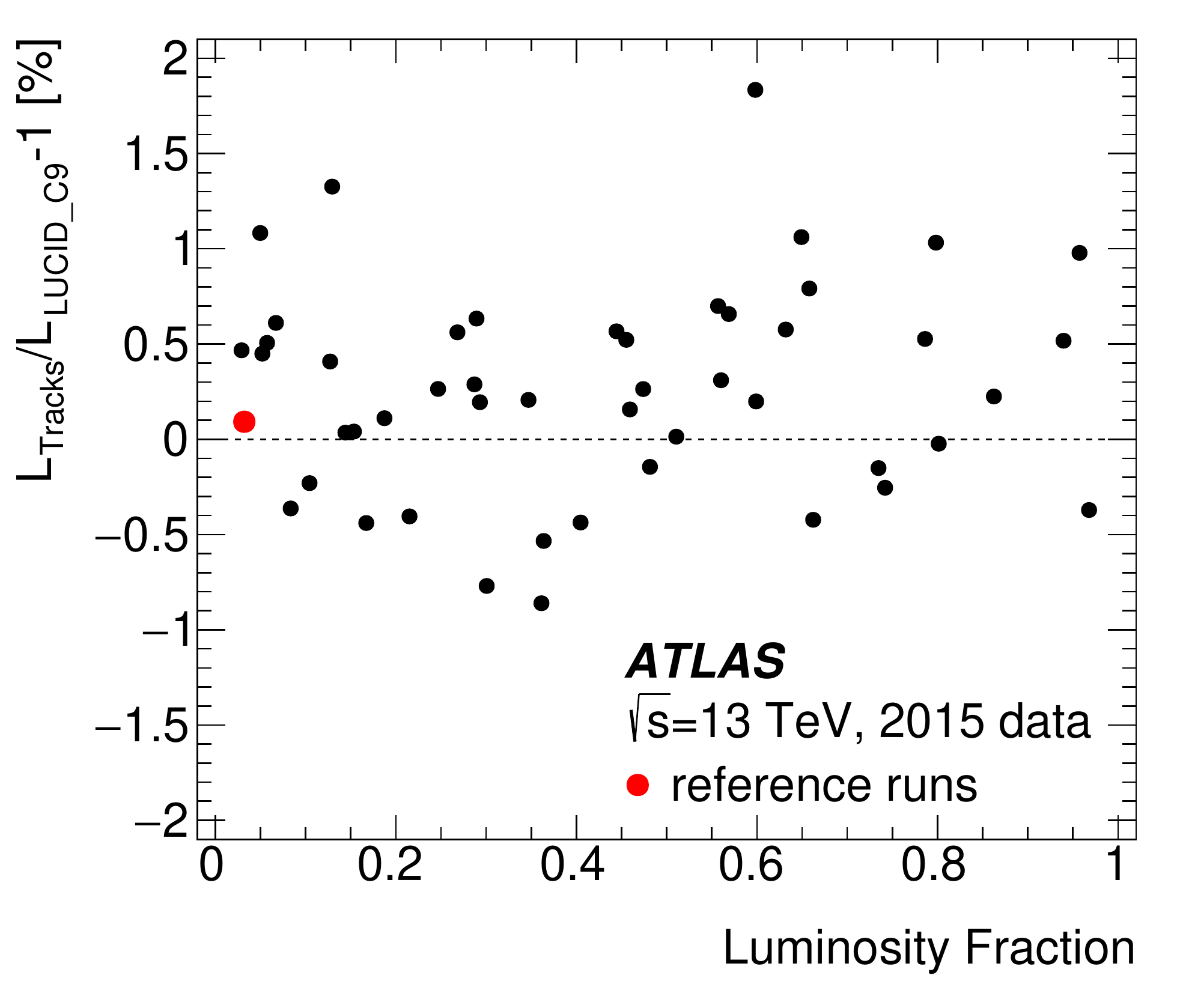}\vspace{-6mm}\center{(a)}}
\parbox{83mm}{\includegraphics[width=78mm]{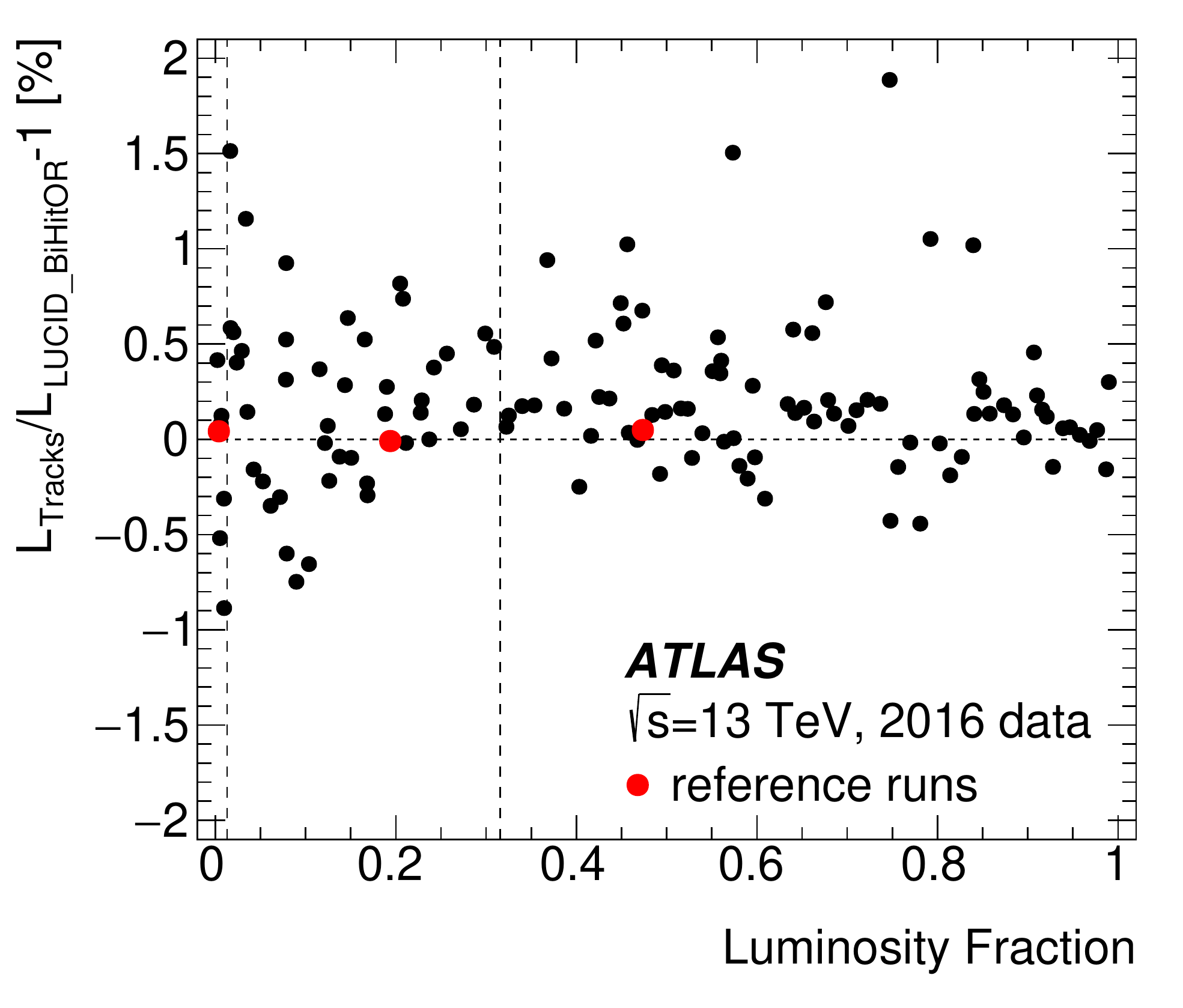}\vspace{-6mm}\center{(b)}}
\parbox{83mm}{\includegraphics[width=78mm]{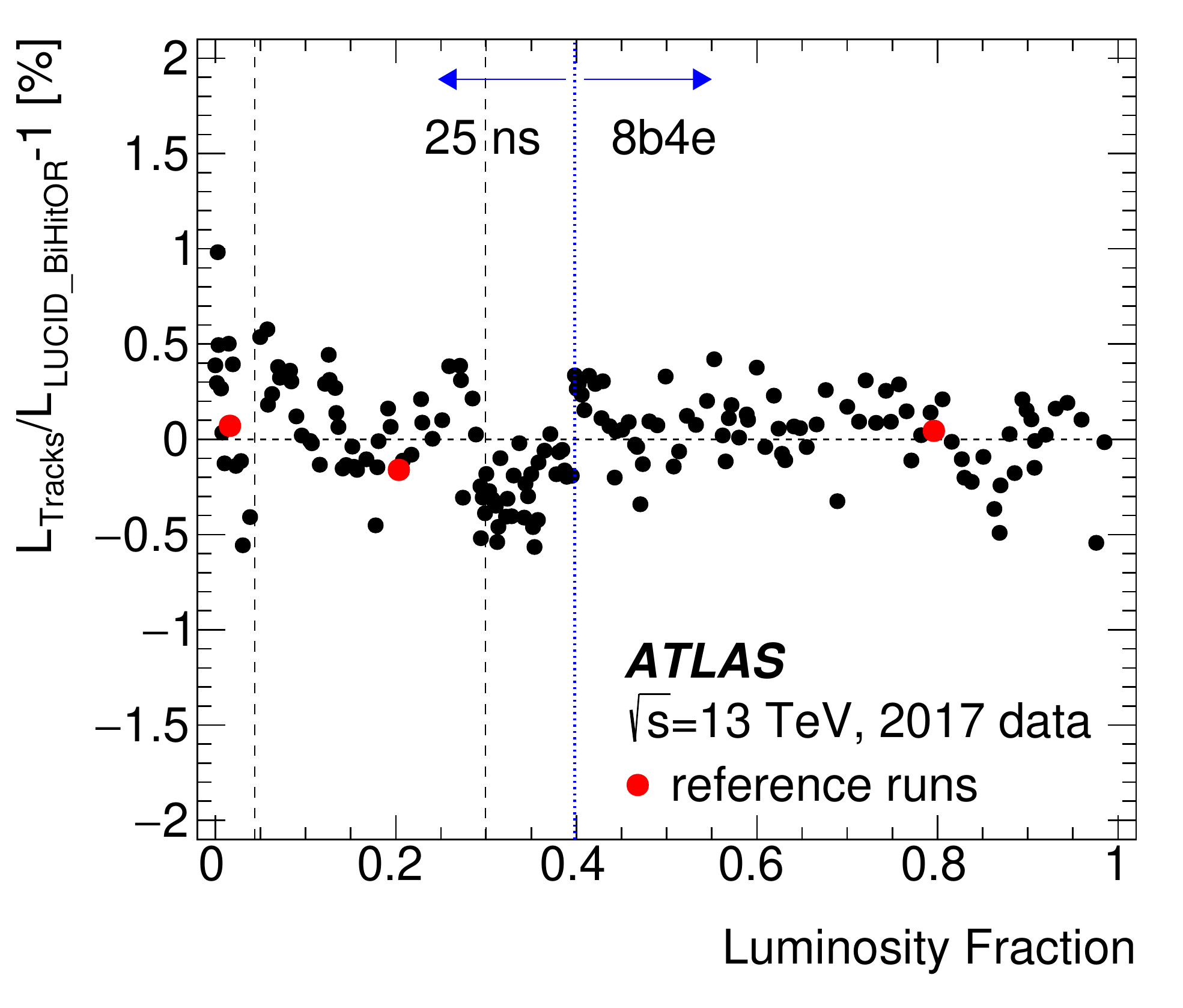}\vspace{-6mm}\center{(c)}}
\parbox{83mm}{\includegraphics[width=78mm]{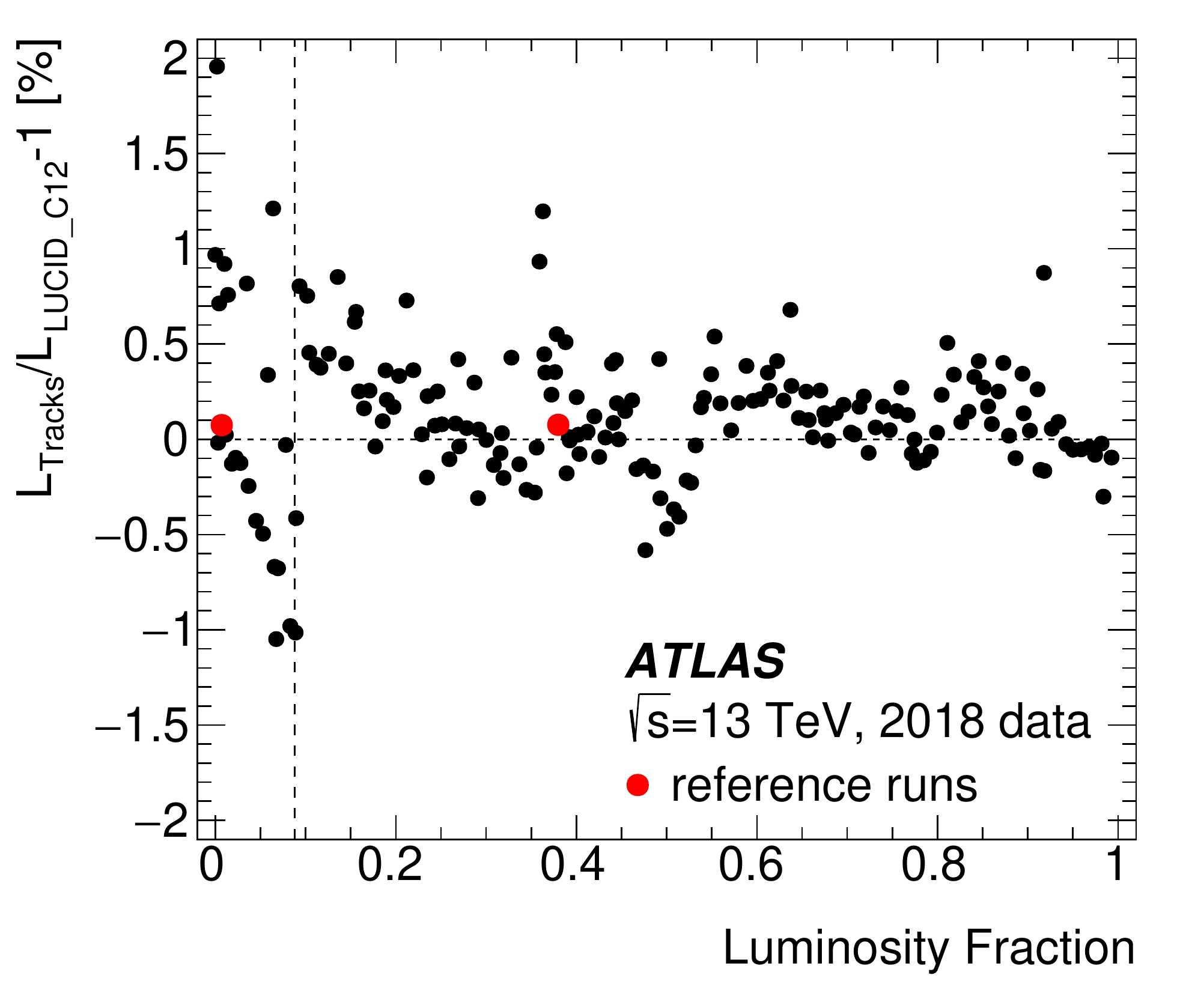}\vspace{-6mm}\center{(d)}}
\caption{\label{f:lucidepoch}Fractional differences in run-integrated luminosity
between the track-counting and corrected LUCID luminosity algorithms, plotted
as functions of cumulative delivered luminosity in each data-taking year.
The division of the data into multiple epochs is shown by the vertical dashed
lines (except in 2015), the separation of 25\,ns and 8b4e running periods in
2017 is shown by the blue dotted line,
and the reference runs used to derive LUCID $\mu$-corrections are shown by
the red points. Only runs with at least 500 colliding-bunch pairs and one hour
of data-taking are shown.}
\end{figure}

The correction of Eq.~(\ref{e:mucorr}) ensures that the LUCID luminosity agrees
well with track-counting throughout the reference fill. If the LUCID
response has no time dependence, a single reference fill and correction
would be adequate for a whole year of data-taking. The regular PMT high-voltage
adjustment based on the LUCID $^{207}$Bi calibration signals ensures that
this is approximately the case, but comparisons with track-counting show
some residual variations, particularly during the first few weeks of
data-taking in each year, when the number of colliding bunches and
instantaneous luminosity increased during the LHC intensity ramp-up phase.
Better agreement with track-counting throughout the year was therefore
obtained by dividing each year into several correction periods, or epochs,
with a long LHC fill representative of each epoch being used as a reference run.
The change from 25\,ns to 8b4e bunch trains in 2017 also required separate
epochs for the two periods (the 8b4e epoch also includes a
period of unstable 25\,ns running with a reduced number
of bunches and many short fills immediately preceding the change to 8b4e).
One epoch in 2015, three epochs in each of 2016
and 2017, and two in 2018 were used in order to achieve good agreement
between LUCID and track-counting throughout each year. Figure~\ref{f:lucidepoch}
shows the resulting level of agreement between track-counting
and LUCID for each data-taking run (normally equivalent to one LHC fill),
in terms of the relative differences of the per-run integrated luminosities
measured by the two algorithms. The differences are plotted as a function
of cumulative integrated luminosity through the year normalised to the per-year
total, giving a fraction which increases
monotonically from zero to one as a function of time within the year.
In all years, the run-integrated LUCID and
track-counting measurements typically agree to better then 0.5\%, and at least
about 70\% of the data in each year is covered by a single correction epoch.
 
% End of text imported from the .//caltrans.tex input file

% The next lines are included from the .//calsyst.tex input file
\section{Calibration transfer uncertainties}\label{s:calsyst}
 
The LUCID correction strategy discussed in Section~\ref{ss:lucidcorr}
implicitly assumes that the track-counting measurement with selection~A
suffers from no
significant non-linearity between the low-luminosity vdM and high-luminosity
physics regimes, in particular as a function of \meanmu, bunch structure or
LHC crossing angle, which all change between the two regimes. The
results shown in Sections~\ref{ss:algcomp} and~\ref{ss:trkperf} suggest that
any such non-linearity is smaller than 1\%, but a more robust validation
using independent detectors is essential for a high-precision luminosity
measurement. Comparisons between track-counting, TileCal D-cells and EMEC
(Figure~\ref{f:mudecayphys}) can probe down to the smallest instantaneous
luminosities exploited in normal physics running, but these luminosities are
still $O(10^3)$ times larger than those in the vdM regime, where the
EMEC and TileCal D6 measurements lose sensitivity. The only other measurements
which have enough dynamic range to span from the vdM to physics regimes are
those from the TileCal E-cell gap scintillators, in particular those from
E3 and E4. Comparisons of the luminosity ratios \rtitk{E}\ in the head-on parts
of the vdM
fills with the equivalent ratios in nearby physics fills quantify any relative
non-linearity between the TileCal E-cell and track-counting luminosity
measurements, and can thus be used to probe potential non-linearity in the
track-counting measurement. Deviations of the double ratio
\begin{equation}\label{e:tiledbl}
\rtdbl{E}=\frac{\rtitk{E}(\mathrm{physics})}{\rtitk{E}(\mathrm{vdM})}
\end{equation}
from unity, where each ratio is integrated over the vdM- or physics-like
running period within a fill,
can therefore be interpreted as a reasonable upper limit on the
change in track-counting luminosity response between the vdM and physics
regimes, and hence as an uncertainty in the calibration transfer correction
applied to LUCID.
 
In principle, these double ratios can be measured using the vdM fills combined
with any suitable physics fill close in time either before or after the vdM
fill, limited by the rapid ageing of the scintillators due to radiation damage
in high-luminosity running. Unfortunately, the TileCal E-cell measurements also
suffer from significant contributions from activation of the surrounding
material. These contributions have magnitudes of up to about 1\% of the primary
luminosity signal, and decay with various time constants from minutes
to tens of hours. The residual activation from a high-luminosity physics
fill produces a TileCal E-cell luminosity signal which is initially of a similar
magnitude to the typical luminosity in a vdM fill, and which takes several
days to decay to a negligible level. Wherever possible, the vdM fills were
scheduled immediately after pauses in the LHC running schedule (e.g.\ after a
technical stop) to minimise this activation, and TileCal luminosity
data was recorded before and after the vdM fills to allow the
activation contributions to be fitted and subtracted. The details of this
activation modelling are described in Section~\ref{ss:actmod}.
 
In 2017 and 2018,
additional `vdM-like' fills with only 140~isolated colliding bunch pairs
were scheduled after periods without  high-luminosity running, to provide
further constraints. These dedicated calibration transfer (CT) fills used
standard low-$\beta^*$ physics optics with a beam crossing angle, but the beams
were partially separated for the first two hours of running to give
$\meanmu\approx 0.5$, the same as in vdM fills. After approximately two hours in
this configuration, a $\mu$-scan up to head-on collisions and back to separated
beams was performed, followed by a period of head-on collisions to give
high-pileup running with isolated bunches. The \rtitk{E}\ ratios in both the low
and high-$\mu$ periods of these runs could then be compared with \rtitk{E}\ in
the following physics fills to give additional constraints on the calibration
transfer systematics. These studies are described in Section~\ref{ss:caltdir}.
 
A further study was performed using data taken in June~2018, during the LHC
intensity ramp-up after a one-week technical stop (TS1).
After the 140~colliding bunch
fill mentioned above, a series of relatively short physics fills with increasing
numbers of bunches took place before the 2018 vdM run, with a further two
fills coming after a one-week period of low-luminosity high-$\beta^*$
running for the LHC forward-physics programme that immediately followed the
vdM run.  This sequence of fills allowed
the evolution of the \rtitk{E}\ ratios to be studied from vdM-like conditions,
through high-pileup data with isolated bunches and then increasing numbers
of bunches in trains, in a step-by-step `ladder' approach discussed in
Section~\ref{ss:caltladder}. Along with some analogous fills in 2017,
it also allowed the calibration transfer uncertainties for $\mu\approx 2$
running with bunch trains to be assessed, as discussed in Section~\ref{s:lowmu}
below.
 
\subsection{Tile calorimeter activation model}\label{ss:actmod}
 
The activation signal in the TileCal E-cells was described using an empirical
model based on multiple components, each representing a contribution from
a distinct radioactive isotope with lifetime $\tau$, produced
in the interaction of primary or secondary collision particles with the
detector material, and decaying some time $t$ later giving a contribution
to the luminosity signal in the detector.
For a short pulse of real instantaneous luminosity \lrefinst\ of duration
$\Delta t$ at time $t=0$, the resulting apparent instantaneous luminosity
\lactinst\ from activation at some later time $t$ is given by
\begin{equation}\label{e:act1}
\lactinst = \Delta t\,\lrefinst\frac{f \mathrm{e}^{-t/\tau}}{\tau} \ ,
\end{equation}
where the fraction $f$ characterises the strength of the activation-induced
luminosity signal with respect to the source luminosity \lrefinst. The model
consists of several such components $i$, each with their own parameters
$f_i$ and $\tau_i$ which can be fitted from data.
 
\begin{figure}[tp]
\parbox{83mm}{\includegraphics[width=76mm]{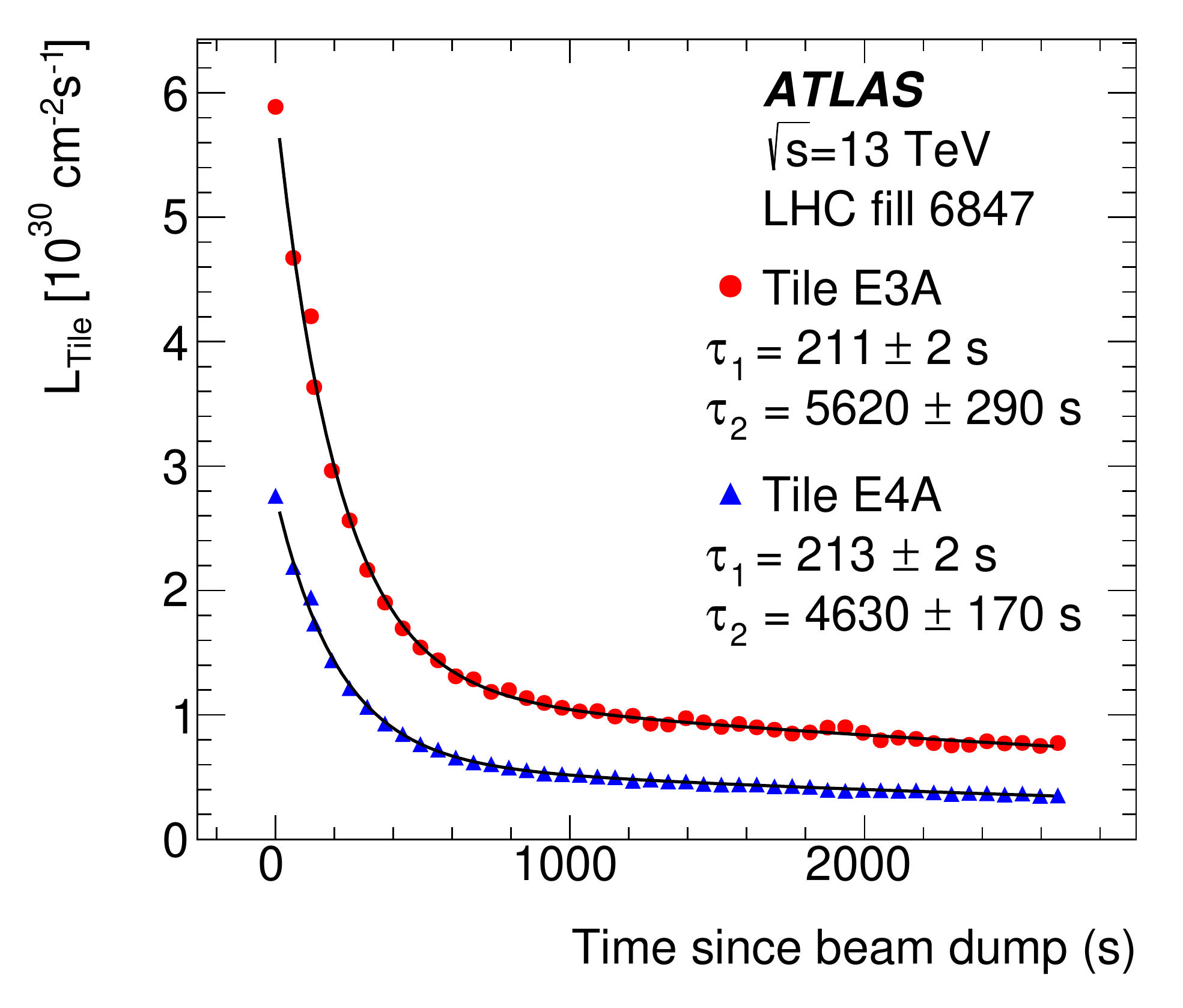}\vspace{-6mm}\center{(a)}}
\parbox{83mm}{\includegraphics[width=76mm]{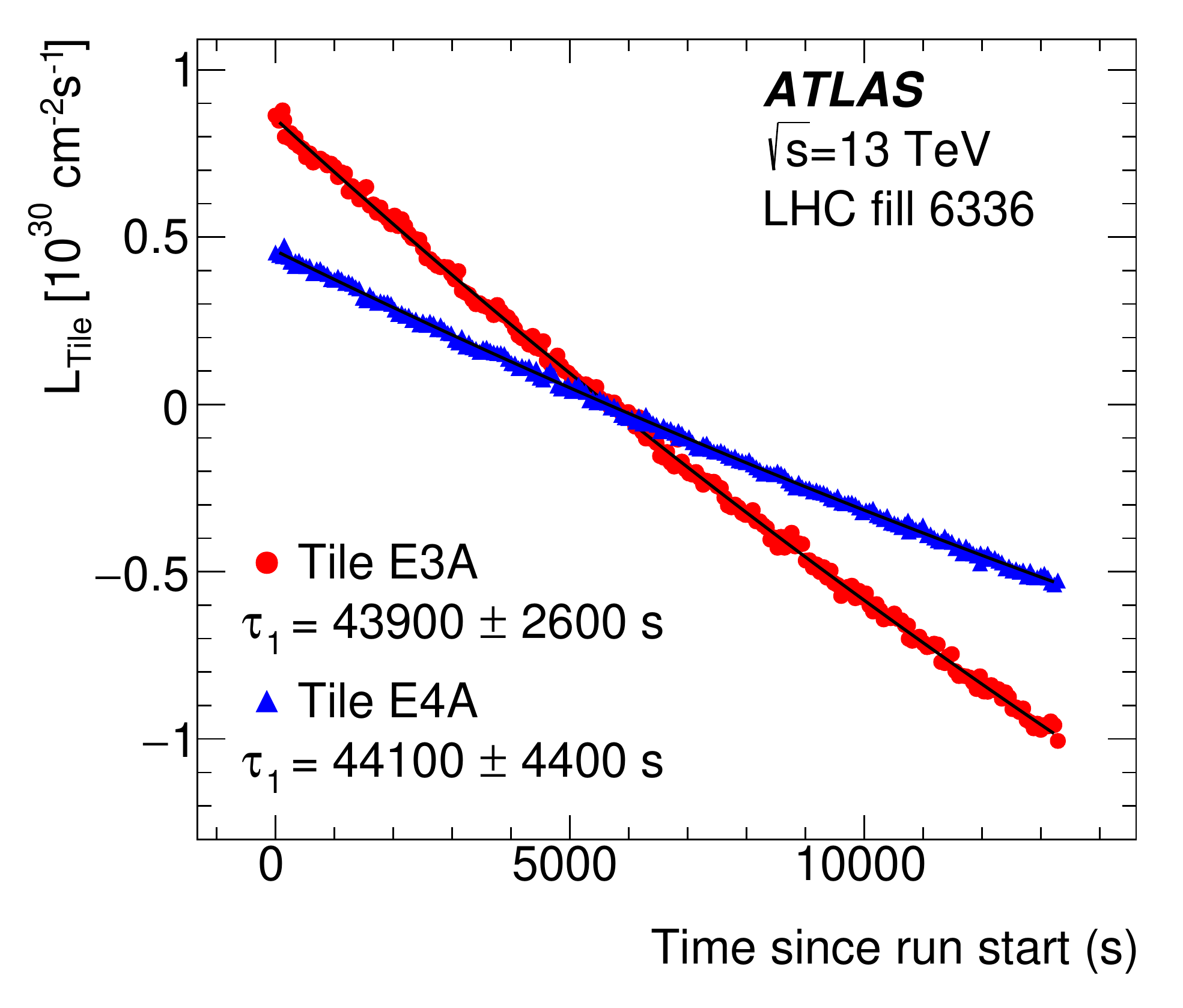}\vspace{-6mm}\center{(b)}}
\caption{\label{f:actdecay}Instantaneous luminosity measured by the A-side
TileCal E3 and E4 cells as functions of time in (a) the period after the beam
dump in fill 6847 from 2018, and (b) the period before collisions in
fill 6336 from 2017.
The measurements are fitted to the sum of two exponential functions in (a), and
one exponential function plus a constant in (b), with the fitted lifetimes and
their uncertainties shown in the legends. The uncertainties in the individual
TileCal measurements are smaller than the marker sizes and were derived from the
RMS of the differences between A- and C-side measurements in each luminosity
block. The pedestal offset in (b) is arbitrary,
and has been set so that the mean of the measurements is zero in the
pre-collision period. The measurements from the C-side cells are similar.}
\end{figure}

Periods without collisions are particularly useful in characterising the
activation contributions to the TileCal E-cell measurements.
Figure~\ref{f:actdecay}(a) shows the instantaneous luminosity measured in the
A-side E3 and E4 cells in the 45~minutes after the beam dump in the
140~colliding bunch fill~6847 in 2018. The instantaneous luminosity just before
the beam dump was about \lval{880}, and immediately afterwards the E3 cells
measured a residual luminosity of about 1\% of this value due to activation,
which subsequently decayed away. The activation fits well to the sum of two
decaying exponential functions as shown.
The measurements from E4 show about a factor
two smaller activation, which decays with similar time constants. The fitted
lifetimes are close to those of $^{28}$Al and $^{39}$Cl ($\tau=194$ and 4934\,s),
two isotopes which are expected to be present in the nearby cryostats and LAr
of the electromagnetic calorimeters. Figure~\ref{f:actdecay}(b) shows
analogous measurements from the period before collisions in fill 6336 from
2017, during the time when the ATLAS data acquisition was running but there
were not yet collisions in the LHC. The data-taking run started 12~hours after
the previous high-luminosity fill was dumped, so the relatively short lifetime
components shown in Figure~\ref{f:actdecay}(a) would have decayed to a
negligible level, but a residual component, with a lifetime of about 12~hours,
can clearly be seen in both E3 and E4 measurements, again with a larger amplitude
in E3 than E4.
 
These studies, together with similar analyses from other LHC fills,  allow the
time constants $\tau_i$ of the activation components in Eq.~(\ref{e:act1}) to be
determined, but not the fractions $f_i$. The latter can only be determined from
knowledge of the luminosity history leading to the activation at any particular
time. If the luminosity history preceding the time $t$ is known, the activation
corresponding to any given set of ($\tau_i,f_i)$ parameters can be calculated
by integration, and compared with data. In practice, the activation is dominated
by the relatively short-lived components shown in Figure~\ref{f:actdecay}(a),
and only their excitation within the fill under consideration needs to be taken
into account. The long-lifetime component shown in Figure~\ref{f:actdecay}(b)
is only relevant for low-luminosity vdM and CT fills which followed soon
after a high-luminosity physics fill, and this component can be normalised with
a free parameter representing the residual activation at the start of the
data-taking run, without knowledge of the luminosity structure in the previous
fill(s) that produced it.
 
This procedure was implemented using the LUCID luminosity measurements
(including the $\mu$-correction of Eq.~(\ref{e:mucorr})) for \lrefinst\ in
Eq.~(\ref{e:act1}) to derive a prediction for the contribution \lactint\ to the
integrated luminosity measured by TileCal in each LB within a
data-taking run. The corrected TileCal luminosity \lcorrint\ is then given by
\begin{equation}\nonumber
\lcorrint=\lcaloint-\lactint-P\Delta t\ ,
\end{equation}
where \lcaloint\ is the uncorrected TileCal measurement and $P$ is a pedestal
correction to the instantaneous luminosity, multiplied by the luminosity
block duration $\Delta t$. This correction was taken to be a constant or a
linear function of time to account
for slow drifts. The activation correction \lactint\ implicitly depends on
the LUCID luminosity history and the parameters $(\tau_i,f_i)$.
 
The relation between the corrected TileCal and track-counting luminosities
was assumed to be a linear function of $\meanmu$:
\begin{equation}\label{e:tiletrk}
\lcorrint=\ltrkint (p_0+\meanmu p_1)\ ,
\end{equation}
where \ltrkint\ is the track-counting integrated luminosity in the LB
and \meanmu\ is measured using track-counting. Finally, a $\chi^2$ was defined
from the comparison of corrected TileCal and track-counting measurements
in each luminosity block, i.e.
\begin{equation}\label{e:actchi}
\chi^2 = \sum\left( \frac{\left(\lcaloint-\lactint-P\Delta t-\ltrkint (p_0+\meanmu p_1)\right)^2}{\sigma^2_{\lcaloint}+\sigma^2_{\ltrkint}} \right)\ ,
\end{equation}
where the terms $\sigma_{\lcaloint}$ and $\sigma_{\ltrkint}$ represent the
per-LB uncertainties in the luminosity measurements, and the sum is taken over
all LBs where both the calorimeter and tracking information are available.
The calorimeter uncertainties were estimated empirically from the spread of
the differences between the A- and C-side measurements in each luminosity block,
and parameterised as linear functions of luminosity. The $\chi^2$ was then
minimised to determine the best values of the activation model parameters
$(\tau_i, f_i)$, together with $p_0$, $p_1$ and the pedestal parameters.
Periods without collisions were also included in the $\chi^2$, assuming the
track-counting luminosity to be zero, but periods with collisions but
no track-counting or calorimeter measurements (e.g.\ before stable beams when
the tracking detectors are not yet active) were excluded. However, the LUCID
luminosity measurements (which are also available outside stable beam periods)
were still used to follow the build-up of calorimeter activation during these
times.
 
\begin{figure}[tp]
\parbox{83mm}{\includegraphics[width=76mm]{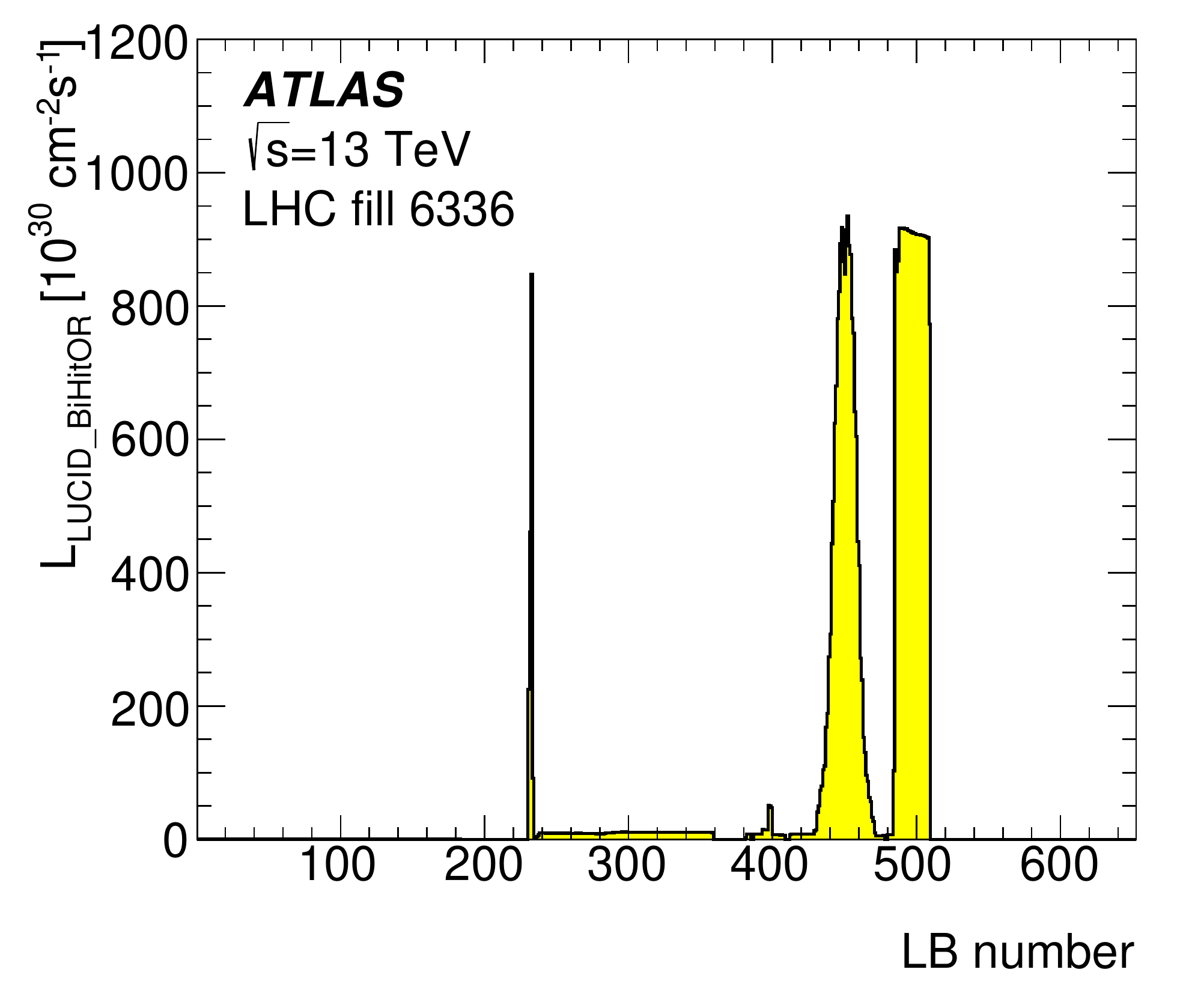}\vspace{-6mm}\center{(a)}}
\parbox{83mm}{\includegraphics[width=76mm]{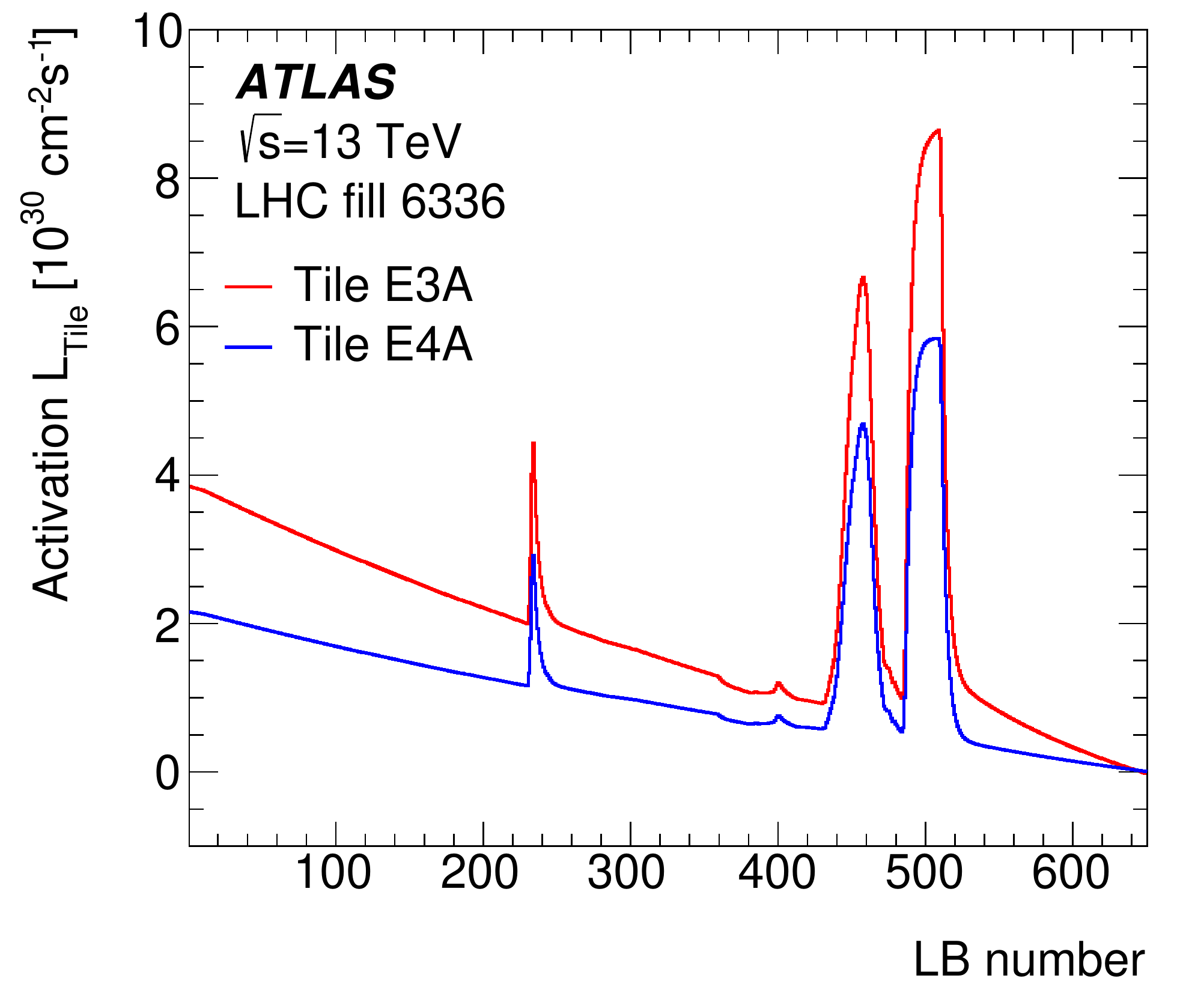}\vspace{-6mm}\center{(b)}}
\parbox{83mm}{\includegraphics[width=76mm]{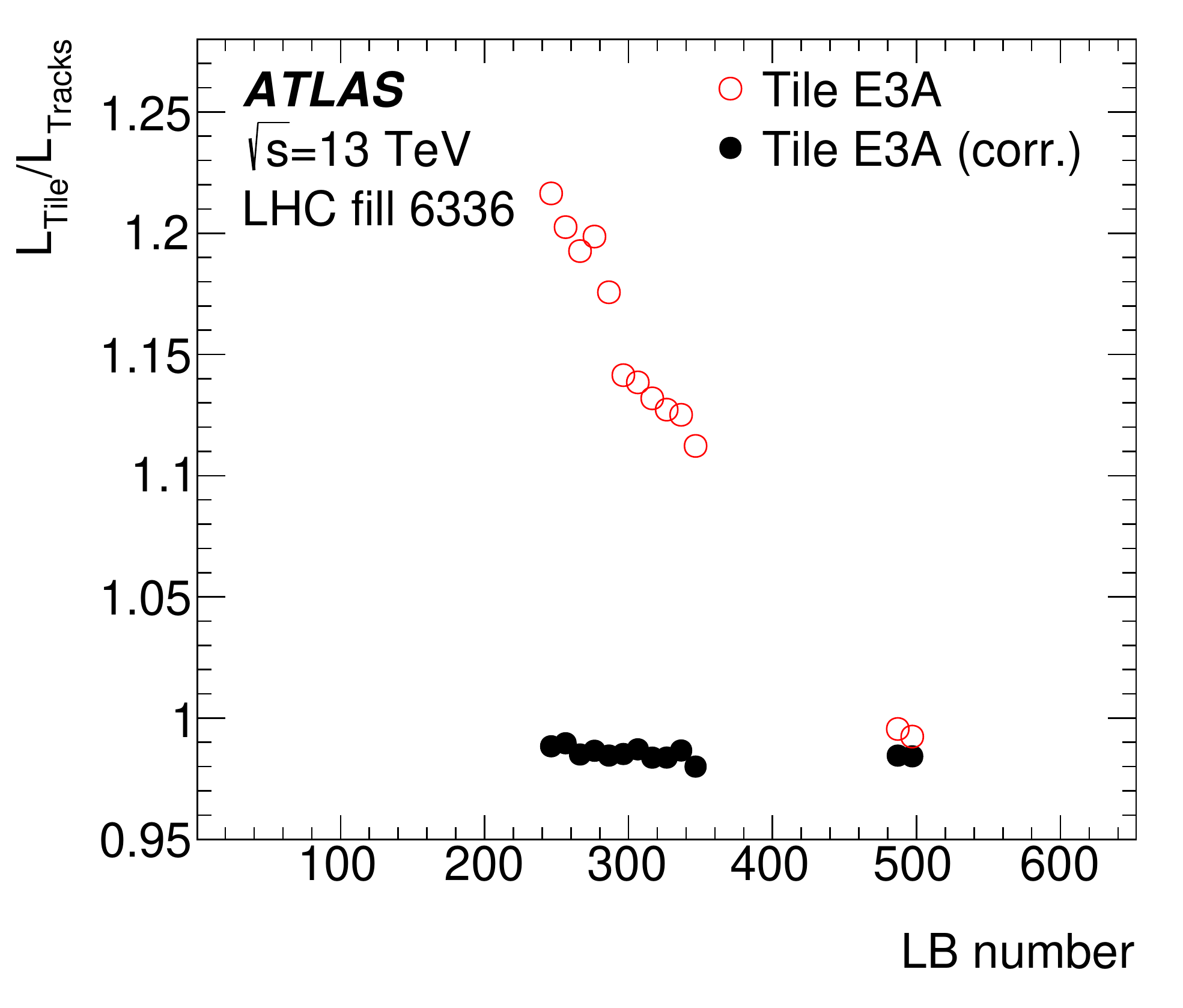}\vspace{-6mm}\center{(c)}}
\parbox{83mm}{\includegraphics[width=76mm]{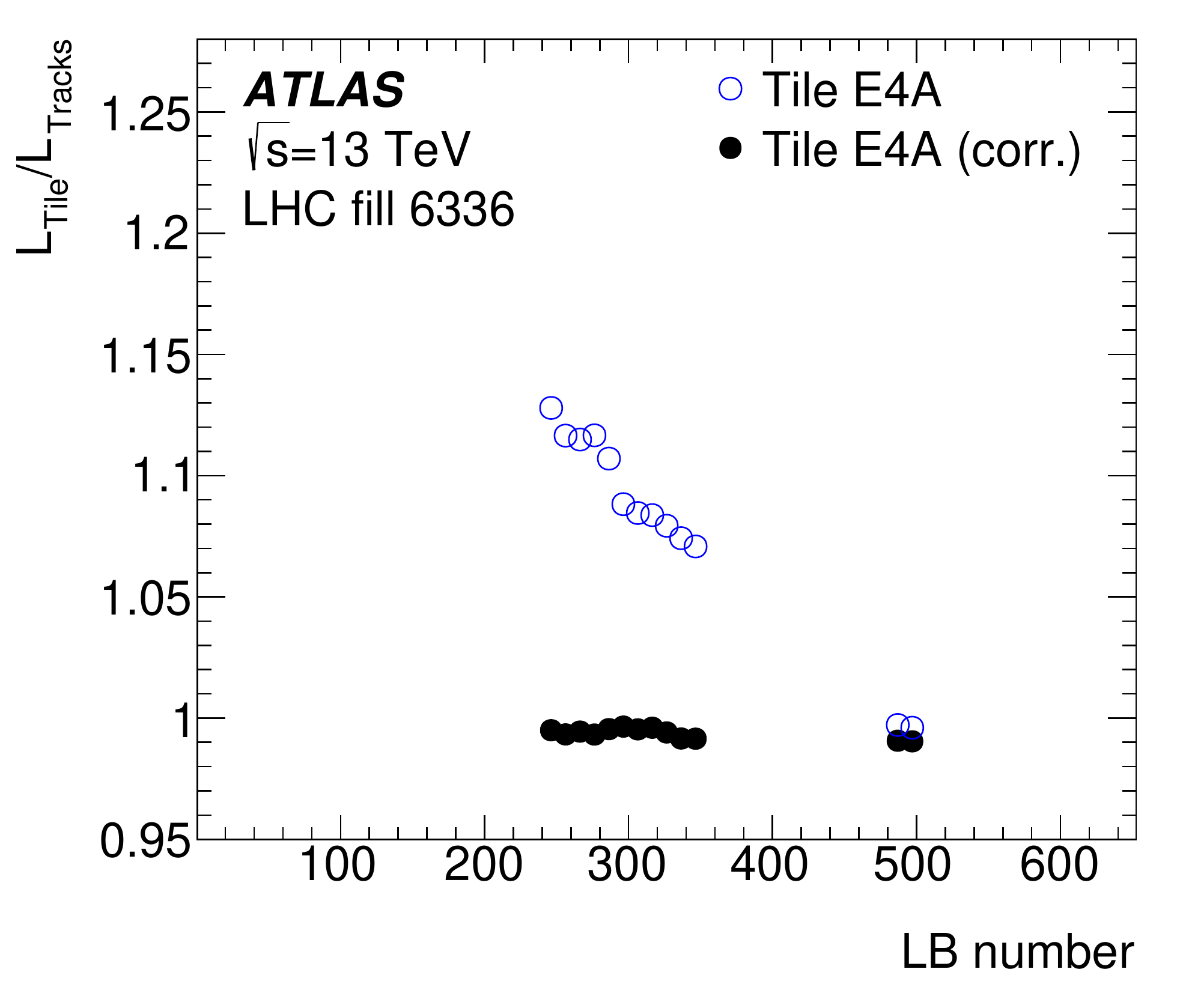}\vspace{-6mm}\center{(d)}}
\caption{\label{f:tilect17}Details of the TileCal E-cell activation analysis for
fill 6336 from 2017 with 140 colliding bunch pairs: (a) luminosity profile as
measured by LUCID as a function of LB number (see text); (b) fitted activation
contributions to the A-side TileCal E3 and E4 measurements; (c, d)
ratios of uncorrected (red or blue open circles) and activation-corrected
(black filled circles) TileCal E-cell luminosity measurements to those from
track-counting for E3 and E4 in the periods with steady-state running
at low and high pileup, excluding the $\mu$-scan. The ratios have been averaged
over ten consecutive luminosity blocks.
The results for the C-side measurements are similar.}
\end{figure}

Figure~\ref{f:tilect17} illustrates this fit procedure for LHC fill 6336, the
140 colliding bunch CT fill recorded in November 2017. Figure~\ref{f:tilect17}(a)
shows the luminosity history of the fill as measured by LUCID: after a four-hour
period before collisions, the beams were brought into head-on collision
at LB~231 before being separated to give two hours at \lval{10} and
$\meanmu=0.5$, a short period with the beams separated in both planes
to give almost-zero luminosity, a $\mu$-scan in LB 428--475 (preceded by
a short aborted $\mu$-scan starting at LB 386), and finally a short period
of head-on running, after which the beams were dumped and the data-taking
continued for another 80~minutes to monitor the TileCal activation decay.
Figure~\ref{f:tilect17}(b) shows the fitted activation contributions
after minimising the $\chi^2$ of Eq.~(\ref{e:actchi}); the lifetimes
of the three fitted components are mainly constrained from the periods
before and after collisions, whilst the fractions are also constrained from
the evolution during collisions. The fractions corresponding to the
$\tau \approx 200$\,s component are around 1\% for E3 and 0.5\% for E4,
and the longer $\tau\approx 5000$\,s component has fitted fractions
of 0.2\% for E3 and less than 0.1\% for E4. For this fit, the long
$\tau\approx 45000$\,s component was only used to fit the initial activation
from previous running, and any activation of this component within the
low-luminosity fill 6336 was neglected (i.e.\ $f$ was fixed to zero).
The offset of the activation contribution is arbitrary, and has been set so
the activation is zero at the end of the data-taking period.
The short-lifetime activation produced by the high-luminosity
periods of the fill (the initial `spike' when the beams were brought into
collision, the $\mu$-scan and the head-on period) is clearly visible above
the background from the slow decay of the residual long-lifetime component.
 
Figures~\ref{f:tilect17}(c) and~\ref{f:tilect17}(d) show the
TileCal/track-counting instantaneous
luminosity ratios during the low-$\mu$ separated-beam and high-$\mu$ head-on
periods of fill~6336. The uncorrected ratios (shown by the open circles)
are far from unity and vary by 5--10\%
during the low-$\mu$ period due to the changing offset caused by the
slowly decaying $\tau\approx 45000$\,s activation component.
The activation-corrected ratios are much closer to unity, although a small
downward slope is still visible for E3A, perhaps due to residual imperfections
in the activation modelling.
During the high-$\mu$ period with $\linst=\lval{900}$, the uncorrected and
corrected ratios are similar and much closer to unity, because
the long-lifetime activation contributions are relatively much less important.
The remaining deviation from unity reflects the absolute normalisation of the
initial TileCal E-cell measurements, which was determined from the following
high-luminosity fill without taking activation effects into account.
The fitted $p_1$ parameters from Eq.~(\ref{e:tiletrk}) for the
four cell families E3A, E3C, E4A and E4C are all small, between
zero and $-1\times 10^{-4}$, and show that the corrected ratios are in good
agreement between the low- and high-$\mu$ periods. These $p_1$ values
correspond to a maximum change of $-0.5$\% in the
TileCal/track-counting ratio between $\meanmu=0.5$ and $\meanmu=50$,
giving a strong constraint on the maximum non-linearity of the
track-counting measurement between low-$\mu$ and high-$\mu$ isolated bunches.
Similar results were obtained from the analogous 140 colliding bunch CT fill
6847 in 2018.

\subsection{Comparisons of TileCal and track-counting measurements: direct approach}\label{ss:caltdir}
 
\begin{figure}[tp]
\parbox{83mm}{\includegraphics[width=76mm]{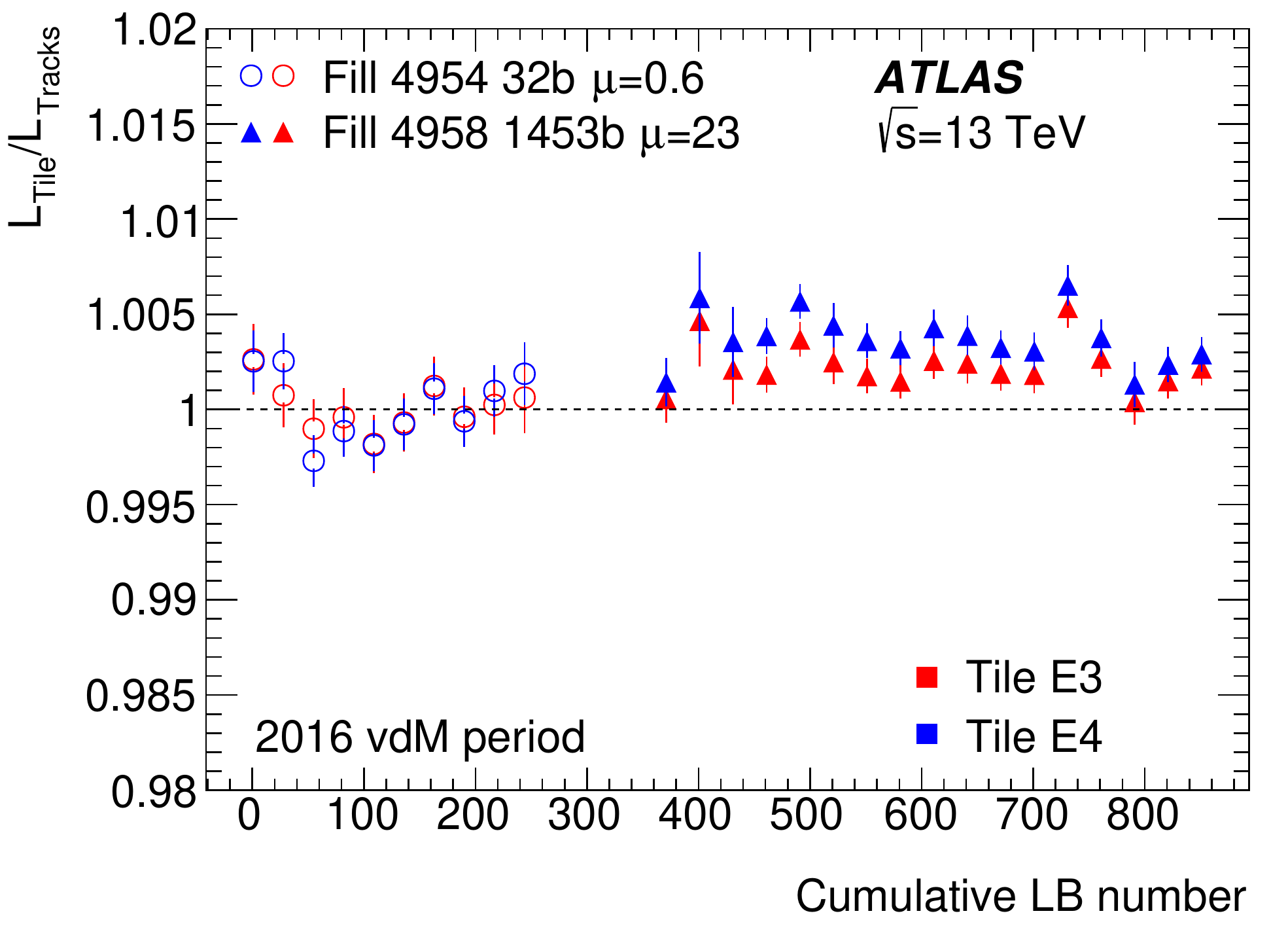}\vspace{-6mm}\center{(a)}}\\
\parbox{83mm}{\includegraphics[width=76mm]{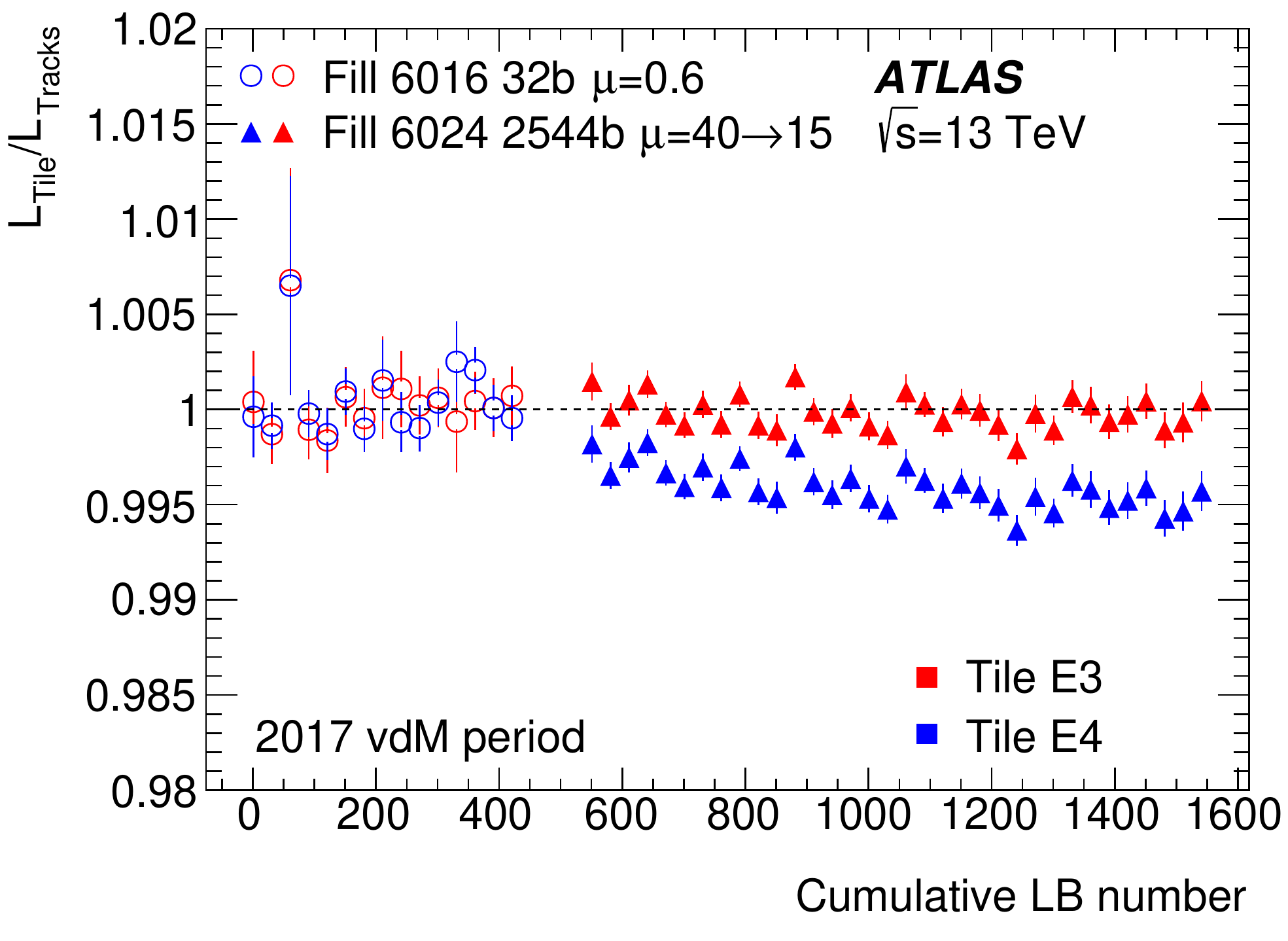}\vspace{-6mm}\center{(b)}}
\parbox{83mm}{\includegraphics[width=76mm]{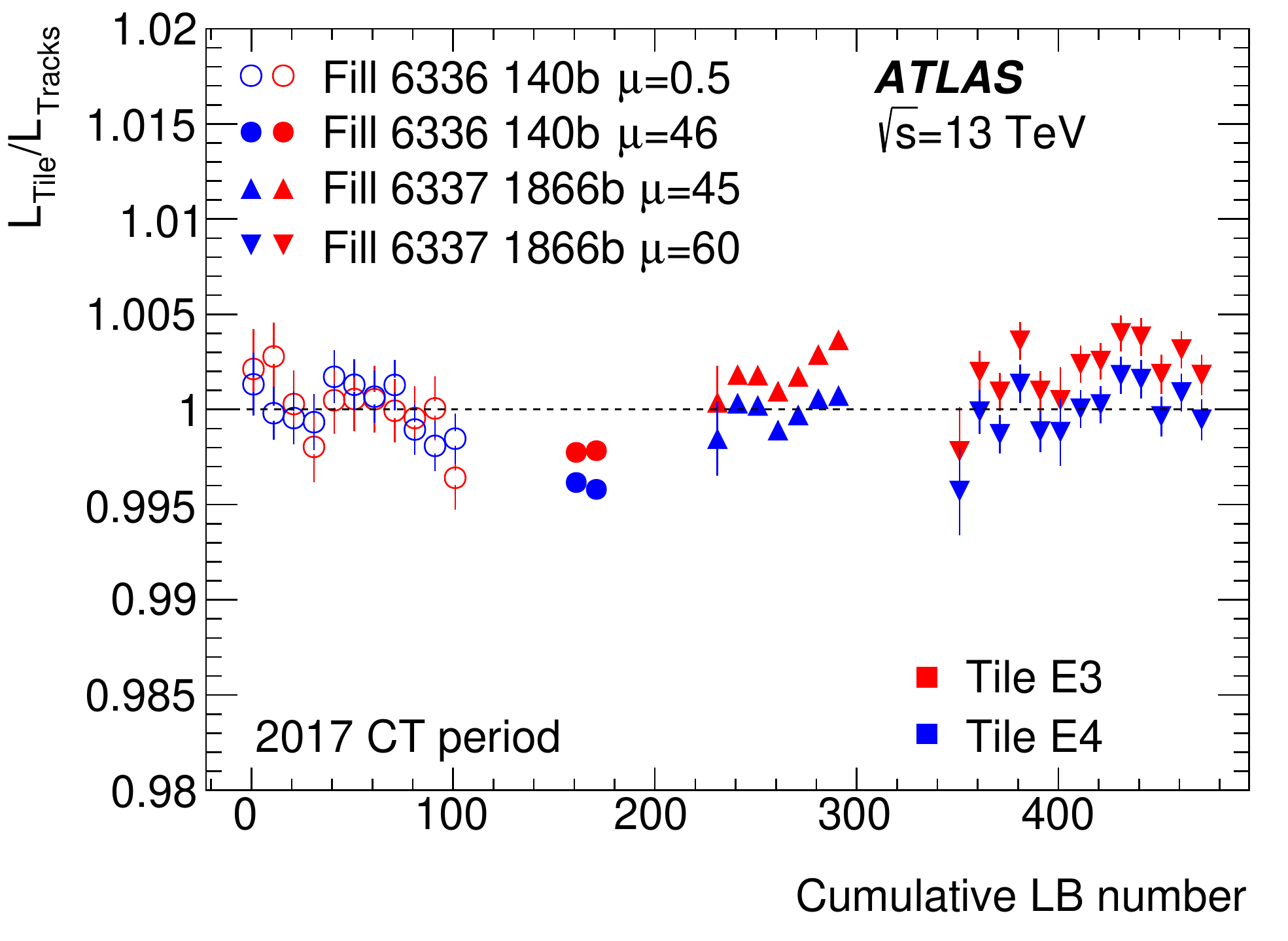}\vspace{-6mm}\center{(c)}}
\parbox{83mm}{\includegraphics[width=76mm]{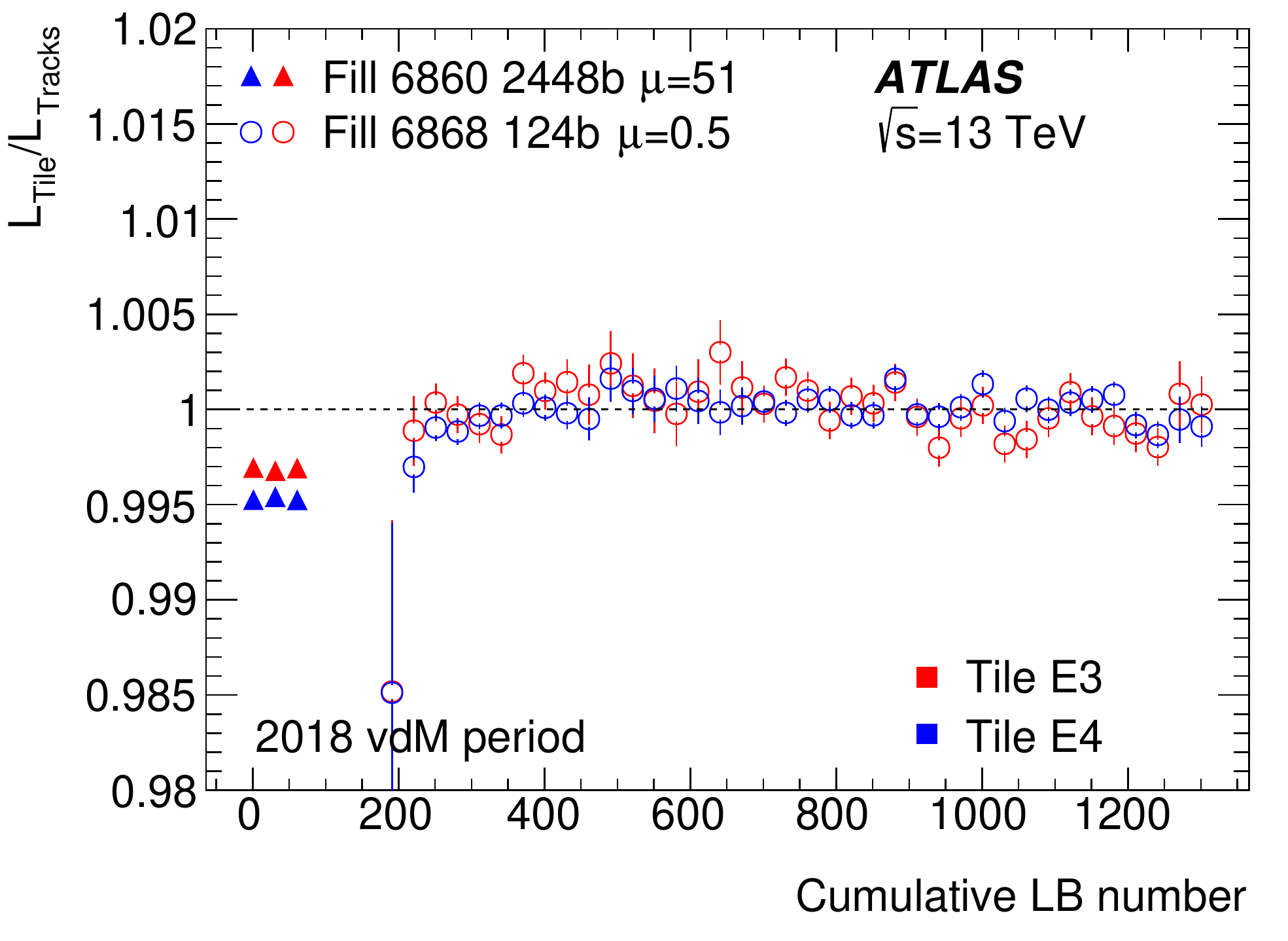}\vspace{-6mm}\center{(d)}}
\parbox{83mm}{\includegraphics[width=76mm]{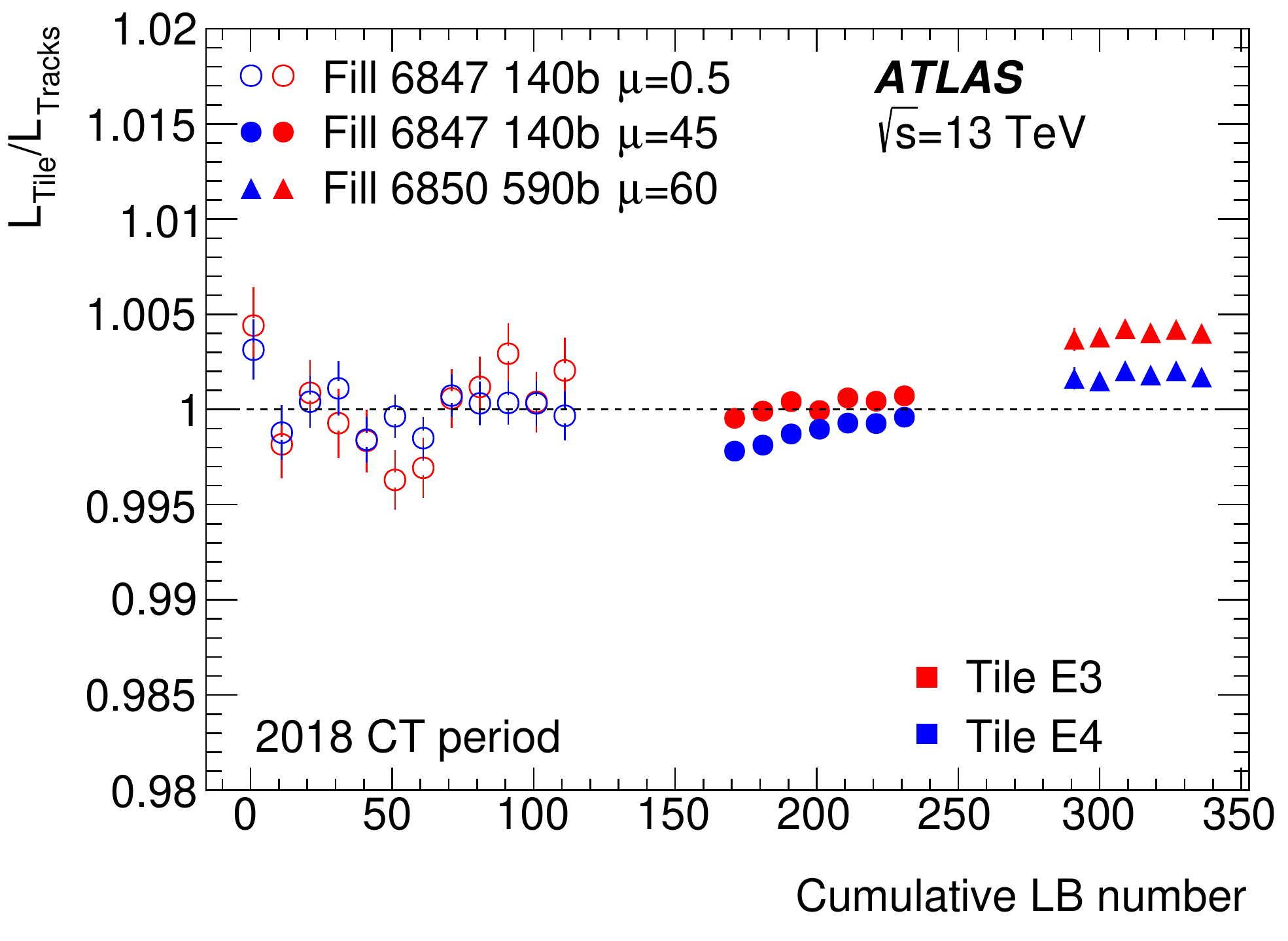}\vspace{-6mm}\center{(e)}}
\caption{\label{f:ctdir}Ratios of luminosity measured by the TileCal E-cells
to that measured by track-counting for sequences of fills  under various LHC
conditions in the 2016--2018 calibration transfer studies.
The ratios are averaged over the
A- and C-side measurements and over~30 (vdM and related fills, left column)
or~10 (CT periods, right column) luminosity blocks, separately for the E3 (red)
and E4 (blue) cells. The uncertainties indicated by the error bars are
dominated by the statistical uncertainties from the track-counting measurements.
Within each plot, the ratios are shown as a function of luminosity block number,
with gaps between fills to give a continuous sequence over related fills.
Periods with no ratio measurements within fills (e.g.\ during vdM scans) are
not shown, and are omitted in the luminosity block numbering. The ratios
have been renormalised separately for E3 and E4 cells in each sequence so that
the integrated ratios in the low-luminosity vdM-like periods are unity. The
numbers of colliding bunch pairs in ATLAS and the typical $\mu$ values are
indicated for each fill, and the various fill periods are plotted with different
marker styles.}
\end{figure}

The activation model described above was used to derive corrected TileCal E3-
and E4-cell measurements for the vdM fills in 2016, 2017 and 2018, and the
dedicated
140~colliding bunch CT fills in 2017 and 2018. Each such fill was paired with
a nearby high-luminosity fill in order to derive the double ratios \rtdbl{E}
defined in Eq.~(\ref{e:tiledbl}). For consistency,
the activation model was also applied to
correct the TileCal luminosity measurements in the high-luminosity fills,
using the $(\tau_i, f_i)$ parameters derived from the corresponding
low-luminosity fill. The datasets and results are summarised in
Table~\ref{t:caltrun}, including comparisons between low-
and high-$\mu$ isolated bunches within each of the two CT fills. The double
ratios are given averaged over the A- and C-sides, but separately for
E3 and E4 cells. More details are given in Figure~\ref{f:ctdir}, which shows the
TileCal E-cell/track-counting single ratios as a function of luminosity block
number in a continuous series for each pair of low- and high-luminosity
runs, normalising the ratios so that the integrals of TileCal and track-counting
luminosities agree in the low-luminosity runs. The stability of the ratios
vs.\ time demonstrates the quality of the activation modelling; as can be seen
from Figures~\ref{f:tilect17}(c) and~\ref{f:tilect17}(d), these corrections can
be of $O(10)$\%, but the corrected TileCal/track-counting ratios for low
and high luminosity agree to within 0.5\% for all the comparisons shown in
Table~\ref{t:caltrun} and Figure~\ref{f:ctdir}. There are some systematic
differences for the ratios involving E3 and E4 cells---the E3/track-counting
ratios, which require larger activation corrections, are generally closer to
unity, but all the E4 double-ratios are also within $\pm 0.5$\% of unity.
There is no systematic difference between the comparisons where the
low-luminosity fill is a vdM fill (without an LHC beam crossing angle) or a
CT fill using physics optics with a crossing angle, suggesting that
the track-counting measurement does not have a crossing-angle dependence.
The 2016 dataset included a second vdM fill (4945) which showed larger
deviations of up to 1\% with respect to nearby physics fills.
However, the TileCal
laser calibration corrections in this period were not fully understood, and
there were inconsistencies between the A- and C-side measurements, so this
dataset was discarded.
 
\begin{table}[tp]
\caption{\label{t:caltrun}Pairs of vdM-like and physics-like fills used to
measure the
TileCal E3- and E4-cell/track-counting double ratios \rtdbl{E} in order to
study the calibration transfer uncertainty in 2016--2018 data. For each fill,
the LHC fill number and number of colliding bunches in ATLAS \nbun\ are given.
The vdM-like fills had only isolated bunches with $\meanmu=0.5$--0.6, with (y)
or without (n) a crossing angle at the ATLAS interaction point.
The physics-like fills all have a crossing angle, and (apart from the
high-$\mu$ parts of fills 6336 and 6847) a filling scheme with bunch trains.
The rightmost columns show the integrated double ratios, including the effects
of activation corrections.}
\centering
 
\begin{tabular}{l|ccc|ccc|rr}\hline
\multicolumn{1}{c|}{Year} & \multicolumn{3}{c|}{vdM-like fill} & \multicolumn{3}{c|}{Physics-like fill} & \multicolumn{2}{c}{Double ratios} \\
& Fill & \nbun & X-angle & Fill & \nbun\ & \meanmu & \rtdbl{E3} & \rtdbl{E4} \\
\hline
2016 vdM & 4954 & 32 & n & 4958 & 1453 & 21 & 1.002 & 1.003 \\
\hline
2017 vdM & 6016 & 32 & n & 6024 & 2544 & 40 & 1.000 & 0.997 \\
2017 CT & 6336 & 140 & y & 6336 & 140 & 46 & 0.998 & 0.996 \\
2017 CT & 6336 & 140 & y & 6337 & 1866 & 45 & 1.002 & 1.000 \\
2017 CT & 6336 & 140 & y & 6337 & 1866 & 58 & 1.002 & 1.000 \\
\hline
2018 CT & 6847 & 140 & y & 6847 & 140 & 45 & 1.000 & 0.999 \\
2018 CT & 6847 & 140 & y & 6850 & 590 & 60 & 1.004 & 1.002 \\
2018 vdM & 6868 & 124 & n & 6860 & 2448 & 51 & 0.998 & 0.995 \\
\hline
\end{tabular}
\end{table}
 
\begin{figure}[tp]
\parbox{83mm}{\includegraphics[width=76mm]{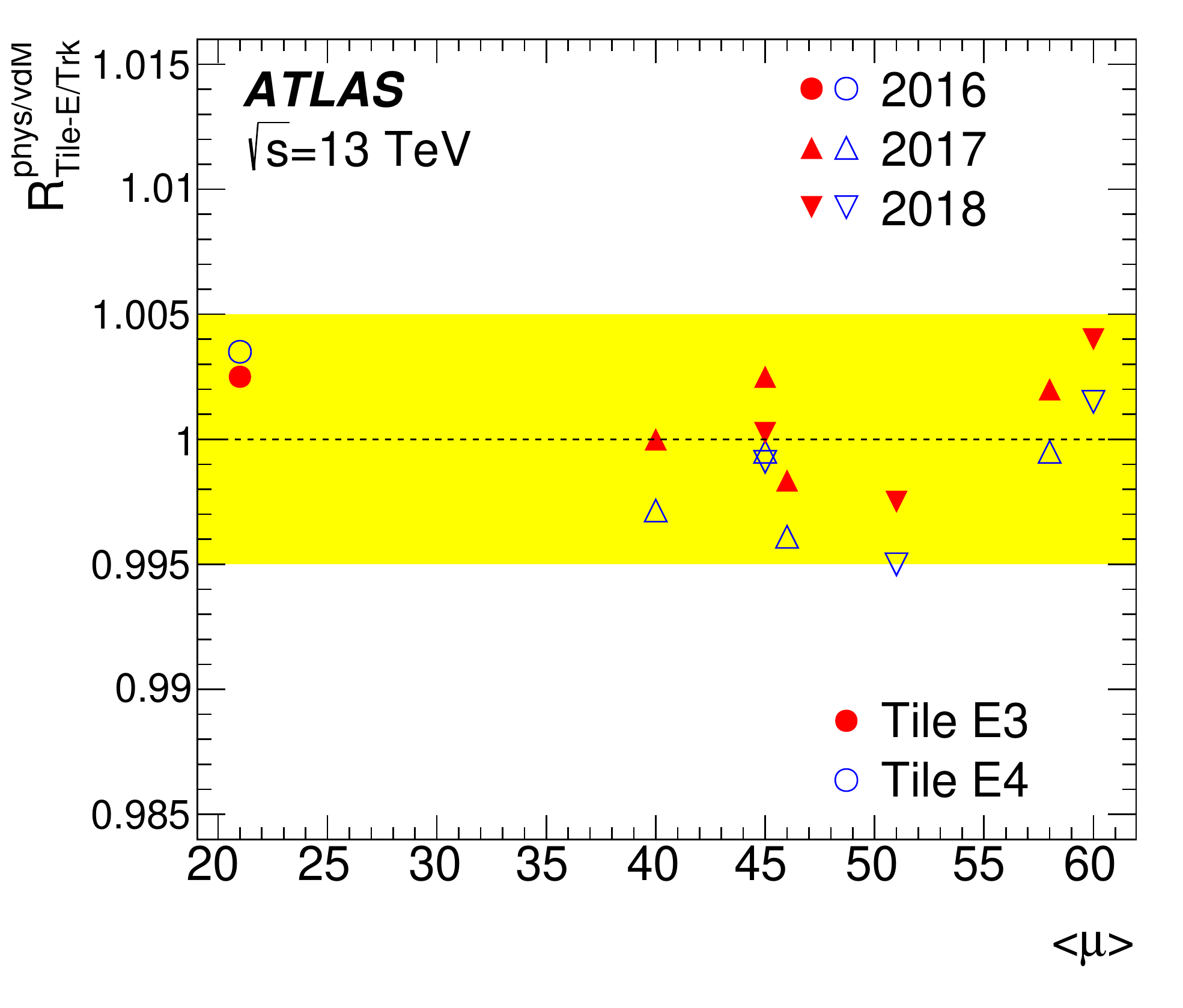}\vspace{-6mm}\center{(a)}}
\parbox{83mm}{\includegraphics[width=76mm]{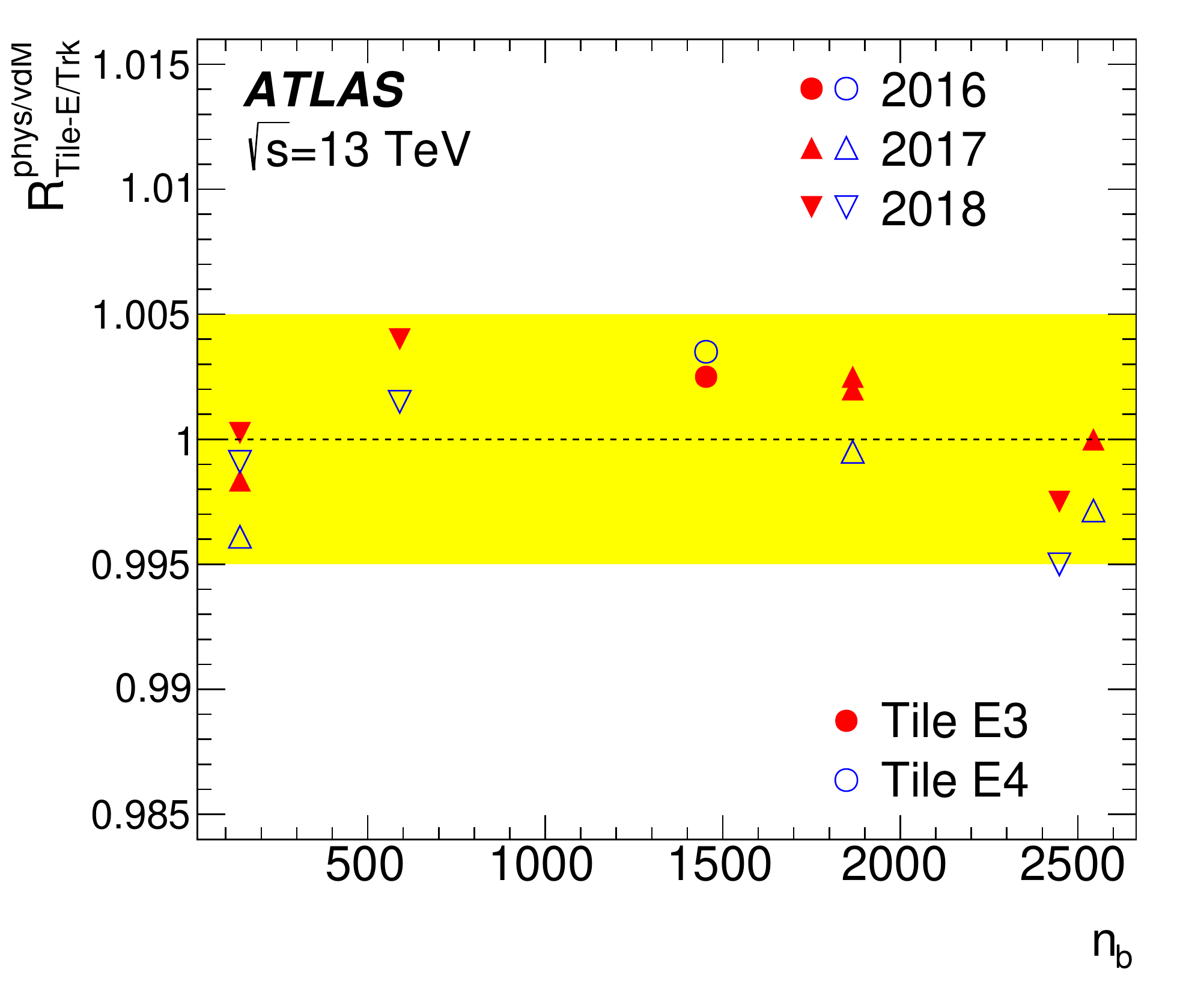}\vspace{-6mm}\center{(b)}}
\caption{\label{f:ctmubcid}Double ratios \rtdbl{E} of integrated luminosity
measured by the TileCal E3 (filled red points) and E4 (open blue points) cells
to that measured by track-counting in physics runs with bunch trains vs.
vdM-like
runs for the pairs of runs shown in Table~\ref{t:caltrun}, shown as a function
of (a) \meanmu\ in the physics run and (b) the number of colliding bunch pairs
\nbun\ in the physics run. The yellow band indicates a range of $\pm 0.5$\%.}
\end{figure}

Figure~\ref{f:ctmubcid} shows the integrated TileCal E-cell/track-counting
double ratios as functions of the \meanmu\ and \nbun\ values in the
physics-like fills, exploiting the variety of conditions sampled over the
three years. No strong correlations
with either quantity are visible, suggesting that the $\pm 0.5$\% maximum
deviation from unity is valid for all the conditions encountered in
high-pileup physics running with bunch trains during Run~2.
 
\begin{figure}[tp]
\parbox{83mm}{\includegraphics[width=76mm]{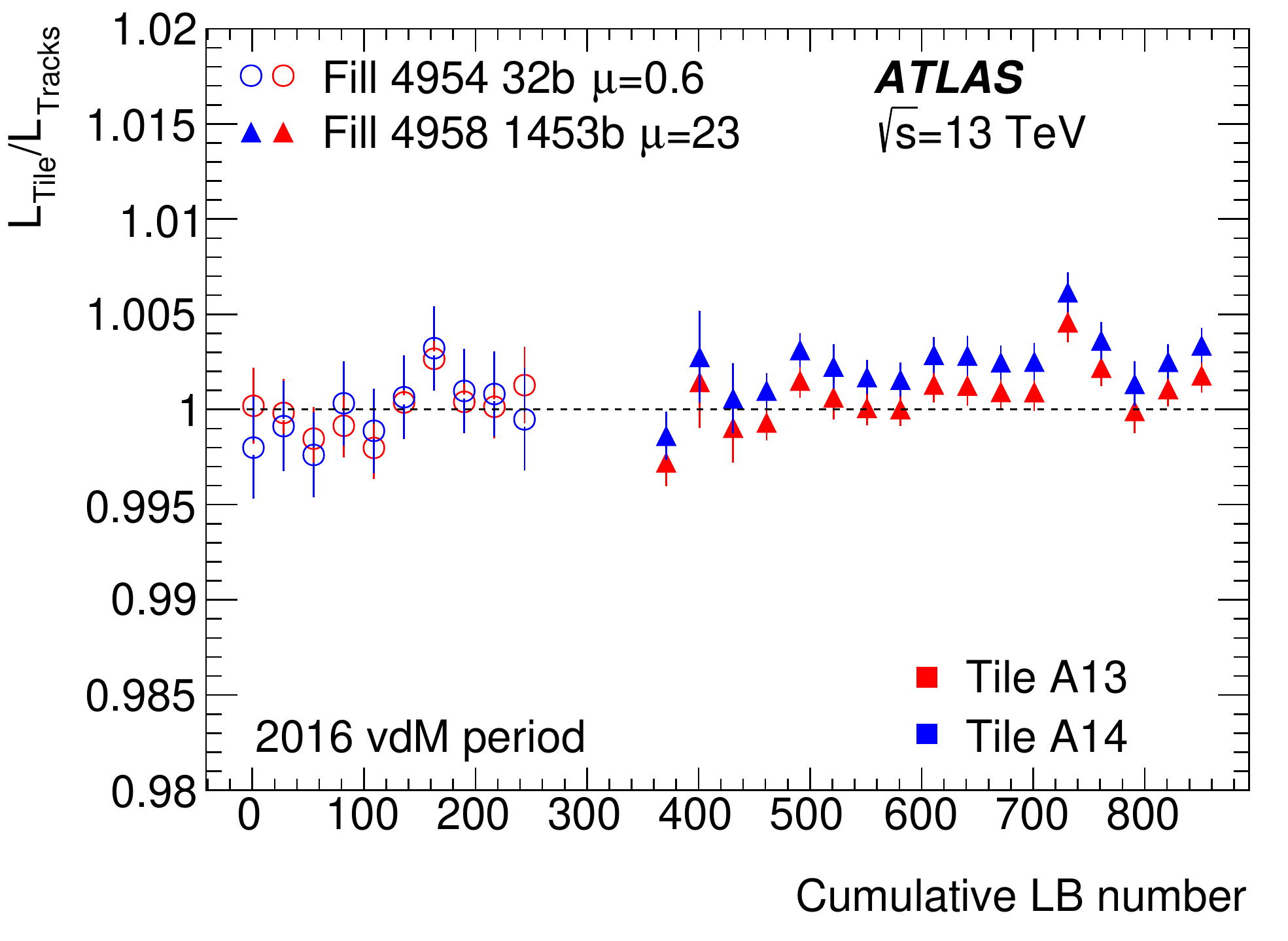}\vspace{-6mm}\center{(a)}}
\parbox{83mm}{\includegraphics[width=76mm]{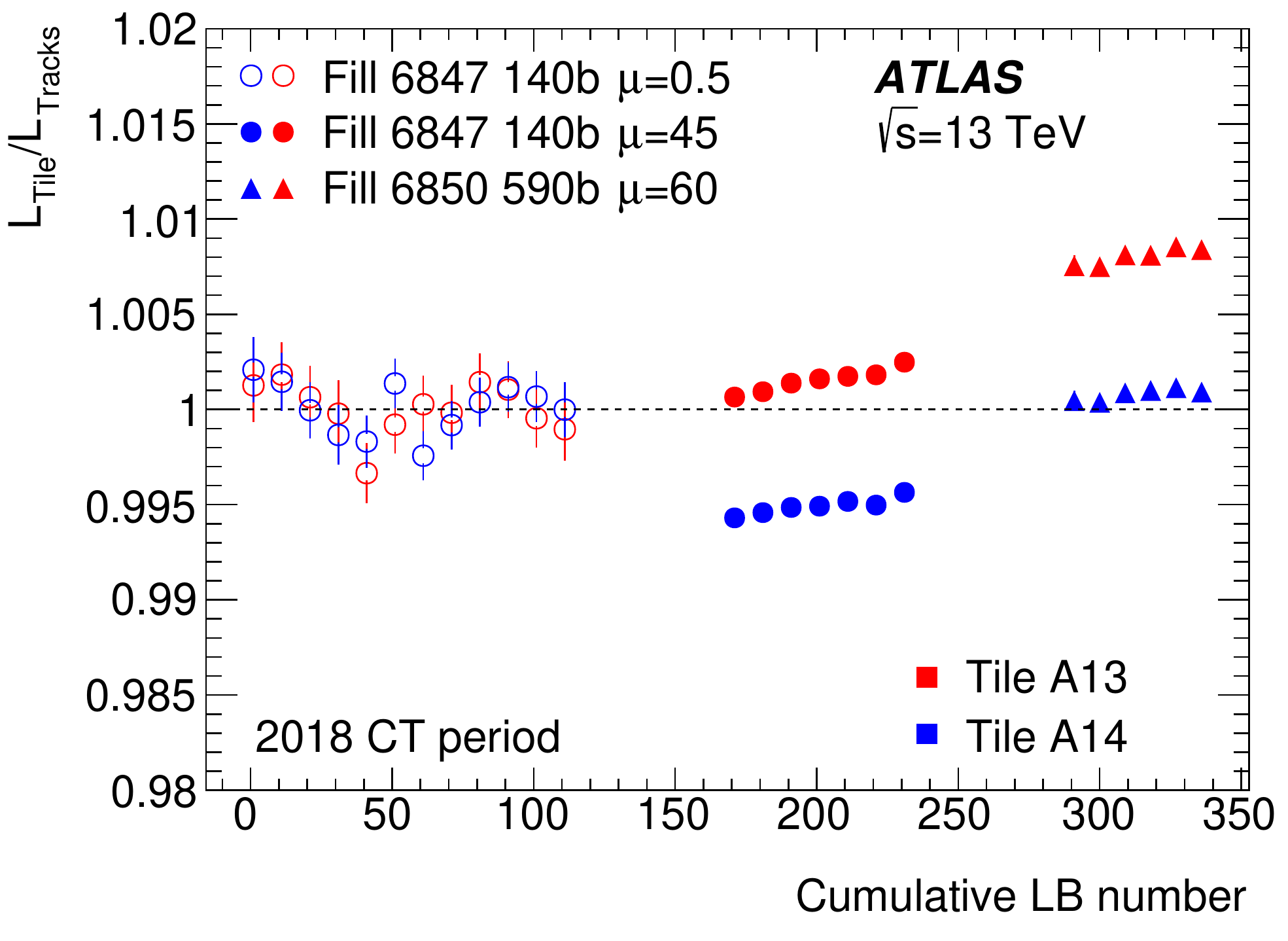}\vspace{-6mm}\center{(b)}}
\caption{\label{f:ctdirA}Ratios of luminosity measured by the TileCal A-cells
to that measured by track-counting for sequences of fills
in the 2016 and 2018 calibration transfer studies.
The ratios are averaged over the
A- and C-side measurements and over~30 (2016 vdM period) or 10 (2018 CT period)
luminosity blocks, separately for the A13 (red)
and A14 (blue) cells. The uncertainties indicated by the error bars are
dominated by the statistical uncertainties from the track-counting measurements.
Within each plot, the ratios are shown as a function of luminosity block number,
with gaps between fills to give a continuous sequence over related fills.
Periods with no ratio measurements within fills (e.g.\ during vdM scans) are
not shown, and are omitted in the luminosity block numbering. The ratios
have been renormalised separately for A13 and A14 cells in each sequence so
that the integrated ratios in the low-luminosity vdM-like periods are unity. The
numbers of colliding bunch pairs in ATLAS and the typical $\mu$ values are
indicated for each fill, and the various fill periods are plotted with different
marker styles.}
\end{figure}

Limited additional studies were performed using the TileCal A13 and A14 cells,
located in the first longitudinal sampling of the extended barrel calorimeter.
These cells have poorer sensitivity in vdM runs than E3 and E4, but suffer
less from radiation-induced ageing and have smaller activation corrections.
The activation was studied and corrections were made using the techniques
described in Section~\ref{ss:actmod}, suggesting activation components with
lifetimes of $\tau\approx 300$, 5800 and 12000\,s, though the lower sensitivity
of the A-cells makes it hard to determine the $(\tau_i,f_i)$ parameters
unambiguously. The 2016 vdM and 2018 CT periods were studied, and the
resulting \rtitk{A}\ ratios are shown in Figure~\ref{f:ctdirA}, to be compared
with Figures~\ref{f:ctdir}(a) and~\ref{f:ctdir}(e) for the E3 and E4 cells.
The 2016 vdM dataset
shows differences below 0.3\% between the physics and vdM runs for both
A13 and A14, whereas the 2018 CT dataset shows larger shifts, and results for
A13 and A14 that differ by around 0.5\% at high-$\mu$, both for 140 isolated
bunches and 590 bunches in trains. However, even in this dataset, the average
of A13 and A14 cells shows shifts within 0.5\% of unity between the low-$\mu$
and high-$\mu$ regimes.
 
\subsection{Comparisons of calorimeter and track-counting measurements: ladder approach}\label{ss:caltladder}
 
\begin{figure}[tp]
\vspace{-5mm} 
 
\parbox{83mm}{\includegraphics[width=76mm]{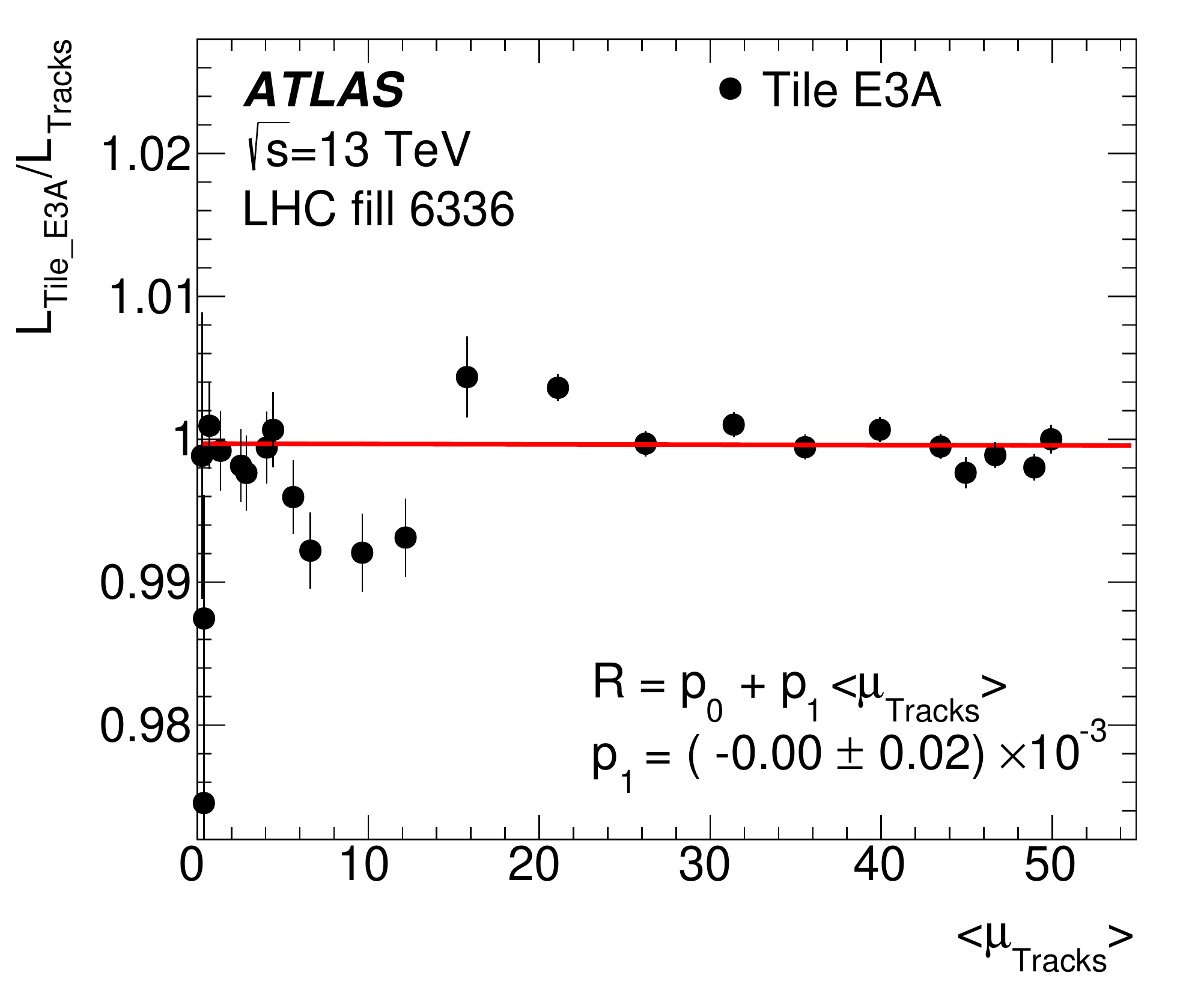}\vspace{-6mm}\center{(a)}}
\parbox{83mm}{\includegraphics[width=76mm]{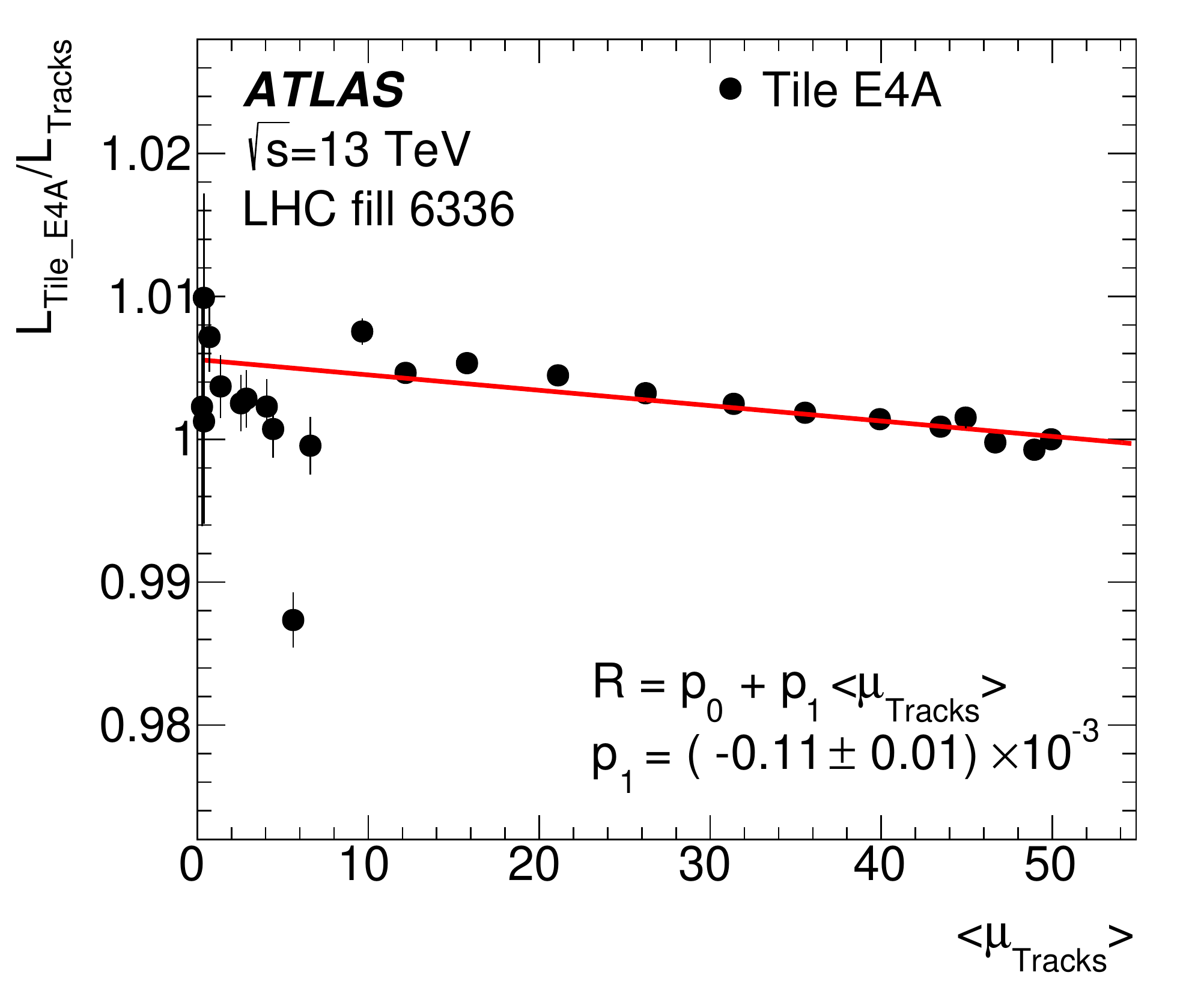}\vspace{-6mm}\center{(b)}}
\parbox{83mm}{\includegraphics[width=76mm]{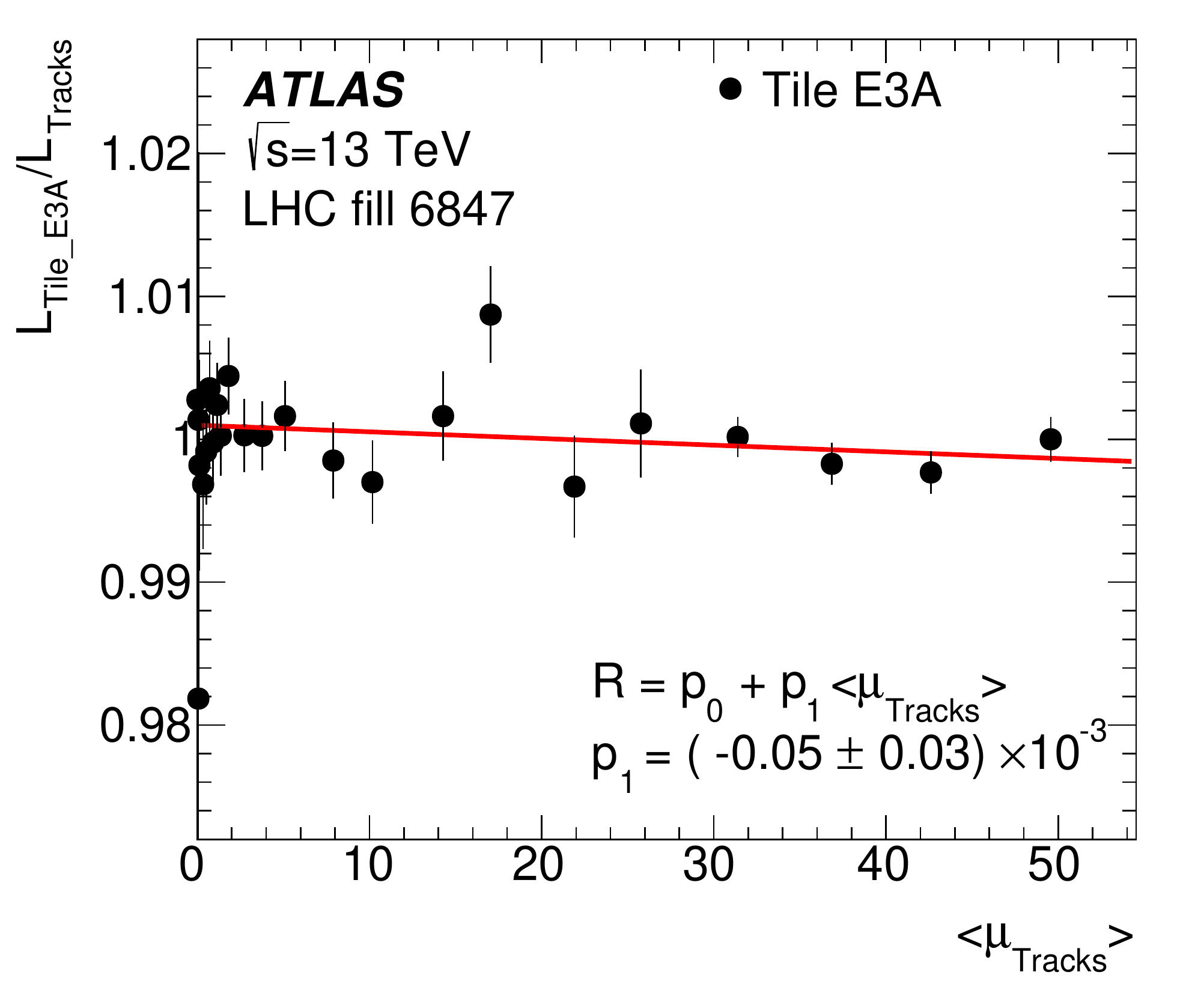}\vspace{-6mm}\center{(c)}}
\parbox{83mm}{\includegraphics[width=76mm]{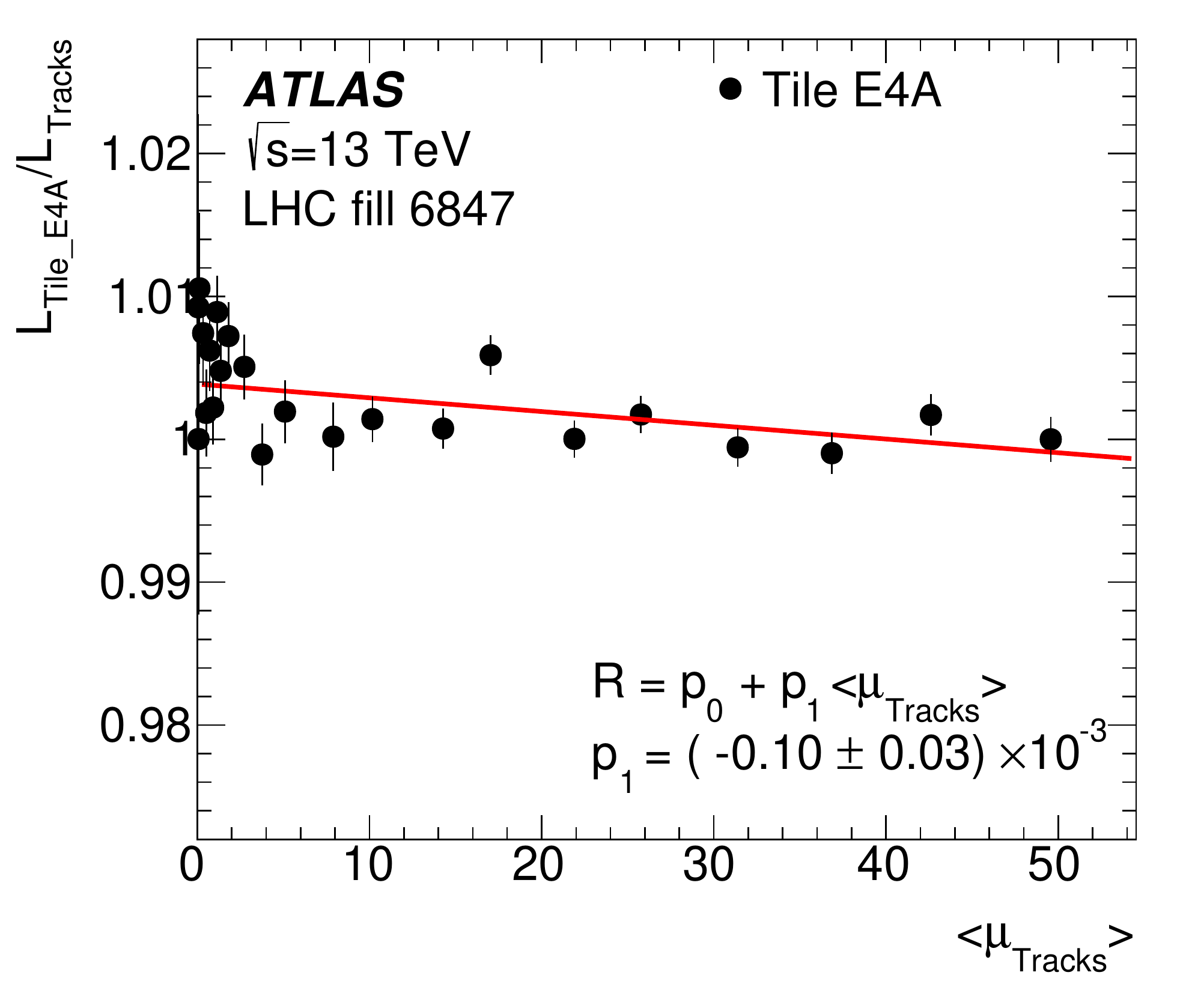}\vspace{-6mm}\center{(d)}}
\parbox{83mm}{\includegraphics[width=76mm]{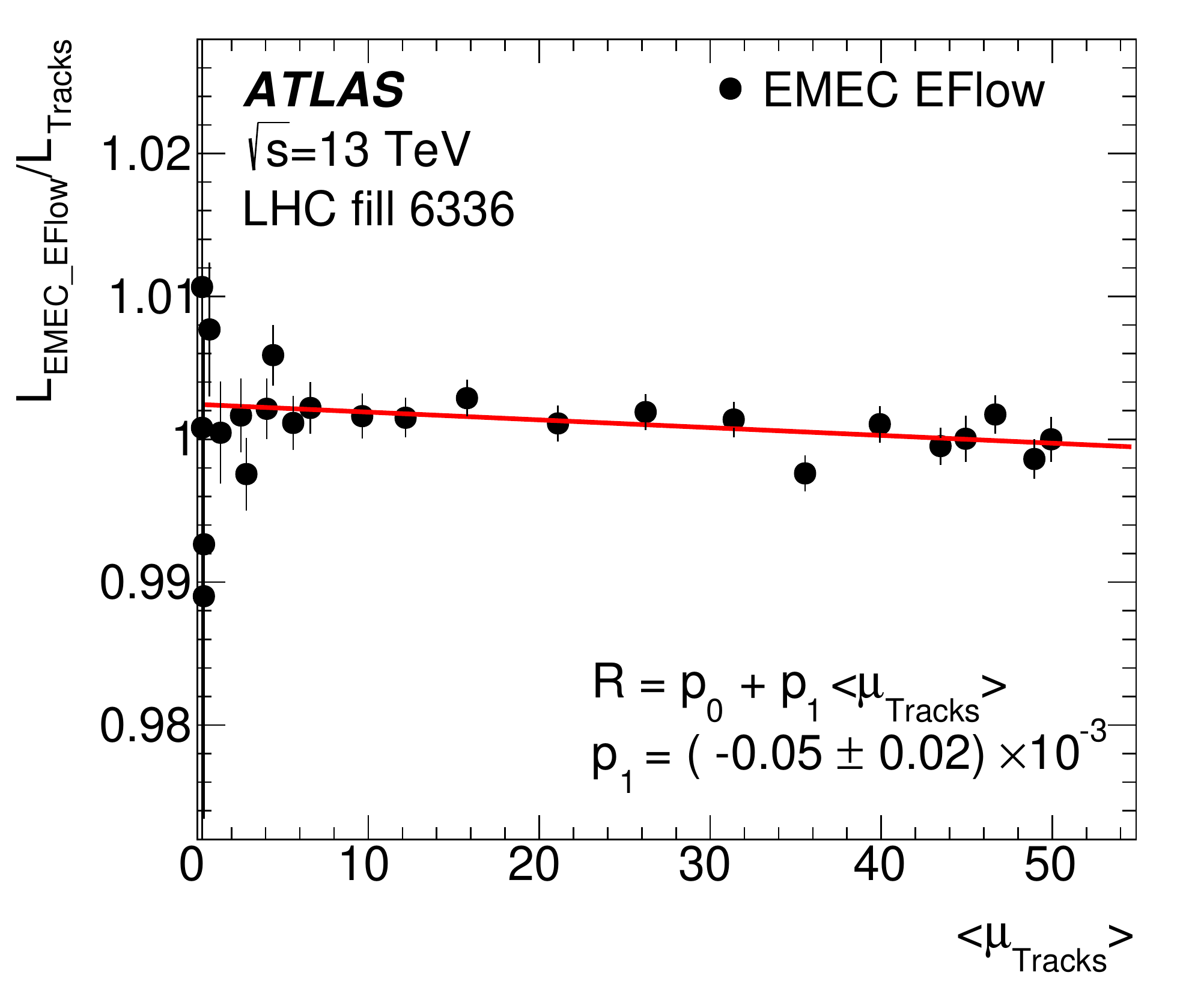}\vspace{-6mm}\center{(e)}}
\parbox{83mm}{\includegraphics[width=76mm]{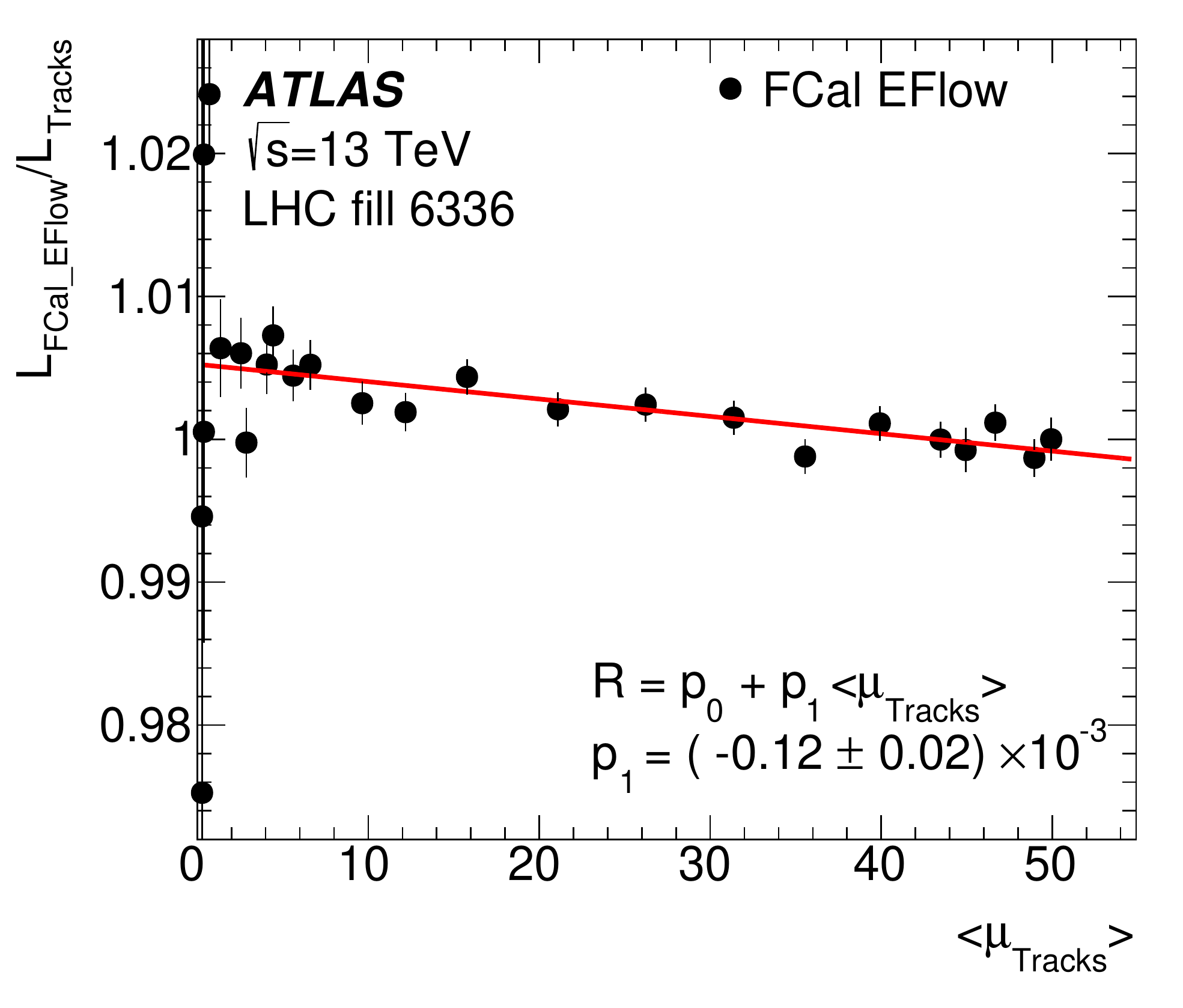}\vspace{-6mm}\center{(f)}}
\caption{\label{f:muscanE}Ratios of luminosity measured by the TileCal E-cells
or LAr energy-flow to that measured by track-counting as a function of \meanmu,
determined from $\mu$-scans recorded in the 140 colliding
bunch fills 6336 in 2017 and 6847 in 2018: (a, b) ratios using the A-side
TileCal E3- and  E4-cells in fill 6336; (c, d) ratios using the A-side TileCal
E3- and E4-cells in fill 6847;
(e) ratios using the EMEC energy-flow measurement in fill 6336, (f)
ratios using the FCal energy-flow measurements in fill 6336.
The ratios are normalised such that the point at highest $\meanmu$ is at
unity, and
the uncertainties are dominated by the statistical uncertainties in the
track-counting measurements. The red lines show linear fits through the points,
and the $p_1$ slope parameters are indicated in the legends. For the TileCal
measurements, the results from the C-side E3 and E4 cells are similar.}
\end{figure}

The ladder approach considers the transition from the vdM to physics regimes in
three steps, low-$\mu$ to high-$\mu$ isolated bunches, high-$\mu$ isolated
bunches to high-$\mu$ bunch trains, then an increasing number of bunches in
trains until the LHC ring is full. The first step was already addressed by
the studies within the 140 colliding bunch fills 6336 and 6847 shown in
Table~\ref{t:caltrun} and Figures~\ref{f:ctdir}(c) and~\ref{f:ctdir}(e),
which suggest a small decrease of at most 0.4\% in \rtitk{E}\ when going from
low- to high-$\mu$
with isolated bunches. This transition was also studied using the $\mu$-scans
carried out in both fills between the low-$\mu$ and high-$\mu$ periods,
as shown for the 2017 fill in Figure~\ref{f:tilect17}(a). The analysis of the
TileCal E-cell data again uses the activation model described in
Section~\ref{ss:actmod}, which is particularly
important in the second part of the $\mu$-scan when the luminosity is decreasing
at each step, increasing the relative importance of activation from
the preceding luminosity blocks with higher instantaneous luminosities.
The ratios \rtitk{E}\ from the A-side E3 and
E4 cells are shown as a function of \meanmu\ for the 2017 140 colliding bunch
fill in Figures~\ref{f:muscanE}(a) and~\ref{f:muscanE}(b), and for the analogous
2018 fill in Figures~\ref{f:muscanE}(c) and~\ref{f:muscanE}(d).
Although the precision is limited
in the low-$\mu$ region by the statistical uncertainties in the track-counting
measurements, and the linear fits are not perfect, the general
trends are clear, and similar between 2017 and 2018. The \rtitk{E}\ ratios
slightly decrease with increasing \meanmu, with negative gradients of up to
$1\times 10^{-4}$, corresponding to a double ratio \rtdbl{E} up to 0.5\%
below unity for $\meanmu\approx 50$, consistent with the results from comparing
the extended low- and high-$\mu$ periods in these fills.
 
The LAr energy-flow luminosity measurements from EMEC and FCal were also studied
in these $\mu$-scans, determining the pedestals from the separated-beam
period with almost-zero luminosity immediately before the $\mu$-scans, and
for the 2018 fill, a second separated-beam period immediately after. The
resulting \rlitk\ ratios for the 2017 fill, where all colliding bunches
were sampled in the special data stream, are shown in Figures~\ref{f:muscanE}(e)
and~\ref{f:muscanE}(f). They show a behaviour very similar to that of the
\rtitk{E}\ ratios, with small negative slopes corresponding to a
negative deviation of up to 0.5\% at $\meanmu\approx 50$.
Similar results were obtained for the 2018 fill, this time only sampling
two of the 140 colliding bunch pairs with higher rate.
These results, which use in-time event-by-event information (thus eliminating
any activation contributions) from an independent detector, provide a powerful
confirmation that the track-counting non-linearities between low- and
high-$\mu$ in isolated bunches are smaller than 0.5\%. Unfortunately the
LAr signal-shaping electronics and the long drift times do not allow the
LAr energy-flow analysis to be extended to fills with bunch trains.
 
\begin{figure}[tp]
\parbox{83mm}{\includegraphics[width=76mm]{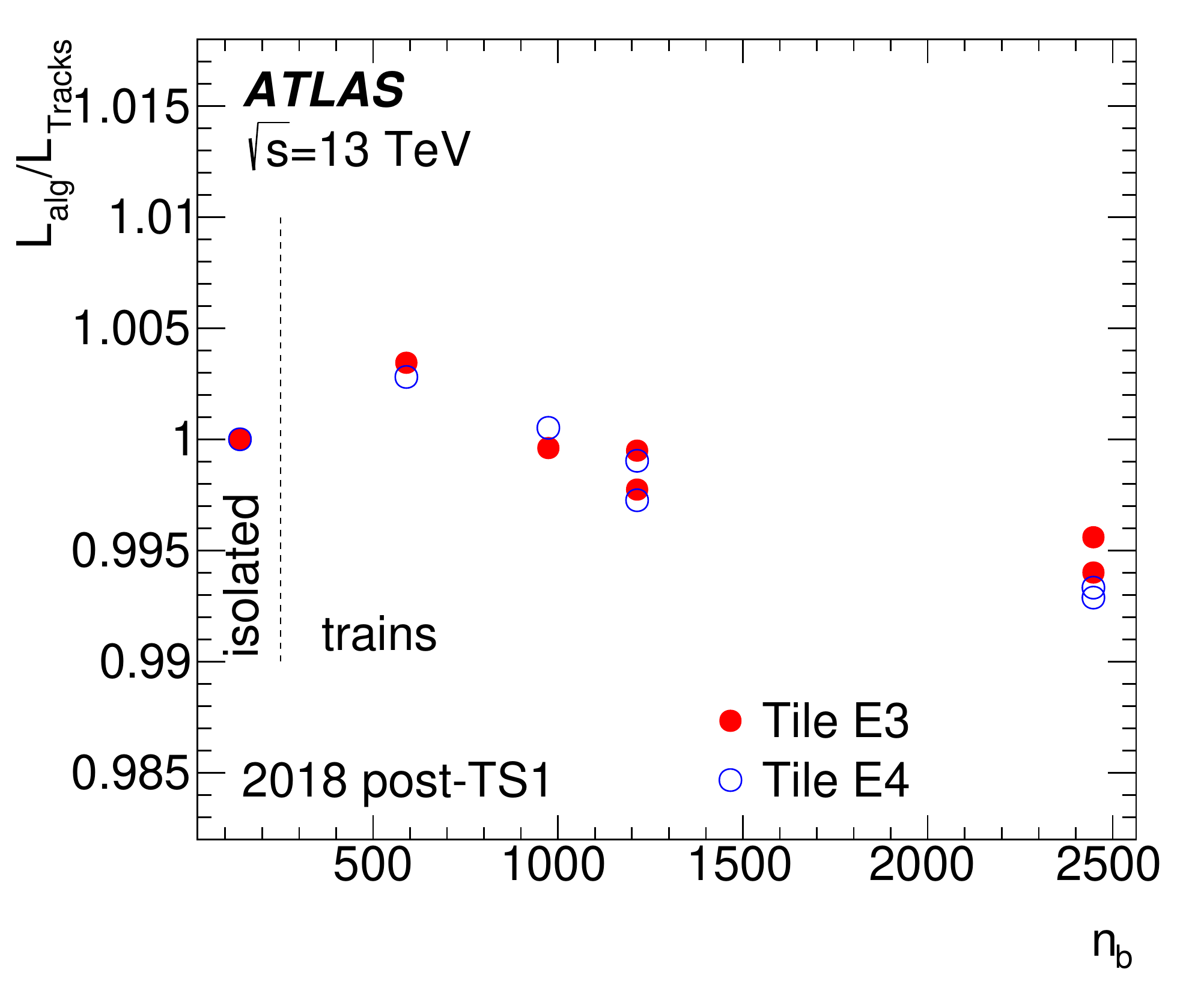}\vspace{-6mm}\center{(a)}}
\parbox{83mm}{\includegraphics[width=76mm]{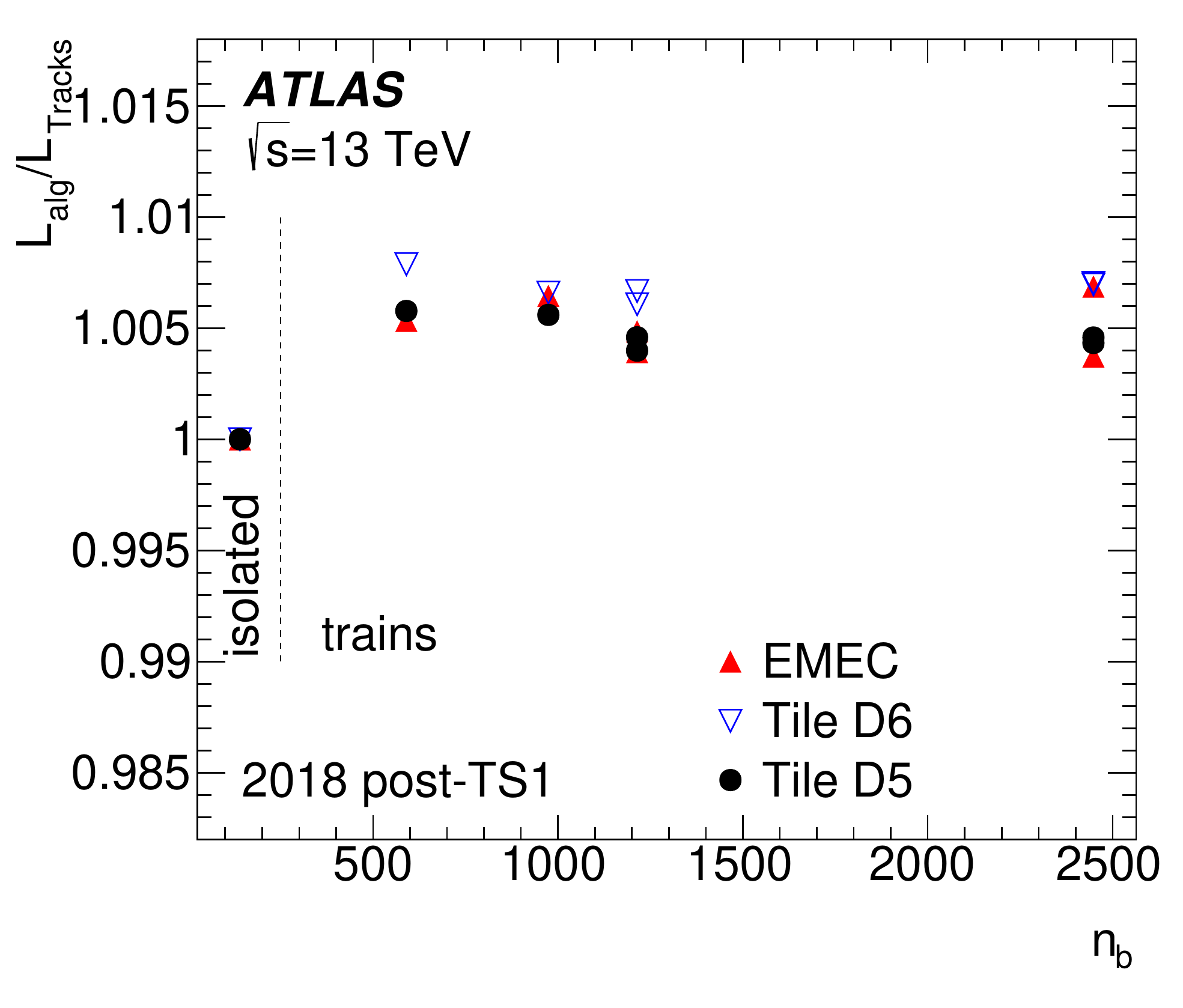}\vspace{-6mm}\center{(b)}}
\caption{\label{f:ctlad}Ratios of integrated luminosity
measured by (a) the TileCal E3 and E4 cells, and (b) the EMEC, TileCal D6
and D5 cells, to that measured by track-counting as a function of the number of
colliding bunch pairs in high-$\mu$ fills in the period after the 2018 technical
stop TS1.
The ratios are averaged over A- and C-side calorimeter measurements and
normalised to the high-$\mu$ part of the 140 isolated-bunch CT fill 6847
(the leftmost point). All the other fills feature bunch trains with 25\,ns
spacing.}
\end{figure}

The second and third steps of the uncertainty ladder were addressed by
comparing calorimeter/track-counting integrated luminosity ratios
in the high-$\mu$ part of the 140 colliding-bunch fill with those in
a sequence of short fills with bunch trains and increasing total numbers of
bunches. The 2018 intensity ramp-up included fills with 590,
1214 (two consecutive fills) and 2448 colliding bunches before the
vdM fill, and 978 and 2448 colliding-bunch fills about one week later. The
resulting ratios are shown as a function of \nbun\ for the TileCal E3 and E4
cells in Figure~\ref{f:ctlad}(a). The luminosity in the 140 colliding-bunch
fill is high enough that the TileCal D cells and EMEC also become sensitive,
and the ratios for TileCal D5, TileCal D6 and EMEC are shown in
Figure~\ref{f:ctlad}(b). All ratios show a positive step of around 0.5\%
when going from isolated bunches to trains.
Given that this step is observed in the ratios obtained using different
calorimeter technologies,
it is most likely to originate from a decrease in the track-counting response
in bunch-train running, e.g.\ because of the time-over-threshold effect in the
pixel detector discussed in Section~\ref{ss:trkperf}.
As the number of bunches increases further, the ratios involving TileCal E-cells
decrease, whereas those involving D-cells or EMEC remain stable within
0.1--0.2\%. This is attributed to scintillator ageing in the TileCal E-cells
as the accumulated luminosity increases through the intensity ramp-up period,
and suggests that the track-counting response does not change significantly
with increasing number of bunches, once the transition from isolated bunches
to trains has been made.
 
Combining all steps, the ladder approach based on comparisons with TileCal D-
and E-cell PMT current measurements, and LAr HV-current and energy-flow
measurements,
implies that the track-counting luminosity response increases by up to 0.5\%
when going from low-$\mu$ to high-$\mu$ isolated bunches, and then decreases
by about 0.5\% going from high-$\mu$ isolated bunches to bunch trains. This
is compatible with the $\pm 0.5$\% overall shift determined from the
direct approach in Section~\ref{ss:caltdir}.
 
Therefore, an overall uncertainty of 0.5\% was assigned to the calibration
transfer correction derived from the track-counting luminosity measurements
and applied to LUCID via Eq.~(\ref{e:mucorr}). Since the
same track-counting selection working point was used for all years, and
no significant year-dependence was seen e.g.\ in Figure~\ref{f:ctmubcid},
the same calibration transfer uncertainty was applied to all the Run~2
datasets (including 2015 where no dedicated studies were performed),
and considered to be correlated between years.

% End of text imported from the .//calsyst.tex input file

% The next lines are included from the .//stab.tex input file
\section{Long-term stability}\label{s:stab}
 
The LUCID correction strategy described in Section~\ref{ss:lucidcorr} results
in a calibrated LUCID luminosity measurement for every luminosity block in
every physics run. However, drifts in the response of LUCID over time
that are not fully corrected by the $^{207}$Bi calibration system, or drifts
in the track-counting measurement used to determine the corrections to LUCID
in the reference runs, could still result in a time-dependent bias in the
baseline LUCID luminosity measurement. These potential biases were
studied by comparing the run-integrated luminosity measurements from LUCID
with independent measurements from the EMEC, FCal and TileCal D6-cell
calorimeter algorithms,
for every physics run in the standard GRL. Since the calorimeter measurements
cannot be calibrated at low luminosity in the vdM run, they were normalised,
or `anchored' to agree with the run-integrated track-counting measurements
in up to ten physics runs close to the vdM fill, such that discrepancies
between the LUCID and calorimeter measurements in physics runs far from the vdM
fill are indicative of long-term drifts in either the LUCID or calorimeter
measurements.
 
Studies of the ratios of the FCal luminosity measurements to those from
track-counting, EMEC and TileCal D6 cells as a function of the bunch-integrated
instantaneous luminosity \linst\
within individual runs, and comparisons of runs with different numbers of
colliding bunches, showed that the FCal measurements suffer from a
reproducible non-linearity that strongly depends on \linst.
Compared to track counting, the FCal response
increases by about 2\% when the luminosity decreases from $15\times 10^{33}$
to $5\times 10^{33}$\,cm$^{-2}$s$^{-1}$;
the lower the luminosity, the steeper the increase in FCal
response. Over the luminosity range of interest for the physics data-taking
(excluding the vdM regime and the low-$\mu$ data discussed in
Section~\ref{s:lowmu}), the evolution of the uncorrected FCal/track-counting
luminosity ratio is well described by a quadratic function of the
uncorrected FCal instantaneous luminosity. Since the studies in
Section~\ref{s:calsyst}
showed that non-linearity in the track-counting measurement is small, the
FCal data were corrected as a function of instantaneous luminosity so
as to agree with track-counting in a single reference fill.
For 2017--18 FCal data, the correction parameters were derived from a single
long physics fill recorded in 2017, and the 2015--16 data were similarly
corrected using a single fill from 2016. These corrections significantly
improved the consistency of FCal and track-counting measurements throughout
the data-taking period, and are applied to all the FCal data shown below.
In addition, significant discrepancies were found between A- and C-side
FCal data in 2015--16, with C-side data agreeing much better with other
luminosity measurements. The A-side data were therefore discarded, and
only the C-side data used in the analysis below. In 2017--18, the two
sides agree well, and their average was used in the comparison with
other luminosity measurements.
 
\begin{figure}[tp]
\parbox{83mm}{\includegraphics[width=78mm]{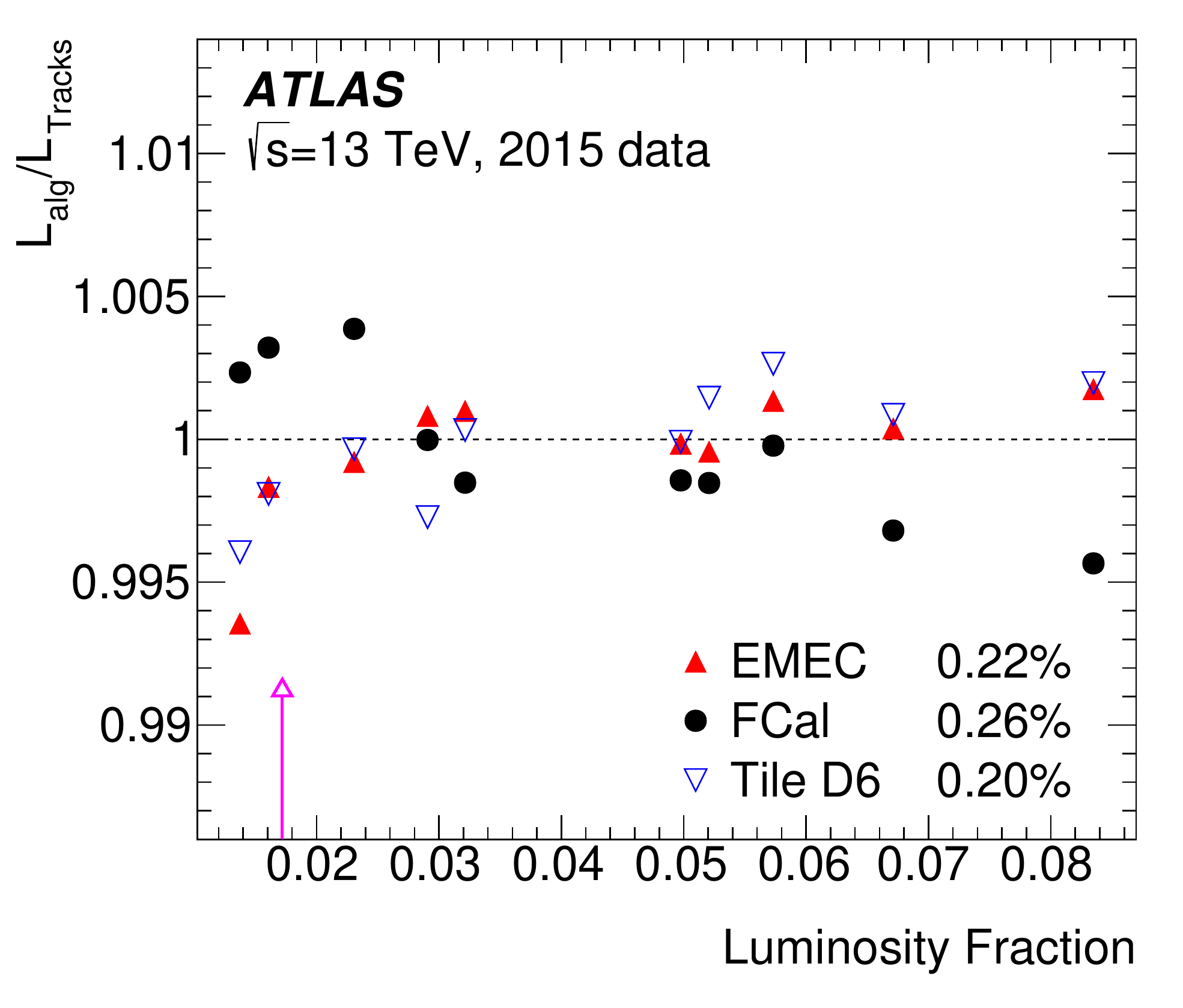}\vspace{-6mm}\center{(a)}}
\parbox{83mm}{\includegraphics[width=78mm]{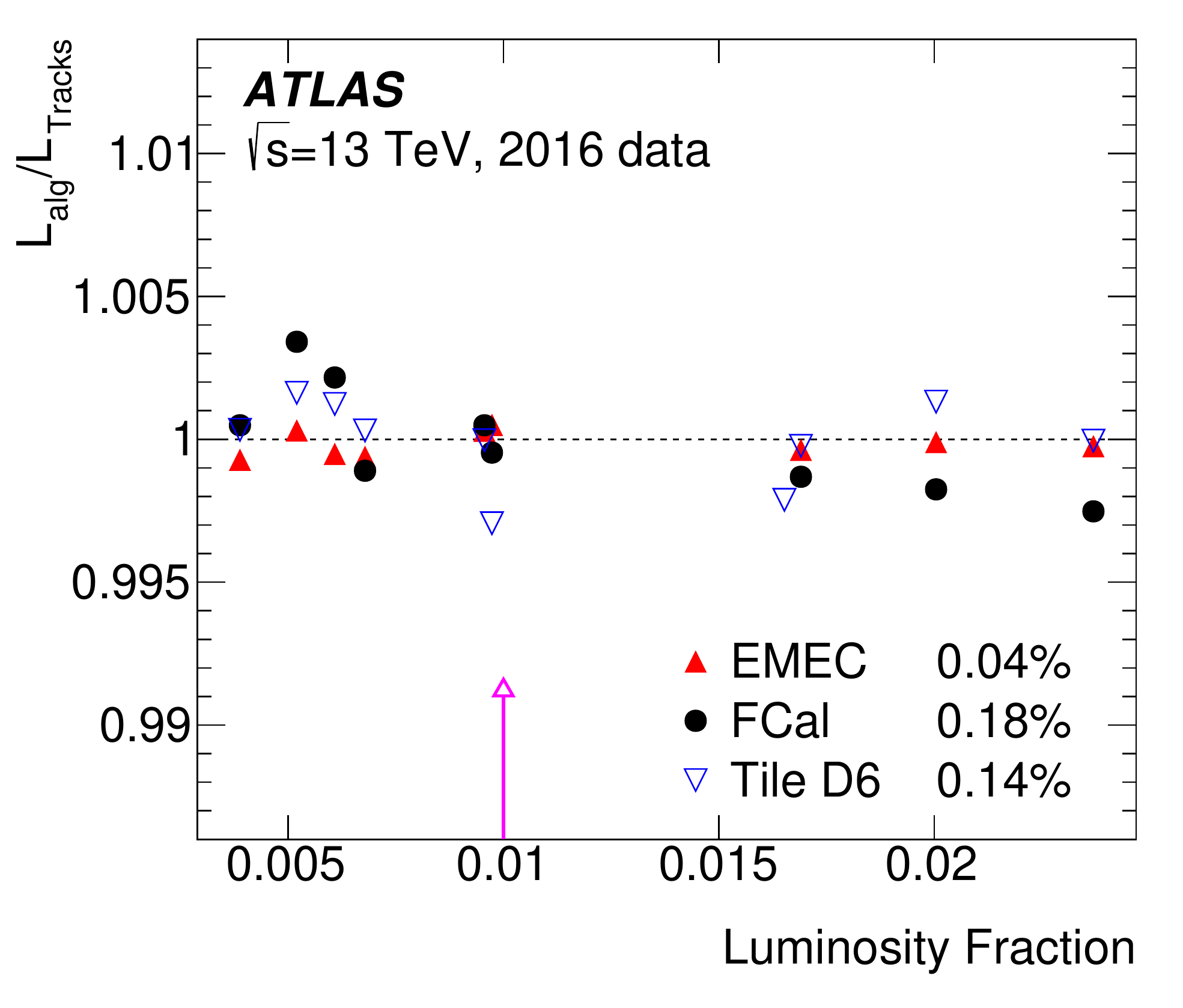}\vspace{-6mm}\center{(b)}}
\parbox{83mm}{\includegraphics[width=78mm]{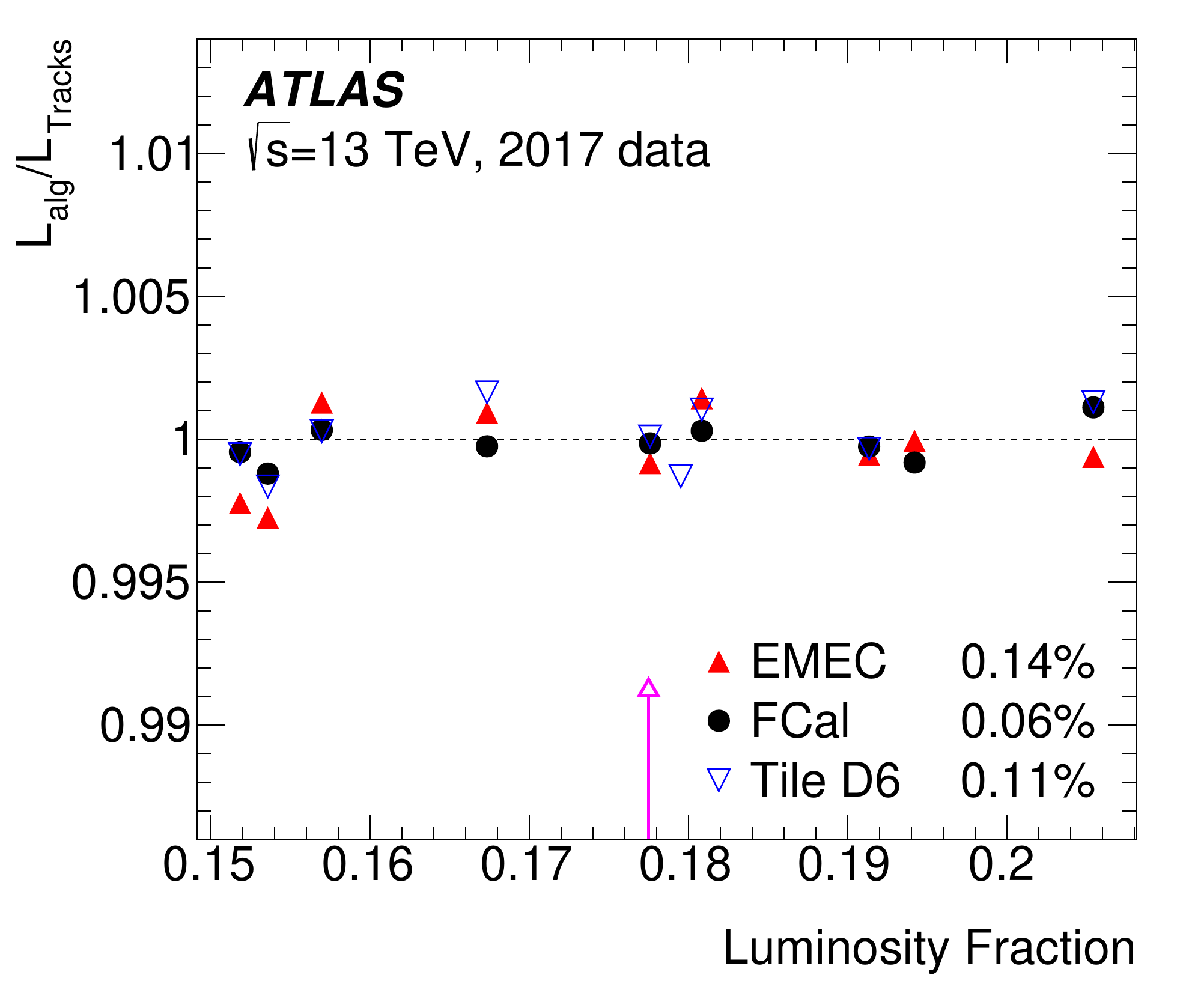}\vspace{-6mm}\center{(c)}}
\parbox{83mm}{\includegraphics[width=78mm]{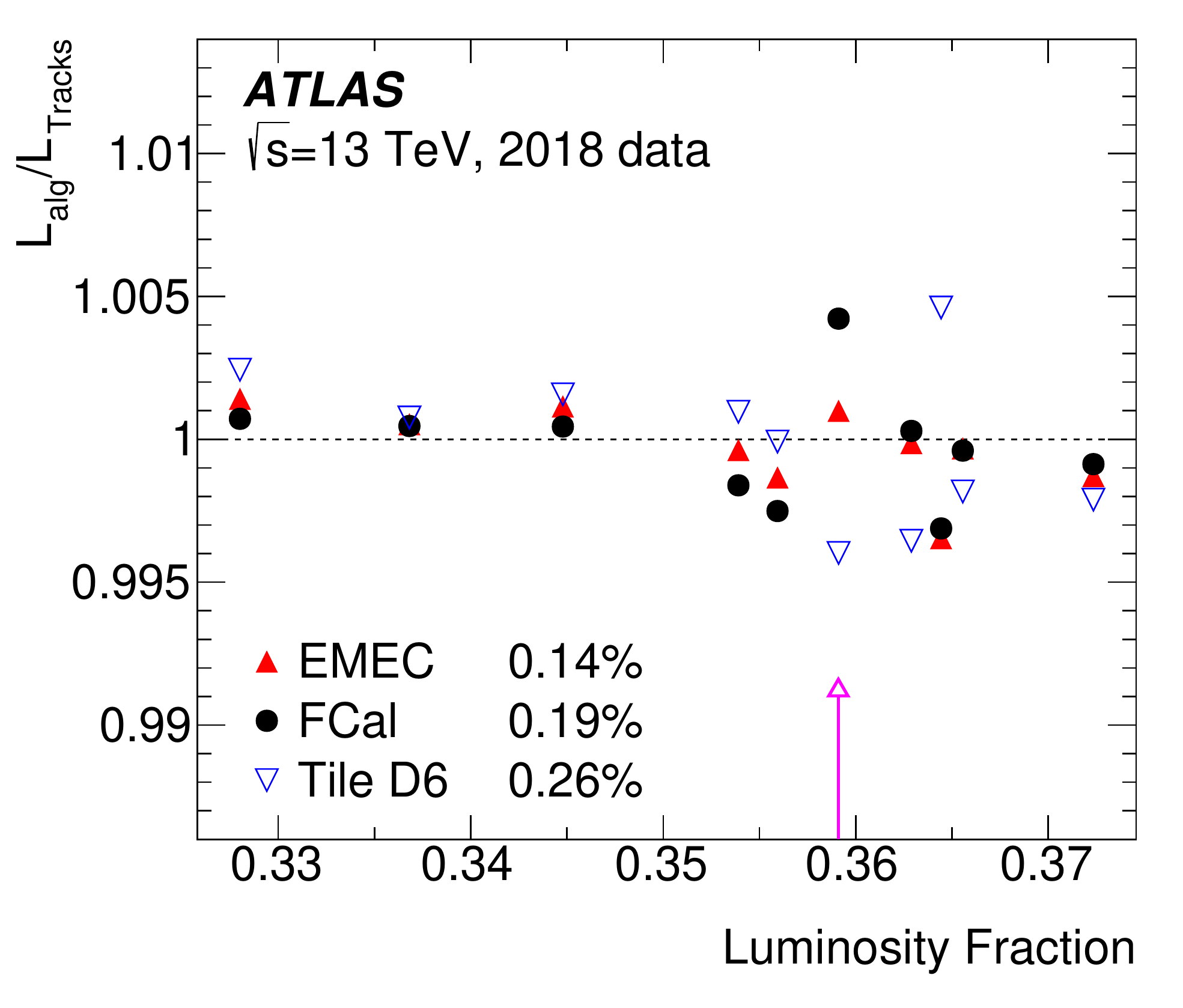}\vspace{-6mm}\center{(d)}}
\caption{\label{f:calanchor}Ratios of the per-run integrated luminosity measured
by the EMEC, FCal and TileCal D6 algorithms to that measured by track-counting
in the ten physics runs surrounding the vdM run, plotted
as functions of the fractional cumulative integrated luminosity in each
data-taking year. The positions of the vdM fills are shown by the purple arrows.
The ratios are each normalised so that the unweighted average
over all ten runs is unity, and the legends show the RMS difference between
the calorimeter and track-counting measurements in each case. Ratios which
are not available for a particular detector and run are not plotted
and removed from the average.}
\end{figure}

The calibration anchoring procedure, which was carried out after correcting
the FCal measurements,
is illustrated in Figure~\ref{f:calanchor}. It shows the
per-run integrated luminosity ratios between calorimeter and track-counting
measurements for the selected physics runs close to the vdM fill,
normalised so that their unweighted
average\footnote{Since the intention is to smooth out possible run-to-run
variations in the calorimeter/track-counting luminosity ratios, the
measurements were not weighted according to the integrated luminosity in each
run.} over all selected runs is unity. Only fills with at least
500 colliding bunch pairs (400 in 2015)
and at least one hour of data-taking were considered. The RMS of each set of
calorimeter/track-counting ratios gives a measure of the short-term
variability for each calorimeter, and is shown in the legend. These variations
are caused partly by differences in the characteristics of each physics fill,
in particular the fill duration, which affects the importance of activation
build up, especially in the EMEC and FCal.
The variations translate into ambiguities in the
calorimeter luminosity normalisation, and were reduced by averaging over
ten fills. In 2015 and 2016, the
vdM fill was performed early in the running period, before the LHC intensity
ramp-up to the full number of colliding bunches had been completed. In 2015,
the first three fills have only 446 or 447  colliding bunches, compared to
1021 at the end of the sequence. The instantaneous luminosity in the first
three fills is atypically low, below the range covered by the FCal intensity
correction derived from 2016 data, and leads to the anomalously high
FCal/track-counting ratios in these fills. In 2016, the number of colliding
bunches increased from 589 to 1740, but the per-bunch instantaneous luminosity
was also higher, leading to smaller variations.
To account for the residual uncertainties in normalising the calorimeter
luminosity measurements, the largest RMS value from
any calorimeter/track-counting calibration seen in each year (excluding the FCal
data in 2015) was taken
as the `calibration anchoring' uncertainty for that year, and is shown in
Table~\ref{t:unc}.
 
\begin{figure}[tp]
\parbox{83mm}{\includegraphics[width=78mm]{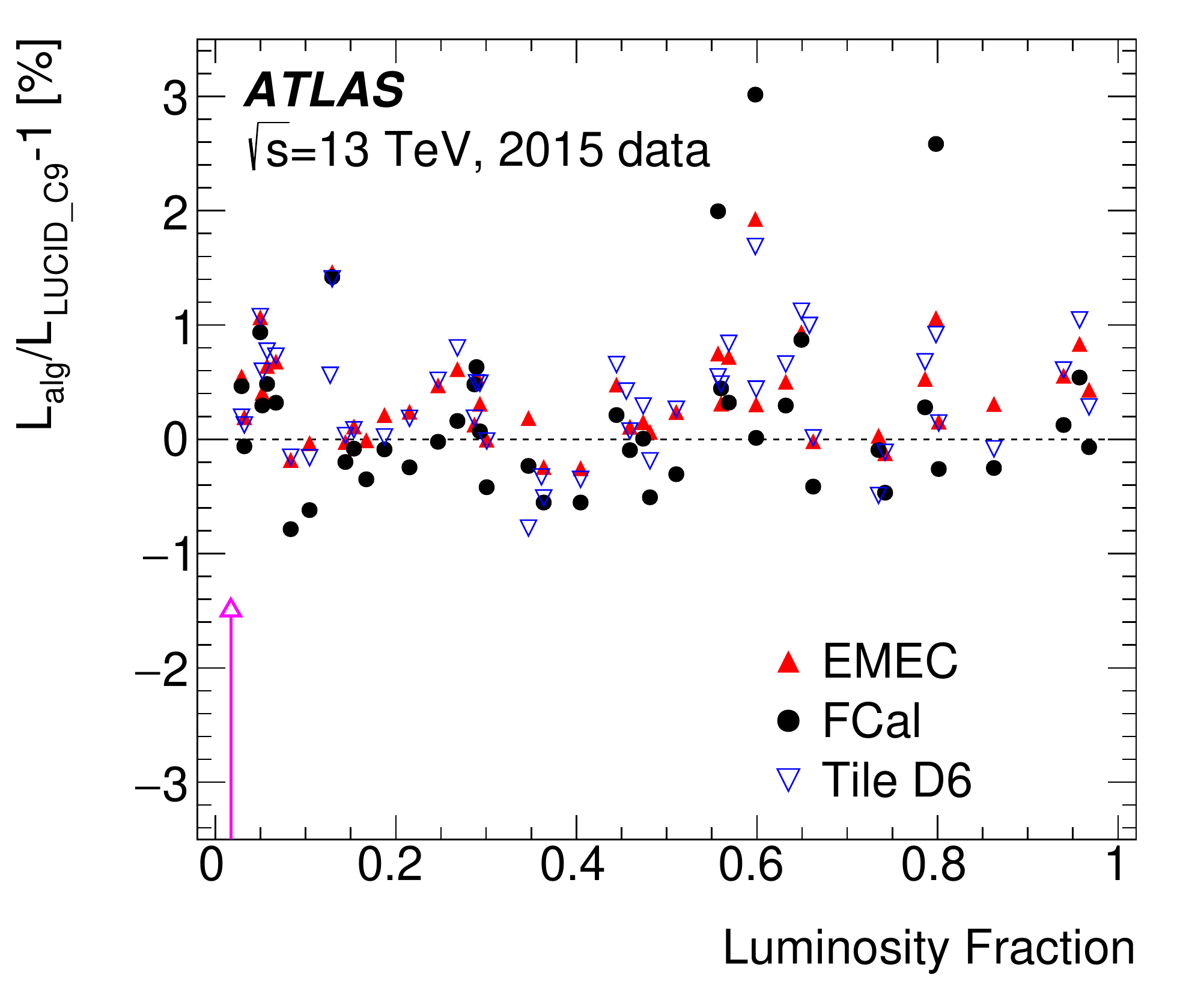}\vspace{-6mm}\center{(a)}}
\parbox{83mm}{\includegraphics[width=78mm]{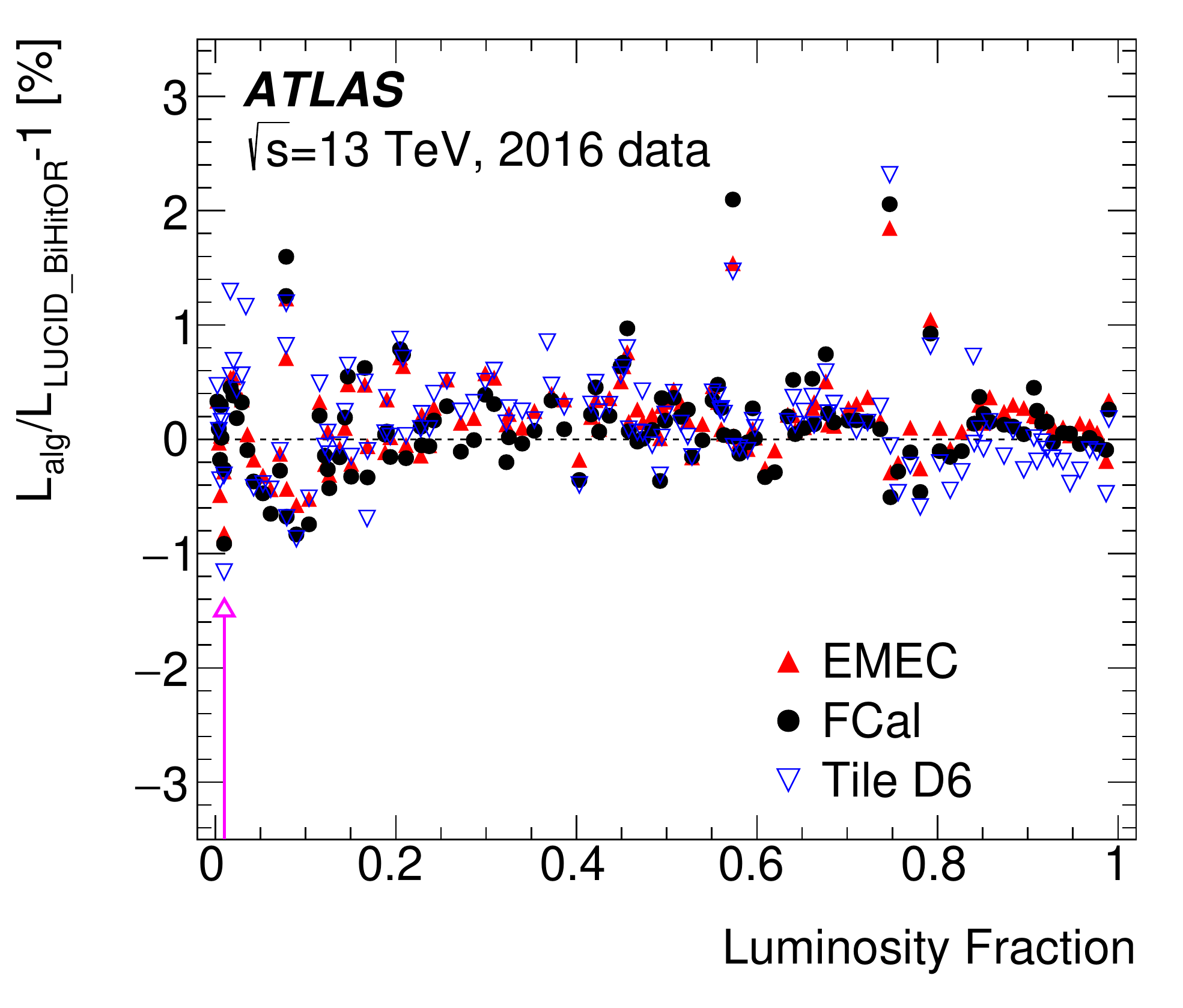}\vspace{-6mm}\center{(b)}}
\parbox{83mm}{\includegraphics[width=78mm]{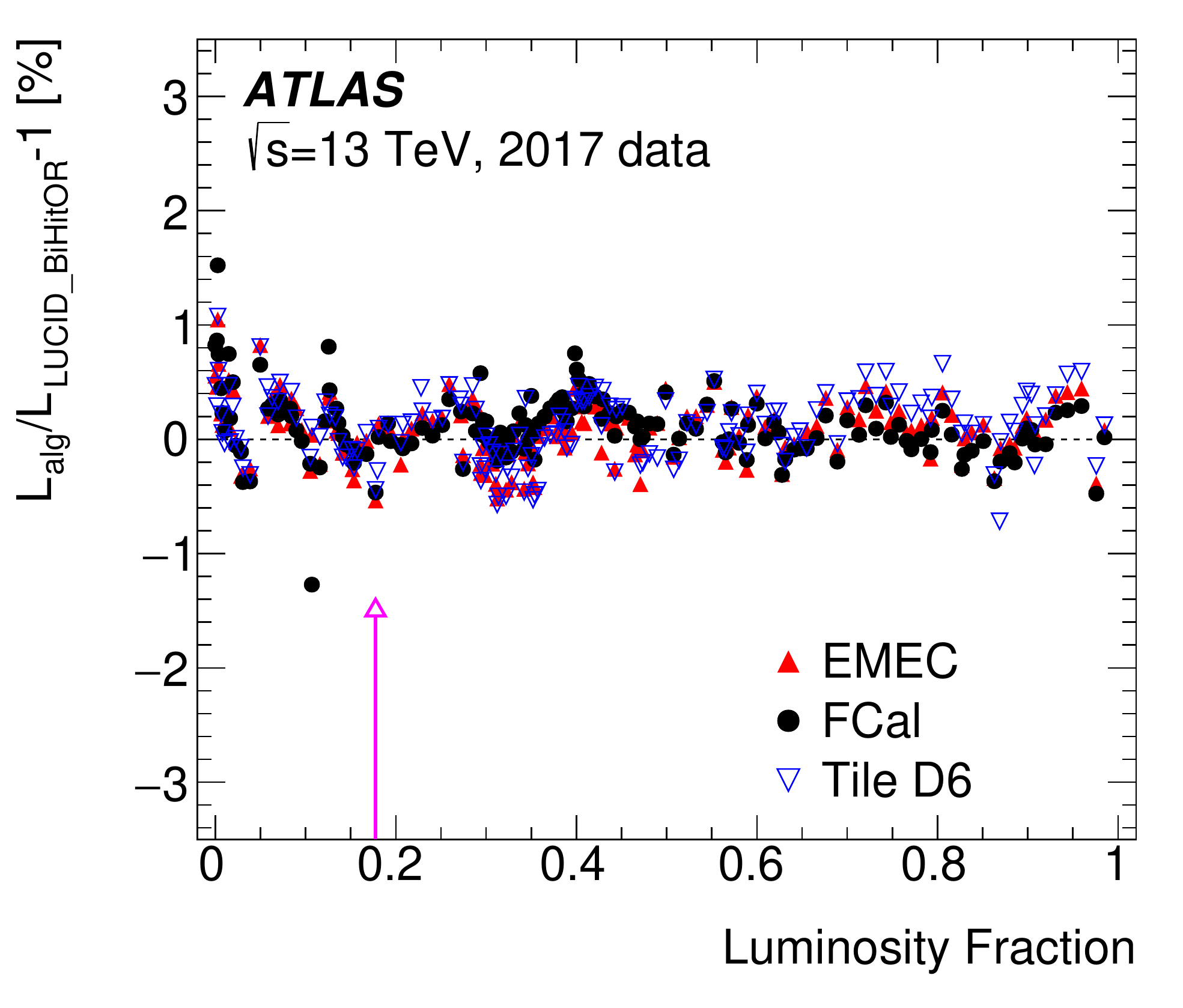}\vspace{-6mm}\center{(c)}}
\parbox{83mm}{\includegraphics[width=78mm]{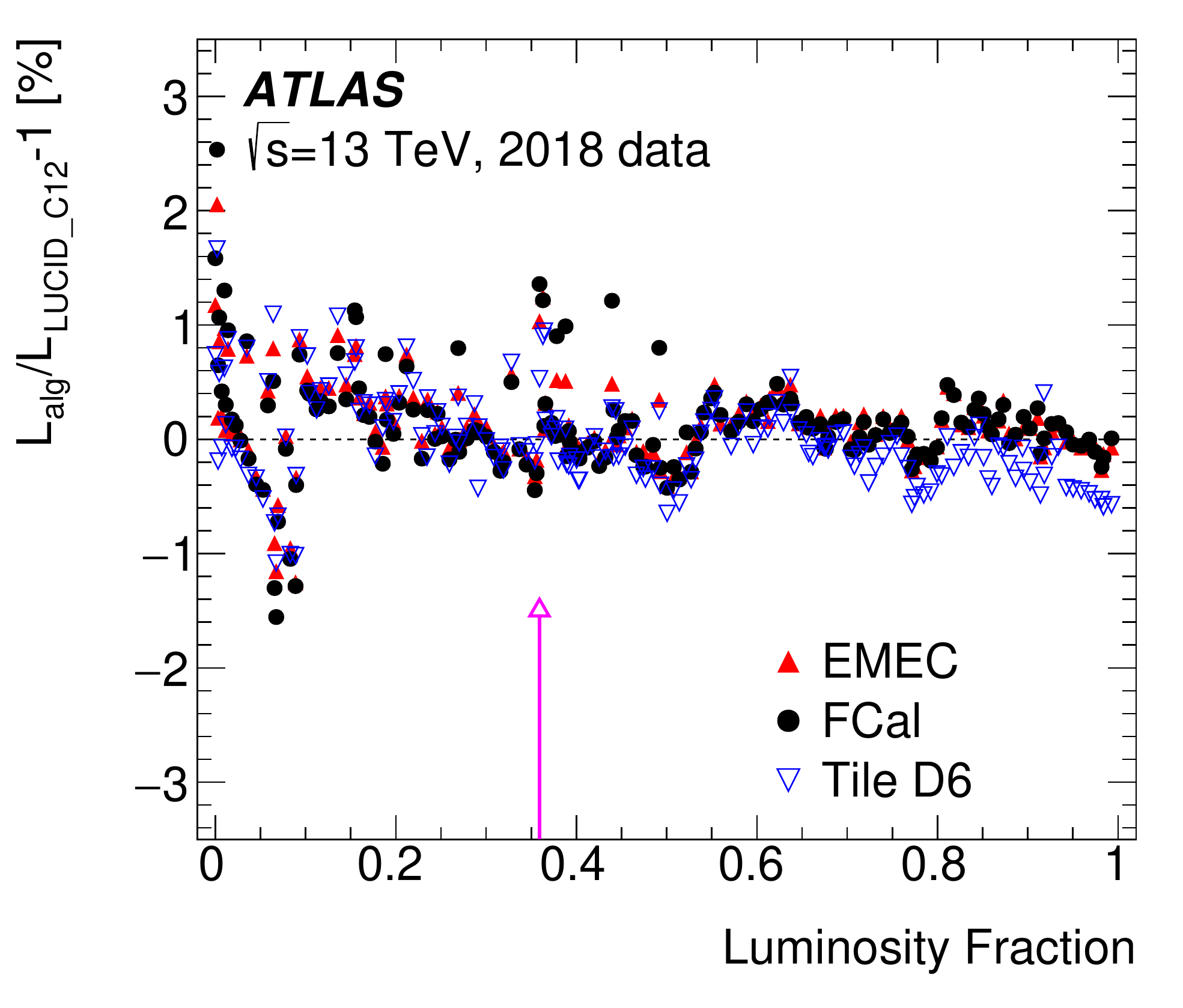}\vspace{-6mm}\center{(d)}}
\caption{\label{f:calostab}Fractional differences in run-integrated luminosity
between the EMEC, FCal and TileCal D6 luminosity measurements and the
baseline LUCID luminosity measurement, plotted as functions
of the fractional cumulative integrated luminosity in each data-taking year.
The positions of the vdM fills are shown by the purple arrows.
Only runs with at least 500 colliding-bunch pairs and one hour
of data-taking are shown.}
\end{figure}

The anchored calorimeter measurements were then compared with
LUCID measurements in each run to assess the long-term stability uncertainty.
Figure~\ref{f:calostab} shows the per-run integrated luminosity differences
between each calorimeter measurement and LUCID as a function of the fractional
cumulative integrated luminosity in each data-taking year, for runs with
at least 500 colliding bunch pairs and one hour of data-taking. Except in
the early parts of the years, most of the variations are well within $\pm 1\%$
in 2015, and within $\pm 0.5$\% in subsequent years. In some periods, correlated
trends common to all of the calorimeter vs. LUCID comparisons are visible,
suggesting that the differences
are caused by variations in the LUCID luminosity measurements.
 
\begin{figure}[tp]
\parbox{83mm}{\includegraphics[width=78mm]{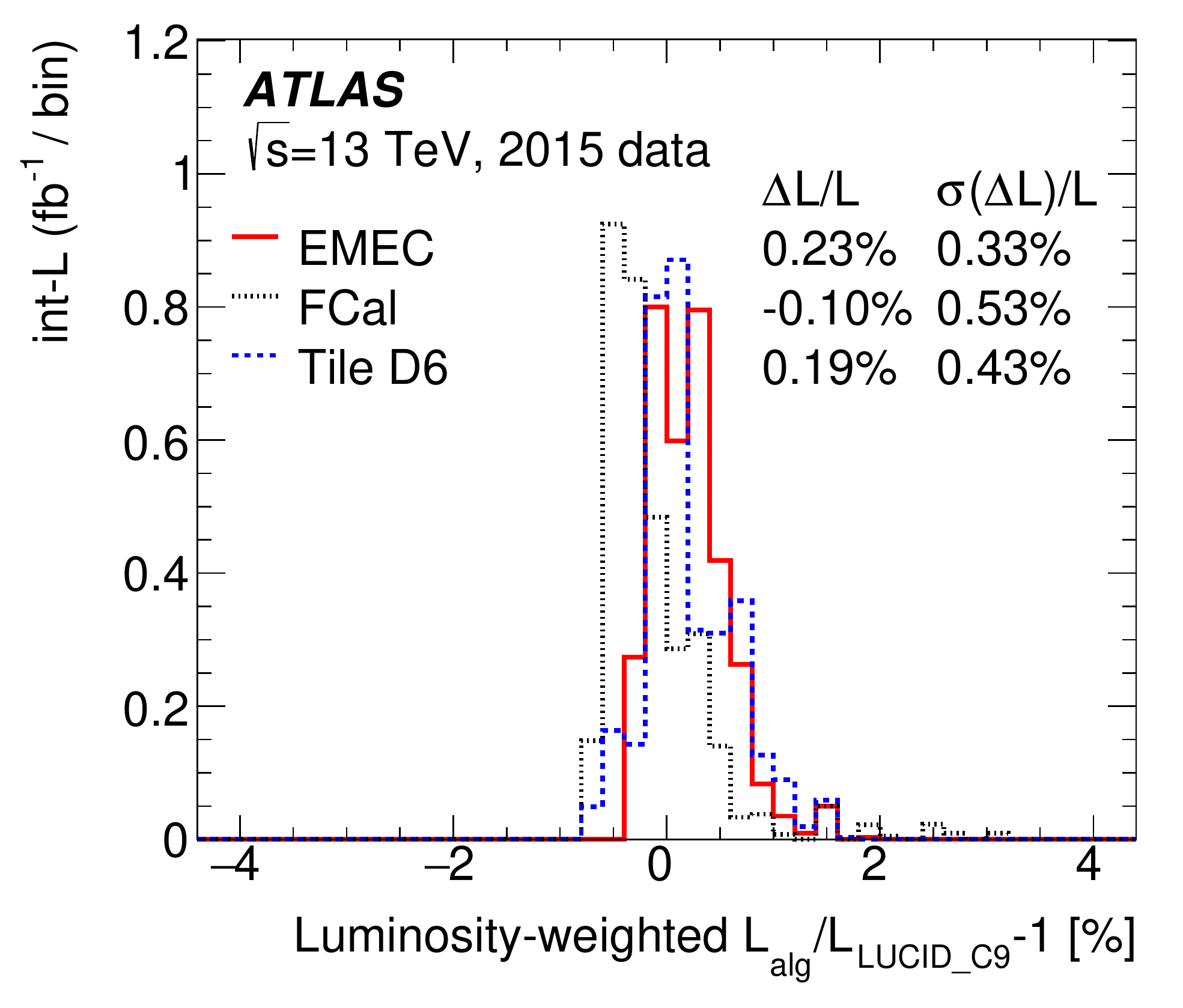}\vspace{-6mm}\center{(a)}}
\parbox{83mm}{\includegraphics[width=78mm]{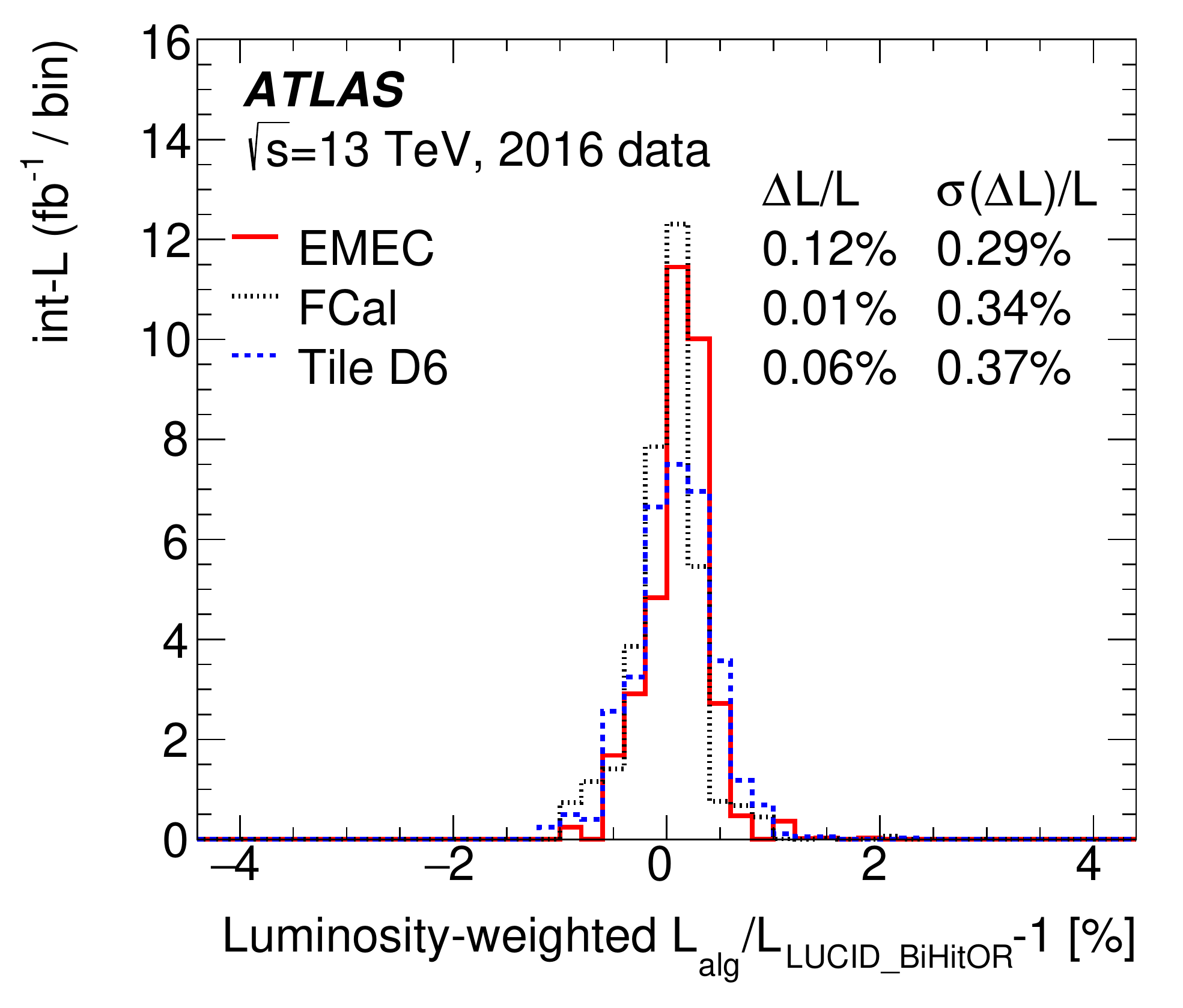}\vspace{-6mm}\center{(b)}}
\parbox{83mm}{\includegraphics[width=78mm]{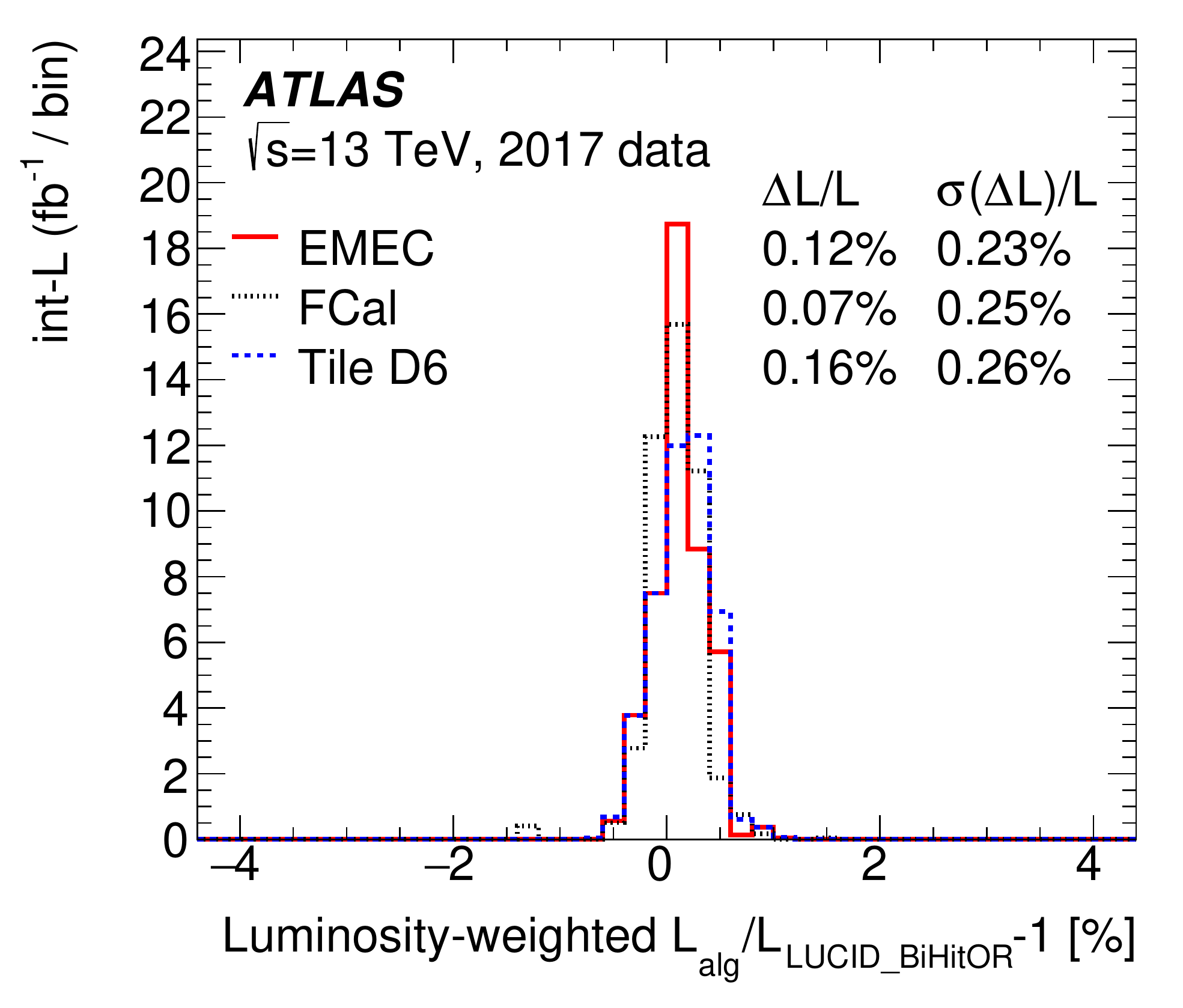}\vspace{-6mm}\center{(c)}}
\parbox{83mm}{\includegraphics[width=78mm]{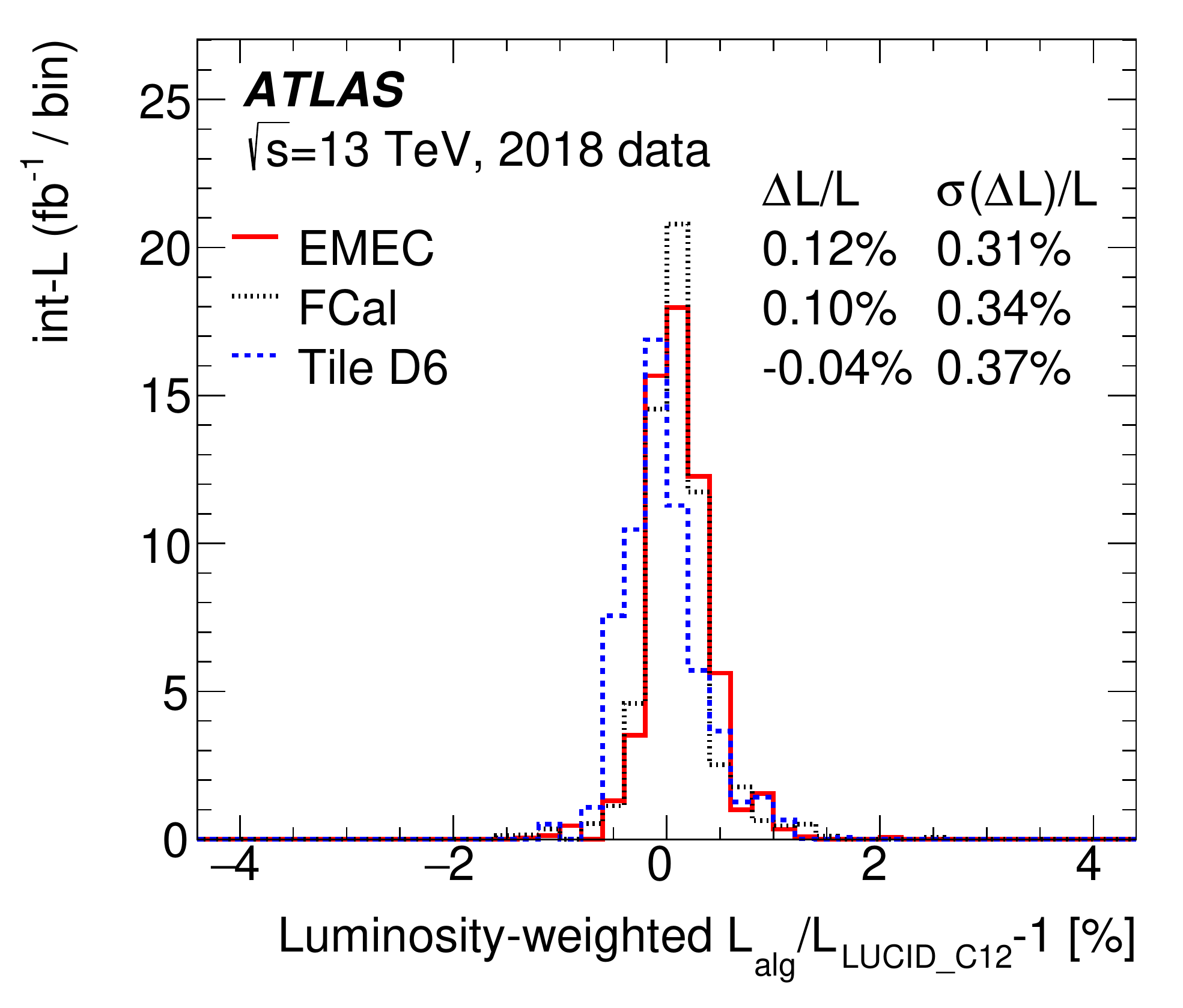}\vspace{-6mm}\center{(d)}}
\caption{\label{f:calorms}Distributions of relative differences
in run-integrated luminosity between the EMEC, FCal and TileCal D6  luminosity
measurements and the baseline LUCID luminosity measurement, weighted by the
integrated luminosity in each run, for each data-taking year. The mean
and RMS of each of the distributions are given in the legend.}
\end{figure}

Figure~\ref{f:calostab} demonstrates the run-to-run spread between
luminosity measurements and their relative trends as a function of time, but
shows all
runs with equal weight. Figure~\ref{f:calorms} shows an alternative presentation
of the same data, showing histograms of the per-run integrated luminosity
differences, weighted by the integrated luminosity of each run. The RMS values
of these distributions reflect the spread also visible in
Figure~\ref{f:calostab}, but with higher weight given to longer runs. The means
represent the fractional difference in integrated luminosity over the entire
year obtained from each calorimeter measurement compared to LUCID. Since
physics analyses require the total integrated luminosity of the dataset,
these differences in integrated luminosity represent the best metric
for assessing the consistency of different luminosity measurements, and hence
the potential error in the baseline LUCID luminosity measurement integrated
over the whole year. The long-term stability uncertainty for each year quoted in
Table~\ref{t:unc} was therefore defined as the largest difference in
year-integrated luminosity between LUCID and any calorimeter measurement. These
differences are in the range 0.1--0.2\%, and originate from comparisons of
LUCID with different detectors in the different years.
 
The calibration anchoring
procedure, including the uncertainty from the RMS over ten runs,
protects against a fortuitous choice of anchoring run leading to an
accidentally small difference in the total integrated luminosities.
The track-counting luminosity
measurements (shown in Figure~\ref{f:lucidepoch}) were not considered in these
comparisons, as the correction of LUCID to track-counting performed with multiple
epochs per year makes them not fully independent. A comparison of the trends
seen in Figure~\ref{f:calostab} with those in Figure~\ref{f:lucidepoch} also
suggests that the residual differences between LUCID and the calorimeters
are rather similar to those between LUCID and track-counting.
 
The calibrations of the different luminosity measurements are in principle
sensitive to shifts of the mean longitudinal position of the luminous region
with respect to the nominal ATLAS interaction point. Studies from a test made
in 2016, where the
mean $z$-position was moved over the range $|z|<300$\,mm, showed that these
effects vary across the different detectors, but that they are all
negligible for the mean $z$-shifts of up to $\pm O(10\,\mathrm{mm})$
seen in physics conditions, especially for algorithms that average measurements
from both sides of the ATLAS detector.

% End of text imported from the .//stab.tex input file

% The next lines are included from the .//uncert.tex input file
\section{Uncertainties and results}\label{s:unccorl}
 
The various sources of uncertainty in the Run~2 luminosity calibration
have been discussed in Sections~\ref{s:vdmcal}--\ref{s:stab}, in terms
of the separate datasets from each data-taking year, each with its own
absolute vdM calibration. However, most ATLAS physics analyses treat the
full Run~2 \sxyt\ $pp$ collision data sample as a single dataset, and
therefore require the uncertainty in the total integrated luminosity of this
combined dataset. The uncertainty correlations between years are discussed in
Section~\ref{ss:corl}, and the final uncertainties are derived and tabulated
in Section~\ref{ss:reshimu}. The consistency between the different years, and
the stability as a function of \meanmu, have been studied via comparisons
with relative luminosity measurements derived from the rates of reconstructed
$Z\rightarrow ee$ and $Z\rightarrow\mu\mu$ events (`$Z$-counting'). These
comparisons are described in Section~\ref{ss:zcount}.
 
\subsection{Uncertainty correlations between years}\label{ss:corl}
 
Since the absolute luminosity scale was calibrated separately for each year
with an independent vdM scan session, a large part of the associated
uncertainty is uncorrelated between years. The bunch-by-bunch consistency,
scan-to-scan reproducibility and reference specific luminosity uncertainties
are driven by the internal consistency of each set of vdM scans, and were
considered to be uncorrelated, as were the beam position jitter
and orbit-drift correction uncertainties, which depend on the specifics of
the jitter and drifts observed in each scan session.
The uncertainties in the FBCT and
DCCT calibration and ghost/satellite measurements are dominated by systematic
instrumental effects and were considered correlated.
The non-factorisation uncertainty was taken to be correlated,
because the same methodology was used to evaluate it in each year, and
a common underlying cause is likely. For the same reasons, the beam--beam
effects, emittance growth correction, background subtraction, fit model and ID
length scale uncertainties were treated as correlated between years.
The results of the non-linear
length scale fits accounting for magnetic non-linearity vary significantly
between years, with no consistent trend or bias being visible in
Figure~\ref{f:magsum}, so the magnetic non-linearity uncertainty
was treated as uncorrelated between years. An alternative treatment, applying
the baseline magnetic non-linearity fit results shown in Figure~\ref{f:magsum}
as correlated corrections to the \sigmavis\ values in each year, gives a shift
in the total Run~2 integrated luminosity which is smaller than the uncertainty
from treating the years as uncorrelated, due to the non-linearity correction
for 2018 having the opposite sign to those for 2015--17. Applying the same
procedure to the alternative magnetic non-linearity fits discussed in
Section~\ref{ss:lscmag} also gives smaller uncertainties than the uncorrelated
treatment. The separate uncertainty in the linear length scale calibration
is statistical in nature and hence uncorrelated between years.
 
The calibration transfer uncertainty was derived from comparisons
of TileCal E-cell and track-counting data at multiple points throughout
2016--2018, and the resulting uncertainty was taken to be correlated between
years, as discussed at the end of  Section~\ref{s:calsyst}.
The calibration anchoring and long-term stability uncertainties
were obtained by taking the largest
difference between a reference measurement from LUCID or track-counting and any
calorimeter-based measurement. These comparisons exhibit significant scatter,
with the largest differences having different signs and coming from different
calorimeter measurements across the years, and the corresponding uncertainties
were taken to be uncorrelated between years.
 
The sensitivity of the final combined uncertainty to the above assumptions was
further investigated by separately considering the beam position jitter,
reference specific luminosity, calibration anchoring and long-term stability
uncertainties to be correlated rather than uncorrelated between years. None
of these variations increased the size of the total Run~2 luminosity
uncertainty by more than 4\% of its original value. Taking the non-factorisation
uncertainty to be uncorrelated rather than correlated reduced the total
uncertainty by 2\%. The total uncertainty is therefore rather insensitive
to the correlations assumed for these  contributions.
 
\subsection{Results for the $\sqrt{s}=13\,\TeV$ Run~2 dataset}\label{ss:reshimu}
 
The integrated luminosities and associated uncertainties for the high-pileup
$pp$ collision data sample at \sxyt\ from each year of Run~2
data-taking are summarised in Table~\ref{t:unc}. The data samples
are those recorded by ATLAS and passing the standard data quality requirements
for ATLAS physics analyses \cite{DAPR-2018-01}, and are significantly smaller
than those for the luminosity delivered to ATLAS shown in Table~\ref{t:lhcpar}.
In 2017 and 2018, they do not include
the \lowmuintl\,\ipb\ of low-pileup data discussed in Section~\ref{s:lowmu}.
 
\begin{table}[tp]
\caption{\label{t:unc}Summary of the integrated luminosities (after standard
data-quality requirements) and uncertainties
for the calibration of each individual year of the Run~2 $pp$
data sample at \sxyt\ and the full combined sample.
As well as the integrated luminosities and total uncertainties,  the table gives
the breakdown and total of contributions to the absolute vdM calibration,
the additional uncertainties for the physics data sample, and the
total relative uncertainty in percent. Contributions
marked $^*$ are considered fully correlated between years,
and the other uncertainties are considered uncorrelated.}
\centering
\begin{tabular}{l|cccc|c}\hline
Data sample  & 2015 & 2016 & 2017 & 2018 & Comb. \\
\hline
Integrated luminosity [fb$^{-1}$]  &    3.24 &   33.40 &   44.63 &   58.79 &  140.07 \\
Total uncertainty [fb$^{-1}$] &   0.04 &   0.30 &   0.50 &   0.64 &     1.17 \\
\hline
Uncertainty contributions [\%]: & & & &  \\
Statistical uncertainty &   0.07 &   0.02 &   0.02 &   0.03 &   0.01 \\
Fit model$^*$ &   0.14 &   0.08 &   0.09 &   0.17 &   0.12 \\
Background subtraction$^*$ &   0.06 &   0.11 &   0.19 &   0.11 &   0.13 \\
FBCT bunch-by-bunch fractions$^*$ &   0.07 &   0.09 &   0.07 &   0.07 &   0.07 \\
Ghost-charge and satellite bunches$^*$ &   0.04 &   0.04 &   0.02 &   0.09 &   0.05 \\
DCCT calibration$^*$ &   0.20 &   0.20 &   0.20 &   0.20 &   0.20 \\
Orbit-drift correction &   0.05 &   0.02 &   0.02 &   0.01 &   0.01 \\
Beam position jitter &   0.20 &   0.22 &   0.20 &   0.23 &   0.13 \\
Non-factorisation effects$^*$ &   0.60 &   0.30 &   0.10 &   0.30 &   0.24 \\
Beam--beam effects$^*$ &   0.27 &   0.25 &   0.26 &   0.26 &   0.26 \\
Emittance growth correction$^*$ &   0.04 &   0.02 &   0.09 &   0.02 &   0.04 \\
Length scale calibration &   0.03 &   0.06 &   0.04 &   0.04 &   0.03 \\
Inner detector length scale$^*$ &   0.12 &   0.12 &   0.12 &   0.12 &   0.12 \\
Magnetic non-linearity &   0.37 &   0.07 &   0.34 &   0.60 &   0.27 \\
Bunch-by-bunch $\sigma_\mathrm{vis}$ consistency &   0.44 &   0.28 &   0.19 &   0.00 &   0.09 \\
Scan-to-scan reproducibility &   0.09 &   0.18 &   0.71 &   0.30 &   0.26 \\
Reference specific luminosity &   0.13 &   0.29 &   0.30 &   0.31 &   0.18 \\
\hline
Subtotal vdM calibration &   0.96 &   0.70 &   0.99 &   0.93 &   0.65 \\
\hline
Calibration transfer$^*$ &   0.50 &   0.50 &   0.50 &   0.50 &   0.50 \\
Calibration anchoring &   0.22 &   0.18 &   0.14 &   0.26 &   0.13 \\
Long-term stability &   0.23 &   0.12 &   0.16 &   0.12 &   0.08 \\
\hline
Total uncertainty [\%] &   1.13 &   0.89 &   1.13 &   1.10 &   0.83 \\
\hline
\end{tabular}
\end{table}
 
The total Run~2 integrated luminosity $\ltot$ is the sum of the integrated
luminosities $\ltoti$ for each individual year:
\begin{equation*}
\ltot = \sum_i \ltoti\ ,
\end{equation*}
and the absolute uncertainty in the total luminosity, $\sigma_{\ltot}$, is given
by standard error propagation as
\begin{equation*}
\sigma^2_{\ltot} = \mathbf{e^T V_L e}\ .
\end{equation*}
Here, $\mathbf{V_L}$ is the covariance matrix of the absolute luminosity
uncertainties for the different years, and $\mathbf{e}$ is a column vector
with unit entries.\footnote{In general, $\sigma^2_{\ltot}=\mathrm{D^T V_L D}$
where the $i$ entries of $\mathrm{D}$ are $\mathrm{d}\ltot/\mathrm{d}\ltoti$, but
since these derivatives are all unity, $\mathrm{D}=\{1,1,1,1\}$.}
The covariance matrix is made up of the sum of terms corresponding to
each uncertainty source in Table~\ref{t:unc}; uncorrelated uncertainties
give rise to terms on the diagonal, whilst correlated
sources are represented by terms with non-zero off-diagonal entries.
 
The results of the combination are shown in the rightmost column in
Table~\ref{t:unc}. The integrated luminosity of the full Run~2 sample
is $\totintl\pm\etotintl\,\ifb$, corresponding to a relative uncertainty
of \totlfrac\%. This uncertainty is smaller than that for any individual
year, reflecting the fact that the total uncertainty is not dominated by effects
that are correlated between years.
The covariance matrix $\mathbf{V_L}$ can also be expressed as
$\mathbf{V_L}=\mathbf{\boldsymbol{\sigma}_L C \boldsymbol{\sigma}_L^T}$, where
the vector $\mathbf{\boldsymbol{\sigma}_L}$
of total absolute uncertainties on $\ltoti$ and (symmetric) correlation
matrix $\mathbf{C}$ are given by
\begin{equation*}
\mathbf{\boldsymbol{\sigma}_L}=\left(
\begin{array}{l}
0.0367 \\
0.296 \\
0.504 \\
0.644 \\
\end{array}
\right)\ \mbox{fb$^{-1}$} \ ,\ \mathbf{C}=\left(\begin{array}{rrrr}
1.000 & &  \\
0.579 & 1.000 &  \\
0.368 & 0.437 & 1.000  \\
0.480 & 0.510 & 0.362 & 1.000  \\
\end{array}
\right)
\ .
\end{equation*}
The off-diagonal elements of the correlation matrix are all smaller
than 0.6, and the relative error of \totlfrac\% in the total luminosity
is significantly smaller than that for any individual year. The largest
single uncertainty in the total luminosity is 0.5\% from calibration
transfer. The total uncertainty from the absolute vdM calibration
is 0.65\%, and the largest contributions come from
non-factorisation corrections, beam--beam effects, magnetic non-linearity
and scan-to-scan reproducibility, all four of which contribute at a comparable
level. The beam--beam correction and the evaluation of potential biases
from magnetic non-linearity have been refined significantly since the
Run~1 luminosity measurements \cite{DAPR-2011-01,DAPR-2013-01}
and the preliminary Run~2 luminosity calibration \cite{ATLAS-CONF-2019-021}
used for most ATLAS Run~2 physics analyses to date. However, the
total uncertainty in the integrated luminosity is significantly smaller
than those achieved previously.
 
The changes to the integrated luminosities of the standard Run~2 physics
samples in each year compared to the preliminary calibration
are listed in Table~\ref{t:lumidel}. The largest
contributions come from the refinements in the vdM calibration, but changes
in the calibration transfer procedure and other effects also contribute.
The increase for
the full Run~2 sample is 0.8\%, and that for the 2015--16 sample is 1.2\%.
The uncertainty has been reduced by a factor two compared with the preliminary
calibration, thanks to improvements in the calibration transfer and long-term
stability analyses, as well as the changes to the vdM calibration.
If the refinements in the absolute
vdM calibration (in particular the revised beam--beam interaction treatment
and the GP4+G fit model) were applied to the Run~1 analyses, the
corresponding integrated luminosities would increase by 0.5--1.0\%, well
within the uncertainties assigned to these measurements.
 
\begin{table}[tp]
\caption{\label{t:lumidel}Increase in the integrated luminosity
(after standard data-quality requirements) of the \sxyt\ $pp$ data sample
recorded in each year of Run~2 ATLAS data-taking, comparing the results of
this analysis with the preliminary calibration reported in
Ref.~\cite{ATLAS-CONF-2019-021}.
}
\centering
 
\begin{tabular}{l|cccc|c}\hline
Year & 2015 & 2016 & 2017 & 2018 & Comb. \\
\hline
$\Delta{\cal L}/{\cal L}$ [\%] & +0.93 & +1.23 & +0.73 & +0.59 & +0.80 \\
\hline
\end{tabular}
\end{table}
 
\subsection{Comparison with $Z$-counting measurements}\label{ss:zcount}
 
The continuous monitoring of high-rate physics processes, such as the production
of $Z$ bosons, provides another potential luminosity measurement. The leptonic
decays $Z\rightarrow ee$ and $Z\rightarrow\mu\mu$ are particularly attractive,
as the distinctive signature is easy to trigger on and reconstruct, has
less than 1\% background \cite{STDM-2018-14},
and the two lepton channels can be compared for internal
consistency checks. The lepton trigger and reconstruction efficiencies can be
measured in-situ using tag-and-probe techniques exploiting the two leptons
in each event \cite{PERF-2017-01,MUON-2018-03}, and the rate of about 10\,Hz
of reconstructed events per lepton flavour at \sxyt\ and
$\linst=10^{34}\,\mathrm{cm}^{-2}\mathrm{s}^{-1}$
gives a statistical precision well below 1\% for runs lasting several hours.
However, the $Z$ production cross-section is only predicted to a precision
of 3--4\% due to uncertainties in the proton parton distribution
functions \cite{STDM-2016-02}, so this method is most useful as a relative
luminosity measurement,
after normalisation to an absolutely calibrated reference. In this mode,
it can be used to study the stability with time, or with respect to other
parameters such as pileup \meanmu.
 
Within ATLAS, $Z$-counting has been used both within the
offline data-quality framework for giving fast feedback on the delivered
luminosity (for example in comparisons with CMS), and for offline
studies. The method is described in detail in Ref.~\cite{ATL-DAPR-PUB-2021-001},
where the results are compared with the preliminary baseline Run~2 luminosity
measurement. A brief overview of the $Z$-counting luminosity measurement is
given here, together with a comparison of the results with the updated
luminosity calibration described in this paper. This serves as a validation
of the stability of the calibration with time and pileup \meanmu.
 
Selected events were required to contain two identified leptons
(electrons or muons) of the same flavour and opposite charge, each having
$\pt>27$\,\GeV\ and $|\eta|<2.4$, and satisfying track-based isolation
criteria, and with at least one lepton matched to a corresponding trigger
signature. The dilepton invariant mass \mll\ was required to lie in the range
$66<\mll<116$\,\GeV. The resulting samples are around 99.5\% pure in
$Z\rightarrow\ell\ell$ events, and the small background was estimated
using simulation and subtracted. Within each luminosity block,
the lepton trigger and reconstruction
efficiencies were estimated from data, automatically accounting for variations
due to transient detector problems or the variation of efficiencies with
pileup, for example. Corrections to account for effects not included in the
tag-and-probe formalism were estimated using simulation. These corrections
amount to 1--2\% for $Z\rightarrow\mu\mu$, and up to about 10\% for
$Z\rightarrow ee$, where a larger fraction of the lepton inefficiencies are not
captured by the tag-and-probe procedure.
The resulting measurements have a statistical precision of 2--5\% per
20 minute period, and show excellent consistency between electron and muon
channels, both as a function of time and \meanmu. The ratios of
same-year-integrated luminosities estimated from $Z\rightarrow ee$ and
$Z\rightarrow\mu\mu$ (where the $Z$ production cross-section cancels out)
are around 0.992--0.993 for all years, consistent with unity within the residual
uncertainties in the lepton reconstruction and trigger efficiencies.
 
Ratios of the run-integrated $Z$-counting luminosity measurements
from $Z\rightarrow ee$ and $Z\rightarrow\mu\mu$ events to the
corresponding ATLAS baseline luminosity measurements are shown in
Figure~\ref{f:zcountyear}. The $Z$-counting measurements are each normalised
to the year-integrated baseline  measurement. The ratios
for the two channels are similar, and show little residual structure,
confirming that the calibration of the baseline luminosity measurement
is stable in time. Figure~\ref{f:zcountrun2} shows a comparison of the
combined $Z$-counting ($Z\rightarrow ee$ and $Z\rightarrow\mu\mu$)
and baseline luminosity measurements over the complete
Run~2 data-taking period, where the $Z$-counting measurement has been
normalised to the baseline integrated luminosity of the full sample. This
comparison is therefore sensitive to differences in the baseline luminosity
calibration between years, and tests the consistency of the absolute
vdM calibrations performed in each year. Year-averaged systematic differences
of $-0.5$\%, 0.2\%, 0.4\% and $-0.4$\% for 2015, 2016, 2017 and 2018
are visible between the $Z$-counting
and baseline luminosity measurements. The consistency of these differences
with the uncertainties detailed in Section~\ref{ss:reshimu} was quantified
using a $\chi^2$ defined as
\begin{equation*}
\chi^2 = \mathbf{\Delta^T} (\mathbf{V_L})^{-1} \mathbf{\Delta}
\end{equation*}
where $\mathbf{\Delta}$ is the vector of per-year differences in the
absolute integrated luminosity measured by $Z$-counting and the baseline
method, and $\mathbf{V_L}$ is the covariance matrix of the baseline
luminosity measurements (neglecting statistical and systematic uncertainties
in the $Z$-counting result). The resulting $\chi^2$ is 0.8 for three degrees of
freedom, confirming the good compatibility of baseline and $Z$-counting
luminosity measurements.
 
\begin{figure}[tp]
\parbox{83mm}{\includegraphics[width=73mm]{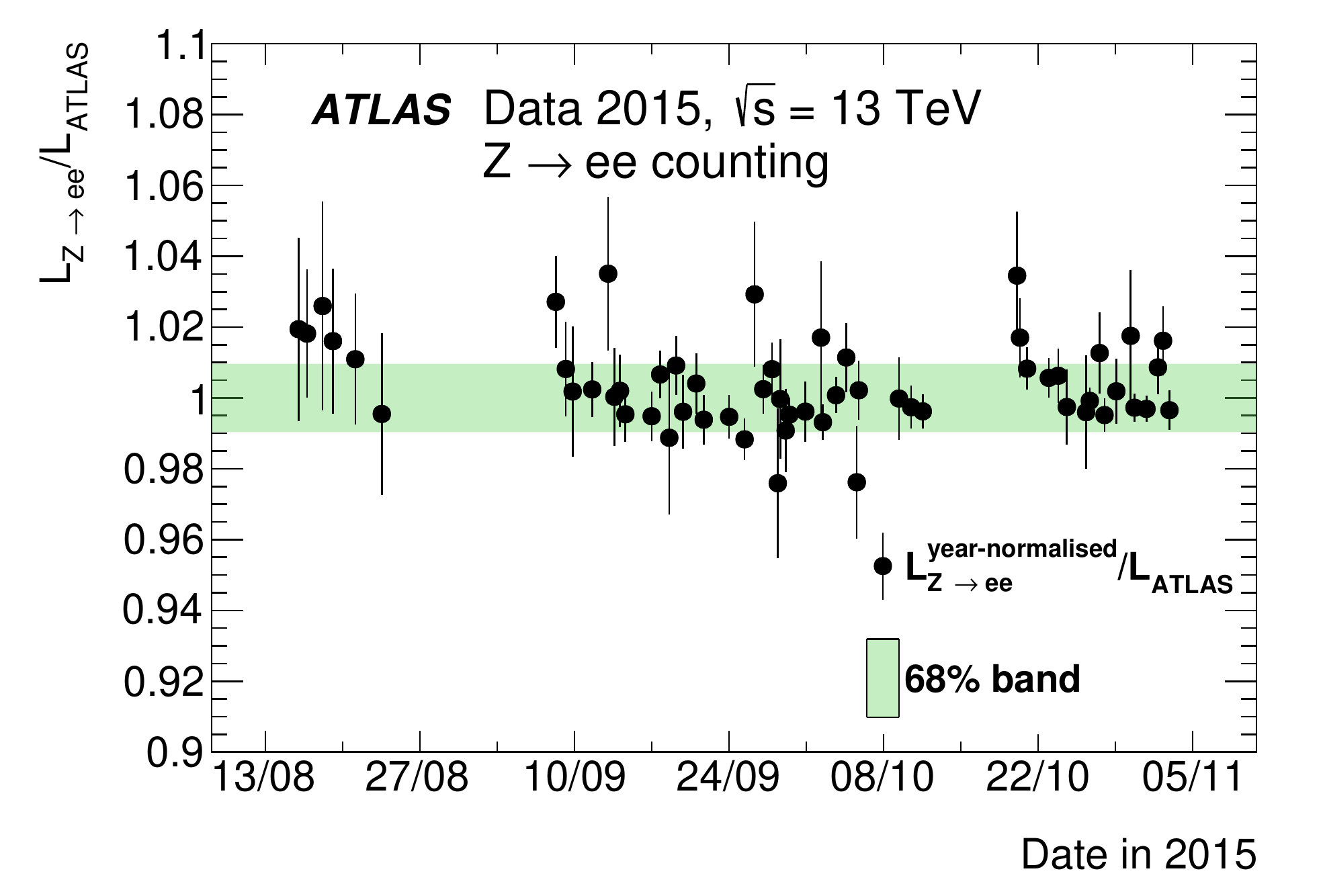}\vspace{-6mm}\center{(a)}}
\parbox{83mm}{\includegraphics[width=73mm]{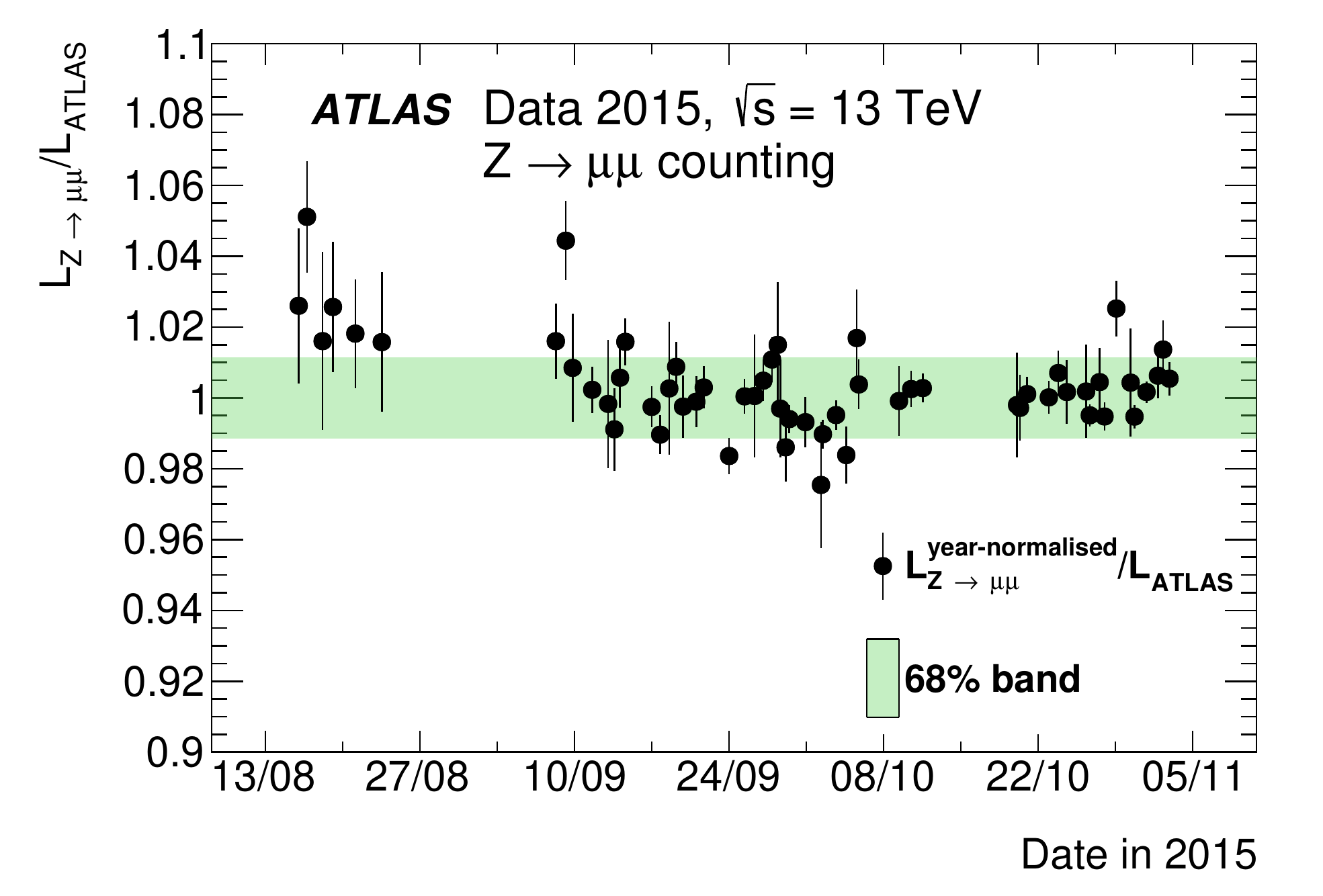}\vspace{-6mm}\center{(b)}}
\parbox{83mm}{\includegraphics[width=73mm]{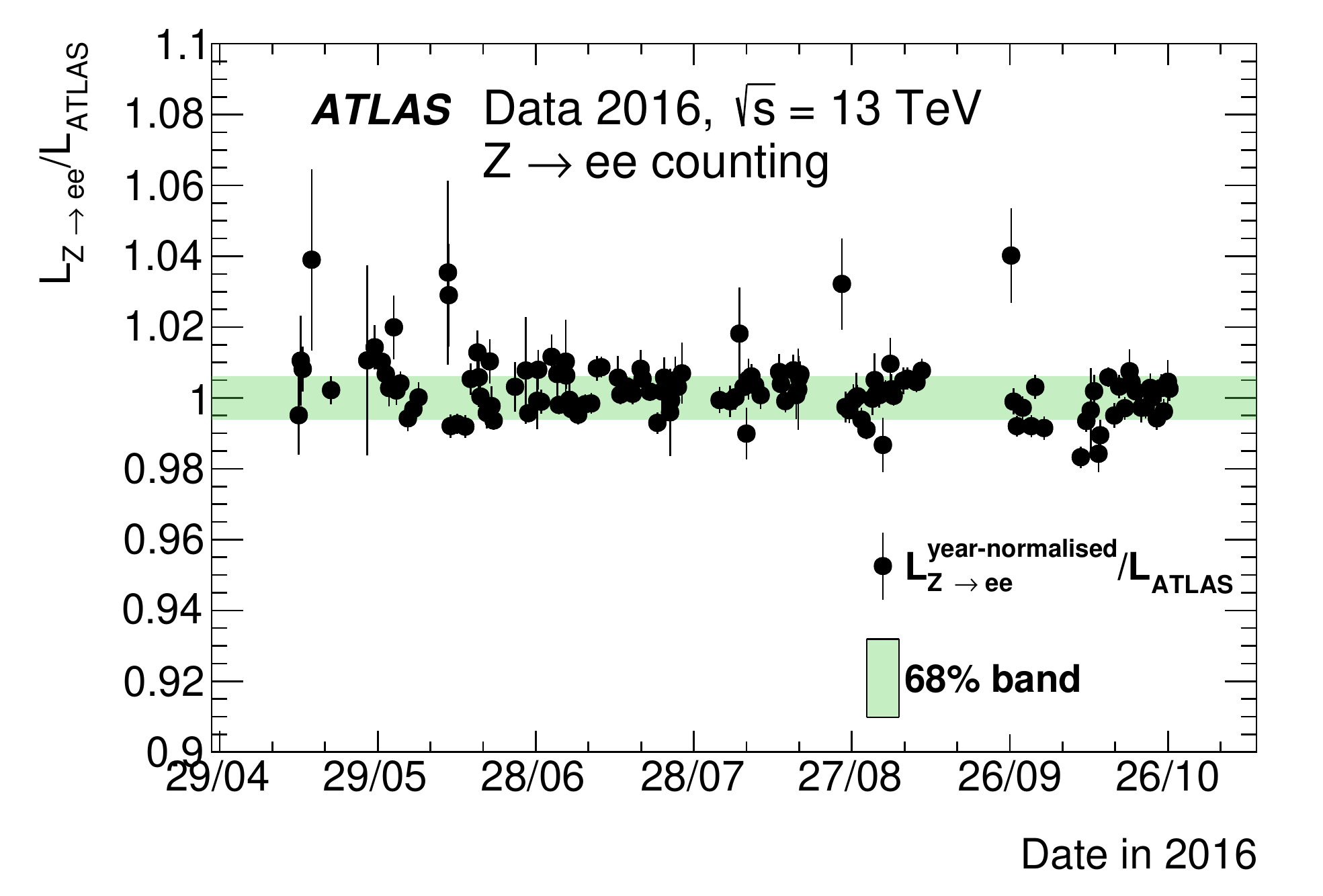}\vspace{-6mm}\center{(c)}}
\parbox{83mm}{\includegraphics[width=73mm]{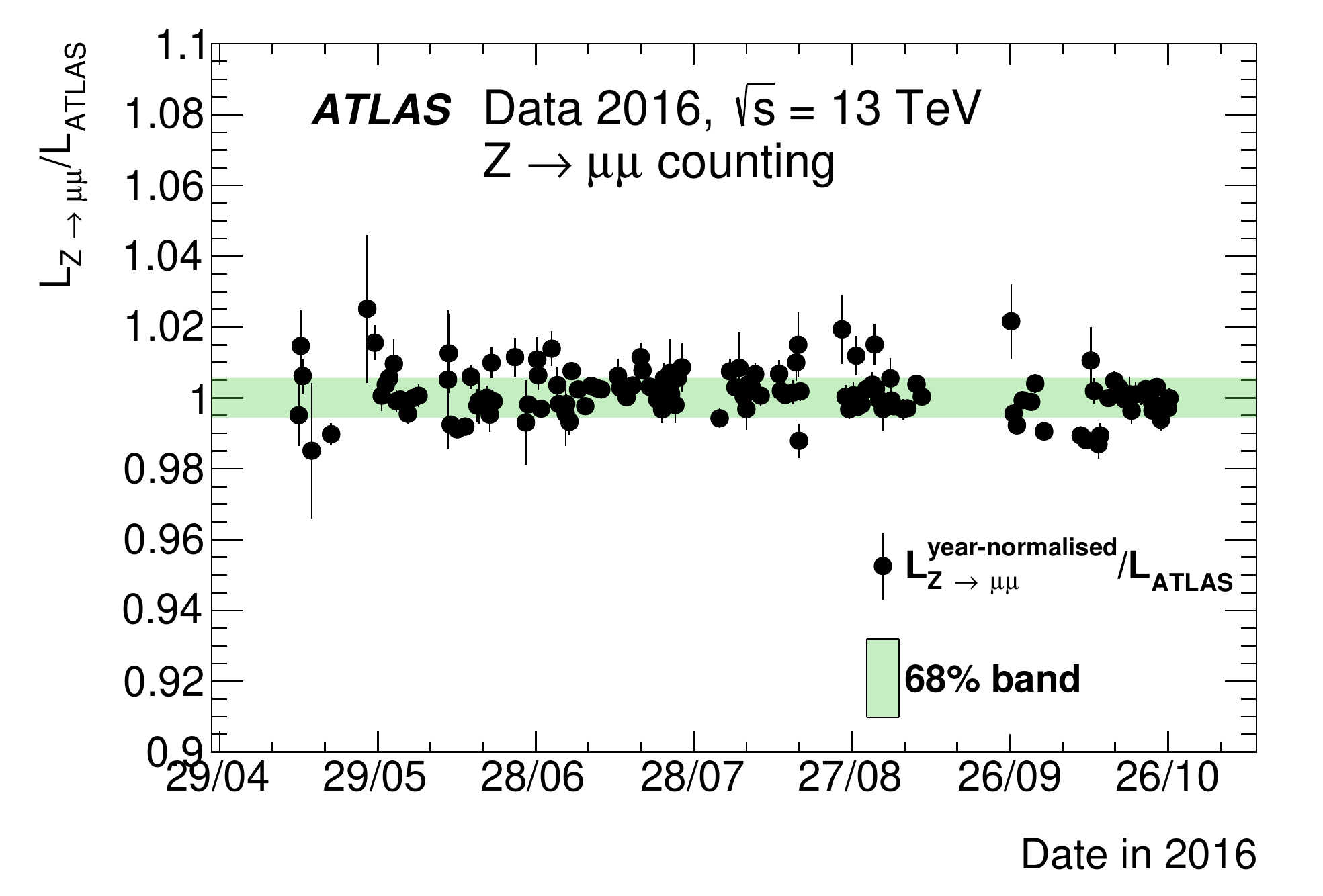}\vspace{-6mm}\center{(d)}}
\parbox{83mm}{\includegraphics[width=73mm]{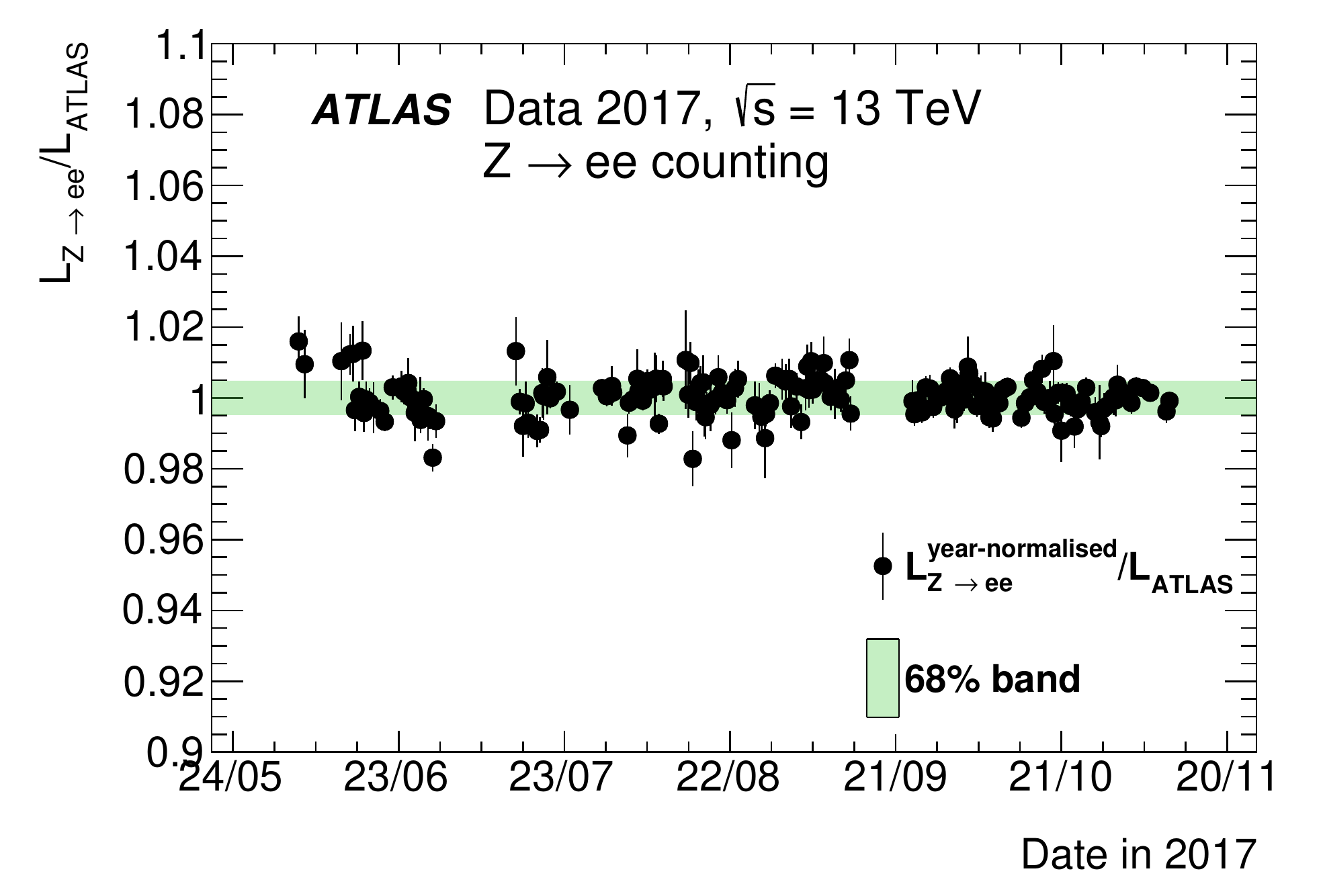}\vspace{-6mm}\center{(e)}}
\parbox{83mm}{\includegraphics[width=73mm]{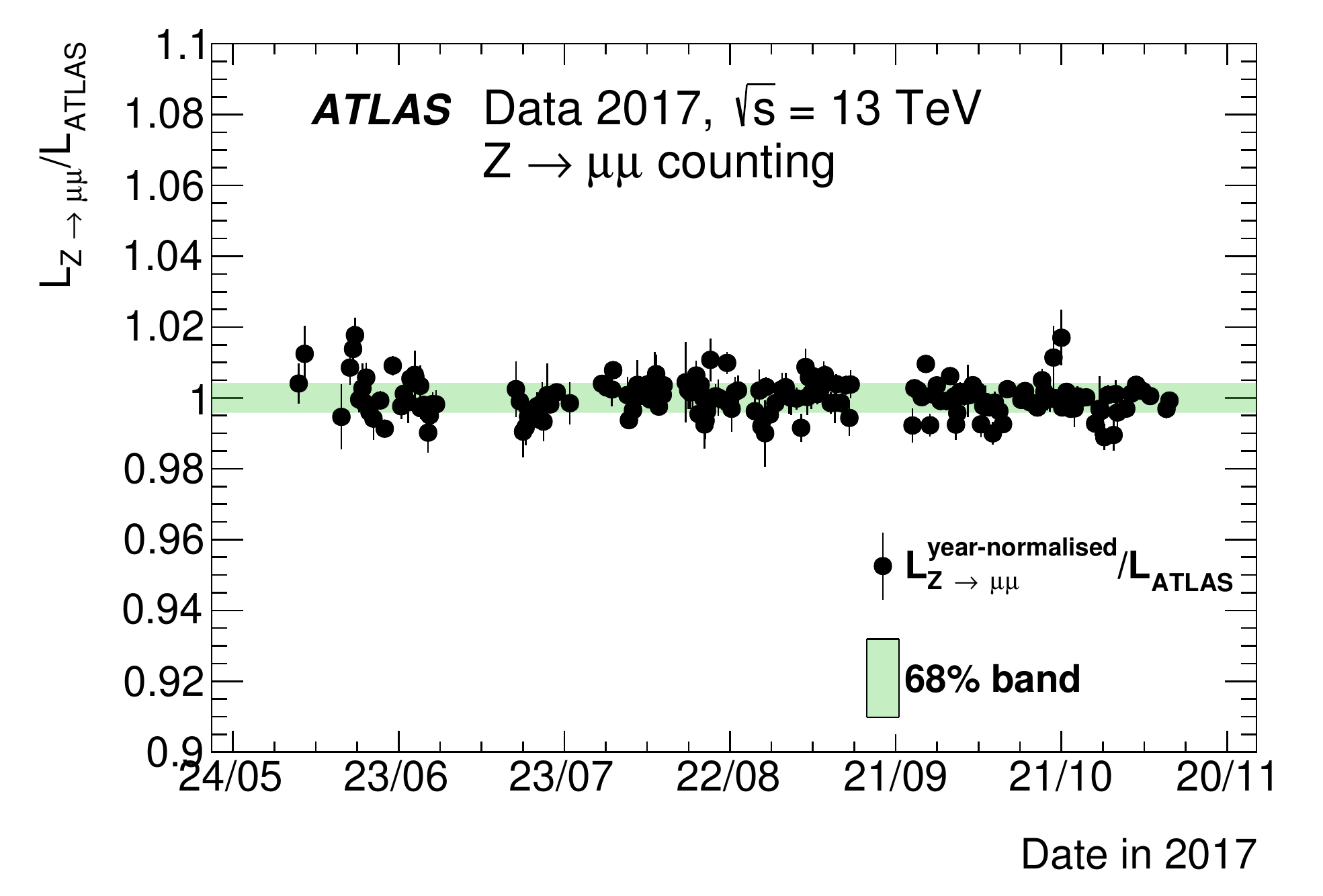}\vspace{-6mm}\center{(f)}}
\parbox{83mm}{\includegraphics[width=73mm]{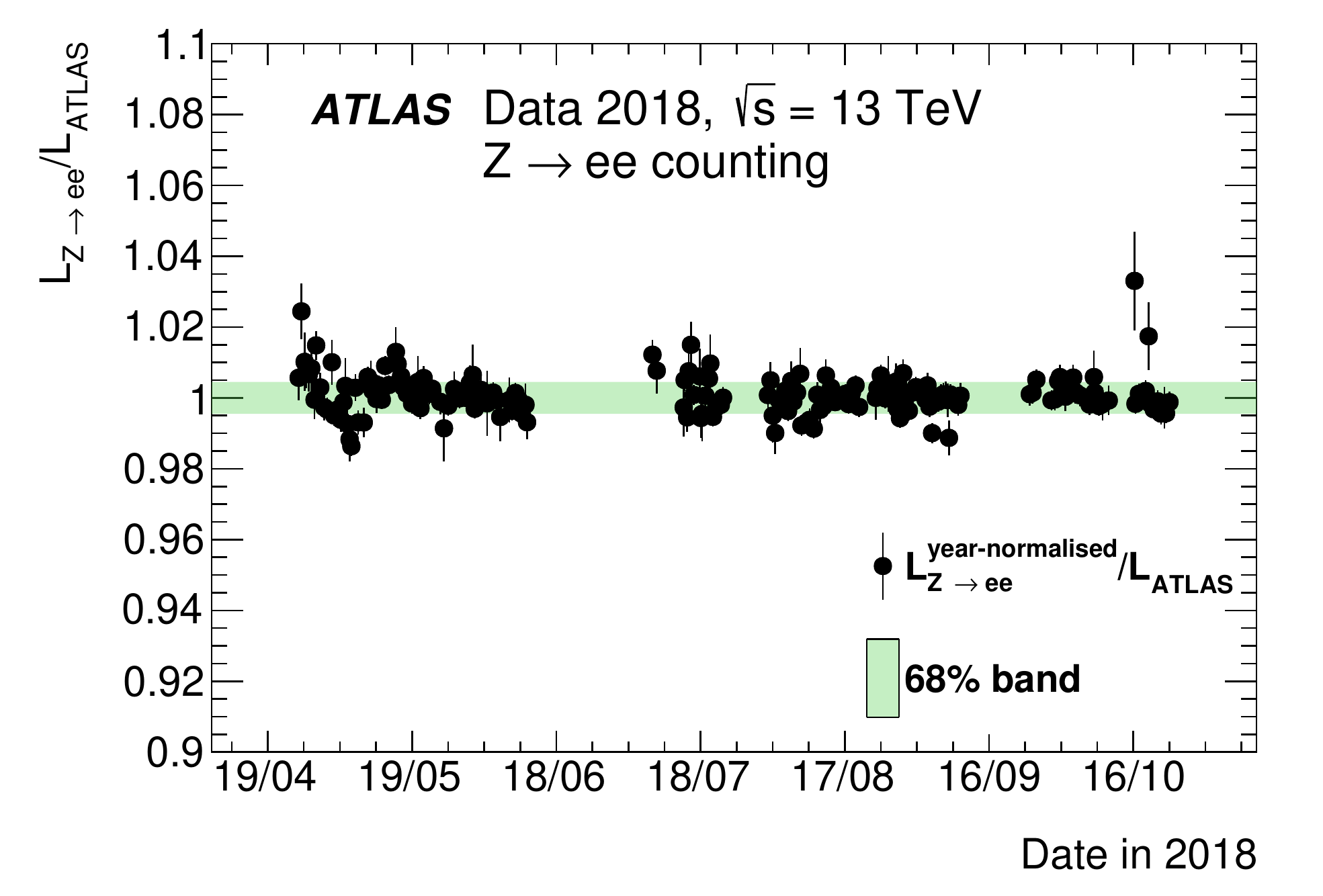}\vspace{-6mm}\center{(g)}}
\parbox{83mm}{\includegraphics[width=73mm]{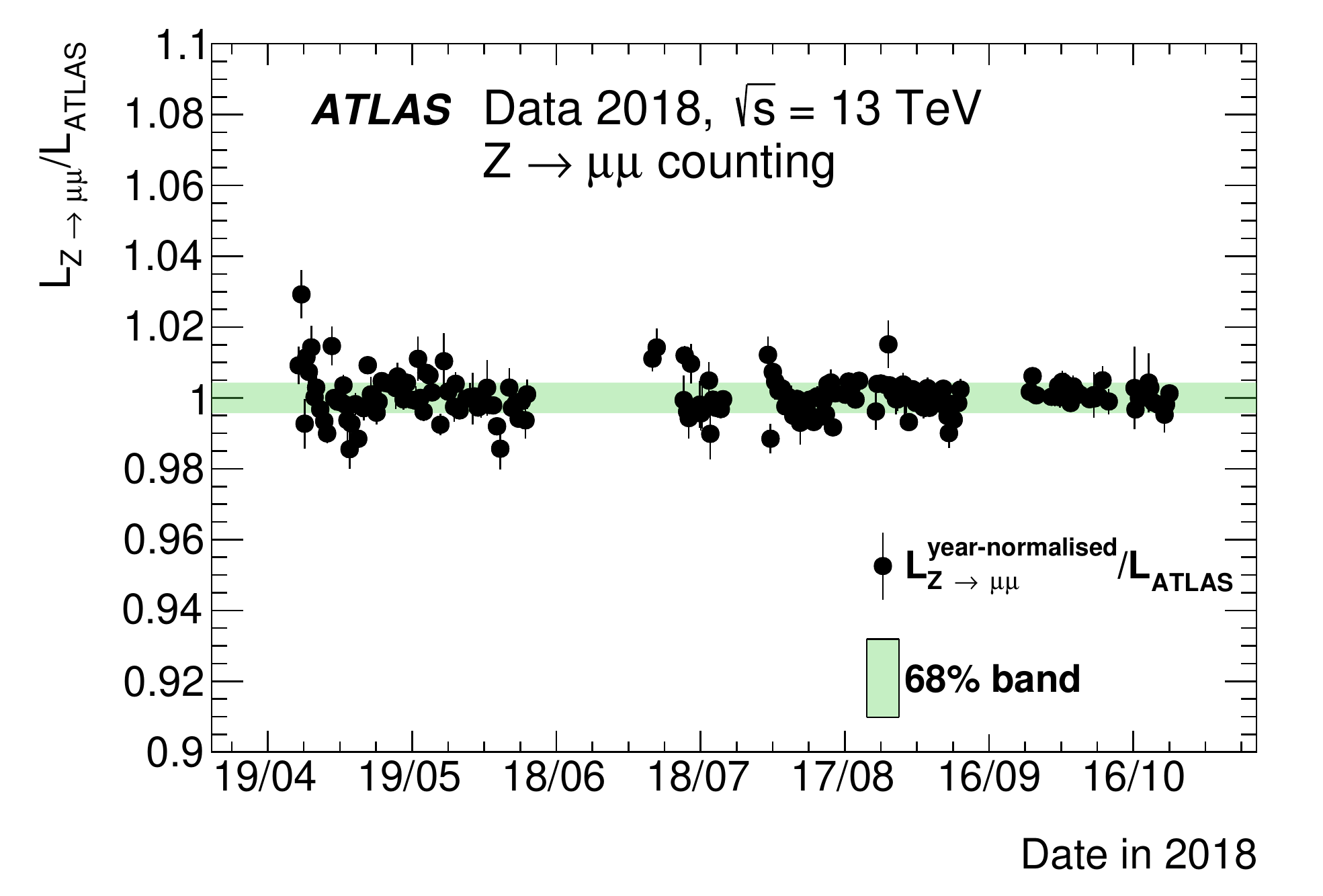}\vspace{-6mm}\center{(h)}}
\caption{\label{f:zcountyear}Ratios of the run-integrated $Z$-counting and
baseline ATLAS luminosity measurements as a function of run date within
each year of Run~2 data-taking, showing (a, c, e, g) $Z\rightarrow ee$ and
(b, d, f, h) $Z\rightarrow\mu\mu$ measurements separately. The $Z$-counting
measurements within each year and lepton channel are each normalised to
the year-integrated baseline luminosity measurement. The error bars show
statistical uncertainties only, and the green bands contain 68\% of all
runs centred around the mean.}
\end{figure}
 
\begin{figure}[tp]
\centering
 
\includegraphics[width=150mm]{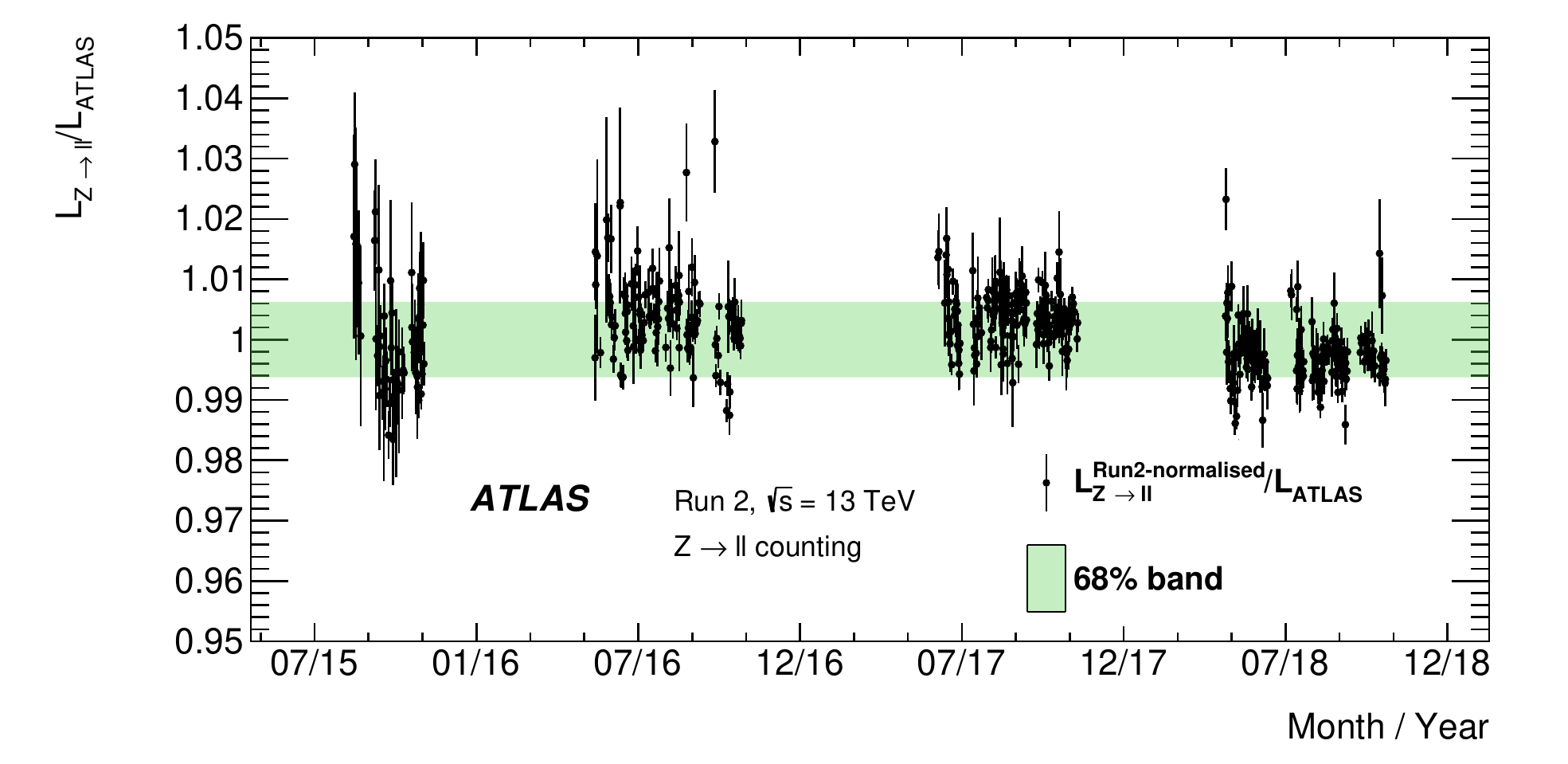}
\caption{\label{f:zcountrun2}Ratios of the run-integrated $Z$-counting
(combining $Z\rightarrow ee$ and $Z\rightarrow\mu\mu$)
and baseline ATLAS luminosity measurements as a function of run date.
The $Z$-counting
measurements are normalised to the baseline luminosity measurement
integrated over the entire Run~2 data-taking period. The error bars show
statistical uncertainties only, and the green bands contain 68\% of all
runs centred around the mean.}
\end{figure}
 
The $Z$-counting and baseline luminosity measurements have also been compared
as a function of pileup \meanmu, as shown in Figure~\ref{f:zcountmu}.
In 2015, 2016 and~2018,
these ratios show very little pileup dependence, except at the extreme
ends of the distributions, which contain only a very small fraction of the data.
In 2017, a trend is visible, with the
$Z$-counting luminosity measurement being higher than the baseline at
low $\mu$ and lower at high $\mu$.
However, the size of these variations does not exceed the
$\pm 0.5$\% calibration transfer uncertainty, and this trend is not visible
in other years.
The distributions are presented separately for each year and not combined,
as the year-dependence visible in Figure~\ref{f:zcountrun2} combined with
the different \meanmu\ profile in each year would produce a spurious
correlation, not connected to $\mu$-dependence within an individual year.
 
\begin{figure}[tp]
\parbox{83mm}{\includegraphics[width=76mm]{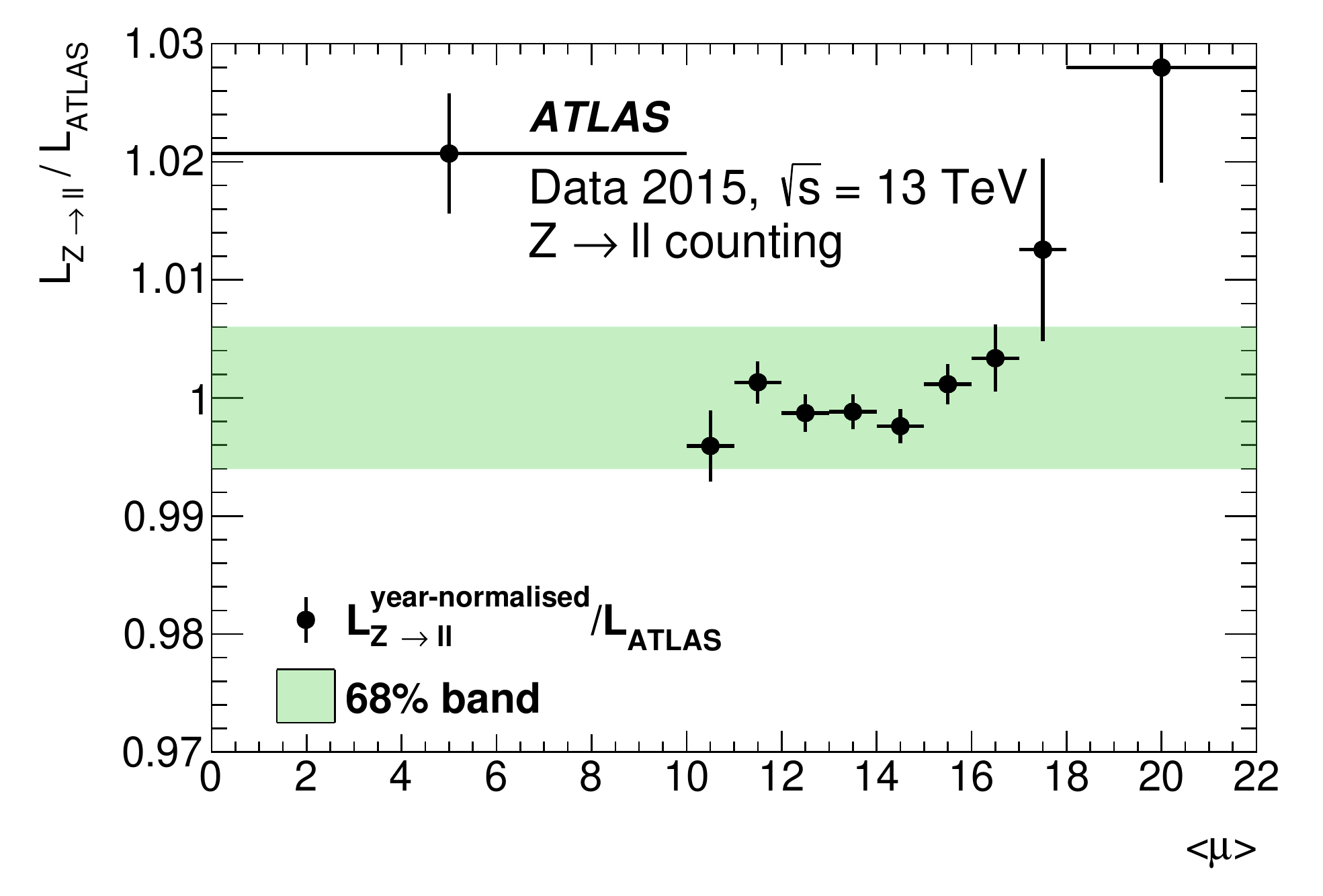}\vspace{-6mm}\center{(a)}}
\parbox{83mm}{\includegraphics[width=76mm]{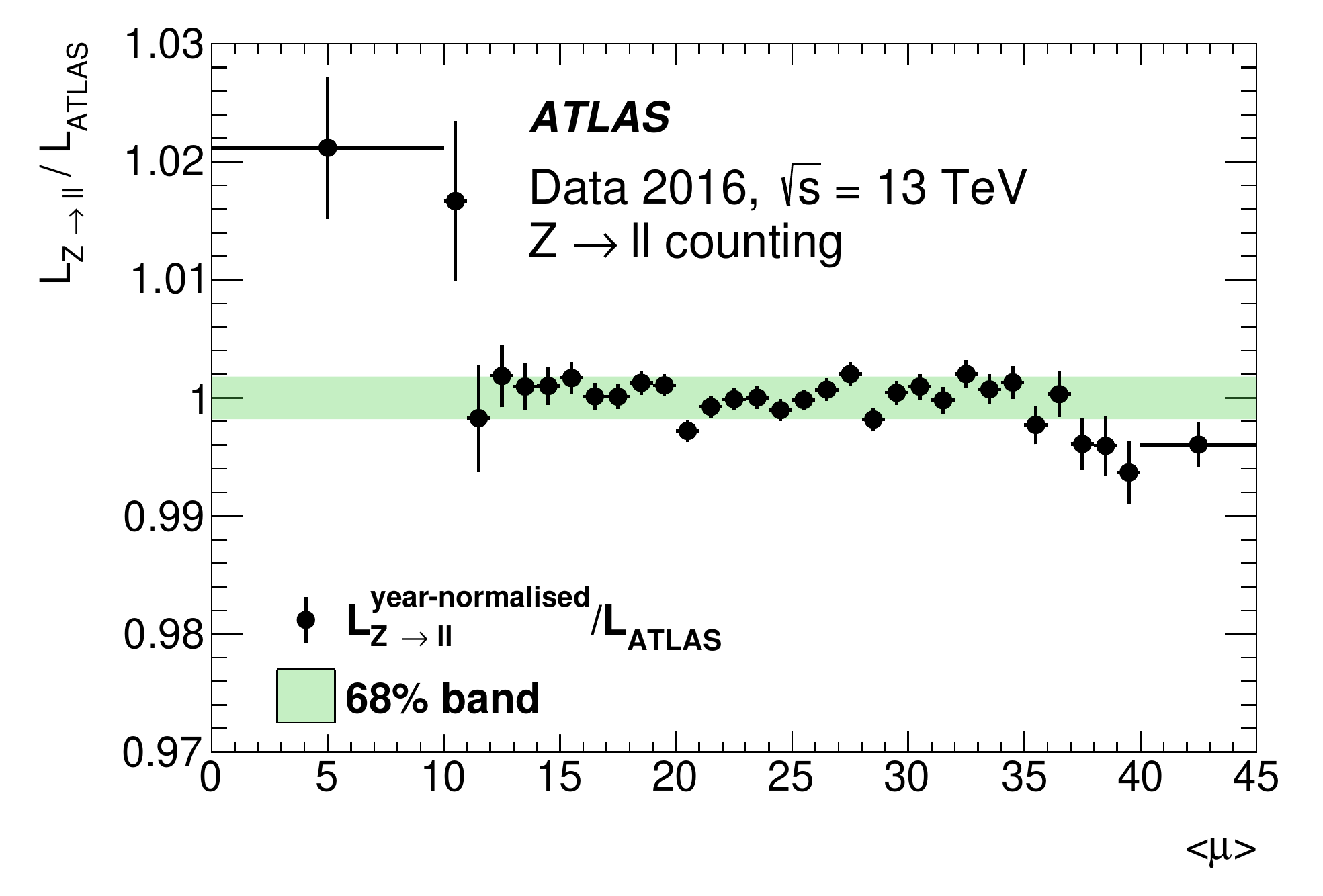}\vspace{-6mm}\center{(b)}}
\parbox{83mm}{\includegraphics[width=76mm]{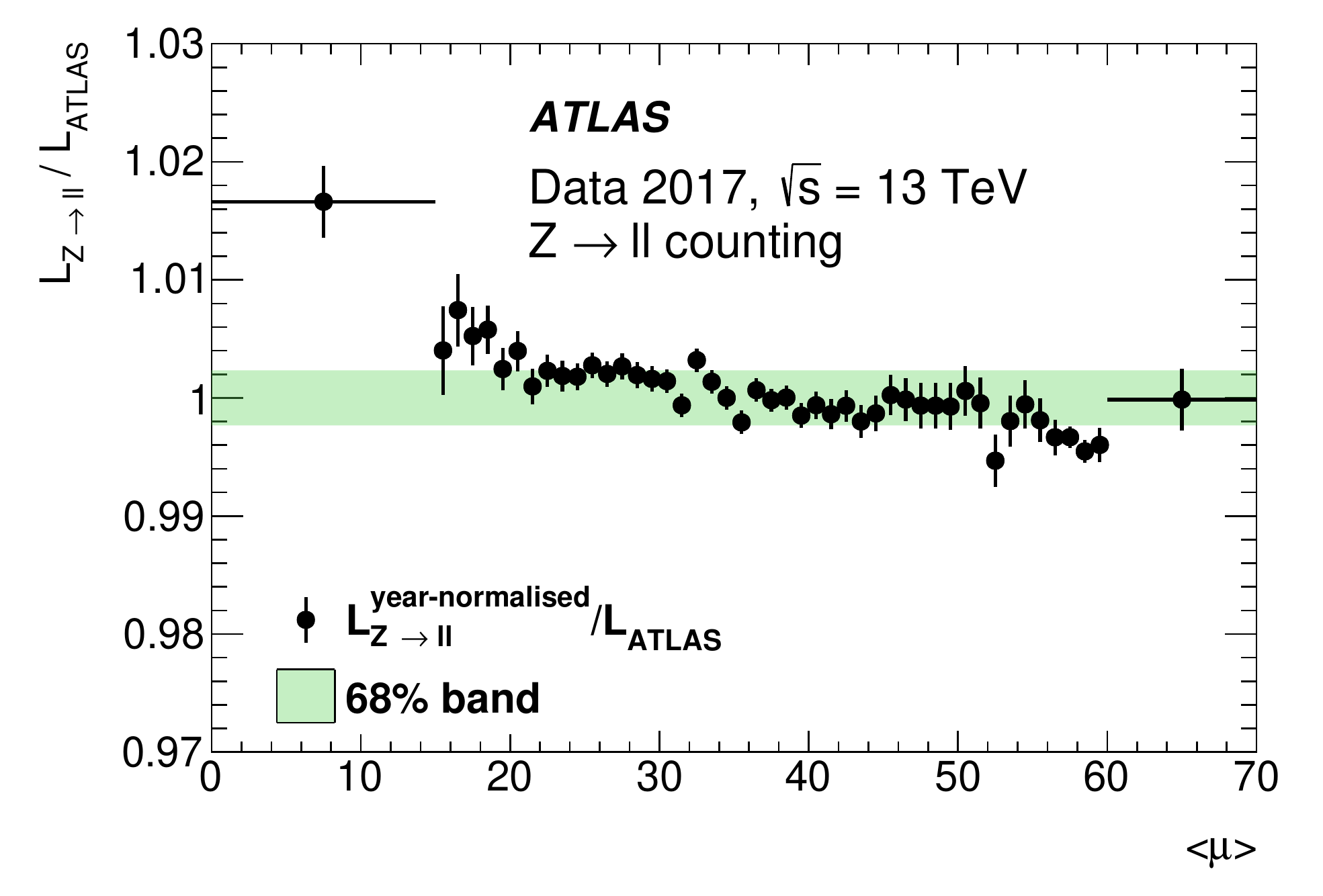}\vspace{-6mm}\center{(c)}}
\parbox{83mm}{\includegraphics[width=76mm]{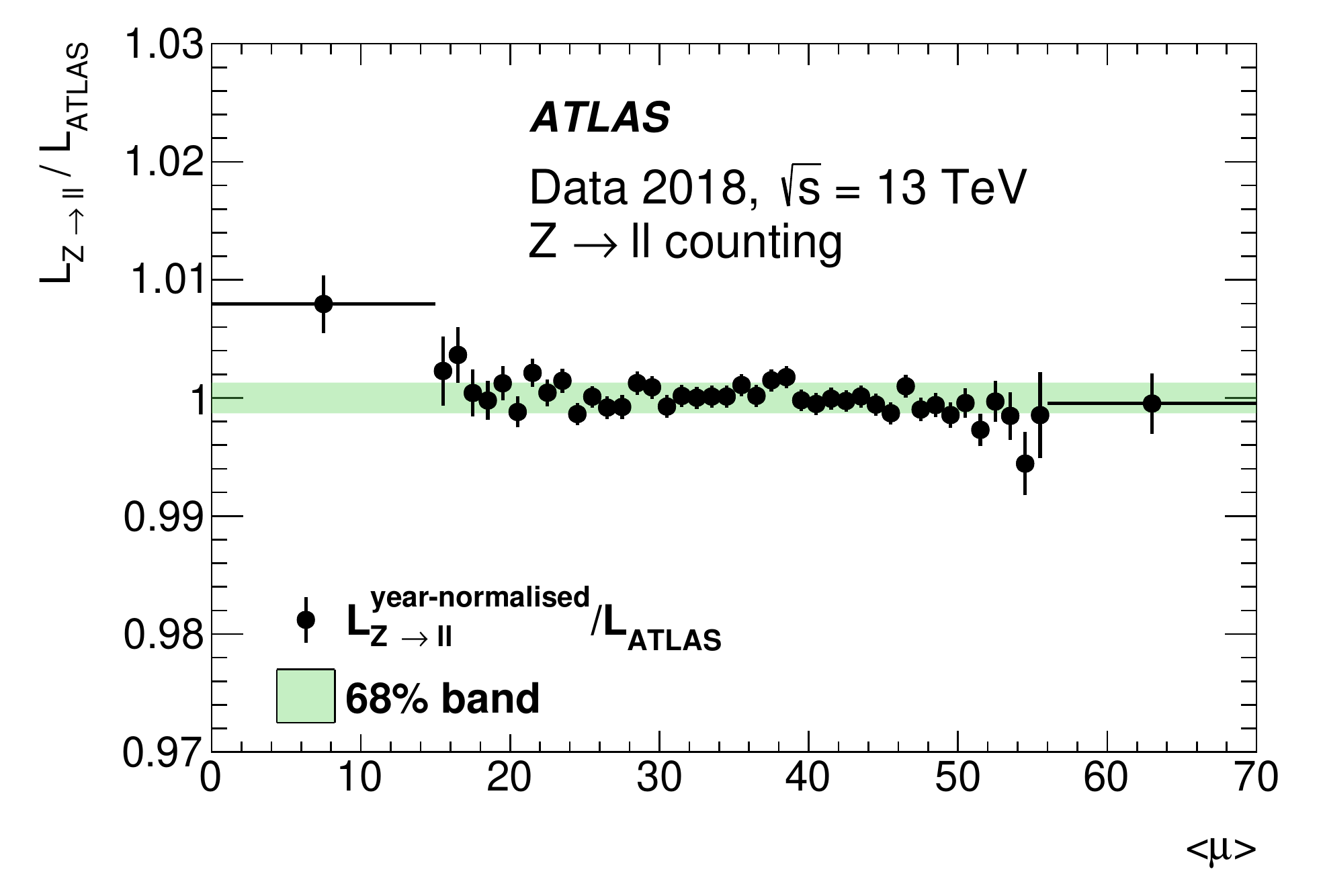}\vspace{-6mm}\center{(d)}}
\caption{\label{f:zcountmu}Ratios of the $Z$-counting
(combining $Z\rightarrow ee$ and $Z\rightarrow\mu\mu$) and baseline ATLAS
luminosity measurements as a function of pileup \meanmu\ determined from the
baseline measurement, separately for each year of Run~2 data-taking. The
$Z$-counting measurements are normalised separately in each year to the
baseline luminosity measurement integrated over the entire year. The error
bars show statistical uncertainties only, and the green bands contain
68\% of the points centred around the mean.}
\end{figure}
 
In summary, the comparisons with $Z$-counting do not reveal any unexpected
time- or \meanmu-dependence in the baseline luminosity calibration, and serve
as an important validation of the calibration transfer procedures and overall
uncertainty estimates presented above.

% End of text imported from the .//uncert.tex input file

% The next lines are included from the .//lowmu.tex input file
\section{Luminosity calibration for low-pileup datasets}\label{s:lowmu}
 
During Run~2, five-day periods in both November~2017 and July~2018 were
devoted to recording \sxyt\ $pp$ collisions with reduced instantaneous
luminosity in ATLAS. In these periods, the beams were partially separated
in the transverse plane at the ATLAS interaction point to give
$\meanmu\approx 2$, and the separation was adjusted continuously to maintain
this level of pileup throughout the fills, which had durations of up to
29~hours.
The lower level of pileup significantly improves the resolution of the ATLAS
missing transverse momentum and hadronic recoil measurements, making these
data samples particularly useful for precision measurements of single $W$/$Z$
boson production \cite{ATL-PHYS-PUB-2017-021}.
 
The luminosity calibration for these data samples was derived in a similar
way as for the \sxyt\ high-pileup data, using the LUCID BiHitOR algorithm
in 2017 and the single-PMT C12 algorithm in 2018. The absolute calibrations
were obtained from the vdM analysis described in Section~\ref{s:vdmcal}.
A dedicated
calibration transfer correction was determined from the ratio of LUCID to
track-counting luminosities in long reference fills within each low-pileup
period. As the \meanmu\ values were approximately constant throughout
these fills, the $p_1$ parameter of Eq.~(\ref{e:mucorr}) was fixed to zero,
reducing the correction to a simple scaling of the LUCID luminosity by $p_0$.
The values of $p_0$ were determined to be 0.987 for the 2017 dataset and 1.009
for 2018. The larger value in 2018 compared to 2017 reflects the LHC beam
crossing angle dependence of the response of individual LUCID PMTs,
which depends on the azimuthal $\phi$-position of the PMT around the beam pipe
(see Section~\ref{ss:lucid}). Since the HitOR algorithm
used in 2017 combines PMTs which are distributed approximately
$\phi$-symmetrically around the beam pipe, the crossing-angle effect mostly
averages out. In 2018, the sign of the vertical boost from the crossing angle
was such that it downwardly biased the response of the single-PMT C12 algorithm,
necessitating a positive correction. The remaining correction is dominated by
the effect of running with bunch trains at $\meanmu\approx 2$, rather than
isolated bunches with $\meanmu\approx 0.5$ as in the vdM fills.
 
This correction procedure assumes that the track-counting luminosity response
does not change between the vdM regime and $\meanmu\approx 2$ physics running
with bunch trains. In the same way as discussed for the high-pileup calibration
transfer correction in Section~\ref{s:calsyst}, this assumption was probed
via comparisons of luminosity measurements from TileCal E3 and E4 cells with
those from track-counting. The last LHC fill of the June 2018 intensity
ramp-up described in Section~\ref{s:calsyst} (fill 6860, with 2448 colliding
bunch pairs) was recorded two days before the 2018 vdM run, and included
six hours of running at $\meanmu\approx2$ after one hour of high-pileup running
and a $\mu$-scan. This high-luminosity running
resulted in a slowly decaying activation contribution to the TileCal
luminosity signal (amounting to 1--2\% of the actual luminosity at the start
of the $\meanmu\approx 2$ period), which was corrected using the
activation model described in Section~\ref{ss:actmod}. The resulting ratios
of TileCal E-cell to track-counting luminosity measurements as a function of
luminosity block number in the $\meanmu\approx 2$ physics and vdM fills are
shown in Figure~\ref{f:ctlowmu}(a). The average ratios are about 0.5\% higher
in the physics fill for E3, and somewhat smaller for E4, suggesting that the
change in track-counting response between these two regimes is at most 0.5\%.
 
\begin{figure}[tp]
\parbox{83mm}{\includegraphics[width=76mm]{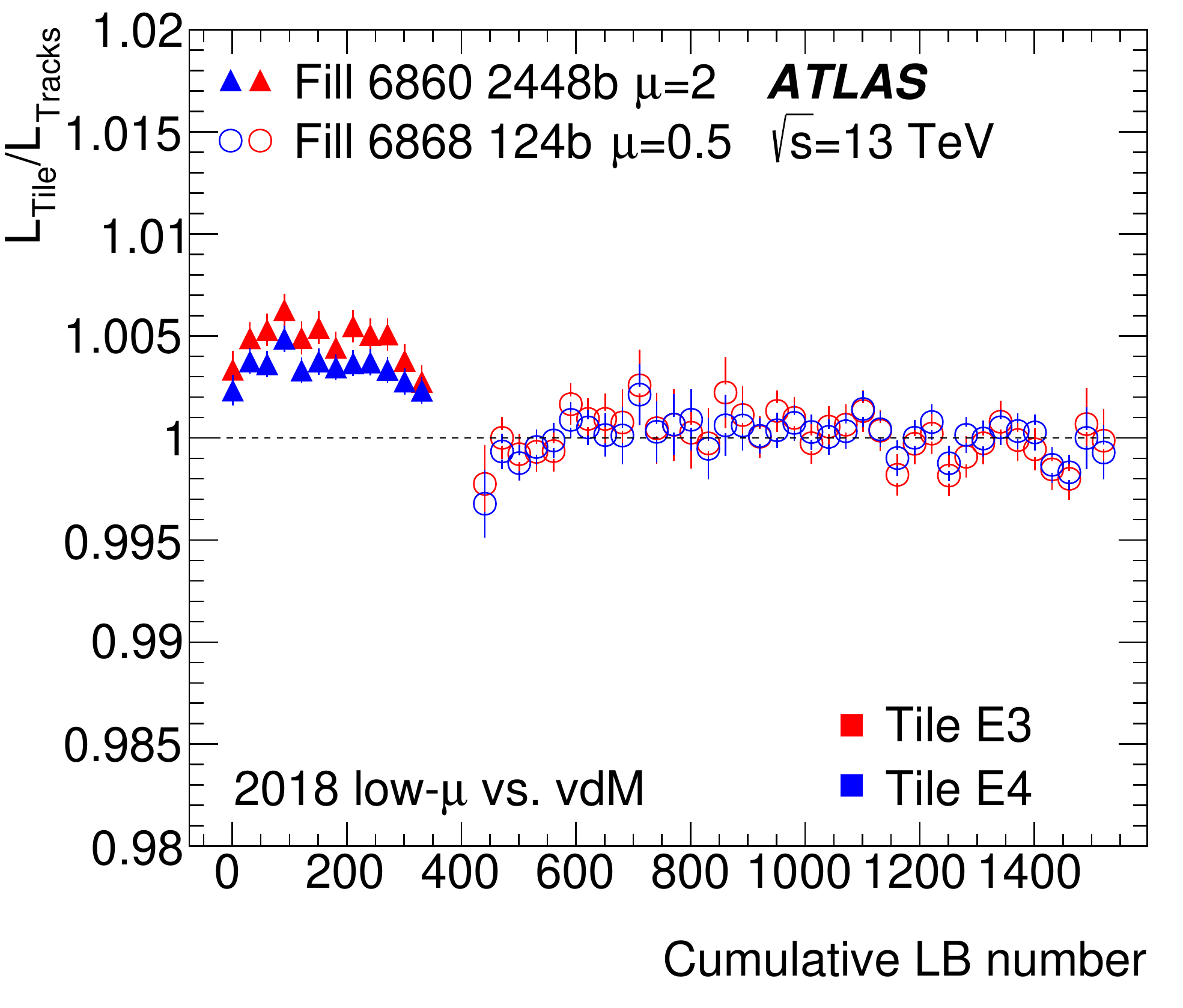}\vspace{-6mm}\center{(a)}}
\parbox{83mm}{\includegraphics[width=76mm]{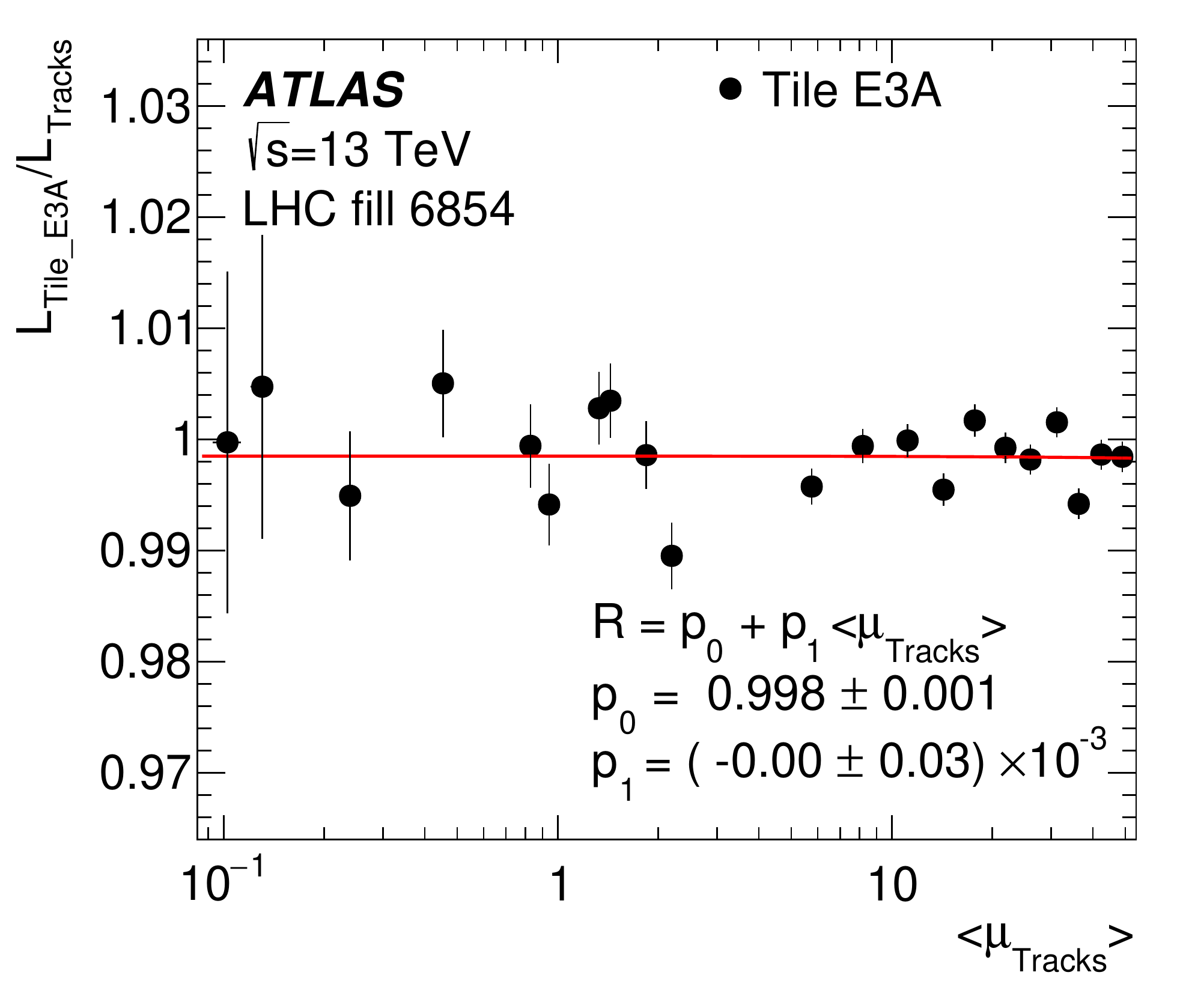}\vspace{-6mm}\center{(b)}}\\
\caption{\label{f:ctlowmu}(a) Ratios of luminosity measured by the TileCal
E-cells to that measured by track-counting for the $\meanmu=2$ part of fill 6860
with 2448 colliding bunch pairs, followed by that in the head-on part of the
vdM fill 6868 with
124 isolated colliding bunch pairs. The ratios are averaged over the
A- and C-side measurements and over 30 luminosity blocks, separately for the
E3 (red) and E4 (blue) cells, and are shown as a function of luminosity block
number, renumbered to give a continuous sequence over both fills.
The ratios have been normalised separately for E3 and E4 cells so that the
integrated ratios in the vdM fill are unity.
(b) Ratios of luminosity measured by the TileCal E3A-cells to that
measured by track-counting as a function of \meanmu\ in the $\mu$-scan
of fill 6854, with 1214 colliding bunch pairs in bunch trains. The ratios
are normalised so that the integrated ratio in the low-$\mu$ part of the nearby
140 colliding bunch fill 6847 is unity, as in Figure~\ref{f:ctdir}(e).
The red line shows a linear fit (as a function of \meanmu)
through the points, and the fit parameters are
shown in the legend. In both plots, the error bars show the uncertainties
in the ratios, dominated by the statistical uncertainties of the track-counting measurements.}
\end{figure}
 
A second constraint was obtained from a $\mu$-scan in the earlier fill 6854
with 1214 colliding bunch pairs, which was recorded two days after the
140 bunch calibration transfer fill 6847 with vdM-like conditions discussed
in Section~\ref{s:calsyst}. Figure~\ref{f:ctlowmu}(b) shows the ratio of
activation-corrected TileCal E3A to track-counting luminosities as a function
of $\meanmu$ in this scan, normalised to the ratio measured in the
$\meanmu=0.5$ part of fill 6847. Although the statistical
precision of the ratios at low $\meanmu$ is limited (as only a few minutes
were spent at each scan point), the intercept $p_0$ of the linear fit
of the ratio as a function of $\meanmu$ can be interpreted as a measurement
of the ratio at low $\meanmu$, giving a ratio 0.2\% below unity for the E3A
TileCal cell family. The corresponding values for E4A and the C-side
measurements
are all within $\pm 0.5$\% of unity, again suggesting that any changes
in track-counting response between the vdM and $\meanmu\approx 2$ physics
regime is at most $\pm 0.5$\%. An uncertainty of 0.5\% was therefore assigned
to the calibration transfer correction of the LUCID luminosity scale in 2018.
Given the similar performance of track-counting seen in the high-pileup
datasets across all years, this uncertainty was also assigned for the 2017
low-pileup dataset.
 
Using a reference run within the low-pileup period in each year also ensures
that the calibration transfer procedure corrects for any time drifts of the
LUCID calibration between the vdM run and the low-pileup period, but it
introduces
a dependency on any drifts of the track-counting measurements. The studies
shown in Sections~\ref{ss:trkperf} and~\ref{s:stab} suggest that these
are small, but an independent check can be obtained from the EMEC measurements,
absolutely calibrated to track-counting in high-pileup physics runs close to
the vdM run using the anchoring procedure described in Section~\ref{s:stab}.
Comparison of the run-integrated luminosity measured by EMEC and LUCID
in each of the low-pileup physics runs is sensitive to time drifts of
either detector, and also further probes the calibration transfer correction
applied to LUCID, assuming that the EMEC luminosity response does not change
between high- and low-pileup data-taking with bunch trains. The FCal
data cannot be used in the same way, because of its strongly non-linear
response with instantaneous luminosity, leading to an offset of several
percent in the FCal to LUCID luminosity ratios at $\meanmu\approx 2$ if the FCal
calibration is taken from the high-pileup anchoring. The FCal luminosity
measurements were therefore anchored in the $\meanmu\approx 2$ reference
runs used to determine the LUCID calibration transfer correction,
and used only for stability checks within each low-pileup dataset. The same
procedure was followed for the TileCal D-cell data; although this detector
is linear in the relevant luminosity range, changes in the pedestal procedures
and TileCal gain settings between high- and low-pileup running meant that the
absolute calibration from the high-pileup anchoring could not be transferred
to the $\meanmu\approx 2$ datasets.
 
The LHC operation sequence followed at the start of physics fills in these
data-taking periods separated the beams at the ATLAS interaction point only
after head-on collisions had been
established and optimised at all four IPs, leading to periods of typically
10--30 minutes at high luminosity before the start of low-pileup data-taking
in each 2018 fill. In 2017, the high-pileup periods were even longer in
the first two fills, but much shorter in the subsequent ones.
Activation from this high-luminosity running biases the EMEC, FCal and TileCal
D-cell luminosity measurements upwards by several percent, in the same way as
discussed above for the TileCal E-cell measurements. The open red points in
Figure~\ref{f:caloactlowmu} illustrate this effect for two fills in 2018,
showing a ratio of EMEC to LUCID instantaneous luminosity which asymptotically
approaches
a constant value as the activation from the initial high-luminosity period
decays away. Assuming that the instantaneous luminosity in the low-pileup
period is constant, the ratio $R(t)$ of calorimeter to LUCID luminosity
as a function of time $t$ can be modelled as
\begin{equation}\label{e:actlowmu}
R(t)=R_0+\sum^{2}_{i=1} a_i \mathrm{e}^{-t/\tau_i} \ ,
\end{equation}
i.e.\ as the sum of a  constant asymptotic value $R_0$ and two
activation terms whose contributions $a_i$ decay exponentially with lifetime
$\tau_i$. The values of $\tau_i$ depend on the lifetimes of the contributing
isotopes, and are constant, whereas the values of $a_i$ depend on the luminosity
profile and duration of the initial high-pileup running, and are thus different
for each fill. This model was fitted simultaneously to the ratios of
EMEC to LUCID luminosities in all five fills in the 2018 low-pileup period,
determining values of $R_0$ for each fill (representing the
activation-corrected run-integrated luminosity ratios for the two detectors),
together with the lifetimes $\tau_1\approx 900$\,s and $\tau_2\approx 8000$\,s
and separate $a_1$ and $a_2$ values for each fill. The black points in
Figure~\ref{f:caloactlowmu} show the EMEC to LUCID ratios after
subtracting the fitted activation contributions; these ratios are much flatter,
indicating that the two-component model of Eq.~(\ref{e:actlowmu})
provides a reasonable description of the activation. In the 2017 dataset,
the fills with large activation contributions are short, making it hard
to constrain the longer-lifetime component, so the six fills were fitted
using fixed lifetimes taken from the 2018 dataset. The same procedure
was applied to the FCal data, which was found to be well described
with two components with $\tau_1\approx 600$\,s and $\tau_2\approx 39000$\,s,
and to the TileCal D6 data, giving components with $\tau_1\approx 1700$\,s and
$\tau_2\approx 29000$\,s.
 
\begin{figure}[tp]
\parbox{83mm}{\includegraphics[width=76mm]{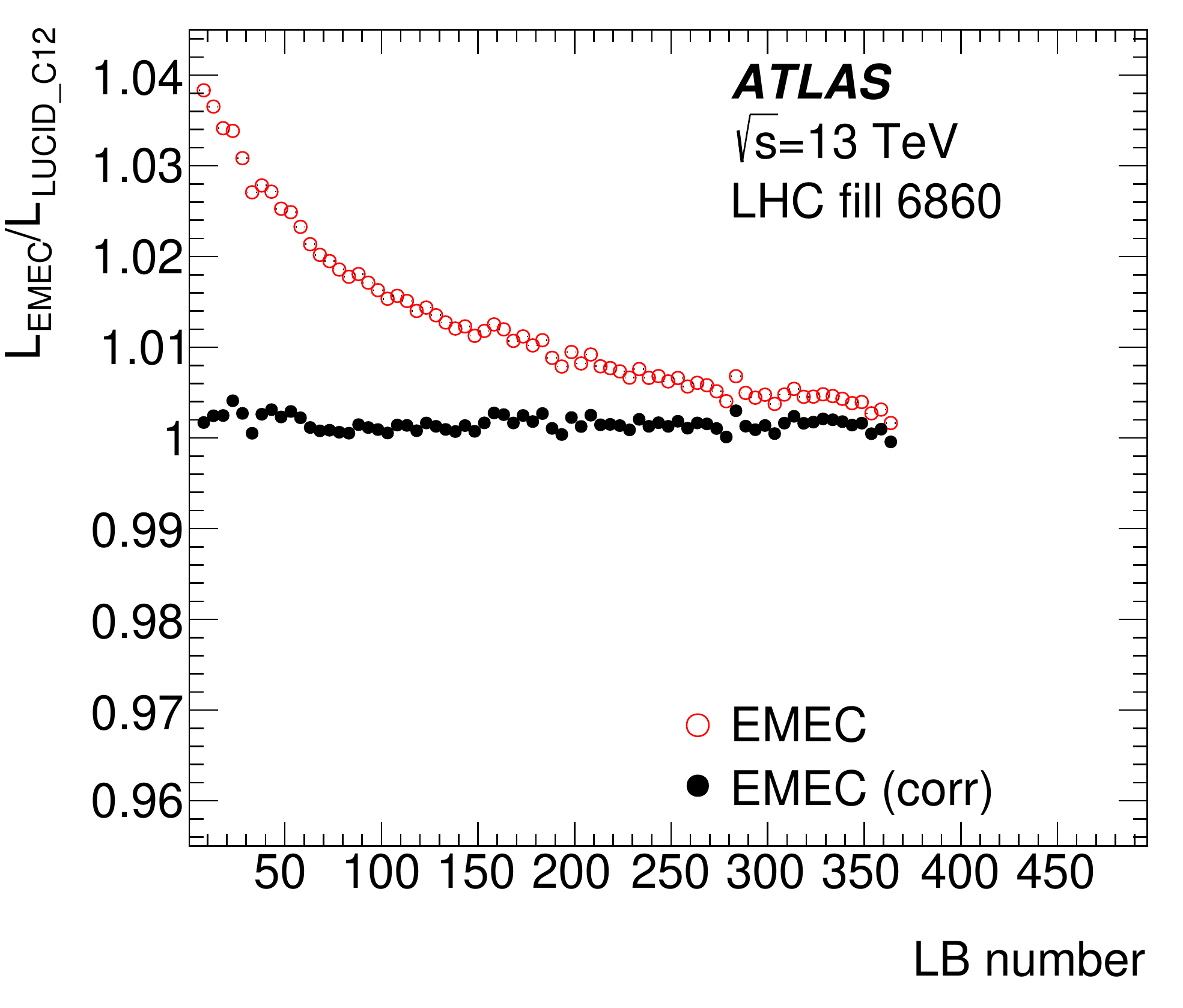}\vspace{-6mm}\center{(a)}}
\parbox{83mm}{\includegraphics[width=76mm]{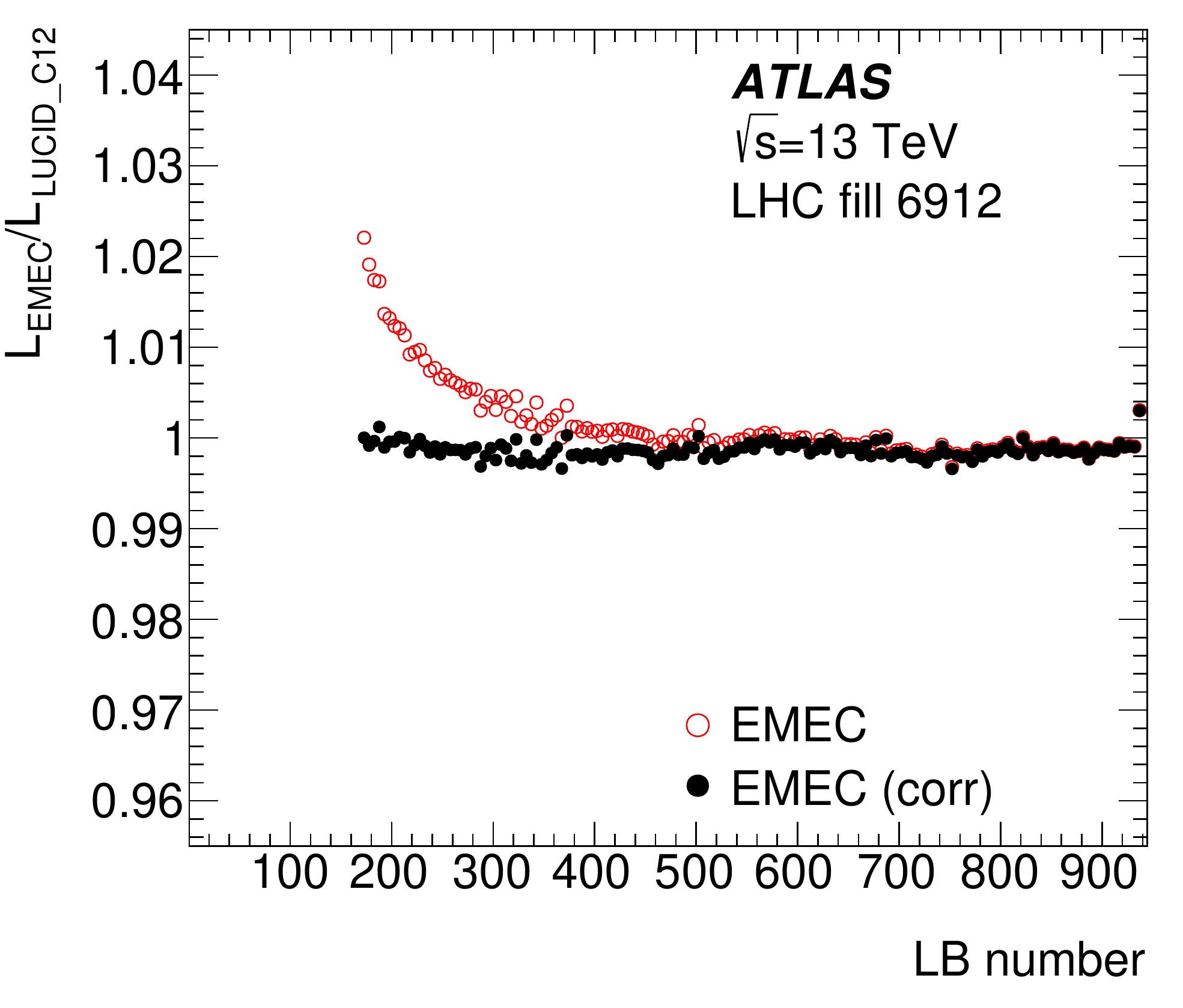}\vspace{-6mm}\center{(b)}}\\
\caption{\label{f:caloactlowmu}Examples of the activation corrections applied
to the EMEC luminosity data in low-pileup runs with $\meanmu\approx 2$ in 2018,
showing LHC fills lasting (a) 6 and (b) 13~hours. The red open points show the
ratios of uncorrected EMEC luminosity measurements to those from LUCID
as a function of luminosity block number, and the black filled points show the
corresponding ratios after correcting for activation effects as discussed
in the text. The ratios have each been averaged over five consecutive luminosity
blocks.}
\end{figure}
 
Figures~\ref{f:lowmustab}(a) and~\ref{f:lowmustab}(b) show the resulting
differences in
run-integrated luminosity between each activation-corrected calorimeter
measurement and LUCID, together with the corresponding track-counting vs.
LUCID differences. The track-counting, FCal and TileCal D6 measurements
agree with LUCID by construction in the two reference runs, shown by the
purple arrows. LUCID and EMEC agree in the reference runs within
0.2\% in 2017 and within 0.05\% in 2018, providing a strong validation
of the calibration transfer correction applied to LUCID. The per-run differences
for other runs and detectors are typically within $\pm 0.5$\% for 2017
and $\pm 0.2$\% for 2018. The FCal measurement for fill 6404 in 2017
is 1.8\% higher than LUCID; however this fill has only 644 colliding bunch
pairs rather than 1866 as for the other fills, and only one third of the
instantaneous luminosity, giving a significantly different FCal response due
to its non-linearity.
In 2018, the activation-corrected TileCal D6 measurement in the 6~hour
fill 6860
is 2\% lower than all the other measurements, but the relatively short
length of this fill compared with the long lifetime of the second activation
component makes it difficult to establish the activation correction precisely.
These two measurements were therefore considered outliers and not used to
assess the run-to-run stability of the LUCID measurement.
 
\begin{figure}[tp]
\parbox{83mm}{\includegraphics[width=76mm]{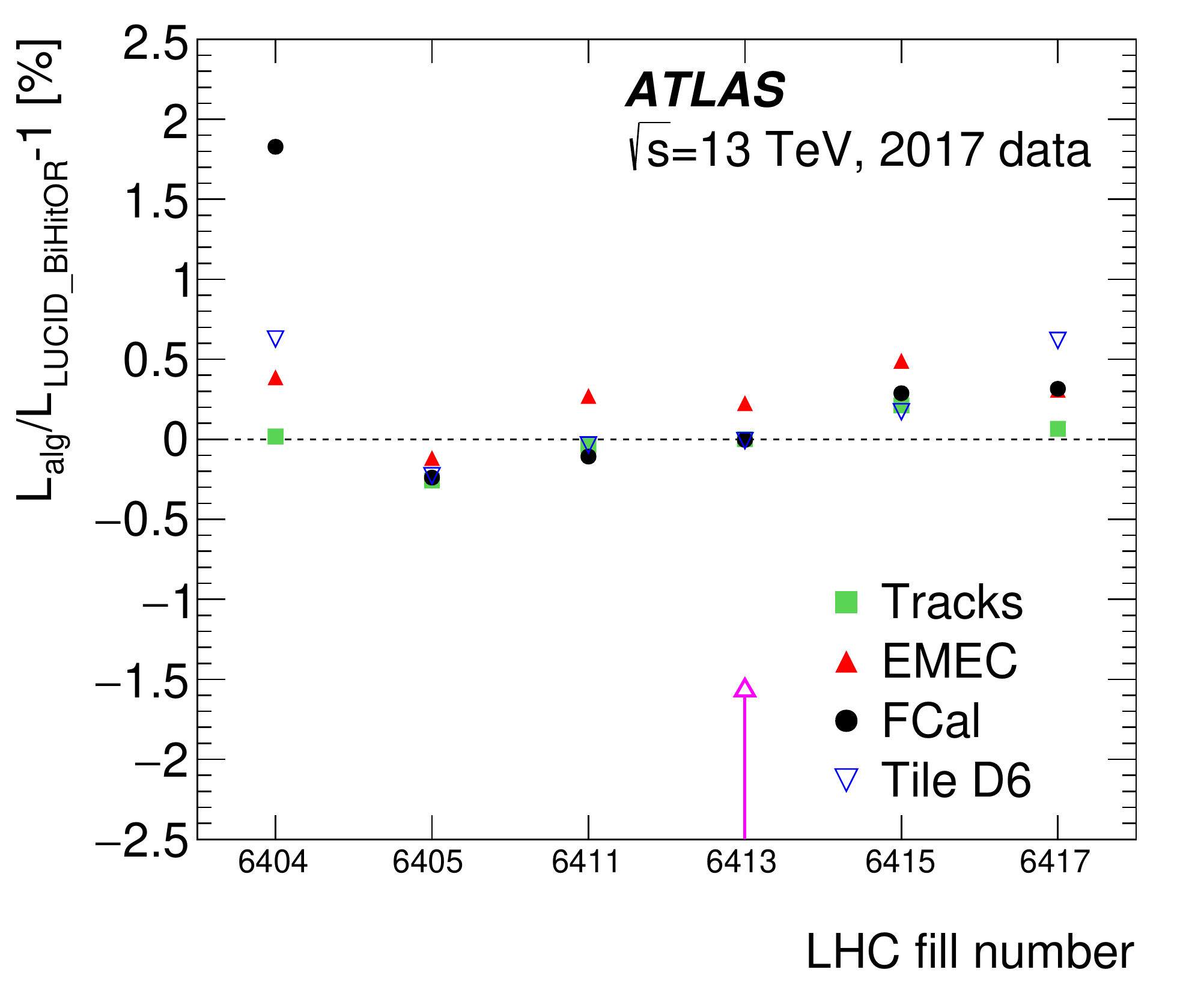}\vspace{-6mm}\center{(a)}}
\parbox{83mm}{\includegraphics[width=76mm]{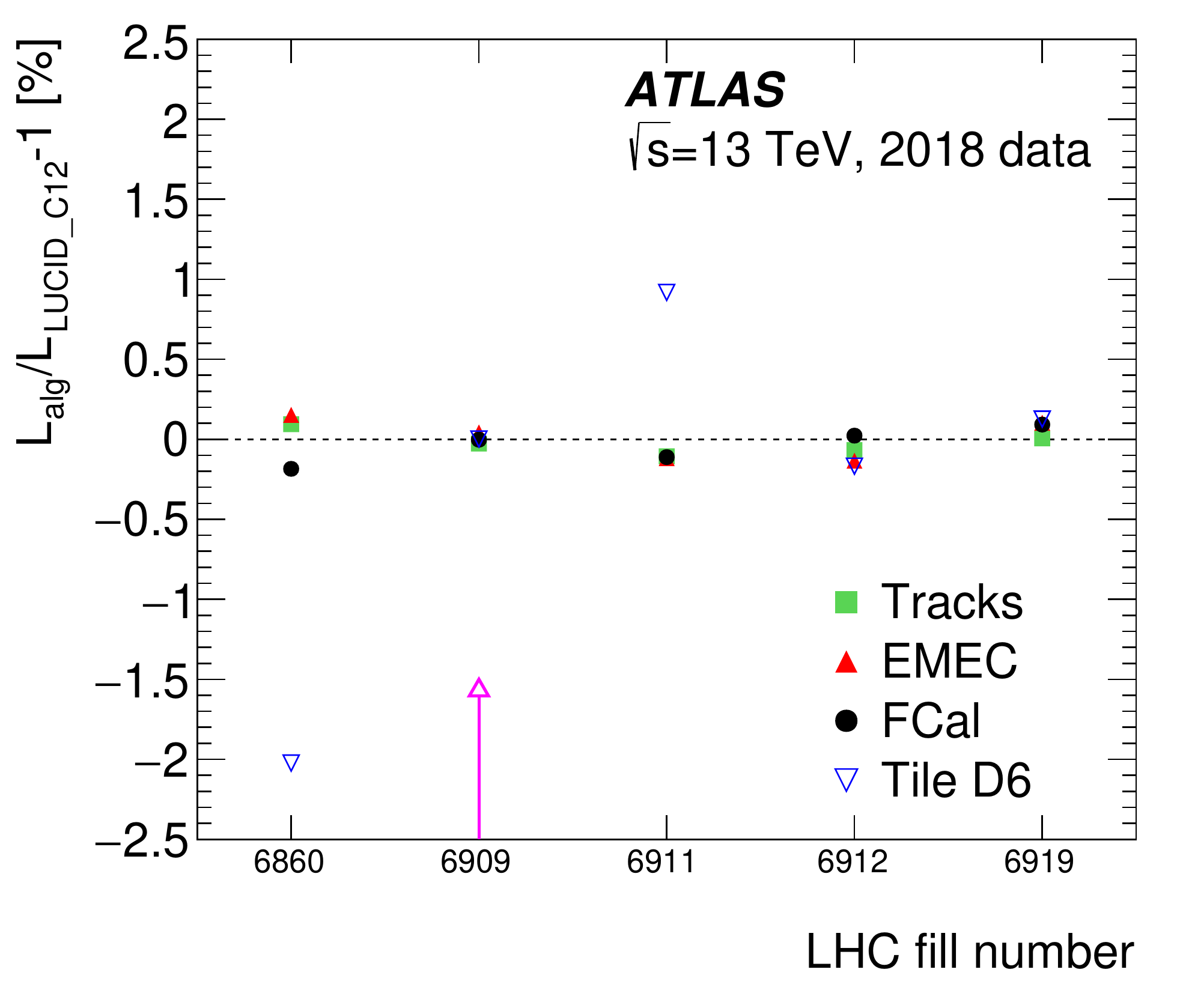}\vspace{-6mm}\center{(b)}}\\
\parbox{83mm}{\includegraphics[width=76mm]{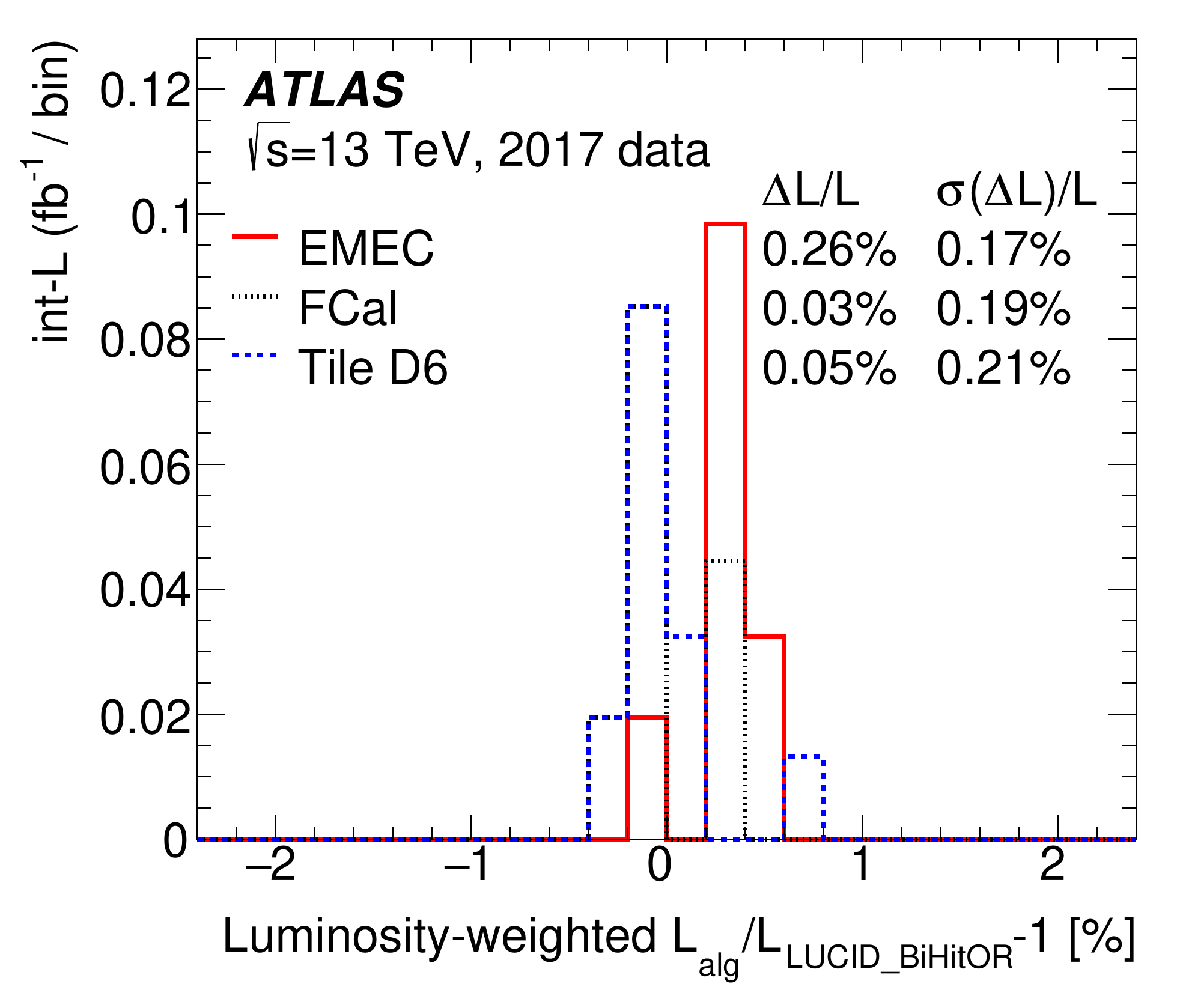}\vspace{-6mm}\center{(c)}}
\parbox{83mm}{\includegraphics[width=76mm]{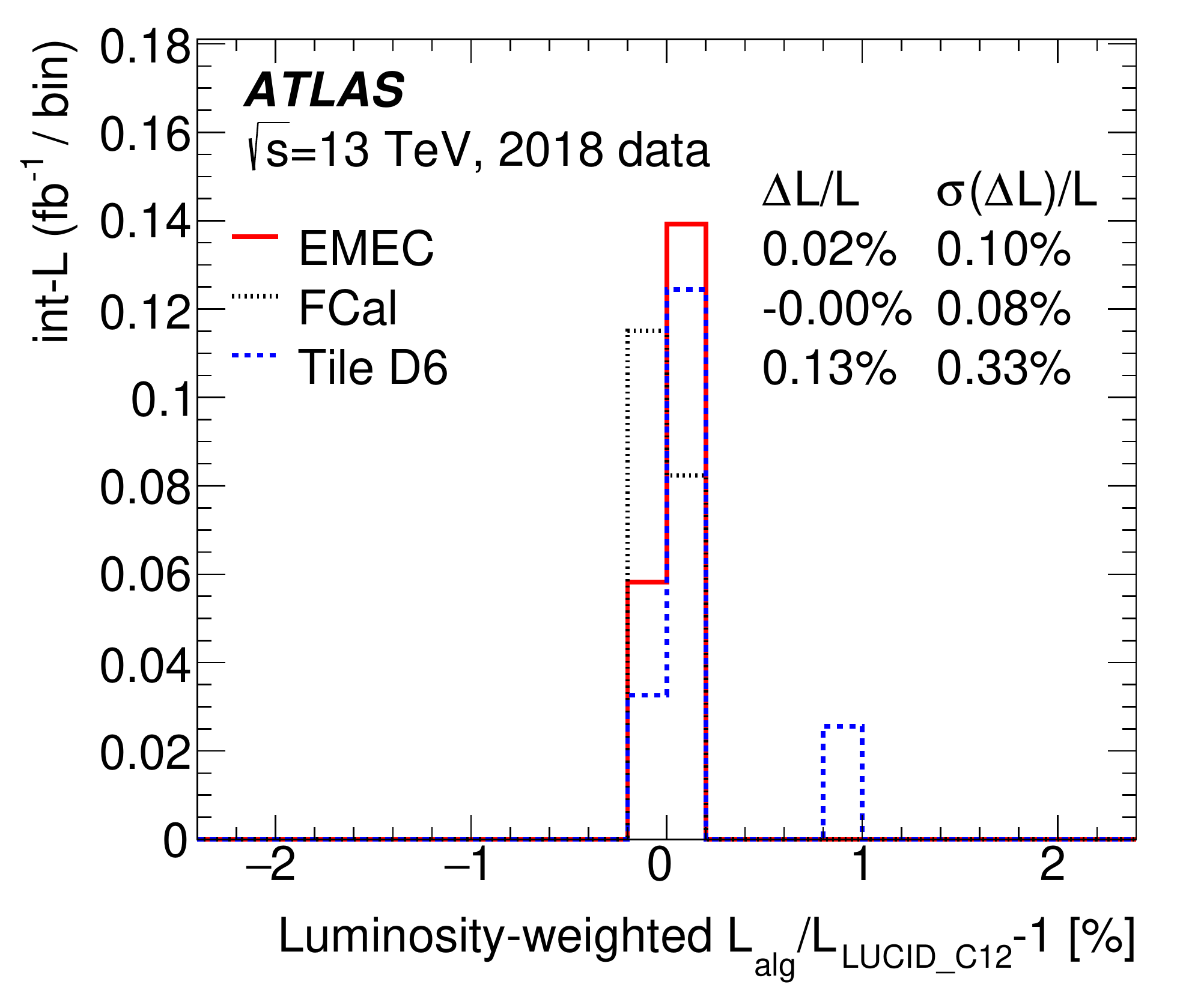}\vspace{-6mm}\center{(d)}}\\
\caption{\label{f:lowmustab}(a, b) Fractional differences in run-integrated
luminosity between the track-counting, EMEC, FCal and TileCal D6 luminosity
measurements and the baseline LUCID luminosity measurements, for each LHC
fill in the 2017 and 2018 $\meanmu\approx 2$ datasets. The reference fills,
used to normalise LUCID to track-counting, and FCal and TileCal to LUCID,
are indicated
by the purple arrows in each year. The EMEC measurements are normalised
in high-pileup runs close to the vdM run, as shown in Figure~\ref{f:calanchor}.
(c, d) Distributions of relative differences in run-integrated luminosity
between the EMEC, FCal and TileCal D6 measurements and the baseline
LUCID measurement, weighted by the integrated luminosity in each run for
the two datasets, excluding fill 6404 for the FCal measurement in 2017 and
fill 6860 for the TileCal measurement in 2018 (see text).
The mean and RMS of each of the distributions are given in the legend.}
\end{figure}
 
Following the approach described for high-pileup data in Section~\ref{s:stab},
the long-term stability uncertainty was defined as the largest difference
in integrated luminosity between LUCID and any calorimeter measurement,
evaluated over the entire low-pileup dataset, separately for 2017 and 2018.
The corresponding plots of fractional differences in per-run integrated
luminosity, weighted by the integrated luminosity of each run, are shown
in Figures~\ref{f:lowmustab}(c) and~\ref{f:lowmustab}(d);
these distributions do not include the two outlier measurements discussed above.
 
The integrated luminosities and uncertainty breakdowns for the 2017 and
2018 low-pileup datasets are summarised in Table~\ref{t:unclow}.
The vdM uncertainties are the same as in Table~\ref{t:unc}, and only
the total vdM uncertainties are shown here.
Since the EMEC luminosity measurements are anchored
in high-pileup runs close to the vdM scans as described in Section~\ref{s:stab}, the EMEC calibration anchoring uncertainties (as shown in
Figures~\ref{f:calanchor}(c) and~\ref{f:calanchor}(d)) are also included.
No additional uncertainty is included for FCal and TileCal measurements, because
they are normalised within the low-pileup periods, and are therefore not
sensitive to drifts between the vdM run and the low-pileup data-taking.
The long-term stability uncertainties are defined from the largest differences in fractional integrated luminosity shown in Figures~\ref{f:lowmustab}(c)
and~\ref{f:lowmustab}(d), and are set by the EMEC measurement in
2017 and TileCal D6 in 2018. The long-term
stability uncertainty is larger in 2017 than 2018, reflecting the long
extrapolation in 2017
between the vdM run (July) and low-pileup runs (November). In 2018,
the last low-pileup run was recorded only two weeks after the vdM run.
 
\begin{table}[tp]
\caption{\label{t:unclow}
Summary of the integrated luminosities (after standard data-quality
requirements) and uncertainties for the calibration of the $\meanmu\approx 2$
low-pileup $pp$ collision data samples at \sxyt\ in 2017 and 2018, and for the
combined sample. As well as the integrated luminosities and total uncertainties,
the table gives the absolute vdM calibration uncertainty and the
additional uncertainties for the physics data sample, and the total relative
uncertainty in percent. The calibration transfer contributions, marked $^*$,
are considered fully
correlated between years, whilst the calibration anchoring and long-term
stability uncertainties are considered
uncorrelated. A detailed breakdown of the vdM calibration uncertainties
for each year is given in Table~\ref{t:unc}.}
\centering
 
\begin{tabular}{l|cc|c}\hline
Data sample  & 2017 & 2018 & Comb. \\
\hline
Integrated luminosity [pb$^{-1}$]  &  147.1 &  191.1 &  338.1 \\
Total uncertainty [pb$^{-1}$] &   1.7 &   2.1 &     3.1 \\
\hline
Uncertainty contributions [\%]: & &  \\
Subtotal vdM calibration &   0.99 &   0.93 &   0.75 \\
\hline
Calibration transfer$^*$ &   0.50 &   0.50 &   0.50 \\
Calibration anchoring &   0.14 &   0.14 &   0.10 \\
Long-term stability &   0.26 &   0.13 &   0.13 \\
\hline
Total uncertainty [\%] &   1.15 &   1.08 &   0.92 \\
\hline
\end{tabular}
\end{table}
 
The rightmost column in Table~\ref{t:unclow} shows the combination of the
2017 and 2018 low-pileup datasets, with the same correlation assumptions
as made for the high-pileup sample. The total integrated luminosity of the
$\meanmu\approx 2$ sample is $\lowmuintl\pm\lowmuetot$\,pb$^{-1}$, corresponding
to a relative uncertainty of \lowmulfrac\%. The covariance matrix is
$\mathbf{V_L}=\mathbf{\boldsymbol{\sigma}_L C \boldsymbol{\sigma}_L^T}$, where
\begin{equation*}
\mathbf{\boldsymbol{\sigma}_L}=\left(
\begin{array}{r}
1.69 \\
2.05 \\
\end{array}
\right)\ \mbox{pb$^{-1}$} \ ,\ \mathbf{C}=\left(\begin{array}{rr}
1.000  \\
0.363 & 1.000  \\
\end{array}
\right)\ .
\end{equation*}
The total luminosity uncertainty for the low-pileup data sample is slightly
larger than for the high-pileup sample, mainly because it contains data from
only two years, normalised using only two vdM calibration sessions.

% End of text imported from the .//lowmu.tex input file

\FloatBarrier

% The next lines are included from the .//conc.tex input file
\section{Conclusions}\label{s:conc}
 
The luminosity scale for the Run~2 \sxyt\ $pp$ collision data recorded
by the ATLAS experiment at the LHC during the years 2015--18
has been calibrated using dedicated van der Meer scans
in each year, and extrapolated to the physics regime using complementary
measurements from several luminosity-sensitive detectors. The total
uncertainties in the integrated luminosities for each individual year
vary from 0.9--1.1\%, and the uncertainty in the combined Run~2 dataset
is $\delta{\cal L}/{\cal L}=\pm\totlfrac$\%. The largest contributions to the
uncertainty come from the extrapolation of the calibration from the
low-luminosity vdM scans to the high-luminosity physics data-taking regime,
followed by the effects of magnetic non-linearity, beam--beam interactions and
scan-to-scan reproducibility on the absolute vdM calibration.
Overall, the uncertainties related to the vdM calibration are larger than
those from other sources in the final uncertainty. This final calibration
implies that previously published ATLAS \sxyt\ $pp$ cross-section values
should be decreased by about 1\%, depending on the data sample used.
 
This calibration is significantly more precise than those achieved by
ATLAS for Run~1 (1.8\% at \sxwt\ \cite{DAPR-2011-01} and 1.9\% at \sxvt\
\cite{DAPR-2013-01}), and is the most
precise luminosity measurement at any hadron collider to date,
apart from some of the second-generation total cross-section
experiments at the CERN ISR which achieved a comparable precision
of 0.9\% using the vdM calibration technique \cite{lumibib,isrtotxsec}.
The 2015 and 2016 calibrations are more precise than the 1.6\% and 1.2\%
achieved by the CMS Collaboration for their corresponding datasets
\cite{CMS-LUM-17-003}.
 
The same techniques have been used to calibrate the small \sxyt\ $pp$ collision
datasets with pileup of $\meanmu\approx 2$ recorded in 2017 and 2018 for
the precision Standard Model physics programme. This sample
has an integrated luminosity of \lowmuintl\,\ipb\ with an uncertainty of
\lowmulfrac\%. This calibration does not apply to the even smaller
samples recorded in Run~2 at high $\beta^*$ for the ALFA forward-physics
programme, which require dedicated analyses appropriate to their specific
experimental conditions.

% End of text imported from the .//conc.tex input file

\section*{Acknowledgements}

% The next lines are included from the .//acknowledgements/Acknowledgements.tex input file

We thank CERN for the very successful operation of the LHC, as well as the
support staff from our institutions without whom ATLAS could not be
operated efficiently.
We also thank the LHC operations team, and the LHC machine and programme
coordinators, for their collaboration in planning and performing the vdM scans
and other dedicated LHC fills used to calibrate the ATLAS luminosity
measurement.
 
We acknowledge the support of
ANPCyT, Argentina;
YerPhI, Armenia;
ARC, Australia;
BMWFW and FWF, Austria;
ANAS, Azerbaijan;
CNPq and FAPESP, Brazil;
NSERC, NRC and CFI, Canada;
CERN;
ANID, Chile;
CAS, MOST and NSFC, China;
Minciencias, Colombia;
MEYS CR, Czech Republic;
DNRF and DNSRC, Denmark;
IN2P3-CNRS and CEA-DRF/IRFU, France;
SRNSFG, Georgia;
BMBF, HGF and MPG, Germany;
GSRI, Greece;
RGC and Hong Kong SAR, China;
ISF and Benoziyo Center, Israel;
INFN, Italy;
MEXT and JSPS, Japan;
CNRST, Morocco;
NWO, Netherlands;
RCN, Norway;
MEiN, Poland;
FCT, Portugal;
MNE/IFA, Romania;
MESTD, Serbia;
MSSR, Slovakia;
ARRS and MIZ\v{S}, Slovenia;
DSI/NRF, South Africa;
MICINN, Spain;
SRC and Wallenberg Foundation, Sweden;
SERI, SNSF and Cantons of Bern and Geneva, Switzerland;
MOST, Taiwan;
TENMAK, T\"urkiye;
STFC, United Kingdom;
DOE and NSF, United States of America.
In addition, individual groups and members have received support from
BCKDF, CANARIE, Compute Canada and CRC, Canada;
PRIMUS 21/SCI/017 and UNCE SCI/013, Czech Republic;
COST, ERC, ERDF, Horizon 2020 and Marie Sk{\l}odowska-Curie Actions, European Union;
Investissements d'Avenir Labex, Investissements d'Avenir Idex and ANR, France;
DFG and AvH Foundation, Germany;
Herakleitos, Thales and Aristeia programmes co-financed by EU-ESF and the Greek NSRF, Greece;
BSF-NSF and MINERVA, Israel;
Norwegian Financial Mechanism 2014-2021, Norway;
NCN and NAWA, Poland;
La Caixa Banking Foundation, CERCA Programme Generalitat de Catalunya and PROMETEO and GenT Programmes Generalitat Valenciana, Spain;
G\"{o}ran Gustafssons Stiftelse, Sweden;
The Royal Society and Leverhulme Trust, United Kingdom.
 
The crucial computing support from all WLCG partners is acknowledged gratefully, in particular from CERN, the ATLAS Tier-1 facilities at TRIUMF (Canada), NDGF (Denmark, Norway, Sweden), CC-IN2P3 (France), KIT/GridKA (Germany), INFN-CNAF (Italy), NL-T1 (Netherlands), PIC (Spain), ASGC (Taiwan), RAL (UK) and BNL (USA), the Tier-2 facilities worldwide and large non-WLCG resource providers. Major contributors of computing resources are listed in Ref.~\cite{ATL-SOFT-PUB-2021-003}.

% End of text imported from the .//acknowledgements/Acknowledgements.tex input file

\printbibliography

@booklet{vdmisr,
      author        = "van der Meer, S",
      title         = "{Calibration of the effective beam height in the ISR}",
      howpublished  = "CERN-ISR-PO-68-31",
      year          = "1968",
      url           = "https://cds.cern.ch/record/296752",
}

@article{atlasibl,
      author         = "Abbott, B. and others",
      title          = "{Production and Integration of the ATLAS Insertable
                        B-Layer}",
      collaboration  = "ATLAS IBL",
      journal        = "JINST",
      volume         = "13",
      year           = "2018",
      number         = "05",
      pages          = "T05008",
      doi            = "10.1088/1748-0221/13/05/T05008",
      eprint         = "1803.00844",
      archivePrefix  = "arXiv",
      primaryClass   = "physics.ins-det",
      SLACcitation   = "%%CITATION = ARXIV:1803.00844;%%"
}

@Booklet{IBLTDR,
     author       = "{ATLAS Collaboration}",
     title        = "ATLAS Insertable B-Layer Technical Design Report",
     howpublished = "ATLAS-TDR-19",
     year         = "2010",
     url          = "https://cds.cern.ch/record/1291633",
     addendum     = "{\textit{ATLAS Insertable B-Layer Technical Design Report Addendum}}, ATLAS-TDR-19-ADD-1, 2012, {\scriptsize{URL}}: \url{https://cds.cern.ch/record/1451888}",
}

@article{lucid2,
  author={G. Avoni and others},
  title={The new LUCID-2 detector for luminosity measurement and monitoring in ATLAS},
  journal={JINST},
  volume={13},
  number={07},
  pages={P07017},
  doi="10.1088/1748-0221/13/07/P07017",
  year={2018},
}

@article{lumibib,
      author         = "Grafström, P. and Kozanecki, W.",
      title          = "{Luminosity determination at proton colliders}",
      journal        = "Prog. Part. Nucl. Phys.",
      volume         = "81",
      year           = "2015",
      pages          = "97-148",
      doi            = "10.1016/j.ppnp.2014.11.002",
      SLACcitation   = "%%CITATION = PPNPD,81,97;%%"
}

@article{conddb,
author="M. B{\"o}hler and others",
title={Evolution of ATLAS conditions data and its management for LHC Run-2},
journal="J.\ Phys. Conf. Ser.",
volume="664",
year="2015",
pages="042005",
doi="10.1088/1742-6596/664/4/042005",
}

@Booklet{weidermann,
author="Wiedemann, H.",
title="Particle Accelerator Physics",
howpublished="Graduate Texts in Physics, Springer",
year="2015",
url="https://www.springer.com/us/book/9783319183169?wt_mc=ThirdParty.SpringerLink.3.EPR653.About_eBook",
}

@article{bcmdet,
    author         = "Cindro, V. and others",
    title          = "{The ATLAS Beam Conditions Monitor}",
    journal        = "JINST",
    volume         = "3",
    year           = "2008",
    pages          = "P02004",
    doi            = "10.1088/1748-0221/3/02/P02004",
    primaryClass   = "hep-ex",
}

@bookletthesis{andersthesis,
      author        = "Anders, Gabriel",
      title         = "{Absolute luminosity determination for the ATLAS experiment}",
      year          = "2013",
      howpublished  = "CERN-THESIS-2013-111",
      url           = "https://cds.cern.ch/record/1595219",
}

@article{lhcblumi,
    author = "{LHCb Collaboration}",
    title = "{Precision luminosity measurements at LHCb}",
    eprint = "1410.0149",
    archivePrefix = "arXiv",
    primaryClass = "hep-ex",
    reportNumber = "LHCB-PAPER-2014-047, CERN-PH-EP-2014-221",
    doi = "10.1088/1748-0221/9/12/P12005",
    journal = "JINST",
    volume = "9",
    number = "12",
    pages = "P12005",
    year = "2014"
}

@Booklet{boccardi,
     author       = "Boccardi, A and Bravin, E and Ferro-Luzzi, M and Mazzoni, S and Palm, M",
     title        = "LHC Luminosity calibration using the Longitudinal Density Monitor",
     howpublished = "CERN-ATS-Note-2013-034 TECH",
     year         = "2013",
     url          = "https://cds.cern.ch/record/1556087",
}

@Booklet{dcctcalib,
     author       = "Barschel, C and Ferro-Luzzi, M and Gras, J-J and Ludwig, M and Odier, P and Thoulet, S",
     title        = "{Results of the LHC DCCT Calibration Studies}",
     howpublished = "CERN-ATS-Note-2012-026 PERF",
     year         = "2012",
     url          = "https://cds.cern.ch/record/1425904",
}

@Booklet{bunchvdm,
     author       = "Bartosik, H and Rumolo, G",
     title        = "Production of single Gaussian bunches for Van der Meer scans in the LHC injector chain",
     howpublished = "CERN-ACC-Note-2013-0008",
     year         = "2013",
     url          = "https://cds.cern.ch/record/1590405",
}

@Booklet{anders,
     author       = "Anders, G. and others",
     title        = "Study of the Relative LHC Bunch Populations for Luminosity Calibration",
     howpublished = "CERN-ATS-Note-2013-028 PERF",
     year         = "2013",
     url          = "https://cds.cern.ch/record/1427726",
}

@article{geant4,
      author         = "Agostinelli, S. and others",
      title          = "{GEANT4---A simulation toolkit}",
      collaboration  = "GEANT4",
      journal        = "Nucl. Instrum. Meth. A",
      volume         = "506",
      pages          = "250",
      doi            = "10.1016/S0168-9002(03)01368-8",
      year           = "2003",
      reportNumber   = "SLAC-PUB-9350, FERMILAB-PUB-03-339",
      SLACcitation   = "%%CITATION = NUIMA,A506,250;%%",
}

@article{pythia8,
      author         = {Sj\"ostrand, Torbjorn and Mrenna, Stephen and Skands, Peter Z.},
      title          = "{A brief introduction to PYTHIA 8.1}",
      journal        = "Comput. Phys. Commun.",
      volume         = "178",
      pages          = "852",
      doi            = "10.1016/j.cpc.2008.01.036",
      year           = "2008",
      eprint         = "0710.3820",
      archivePrefix  = "arXiv",
      primaryClass   = "hep-ph",
      reportNumber   = "CERN-LCGAPP-2007-04, LU-TP-07-28,
                        FERMILAB-PUB-07-512-CD-T",
      SLACcitation   = "%%CITATION = ARXIV:0710.3820;%%",
}

@booklet{bassetti,
  author="Bassetti, M. and Erskine, G.",
  title="Closed Expression for the Electrical Field of a Two-dimensional Gaussian Charge",
  year="1980",
  howpublished="CERN-ISR-TH-80-06",
  url="https://cds.cern.ch/record/122227",
}

@article{bambade,
  author="Bambade, P. and others",
  title="Observation of beam-beam deflections at the interaction point of the SLAC Linear Collider",
  journal="Phys. Rev. Lett.",
  volume="62",
  year="1989",
  pages="2949",
  doi="10.1103/PhysRevLett.62.2949",
}

@article{magmeas,
    author = "Chmieli\'nska, Agnieszka and Fiscarelli, Lucio and Hostettler, Michi and Kozanecki, Witold and Russenschuck, Stephan and Todesco, Ezio",
    title = "{Magnetization in superconducting corrector magnets and impact on luminosity-calibration scans in the Large Hadron Collider}",
    journal = "Eur. Phys. J. Plus",
    volume = "138",
    year   = "2023",
    pages  = "796",
    doi = "10.1140/epjp/s13360-023-04427-x",
    eprint = "2304.06559",
    archivePrefix = "arXiv",
    primaryClass = "physics.acc-ph",
    reportNumber = "CERN-TE-2023-001",
    year = "2023"
}

@booklet{madx,
  author="{CERN Accelerator Beam Physics Group}",
  title="MAD - Methodical Accelerator Design",
  url="https://mad.web.cern.ch/mad",
}

@article{bstarb,
    author         = "Balagura, Vladislav",
    title          = "{Van der Meer scan luminosity measurement and beam-beam correction}", 
    journal        = "Eur. Phys. J. C",
    volume         = "81",
    year           = "2021",
    pages          = "26",
    doi            = "10.1140/epjc/s10052-021-08837-y",
    eprint         = "2012.07752",
    archivePrefix  = "arXiv",
    primaryClass   = "hep-ex",
}

@booklet{combi,
      author        = "Jones, F W and Herr, Werner and Pieloni, T",
      title         = "{Parallel Beam-Beam Simulation Incorporating Multiple
                       Bunches and Multiple Interaction Regions}",
      howpublished  = "CERN-LHC-PROJECT-Report-1038",
      year          = "2007",
      url           = "https://cds.cern.ch/record/1058519",
}

@article{newbb,
  author="Babaev, A. and others",
  title="{Impact of Beam-Beam Effects on Absolute Luminosity Calibrations at the CERN Large Hadron Collider}",
  eprint = "2306.10394",
  archivePrefix = "arXiv",
  primaryClass = "physics.acc-ph",
  month = "6",
  year = "2023"
}

@article{isrtotxsec,
  author="Amaldi, U. and others",
  title="{Precision measurement of proton--proton total cross section at the CERN intersecting storage rings}",
  journal="Nucl. Phys. B.",
  volume="145",
  year="1978",
  pages="367",
  doi="10.1016/0550-3213(78)90090-1",
}

@Article{PERF-2007-01,
    author         = "{ATLAS Collaboration}",
    title          = "{The ATLAS Experiment at the CERN Large Hadron Collider}",
    journal        = "JINST",
    volume         = "3",
    year           = "2008",
    pages          = "S08003",
    doi            = "10.1088/1748-0221/3/08/S08003",
    primaryClass   = "hep-ex",
}

@Article{LARG-2009-02,
    author         = "{ATLAS Collaboration}",
    title          = "{Drift Time Measurement in the ATLAS Liquid Argon Electromagnetic Calorimeter using Cosmic Muons}",
    journal        = "Eur. Phys. J. C",
    volume         = "70",
    year           = "2010",
    pages          = "755",
    doi            = "10.1140/epjc/s10052-010-1403-6",
    eprint         = "1002.4189",
    archivePrefix  = "arXiv",
    primaryClass   = "hep-ex",
}

@Article{DAPR-2010-01,
    author         = "{ATLAS Collaboration}",
    title          = "{Luminosity determination in \(pp\) collisions at \(\sqrt{s} = 7\,\text{TeV}\) using the ATLAS detector at the LHC}",
    journal        = "Eur. Phys. J. C",
    volume         = "71",
    year           = "2011",
    pages          = "1630",
    doi            = "10.1140/epjc/s10052-011-1630-5",
    reportNumber   = "CERN-PH-EP-2010-069",
    eprint         = "1101.2185",
    archivePrefix  = "arXiv",
    primaryClass   = "hep-ex",
}

@Article{SOFT-2010-01,
    author         = "{ATLAS Collaboration}",
    title          = "{The ATLAS Simulation Infrastructure}",
    journal        = "Eur. Phys. J. C",
    volume         = "70",
    year           = "2010",
    pages          = "823",
    doi            = "10.1140/epjc/s10052-010-1429-9",
    eprint         = "1005.4568",
    archivePrefix  = "arXiv",
    primaryClass   = "physics.ins-det",
}

@Article{TCAL-2010-01,
    author         = "{ATLAS Collaboration}",
    title          = "{Readiness of the ATLAS Tile Calorimeter for LHC collisions}",
    journal        = "Eur. Phys. J. C",
    volume         = "70",
    year           = "2010",
    pages          = "1193",
    doi            = "10.1140/epjc/s10052-010-1508-y",
    reportNumber   = "CERN-PH-EP-2010-024",
    eprint         = "1007.5423",
    archivePrefix  = "arXiv",
    primaryClass   = "hep-ex",
}

@Article{DAPR-2011-01,
    author         = "{ATLAS Collaboration}",
    title          = "{Improved luminosity determination in \(pp\) collisions at \(\sqrt{s} = 7\,\text{TeV}\) using the ATLAS detector at the LHC}",
    journal        = "Eur. Phys. J. C",
    volume         = "73",
    year           = "2013",
    pages          = "2518",
    doi            = "10.1140/epjc/s10052-013-2518-3",
    reportNumber   = "CERN-PH-EP-2013-026",
    eprint         = "1302.4393",
    archivePrefix  = "arXiv",
    primaryClass   = "hep-ex",
}

@Article{DAPR-2013-01,
    author         = "{ATLAS Collaboration}",
    title          = "{Luminosity determination in \(pp\) collisions at \(\sqrt{s} = 8\,\text{TeV}\) using the ATLAS detector at the LHC}",
    journal        = "Eur. Phys. J. C",
    volume         = "76",
    year           = "2016",
    pages          = "653",
    doi            = "10.1140/epjc/s10052-016-4466-1",
    eprint         = "1608.03953",
    archivePrefix  = "arXiv",
    primaryClass   = "hep-ex",
}

@Article{PERF-2015-01,
    author         = "{ATLAS Collaboration}",
    title          = "{Reconstruction of primary vertices at the ATLAS experiment in Run~1 proton--proton collisions at the LHC}",
    journal        = "Eur. Phys. J. C",
    volume         = "77",
    year           = "2017",
    pages          = "332",
    doi            = "10.1140/epjc/s10052-017-4887-5",
    reportNumber   = "CERN-EP-2016-150",
    eprint         = "1611.10235",
    archivePrefix  = "arXiv",
    primaryClass   = "hep-ex",
}

@Article{PERF-2015-10,
    author         = "{ATLAS Collaboration}",
    title          = "{Muon reconstruction performance of the ATLAS detector in proton--proton collision data at \(\sqrt{s} = 13\,\text{TeV}\)}",
    journal        = "Eur. Phys. J. C",
    volume         = "76",
    year           = "2016",
    pages          = "292",
    doi            = "10.1140/epjc/s10052-016-4120-y",
    reportNumber   = "CERN-EP-2016-033",
    eprint         = "1603.05598",
    archivePrefix  = "arXiv",
    primaryClass   = "hep-ex",
}

@Article{STDM-2016-02,
    author         = "{ATLAS Collaboration}",
    title          = "{Measurements of top-quark pair to \(Z\)-boson cross-section ratios at \(\sqrt{s} = 13, 8, 7\,\text{TeV}\) with the ATLAS detector}",
    journal        = "JHEP",
    volume         = "02",
    year           = "2017",
    pages          = "117",
    doi            = "10.1007/JHEP02(2017)117",
    reportNumber   = "CERN-EP-2016-271",
    eprint         = "1612.03636",
    archivePrefix  = "arXiv",
    primaryClass   = "hep-ex",
}

@Article{PERF-2017-01,
    author         = "{ATLAS Collaboration}",
    title          = "{Electron reconstruction and identification in the ATLAS experiment using the 2015 and 2016 LHC proton--proton collision data at \(\sqrt{s} = 13\,\text{TeV}\)}",
    journal        = "Eur. Phys. J. C",
    volume         = "79",
    year           = "2019",
    pages          = "639",
    doi            = "10.1140/epjc/s10052-019-7140-6",
    reportNumber   = "CERN-EP-2018-273",
    eprint         = "1902.04655",
    archivePrefix  = "arXiv",
    primaryClass   = "hep-ex",
}

@Article{DAPR-2018-01,
    author         = "{ATLAS Collaboration}",
    title          = "{ATLAS data quality operations and performance for 2015--2018 data-taking}",
    journal        = "JINST",
    volume         = "15",
    year           = "2020",
    pages          = "P04003",
    doi            = "10.1088/1748-0221/15/04/P04003",
    reportNumber   = "CERN-EP-2019-207",
    eprint         = "1911.04632",
    archivePrefix  = "arXiv",
    primaryClass   = "physics.ins-det",
}

@Article{MUON-2018-03,
    author         = "{ATLAS Collaboration}",
    title          = "{Muon reconstruction and identification efficiency in ATLAS using the full Run~2 \(pp\) collision data set at \(\sqrt{s} = 13\,\text{TeV}\)}",
    journal        = "Eur. Phys. J. C",
    volume         = "81",
    year           = "2021",
    pages          = "578",
    doi            = "10.1140/epjc/s10052-021-09233-2",
    reportNumber   = "CERN-EP-2020-199",
    eprint         = "2012.00578",
    archivePrefix  = "arXiv",
    primaryClass   = "hep-ex",
}

@Article{STDM-2018-14,
    author         = "{ATLAS Collaboration}",
    title          = "{Measurement of the transverse momentum distribution of Drell--Yan lepton pairs in proton--proton collisions at \(\sqrt{s} = 13\,\text{TeV}\) with the ATLAS detector}",
    journal        = "Eur. Phys. J. C",
    volume         = "80",
    year           = "2020",
    pages          = "616",
    doi            = "10.1140/epjc/s10052-020-8001-z",
    reportNumber   = "CERN-EP-2019-223",
    eprint         = "1912.02844",
    archivePrefix  = "arXiv",
    primaryClass   = "hep-ex",
}

@Article{IDTR-2019-05,
    author         = "{ATLAS Collaboration}",
    title          = "{Alignment of the ATLAS Inner Detector in Run-2}",
    journal        = "Eur. Phys. J. C",
    volume         = "80",
    year           = "2020",
    pages          = "1194",
    doi            = "10.1140/epjc/s10052-020-08700-6",
    reportNumber   = "CERN-EP-2020-108",
    eprint         = "2007.07624",
    archivePrefix  = "arXiv",
    primaryClass   = "hep-ex",
}

@Article{TRIG-2019-04,
    author         = "{ATLAS Collaboration}",
    title          = "{Operation of the ATLAS trigger system in Run~2}",
    journal        = "JINST",
    volume         = "15",
    year           = "2020",
    pages          = "P10004",
    doi            = "10.1088/1748-0221/15/10/P10004",
    reportNumber   = "CERN-EP-2020-109",
    eprint         = "2007.12539",
    archivePrefix  = "arXiv",
    primaryClass   = "hep-ex",
}

@Article{DAPR-2021-01,
    author         = "{ATLAS Collaboration}",
    title          = "{Luminosity determination in \(pp\) collisions at \(\sqrt{s} = 13\,\text{TeV}\) using the ATLAS detector at the LHC}",
    year           = "2022",
    reportNumber   = "CERN-EP-2022-281",
    eprint         = "2212.09379",
    archivePrefix  = "arXiv",
    primaryClass   = "hep-ex",
}

@Booklet{ATL-SOFT-PUB-2021-001,
    author         = "{ATLAS Collaboration}",
    title          = "{The ATLAS Collaboration Software and Firmware}",
    howpublished   = "{ATL-SOFT-PUB-2021-001}",
    url            = "https://cds.cern.ch/record/2767187",
    year           = "2021",
}

@Booklet{ATL-SOFT-PUB-2021-003,
    author         = "{ATLAS Collaboration}",
    title          = "{ATLAS Computing Acknowledgements}",
    howpublished   = "{ATL-SOFT-PUB-2021-003}",
    url            = "https://cds.cern.ch/record/2776662",
    year           = "2021",
}

@Article{CMS-LUM-17-003,
    author         = "{CMS Collaboration}",
    title          = "{Precision luminosity measurement in proton--proton collisions at \(\sqrt{s} = 13\,\text{TeV}\) in 2015 and 2016 at CMS}",
    journal        = "Eur. Phys. J. C",
    volume         = "81",
    year           = "2021",
    pages          = "800",
    doi            = "10.1140/epjc/s10052-021-09538-2",
    reportNumber   = "CERN-EP-2021-033",
    eprint         = "2104.01927",
    archivePrefix  = "arXiv",
    primaryClass   = "hep-ex",
}

@Booklet{ATLAS-CONF-2019-021,
    author         = "{ATLAS Collaboration}",
    title          = "{Luminosity determination in \(pp\) collisions at \(\sqrt{s} = 13\,\text{TeV}\) using the ATLAS detector at the LHC}",
    howpublished   = "{ATLAS-CONF-2019-021}",
    url            = "https://cds.cern.ch/record/2677054",
    year           = "2019",
}

@Booklet{ATL-PHYS-PUB-2015-051,
    author         = "{ATLAS Collaboration}",
    title          = "{Early Inner Detector Tracking Performance in the 2015 Data at \(\sqrt{s} = 13~\text{TeV}\)}",
    howpublished   = "{ATL-PHYS-PUB-2015-051}",
    url            = "https://cds.cern.ch/record/2110140",
    year           = "2015",
}

@Booklet{ATL-PHYS-PUB-2016-017,
    author         = "{ATLAS Collaboration}",
    title          = "{The Pythia~8 A3 tune description of ATLAS minimum bias and inelastic measurements incorporating the Donnachie--Landshoff diffractive model}",
    howpublished   = "{ATL-PHYS-PUB-2016-017}",
    url            = "https://cds.cern.ch/record/2206965",
    year           = "2016",
}

@Booklet{ATL-PHYS-PUB-2017-021,
    author         = "{ATLAS Collaboration}",
    title          = "{Prospects for the measurement of the \(W\)-boson transverse momentum with a low pileup data sample at \(\sqrt{s} = 13~\text{TeV}\) with the ATLAS detector}",
    howpublished   = "{ATL-PHYS-PUB-2017-021}",
    url            = "https://cds.cern.ch/record/2298152",
    year           = "2017",
}

@Booklet{ATL-DAPR-PUB-2021-001,
    author         = "{ATLAS Collaboration}",
    title          = "{Luminosity monitoring using \(Z\to{\ell^+\ell^-}\) events at \(\sqrt{s} = 13\,\text{TeV}\) with the ATLAS detector}",
    howpublished   = "{ATL-DAPR-PUB-2021-001}",
    url            = "https://cds.cern.ch/record/2752951",
    year           = "2021",
}
 
\clearpage
% ATLAS Collaboration author list
% Reference date of DAPR-2021-01 is 2022-10-10
% Author list last updated on date 03-NOV-23
% Data extracted on 03-Nov-2023 for paper reference DAPR-2021-01
% at 2:54pm
 
\begin{flushleft}
\hypersetup{urlcolor=black}
{\Large The ATLAS Collaboration}

\bigskip

\AtlasOrcid[0000-0002-6665-4934]{G.~Aad}$^\textrm{\scriptsize 102}$,
\AtlasOrcid[0000-0002-5888-2734]{B.~Abbott}$^\textrm{\scriptsize 120}$,
\AtlasOrcid[0000-0002-1002-1652]{K.~Abeling}$^\textrm{\scriptsize 55}$,
\AtlasOrcid[0000-0002-8496-9294]{S.H.~Abidi}$^\textrm{\scriptsize 29}$,
\AtlasOrcid[0000-0002-9987-2292]{A.~Aboulhorma}$^\textrm{\scriptsize 35e}$,
\AtlasOrcid[0000-0001-5329-6640]{H.~Abramowicz}$^\textrm{\scriptsize 151}$,
\AtlasOrcid[0000-0002-1599-2896]{H.~Abreu}$^\textrm{\scriptsize 150}$,
\AtlasOrcid[0000-0003-0403-3697]{Y.~Abulaiti}$^\textrm{\scriptsize 117}$,
\AtlasOrcid[0000-0003-0762-7204]{A.C.~Abusleme~Hoffman}$^\textrm{\scriptsize 137a}$,
\AtlasOrcid[0000-0002-8588-9157]{B.S.~Acharya}$^\textrm{\scriptsize 69a,69b,p}$
% ...
\bigskip
\\

% ...
$^{29}$Physics Department, Brookhaven National Laboratory, Upton NY; United States of America.\\
% ...
$^{35}$$^{(a)}$Facult\'e des Sciences Ain Chock, R\'eseau Universitaire de Physique des Hautes Energies - Universit\'e Hassan II, Casablanca;$^{(b)}$Facult\'{e} des Sciences, Universit\'{e} Ibn-Tofail, K\'{e}nitra;$^{(c)}$Facult\'e des Sciences Semlalia, Universit\'e Cadi Ayyad, LPHEA-Marrakech;$^{(d)}$LPMR, Facult\'e des Sciences, Universit\'e Mohamed Premier, Oujda;$^{(e)}$Facult\'e des sciences, Universit\'e Mohammed V, Rabat;$^{(f)}$Institute of Applied Physics, Mohammed VI Polytechnic University, Ben Guerir; Morocco.\\
% ...
$^{55}$II. Physikalisches Institut, Georg-August-Universit\"{a}t G\"ottingen, G\"ottingen; Germany.\\
% ...
$^{69}$$^{(a)}$INFN Gruppo Collegato di Udine, Sezione di Trieste, Udine;$^{(b)}$ICTP, Trieste;$^{(c)}$Dipartimento Politecnico di Ingegneria e Architettura, Universit\`a di Udine, Udine; Italy.\\
% ...
$^{102}$CPPM, Aix-Marseille Universit\'e, CNRS/IN2P3, Marseille; France.\\
% ...
$^{117}$Department of Physics, New York University, New York NY; United States of America.\\
% ...
$^{120}$Homer L. Dodge Department of Physics and Astronomy, University of Oklahoma, Norman OK; United States of America.\\
% ...
$^{137}$$^{(a)}$Departamento de F\'isica, Pontificia Universidad Cat\'olica de Chile, Santiago;$^{(b)}$Millennium Institute for Subatomic physics at high energy frontier (SAPHIR), Santiago;$^{(c)}$Instituto de Investigaci\'on Multidisciplinario en Ciencia y Tecnolog\'ia, y Departamento de F\'isica, Universidad de La Serena;$^{(d)}$Universidad Andres Bello, Department of Physics, Santiago;$^{(e)}$Instituto de Alta Investigaci\'on, Universidad de Tarapac\'a, Arica;$^{(f)}$Departamento de F\'isica, Universidad T\'ecnica Federico Santa Mar\'ia, Valpara\'iso; Chile.\\
% ...
$^{150}$Department of Physics, Technion, Israel Institute of Technology, Haifa; Israel.\\
$^{151}$Raymond and Beverly Sackler School of Physics and Astronomy, Tel Aviv University, Tel Aviv; Israel.\\
% ...
$^{a}$ Also Affiliated with an institute covered by a cooperation agreement with CERN.\\
$^{b}$ Also at An-Najah National University, Nablus; Palestine.\\
$^{c}$ Also at Borough of Manhattan Community College, City University of New York, New York NY; United States of America.\\
$^{d}$ Also at Bruno Kessler Foundation, Trento; Italy.\\
$^{e}$ Also at Center for High Energy Physics, Peking University; China.\\
$^{f}$ Also at Center for Interdisciplinary Research and Innovation (CIRI-AUTH), Thessaloniki; Greece.\\
% ...
$^{p}$ Also at Department of Physics, King's College London, London; United Kingdom.\\
% ...

\end{flushleft}

% Created with Glance <Atlas.Glance@cern.ch>

\end{document}